\documentclass[a4paper,12pt]{report}
\usepackage{amssymb}
\usepackage{graphicx,color}

\begin{document}

\title{{\bf Chiral Phase Transition in QCD}
\\
{\bf and Vector Manifestation}}
\author{Chihiro Sasaki
\vspace*{0.4cm}
\\
{\it Department of Physics, Nagoya University,} 
\\
{\it Nagoya, 464-8602, Japan}}
\date{\today}
\maketitle

\newpage

\begin{center}
\begin{minipage}{14cm}
\begin{abstract}

\qquad
Spontaneous chiral symmetry breaking is one of
the most important features in low-energy QCD.
The chiral symmetry is expected to be restored 
at very high temperature and/or density.
Accompanied by the chiral phase transition,
properties of hadrons will be changed 
especially near the critical point.
The study of the phenomena associated with the chiral phase transition
will give us some clues on the connection between the chiral symmetry
and the low-energy hadron dynamics.

\qquad
We develop the theory based on the hidden local symmetry (HLS)
at finite temperature, 
which is an effective field theory of QCD and includes pions and 
vector mesons as the dynamical degrees of freedom,
and study the chiral phase transition in hot matter.
We show that the chiral symmetry is restored 
as the vector manifestation (VM),
in which the massless degenerate pion (and its flavor partners) 
and the longitudinal $\rho$ meson (and its flavor partners) 
as the chiral partner.
We also present several predictions based on the VM.
We estimate the critical temperature $T_c$ and
show the following phenomena near $T_c$:
the vector charge susceptibility becomes equal to the axial-vector
charge susceptibility; 
the vector dominance of the electromagnetic form factor of the pion
is largely violated;
the pion velocity is close to the speed of light.
Furthermore,
we show that the remnant of the VM can be clearly seen
in the system of heavy mesons. 
We expect that the VM and its predictions are testable 
by current and future experiments and the lattice analysis.

\end{abstract}

\end{minipage}
\end{center}

\newpage

\tableofcontents

\newpage

\chapter{Introduction}
\label{ch:Intro}


Quantum Chromo Dynamics (QCD) is the theory which governs
the dynamics of quarks and gluons, and eventually the hadron dynamics.
The QCD Lagrangian has an approximate global chiral symmetry
in the light quark sector.
The QCD vacuum, however, holds no longer the chiral symmetry
which is spontaneously broken down
caused by its strongly interacting dynamics.
Spontaneous chiral symmetry breaking is one of the 
most important features in low-energy QCD.
The phenomena of the light pseudoscalar mesons 
(the pion and its flavor partner), 
which are regarded as the approximate Nambu-Goldstone bosons  
associated with the symmetry breaking, are well described
by the symmetry property in the low-energy region.

For the purpose to clarify
the connection between the low-energy phenomena
and the chiral symmetry,
one of the most important techniques to study them 
is using effective field theories (EFTs)
in which a systematic low-energy expansion is done, 
as in the chiral perturbation theory (ChPT).
The EFTs are constructed based on the chiral symmetry,
in which the non-perturbative effects are ``dressed''
around the (bare) quarks and gluons
and the fundamental degrees of freedom
are replaced with the relevant ones 
in the low-energy region, i.e., hadrons.
In general,
the EFTs have many parameters like hadron masses and coupling constants.
Some of them can be determined in low-energy limit by the chiral
symmetry, however others are not completely restricted 
by only the symmetry.
When one performs the matching of the EFT to QCD
to determine the parameters of the EFT,
one can clarify how the hadron physics succeeds 
the dynamics of underlying QCD.
Furthermore,
including quantum effects into the parameters,
we can understand the hadron physics based on QCD
from many different angles
and clarify the non-perturbative aspects of QCD.

\vspace*{0.5cm}

However,
the mechanism how hadrons acquire their masses
as a result of the dynamics
is still an open problem of QCD.
In order to find out some clues on this issue,
it is efficient to consider a world where the QCD vacuum
holds the chiral symmetry.
Although the chiral symmetry is spontaneously broken in the real world,
it is expected to be restored under some extreme conditions, 
e.g., very high temperature and/or density.
Properties of hadrons will be changed at finite temperature/density,
especially near the critical point
of the chiral symmetry restoration
~\cite{HatsudaKunihiro,Pisarski:95,%
Brown-Rho:96,Brown:2001nh,HatsudaShiomiKuwabara,%
Rapp-Wambach:00,Wilczek}.


Both theoretically and experimentally,
it is important to investigate the physics 
near the phase transition point.
In fact, the CERN Super Proton Synchrotron (SPS) observed
an enhancement of dielectron ($e^+e^-$) mass spectra
below the $\rho / \omega$ resonance~\cite{Agakishiev:1995xb}.
This can be explained by the dropping mass of the $\rho$ meson
(see, e.g., Refs.~\cite{Li:1995qm, Brown-Rho:96, Rapp-Wambach:00})
associated with the chiral symmetry restoration
following the Brown-Rho scaling proposed in Ref.~\cite{BR}.
The Relativistic Heavy Ion Collider (RHIC) has started
to measure several physical processes in hot matter
which include the dilepton energy spectra.
In near future,
the CERN Large Hadron Collider (LHC) and GSI
are planning to measure them.
These experiments will further clarify the properties 
of vector mesons in hot matter.

\vspace*{0.5cm}

As indicated by the SPS data,
the vector meson is the one of important degrees of freedom
in studying the phenomena near the QCD phase transition.
The theory based on the hidden local symmetry~\cite{BKUYY,BKY:PRep}
is an EFT including both pions and vector mesons,
where the low-energy expansion is systematically done
(for details, see Ref.~\cite{HY:PRep}).
The HLS theory has a wider energy range of the validity than
the ordinary ChPT.
In this thesis,
we review the construction of the HLS theory in hot matter
and its application to the chiral phase transition
following Refs.~\cite{HS:VM,HKRS:SUS,HS:VVD,Sasaki:2003qj,HKRS:PV}.

\vspace*{0.5cm}

There is no strong restriction concerning vector meson masses
in the standard scenario of chiral symmetry restoration, where
the pion joins with the scalar meson in the same chiral representation
[see for example, Refs.~\cite{HatsudaKunihiro,Bochkarev:1995gi,PT:96}].
However there is a scenario which certainly requires the dropping mass 
of the vector meson and supports the Brown-Rho scaling:
In Ref.~\cite{HY:VM}, it was proposed that
there can be another possibility for the pattern 
of chiral symmetry restoration, the vector manifestation (VM).
The VM was proposed as a novel manifestation of 
the chiral symmetry in the Wigner realization, 
in which the chiral symmetry is restored by 
the massless degenerate pion (and its flavor partners) and the longitudinal 
$\rho$ meson (and its flavor partners) as the chiral partner.
In terms of the chiral representations of the low-lying mesons,
there is a representation mixing in the vacuum~\cite{HY:VM,HY:PRep}.
Approaching the critical point, we find that there are two possibilities
for the pattern of the chiral symmetry restoration
(see chapter~\ref{ch:VMHM}):
One possible pattern is the standard scenario and another is the VM.
Both of them are on an equal footing with each other
in terms of the chiral representations.
It is worthwhile to study the physics associated with the VM as well as
that with the standard scenario of the chiral symmetry restoration.


It has been shown that the VM is formulated at a large number of
flavor~\cite{HY:VM}, critical temperature~\cite{HS:VM} and
critical density~\cite{HKR:VM} by using the HLS theory,
where a second order or weakly first order phase transition was assumed.
In the VM at finite temperature and/or density, 
the {\it intrinsic temperature and/or density dependences} 
of the {\it bare} parameters of the HLS Lagrangian played important roles
to realize the chiral symmetry restoration consistently
(see chapters~\ref{ch:WMFT} and \ref{ch:VMHM}):
Since the HLS theory is an EFT of underlying QCD,
the parameters of the HLS Lagrangian do depend on the temperature
and/or density.
In the framework of the HLS the equality between 
the axial-vector and vector current correlators at the critical point
can be satisfied only if the intrinsic thermal and/or density effects are
included.
The intrinsic temperature and/or density dependences are 
nothing but the information integrated out 
which is converted from the underlying QCD 
through the Wilsonian matching~\cite{HY:PRep,HS:VM}.
In other words,
the intrinsic effects are
the signature that the hadron has an internal structure 
constructed from the quarks and gluons.

The VM explains an anomalous enhancement of 
dielectron mass spectra observed at SPS.
This is a theoretical support of
the dropping mass of $\rho$ meson following the Brown-Rho scaling,
which can explain the SPS data.

\vspace*{0.5cm}
In this thesis, we assume that only the pions and vector mesons are 
the relevant (light) degrees of freedom
until near the critical temperature, say $T_c - \epsilon$,
and that the chiral symmetry is partially restored already 
at $T_c - \epsilon$.
Then we study whether
the dropping $\rho$ can be formulated in some EFT.
We should note that, although we assume the dropping $\rho$ meson mass 
$m_\rho \sim {\cal O}(\epsilon^\prime)$ 
as an input, it is non-trivial that the dropping $\rho$ can actually be
formulated in the framework of quantum field theory.

\begin{figure}
 \begin{center}
  \includegraphics[width = 11cm]{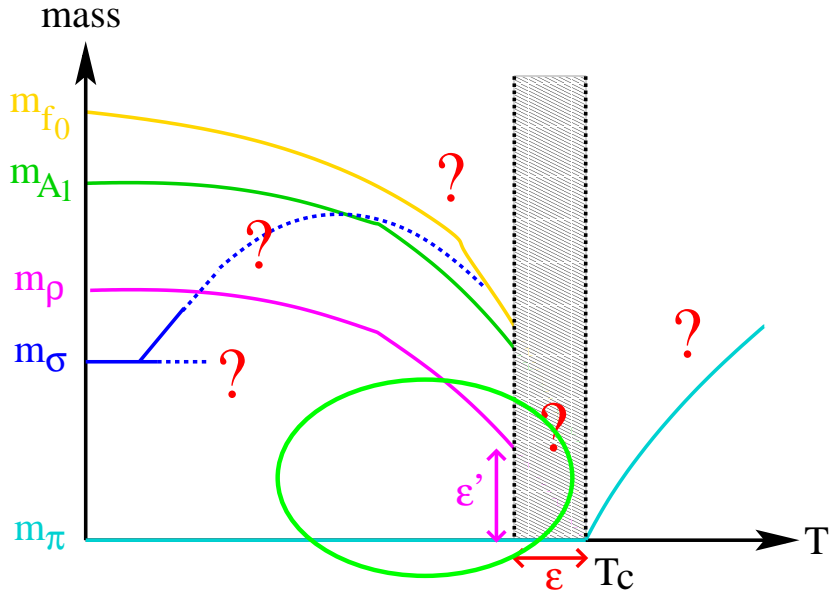}
 \end{center}
 \caption{Our picture. The solid lines denote the temperature
 dependences of light mesons, 
 $\pi, \sigma, \rho, A_1$ and $f_0 (1370)$.
 Throughout this thesis, we concentrate on the region 
 near $T_c - \epsilon$ (denoted by the green ellipsis)
 where we assume that only the pions
 and vector mesons are relevant degrees of freedom.}
 \label{fig:our picture}
\end{figure}

Now, what happens to other mesons?
In Fig.~\ref{fig:our picture}, 
we sketch our view of the temperature dependences of 
the light mesons masses.
We assume that the $A_1$ meson is still heavy at $T_c - \epsilon$.
The scalar mesons and their fate near $T_c$ may be more controversial.
Several pictures on the construction of the scalar meson are proposed
[see e.g., Refs.~\cite{Jaffe,Tornqvist:1995kr,Black:1998wt,Hagiwara:fs}].
In our present picture, we do not introduce the scalar mesons as 
the dynamical degrees of freedom into the EFT because of 
the following naive speculations
based on two kinds of picture on the quark content of scalar mesons.
\begin{description}
\item[Two-quark state :]
The light scalar meson is a bound state of a quark and an anti-quark.
The sigma meson mass $m_\sigma \sim 600\,\mbox{MeV}$ 
in the real world can be explained by the existence of the disconnected
quark diagram joined by the gluon exchange~\cite{Kunihiro:2003yj}.
At finite temperature, we expect that the effective QCD coupling
constant becomes weaker than that in the vacuum and that
$m_\sigma$ may go up.
In that case, the $m_\sigma$ may go to zero in the limit $T \to T_c$
although it might be still heavy until around $T_c - \epsilon$.
There may be a possibility that the $f_0 (1370)$ 
(or the mixture of $\sigma$ and $f_0(1370)$) goes down
at $T \to T_c$.
\item[Four-quark state :]
The light scalar meson is a bound state of a diquark and an anti-diquark.
The sigma meson may be melted at finite temperature less than $T_c$
since the binding energy of $qq$ is smaller than that of $\bar{q}q$.
When there exists the light scalar meson near $T_c$
(above $T_c - \epsilon$), such scalar meson might be related to
the $f_0(1370)$ at $T=0$.
\end{description}
In order to understand which universality class the VM belongs to,
we may need to include the scalar and $A_1$ mesons other than
pions and vector mesons for the analysis of critical exponents.
In the chiral symmetric phase, on the other hand,
an existence of some mesonic (soft) modes is proposed
[see e.g, Refs.~\cite{Hatsuda:1985eb,deForcrand:2000jx,
Shuryak:2003ty,Brown:2003km,nemoto}].

Throughout this thesis, we concentrate on the region
out of the window
near $T_c - \epsilon$ where we assume that only the pions and
vector mesons are relevant degrees of freedom.
In that region, we will see that the dropping $\rho$ can be formulated
within the EFT including both pions and vector mesons, 
related to the chiral symmetry restoration.
Furthermore, the present framework gives several physical predictions
which are closely related to the dropping $\rho$.
These predictions can be tested in future experiments and numerical
simulations on the lattice.

\vspace*{0.5cm}

Now let us assume that the chiral symmetry of QCD is restored as the VM.
Then can we see some remnant of the VM in the real world?
In order to answer this question,
in chapter~\ref{ch:CDHLM},
we study the heavy meson system which consists of
one heavy quark and one light anti-quark
following Ref.~\cite{HRS:CD}.
We construct an effective Lagrangian based on the VM
and heavy quark symmetry at the restoration point,
and then we introduce the effects of the spontaneous chiral 
symmetry breaking to go back to the real world.
We will show that the EFT can describe the recent experimental data.


\vspace*{0.6cm}

This paper is organized as follows:

In chapter~\ref{ch:EFTRG},
we give a general concept of EFTs
and the matching between an EFT and the underlying theory 
in the Wilsonian sense.
Then
we apply the Wilsonian matching to the QCD in hot and/or dense matter.
We give an account of the general idea of the intrinsic 
temperature and/or density dependences of the parameters,
which are inevitably introduced as a result 
of the Wilsonian matching.

In chapter~\ref{ch:THLS},
we briefly review the HLS and the ChPT with HLS 
in hot matter following Refs.~\cite{HS:VM,HS:VVD}:
In section~\ref{sec:HLS},
we review the theory based on the HLS
and in section~\ref{sec:TPFBFG},
we show details of the 
calculation of the hadronic thermal corrections 
as well as the quantum corrections
in the background field gauge.
In section~\ref{sec:CC},
we construct the axial-vector and vector current correlators.
Then in section~\ref{sec:TPpi-rhoP},
we study the temperature dependences of the vector meson mass,
the pion decay constant and the pion velocity.
We also study the validity of the vector dominance (VD) in hot matter.
In section~\ref{sec:LNBL}, 
we show the bare HLS Lagrangian without Lorentz invariance.

In chapter~\ref{ch:WMFT},
we briefly review the Wilsonian matching at finite temperature
following Refs.~\cite{HS:VM,HKRS:SUS,HS:VVD,Sasaki:2003qj,HKRS:PV}:
We extend the Wilsonian matching at zero temperature
to the version at non-zero temperature.
We also study in detail the effect of Lorentz non-invariance of
the bare pion decay constants.

In chapter~\ref{ch:VMHM},
we show how the VM is realized
at the critical point in the framework of the HLS theory.
We also study the predictions of the VM 
in hot matter:
In section~\ref{sec:CBP},
we present the conditions for the bare parameters
through the Wilsonian matching at the critical temperature.
Further, 
we take into account the Lorentz non-invariance in the bare HLS theory
and present the extended conditions.
Then we show that the conditions are still protected 
as the fixed point of the RGEs.
In section~\ref{sec:VMM},
we review the formulation of the VM 
in hot matter following Ref.~\cite{HS:VVD}.
Then in section~\ref{sec:PDCPV}, 
we show how the pion decay constants 
vanish at the critical temperature following Ref.~\cite{HKRS:SUS}.
Next,
in section~\ref{sec:PVM},
we summarize the predictions of the VM in hot matter
studied in Refs.~\cite{HS:VM,HKRS:SUS,HS:VVD,Sasaki:2003qj,HKRS:PV}:
In subsection~\ref{ssec:Tc}, 
we estimate the value of the critical temperature.
In subsection~\ref{sec:SUS},
we focus on the vector and axial-vector susceptibilities 
very near the critical temperature $T_c$.
Then we address the issue of what the relevant degrees
of freedom can be at the critical point. 
In subsection~\ref{sec:PaVVD},
we shed some light on the validity of the vector dominance (VD)
of electromagnetic form factor of the pion 
near the critical temperature.
We find that {\it the VM predicts a large violation of the VD at
the critical temperature}.
This indicates that the assumption of 
the VD may need to be weakened, at least in some amounts,
for consistently including the
effect of the dropping mass of the vector meson.
In subsection~\ref{sec:PV}, 
we study the pion velocity at the critical temperature
based on the VM.
There we show the non-renormalization property on the pion velocity
at the critical temperature, which is protected by the VM.
Then using this property and through matching to the OPE,
we estimate the value of the pion velocity.
In section~\ref{sec:SPVMM}, 
we discuss the critical behavior of $m_\rho (T)$ near
the critical temperature.

In chapter~\ref{ch:CDHLM},
we study the system of heavy mesons like $D$ meson,
following Ref.~\cite{HRS:CD}.
We construct an EFT based on the VM and the heavy quark symmetry
motivated by the recent discovery of new $D$ mesons ($J^P = 0^+$ and $1^+$)
in Babar, CLEO and Belle~\cite{BABAR,CLEO,Belle}.
We show that the mass splitting between new $D$ mesons and 
the  existing $D$ mesons ($J^P = 0^-$ and $1^-$)
is directly proportional to the light quark condensate.
Further we study the characteristic decay modes and 
give the predictions on the decay widths and the branching ratios.

We give a brief
summary and discussions in chapter~\ref{ch:SD}.

\newpage

\chapter{Effective Field Theory and Renormalization Group}
\label{ch:EFTRG}

Use of effective field theories (EFTs) is one of important implements
to study strongly interacting system like low-energy QCD.
An EFT works only in low energy below a characteristic scale,
which is dependent on dynamics of the EFT,
and is expanded in powers of momentum.
At a given order, the theory has a finite number of counter terms
and thus we can perform {\it an order-by-order renormalization}
in the sense of momentum expansion
although the EFT is conventionally non-renormalized.

In this chapter,
we give a general idea of EFTs
and the matching between an EFT and the underlying theory 
in the Wilsonian sense.

\section{General Concept of Effective Field Theory}
\label{sec:GCEFT}

We consider the QCD Lagrangian with $N_f$ massless quarks:
\begin{equation}
{\cal L}_{\rm QCD}^{(0)} 
 = \bar{q}i\gamma^\mu (\partial_\mu - i g_s G_\mu)q
  {}- \frac{1}{2}\mbox{tr}\bigl[ G_{\mu\nu}G^{\mu\nu} \bigr]\,,
\end{equation}
with $G_\mu$ and $g_s$ being the gluon field and the QCD gauge 
coupling constant.
The field strength $G_{\mu\nu}$ is defined by
\begin{equation}
G_{\mu\nu} = \partial_\mu G_\nu - \partial_\nu G_\mu 
 {}- i g_s \bigl[ G_\mu, G_\nu \bigr]\,.
\end{equation}
We define the left- and right-handed quarks by
\begin{equation}
 q_L = \frac{1}{2}(1 - \gamma_5)q,
\quad
 q_R = \frac{1}{2}(1 + \gamma_5)q.
\end{equation}
The interacting term between $q_L$ and $q_R$ never appear
in the Lagrangian.
It implies that
the QCD Lagrangian has an 
$SU(N_f)_L \times SU(N_f)_R \times U(1)_V \times U(1)_A$ symmetry.
In the real world, however,
the chiral symmetry is spontaneously broken down 
by the strongly interacting dynamics,
which makes pions to be much lighter than other hadrons.
The pion are identified as massless Nabmu-Goldstone (NG) bosons 
associated with the spontaneous chiral symmetry breaking.

In the construction of an EFT valid in some energy region,
it is important to specify the relevant degrees of freedom 
in the energy scale 
as well as to respect symmetries of the fundamental theory.
The relevant degrees of freedom may be changed according to
the energy scale that one is interested in.
We show a schematic view of the relation between EFTs and their
underlying QCD in Fig.~\ref{fig:EFT}: 
\begin{figure}
 \begin{center}
  \includegraphics[width = 10cm]{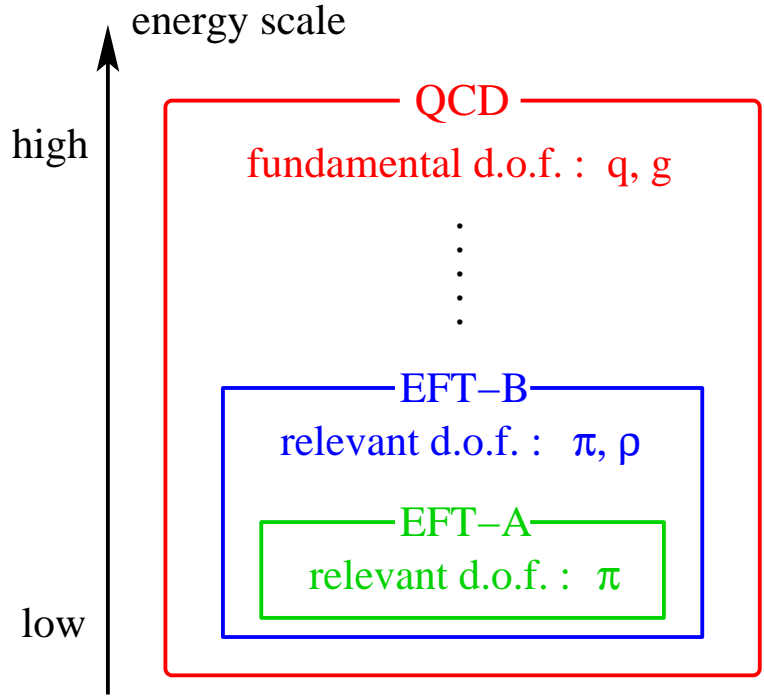}
 \end{center}
 \caption{Schematic view of effective field theories (EFTs).
          Here $q$ and $g$ denote quarks and gluons respectively.}
 \label{fig:EFT}
\end{figure} 
In low energies, the relevant degrees of freedom are hadrons.
Especially pion is the lightest hadron, which is the most relevant
degree of freedom in low-energy limit,
and its dynamics is well described
by EFTs based on the chiral symmetry of QCD like
the chiral perturbation theory (ChPT)~\cite{Weinberg:1978kz,Gas:84,Gas:85b}.
Here we consider an EFT, say EFT-A, which includes only pions as
the relevant degrees of freedom.
In the energy region where the EFT-A works well,
other hadrons, e.g., vector mesons, nucleons and excited states,
are sitting as the background.
However at some energy scale, the EFT-A breaks down, and then
one has to go to a more suitable theory which might be an EFT:
As the energy scale increases, 
other mesons like $\rho$ mesons are elevated from the background
and become relevant degrees of freedom as well as pions.
The background matter now consists of other hadrons except for
pions and $\rho$ mesons.
Thus one can switch over the EFT-A to new EFT, EFT-B,
which includes both pions and $\rho$ mesons as 
the relevant degrees of freedom.

Let us consider the QCD Lagrangian with external source fields:
\begin{equation}
{\cal L}_{\rm QCD} = {\cal L}_{\rm QCD}^{(0)} 
 {}+ \bar{q}(\gamma_\mu {\cal V}^\mu + \gamma_5\gamma_\mu {\cal A}^\mu)q
 {}- \bar{q}({\cal S} - i{\cal P})q\,,
\end{equation}
where the external vector (${\cal V_\mu}$),
axial-vector (${\cal A_\mu}$), scalar (${\cal S}$)
and pseudoscalar (${\cal P}$) fields are the hermitian $N_f \times N_f$
matrices in flavor space.
If there is an explicit chiral symmetry breaking,
it is introduced as the vacuum expectation value (VEV) of ${\cal S}$.
In the case of $N_f = 3$,
\begin{equation}
 \langle {\cal S} \rangle = {\cal M} = \mbox{diag}(m_u, m_d, m_s).
\end{equation}
The following functional of a set of source fields denoted by $J$,
\begin{equation}
 Z[J] = \int {\cal D}q {\cal D}\bar{q} {\cal D}G
        \exp\Biggl[ i\int d^4x {\cal L}_{\rm QCD}
            \bigl[ q,\bar{q},G;J \bigr] \Biggr]\,,
\label{gf}
\end{equation}
generates Green's functions which have important information of the system.
The basic concept of the EFT is based on the following assumption:
The effective Lagrangian,
which has the most general form constructed from the chiral symmetry,
can give the same generating functional as in Eq.~(\ref{gf}):
\begin{equation}
 Z[J] = \int {\cal D}U \exp\Biggl[ i\int d^4x {\cal L}_{\rm eff}
                           \bigl[ U;J \bigr]\Biggr],
\label{gf-eff}
\end{equation}
where $U$ denotes the relevant hadronic fields such as the pion fields
and ${\cal L}_{\rm eff}$ is an Lagrangian 
expressed in terms of these hadrons.
In the ChPT,
$N_f \times N_f$ special-unitary matrix $U$ is
parametrized by the pseudoscalar fields $\pi = \pi^a T_a$ as
\begin{equation}
 U = e^{2i\pi / F_\pi},
\label{U decomp}
\end{equation}
where $F_\pi$ denotes the pion decay constant.
The leading Lagrangian in the chiral limit is 
given by~\cite{Gas:84,Gas:85a}
\begin{equation}
 {\cal L}_{(2)}^{\rm ChPT}
 = \frac{F_\pi^2}{4}\mbox{tr}\bigl[ D_\mu U^\dagger D^\mu U\bigr],
\label{chiral L}
\end{equation}
where $D_\mu$ is the covariant derivative acting on $U$.

Corresponding to each Green's function derived from Eq.~(\ref{gf}),
we have the same Green's function obtained from the EFT
through Eq.~(\ref{gf-eff}).
We should note that according to the energy scale,
the relevant degrees of freedom are different
although both fundamental ($q$ and $g$) and hadronic ($U$) degrees of
freedom are integrated over all configuration space.

The idea mentioned above is the most general construction of EFTs,
where only the chiral symmetry gives constraints for the forms of
the effective Lagrangian.
Thus various EFTs consistent with the chiral symmetry
form an ``EFT space'' and the real QCD holds a some region
in the EFT space.


\section{Matching in the Wilsonian Sense}
\label{sec:MWS}

In general,
an EFT has many parameters like masses and 
coupling constants among hadrons.
We can determine them by performing the matching 
between two Green's functions, Eqs.~(\ref{gf}) and (\ref{gf-eff})
at a scale.

In some matching schemes, the renormalized parameters of the EFT
are determined from QCD.
On the other hand,
the matching in the Wilsonian sense is performed based on the 
following general idea:
\begin{description}
\item[Step 1]
We imagine a some high-energy scale $\Lambda$ 
which separates the perturbative (hard) region 
from the non-perturbative (soft) region [see Fig.~\ref{fig:wm}].
\begin{figure}
 \begin{center}
  \includegraphics[width = 7cm]{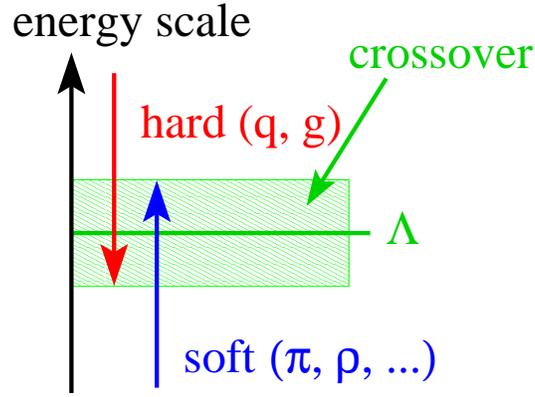}
 \end{center}
 \caption{Separation over two regions (hard and soft).
   A crossover region where both theories are available is assumed. 
   In this region, we can perform the matching between two theories.}
 \label{fig:wm}
\end{figure}  
Although the quarks and gluons are relevant degrees of freedom
in the hard region, they are not relevant anymore in the soft region.
Then we assume that one can replace the quarks and gluons with hadrons
since ``bare'' quarks and gluons in the QCD Lagrangian become
``dressed'' ones by integrating out the high-energy quarks and gluons
(i.e., the quarks and gluons above $\Lambda$),
which eventually form hadrons.
\item[Step 2]
Under the above replacement of the degrees of freedom at $\Lambda$,
we define the EFT at $\Lambda$ as the {\it bare} EFT.
Thus the generating functional derived from the bare EFT 
leads to the same Green's function as that derived from Eq.~(\ref{gf})
at $\Lambda$.
The {\it bare} parameters of the EFT are determined 
after integrating out the quarks and gluons above $\Lambda$,
and then carry information of the underlying QCD.
\end{description}

Once the bare theory is determined through the matching
following the above steps,
we can obtain the physical quantities in low-energy region
within the EFT including quantum corrections,
for instance, by using
renormalization group equations (RGEs).

It is worthwhile mentioning the difference from the 
way to determine the renormalized parameters of the EFT 
through the matching to QCD.
In this scheme, the theory at the matching scale $\Lambda$ is 
{\it full quantum} (i.e., renormalized) theory which is achieved 
by starting over in low energies.
While in the Wilsonian matching, 
one starts from the {\it bare} theory defined at $\Lambda$,
and in order-by-order renormalizes quantum effects into the parameters
to go down to low-energy scales.
Both of them are almost equivalent in the case that
$\Lambda$ is much larger than a characteristic scale $M$ of the EFT,
e.g, energy, temperature and density.
In this case, corrections ${\cal O}(M/\Lambda)$ are neglected
and we can replace the bare parameter with the renormalized one
at $\Lambda$ and vice versa.


\section{Application to Hot and/or Dense Matter}
\label{sec:AHDM}

In this section,
we apply the Wilsonian matching to the QCD in hot and/or dense matter.
We first give an account of the general idea of the intrinsic 
temperature and/or density dependences of the bare parameters,
which are inevitably introduced as a result 
performed the Wilsonian matching.
We also discuss the effect of Lorentz symmetry violation
at bare level.

In the following,
we consider that one performs the Wilsonian matching of the EFT-B
shown in Fig.~\ref{fig:EFT} 
to perturbative QCD at finite temperature and/or density:
In order to obtain the bare EFT-B,
we integrate out high frequency modes in hot and/or dense matter.
The hard modes integrated out are converted into the bare parameters
and thus the temperature and/or density carried by them become 
temperature and/or density dependences of the bare parameters. 
We call the thermal and/or dense effects as {\it intrinsic} 
temperature and/or density dependences.
They carry the information without (or behind) the EFT-B,
i.e., the information of the underlying theory QCD.
Ordinary hadronic thermal and/or dense effects are carried 
by the relevant degrees of freedom below $\Lambda$, $\pi$ and $\rho$, 
which are generated by dynamics of the soft region.
Thus in the Wilsonian matching, we can for convenience distinguish
two kinds of temperature and/or density dependences:
\begin{description}
\item[Intrinsic temperature and/or density dependences :]
Carried by the relevant degrees of freedom above $\Lambda$ 
(hard region). Such degrees of freedom are integrated out
in the EFT-B and contribute to bare parameters.
\item[Hadronic temperature and/or density dependences :]
Carried by the relevant degrees of freedom below $\Lambda$
(soft region). Such degrees of freedom interact to heat/particle bath
and generate thermal and/or dense effects 
which appear in the Bose/Fermi distribution functions.
\end{description}
In the present case, the hard modes integrated out are
all hadronic degrees of freedom except for $\pi$ and $\rho$ 
relevant below $\Lambda$.
It should be noted that the intrinsic effects never appear
in the bare parameters if one does not take into account
these ``residual'' degrees of freedom.
However, temperature and/or density of the system are provided by
the hadronic matter formed by {\it all} hadronic degrees of freedom.
The matter including only $\pi$ and$\rho$ is not identical with
the QCD matter.
Thus we need to take into account ``all the others'' never
incorporated in system within the EFT,
which enable us to include the intrinsic effects.

Further, according to Step 1 mentioned in the previous section, 
we replace the hard modes with the quarks and gluons above $\Lambda$.
The replacement indicates the following:
The intrinsic temperature and/or density dependences are 
nothing but the signature that the hadron has an internal structure 
constructed from the quarks and gluons.
This is similar to the situation where the coupling constants
among hadrons are replaced with the momentum-dependent form factor
in higher energy region
since the hadronic picture becomes irrelevant.
Thus the intrinsic thermal and/or dense effects play more important
roles in the higher temperature region, 
especially near the phase transition point.

Intrinsic thermal/dense effects also cause
Lorentz non-invariance in the bare theory:
At finite temperature and/or density,
theory has no longer the Lorentz invariance
since we specify the rest frame of medium.
The Lorentz non-invariance of hot and/or dense QCD is
generated through the dynamics of quarks and gluons.
Then the QCD Lagrangian takes the Lorentz invariant form.
On the other hand,
parameters of EFT have implicitly the information of
the fundamental theory as the intrinsic effects.
Thus the intrinsic temperature and/or density dependences carry 
the Lorentz non-invariance of hot/dense QCD,
and eventually they cause the Lorentz symmetry breaking in the bare EFT.

\newpage

\chapter{Hidden Local Symmetry Theory at Finite Temperature}
\label{ch:THLS}

The hidden local symmetry (HLS) theory~\cite{BKUYY,BKY:PRep}
is a natural extension of the non-linear chiral Lagrangian
and it can describe a system including both
pseudoscalars (pions) and vector mesons.
In the present framework,
the vector meson is introduced as the gauge boson of the HLS.
\begin{figure}
 \begin{center}
  \includegraphics[width = 12cm]{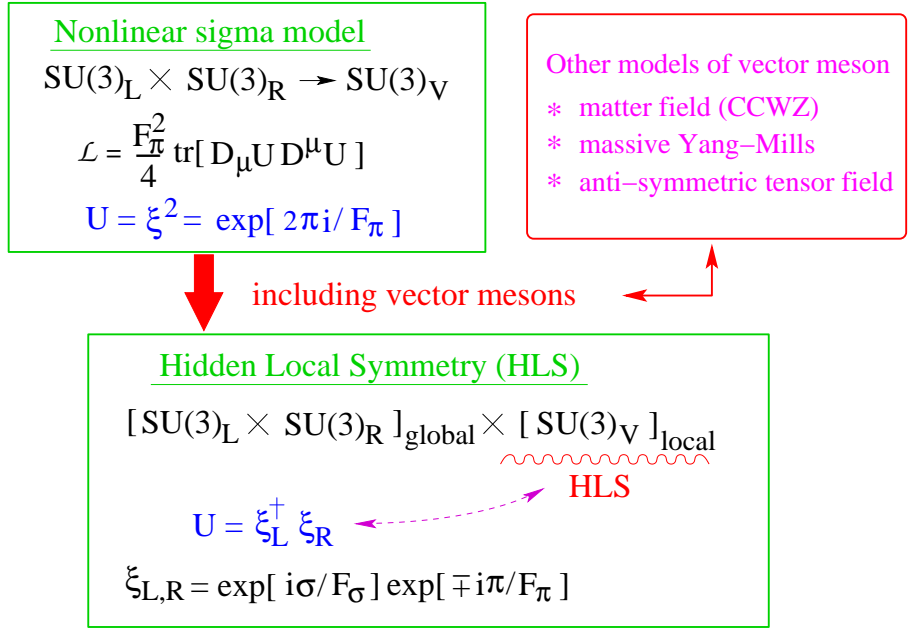}
 \end{center}
 \caption{Non-linear chiral Lagrangian and the HLS theory.}
 \label{fig:HLS}
\end{figure}
We can introduce the vector mesons in other frameworks,
e.g., matter field~\cite{Ecker:1989yg}, 
massive Yang-Mills~\cite{Schwinger:1967tc,Wess:1967jq,Gasiorowicz:1969kn,
Kaymakcalan:1983qq,Meissner:1987ge} 
and anti-symmetric tensor field~\cite{Gas:84,Ecker:1988te}.
It was shown that these models are equivalent with the HLS theory
at tree level~\cite{Ecker:1989yg,Birse:1996hd,Schechter:1986vs,
Yamawaki:1986zz,Golterman:1986cz,Meissner:1986tc,Tanabashi:1995nz}.
However it is the most important advantage using the HLS theory
that the systematic low-energy expansion can be performed.

In this chapter, we first briefly review 
the HLS and the chiral perturbation theory (ChPT) with HLS.
(For details, see Ref.~\cite{HY:PRep}.)
Next we apply the HLS theory to QCD at finite temperature
following Refs.~\cite{HS:VM,HKRS:SUS,HS:VVD,Sasaki:2003qj,HKRS:PV}.
We present the predictions in the low temperature region:
We show the temperature dependences
of the vector meson mass and the pion parameters,
and the validity of the vector dominance of the electromagnetic
form factor of the pion.

\section{Hidden Local Symmetry}
\label{sec:HLS}

The HLS theory~\footnote{
 In the modern interpretation~\cite{georgi},
 implementing HLS in the chiral Lagrangian can be associated with
 the ``ultraviolet completion'' to the fundamental theory of strong
 interactions, i.e., QCD. The matching to QCD at a matching scale
 is therefore a crucial ingredient of the approach.
}
is based on 
the $G_{\rm{global}} \times H_{\rm{local}}$ symmetry,
where $G=SU(N_f)_L \times SU(N_f)_R$ is the chiral symmetry
and $H=SU(N_f)_V$ is the HLS. 
The basic quantities are 
the HLS gauge boson and two matrix valued
variables $\xi_L(x)$ and $\xi_R(x)$
which transform as
 \begin{equation}
  \xi_{L,R}(x) \to \xi^{\prime}_{L,R}(x)
  =h(x)\xi_{L,R}(x)g^{\dagger}_{L,R}\ ,
 \end{equation}
where $h(x)\in H_{\rm{local}}\ \mbox{and}\ g_{L,R}\in
[\mbox{SU}(N_f)_{\rm L,R}]_{\rm{global}}$.
These variables are parameterized as
 \begin{equation}
  \xi_{L,R}(x)=e^{i\sigma (x)/{F_\sigma}}e^{\mp i\pi (x)/{F_\pi}}\ ,
 \end{equation}
where $\pi = \pi^a T_a$ denotes the pseudoscalar Nambu-Goldstone bosons
associated with the spontaneous symmetry breaking of
$G_{\rm{global}}$ chiral symmetry, 
and $\sigma = \sigma^a T_a$ denotes
the Nambu-Goldstone bosons associated with 
the spontaneous breaking of $H_{\rm{local}}$~\footnote{
The $\sigma$ is not to be confused with the scalar that figures in
two-flavor linear sigma model.}.
This $\sigma$ is absorbed into the HLS gauge 
boson through the Higgs mechanism. 
$F_\pi \ \mbox{and}\ F_\sigma$ are the decay constants
of the associated particles.
The phenomenologically important parameter $a$ is defined as 
 \begin{equation}
  a = \frac{{F_\sigma}^2}{{F_\pi}^2}\ .
 \end{equation}
The covariant derivatives of $\xi_{L,R}$ are given by
\begin{eqnarray}
 D_\mu \xi_L &=& \partial_\mu\xi_L - iV_\mu \xi_L + i\xi_L{\cal{L}}_\mu,
 \nonumber\\
 D_\mu \xi_R &=& \partial_\mu\xi_R - iV_\mu \xi_R + i\xi_R{\cal{R}}_\mu,
\end{eqnarray}
where $V_\mu$ is the gauge field of $H_{\rm{local}}$, and
${\cal{L}}_\mu \ \mbox{and}\ {\cal{R}}_\mu$ are the external
gauge fields introduced by gauging $G_{\rm{global}}$ symmetry.

The basic quantities in constructing the Lagrangian are the
following two 1-forms:
\begin{eqnarray}
  \hat{\alpha}_{\parallel\mu} =
  \frac{1}{2i} \left(
    D_\mu \xi_{\rm R} \cdot \xi_{\rm R}^\dag +
    D_\mu \xi_{\rm L} \cdot \xi_{\rm L}^\dag
  \right)
\ ,
\nonumber\\
  \hat{\alpha}_{\perp\mu} =
  \frac{1}{2i} \left(
    D_\mu \xi_{\rm R} \cdot \xi_{\rm R}^\dag -
    D_\mu \xi_{\rm L} \cdot \xi_{\rm L}^\dag
  \right)
\ .
\label{alpha}
\end{eqnarray}
They transform as
\begin{eqnarray}
  \hat{\alpha}_{\perp,\parallel}^{\mu} \rightarrow
  h(x) \cdot \hat{\alpha}_{\perp,\parallel}^{\mu} \cdot h^\dag(x)
\ .
\label{transform}
\end{eqnarray}
When HLS is gauge-fixed to the unitary gauge, $\sigma=0$,
$\xi_{\rm L}$ and $\xi_{\rm R}$ are related with each other by
\begin{equation}
\xi_{\rm L}^\dag = \xi_{\rm R} \equiv
\xi = e^{i\pi/F_\pi} \ .
\end{equation}
This unitary gauge is not preserved under the $G_{\rm global}$
transformation, which in general has the following form
\begin{eqnarray}
 G_{\rm global} \ : \  \xi \ \rightarrow \ \xi^\prime
&=&
 \xi \cdot g_{\rm R}^\dag = g_{\rm L} \cdot \xi
\nonumber\\
&=&
 \exp \left[i \sigma^\prime (\pi,g_{\rm R},g_{\rm L})/F_\sigma\right]
 \exp \left[ i\pi^\prime/F_\pi\right]
\nonumber\\
&=&
 \exp \left[ i\pi^\prime/F_\pi\right]
 \exp \left[-i\sigma^\prime (\pi,g_{\rm R},g_{\rm L})/F_\sigma\right]
\ .
\end{eqnarray}
The unwanted factor
$\exp \left[i \sigma^\prime (\pi,g_{\rm R},g_{\rm L})/F_\sigma\right]$
can be eliminated if we simultaneously perform the
$H_{\rm local}$ gauge transformation with
\begin{equation}
H_{\rm local} \ : \
h = \exp
  \left[i \sigma^\prime (\pi,g_{\rm R},g_{\rm L})/F_\sigma\right]
\equiv h(\pi,g_{\rm R},g_{\rm L})
\ .
\end{equation}
Then the system has a global symmetry
$G=SU(3)_{\rm L} \times SU(3)_{\rm R}$
under the following combined transformation:
\begin{equation}
G \ : \
 \xi \ \rightarrow\
 h(\pi,g_{\rm R},g_{\rm L}) \cdot \xi \cdot g_{\rm R}^\dag
 = g_{\rm L} \cdot \xi \cdot h(\pi,g_{\rm R},g_{\rm L})
\ .
\end{equation}
Under this transformation the HLS gauge boson field $V_\mu$
in the unitary gauge transforms as
\begin{equation}
G \ : \
 V_\mu \ \rightarrow\
 h(\pi,g_{\rm R},g_{\rm L}) \cdot  V_\mu
 \cdot h^\dag(\pi,g_{\rm R},g_{\rm L})
 - i \partial h(\pi,g_{\rm R},g_{\rm L}) \cdot
  h^\dag(\pi,g_{\rm R},g_{\rm L})
\ ,
\end{equation}
which is nothing but the transformation property of
Weinberg's ``$\rho$ meson''~\cite{Wei:68}.
The two 1-forms $\hat{\alpha}_{\parallel}^{\mu}$ and
$\hat{\alpha}_{\perp}^{\mu}$ transform as
\begin{eqnarray}
  \hat{\alpha}_{\perp,\parallel}^{\mu} \rightarrow
  h(\pi,g_{\rm R},g_{\rm L}) \cdot
  \hat{\alpha}_{\perp,\parallel}^{\mu} \cdot
  h^\dag(\pi,g_{\rm R},g_{\rm L})
\ .
\end{eqnarray}
Then, we can regard these 1-forms
as the fields belonging to
the chiral representations
$(1,8) + (8,1)$ and $(1,8) - (8,1)$
under $SU(3)_{\rm L} \times SU(3)_{\rm R}$.

The HLS Lagrangian with lowest derivative terms at the chiral limit
is given by~\cite{BKUYY,BKY:PRep}
 \begin{equation}
  {\cal{L}}_{(2)} = {F_\pi}^2\mbox{tr}\bigl[ \hat{\alpha}_{\perp\mu}
                                      \hat{\alpha}_{\perp}^{\mu}
                                   \bigr] +
       {F_\sigma}^2\mbox{tr}\bigl[ \hat{\alpha}_{\parallel\mu}
                  \hat{\alpha}_{\parallel}^{\mu}
                  \bigr] -
        \frac{1}{2g^2}\mbox{tr}\bigl[ V_{\mu\nu}V^{\mu\nu}
                   \bigr]
\ , \label{eq:L(2)}
 \end{equation}
where $g$ is the HLS gauge coupling,
$V_{\mu\nu}$ is the field strength
of $V_\mu$ and
 \begin{eqnarray}
  \hat{\alpha}_{\perp }^{\mu}
     &=& \frac{1}{2i}\bigl[ D^\mu\xi_R \cdot \xi_R^{\dagger} -
                          D^\mu\xi_L \cdot \xi_L^{\dagger}
                   \bigr] \ ,
\nonumber\\
  \hat{\alpha}_{\parallel}^{\mu}
     &=& \frac{1}{2i}\bigl[ D^\mu\xi_R \cdot \xi_R^{\dagger}+
                          D^\mu\xi_L \cdot \xi_L^{\dagger}
                   \bigr]
\ .
 \end{eqnarray}
We expand the Lagrangian~(\ref{eq:L(2)}) in terms of 
the $\pi$ field taking the unitary gauge of the 
HLS ($\sigma = 0$) to obtain
\begin{eqnarray}
{\cal{L}}_{(2)} 
&=&
\mbox{tr} \left[ \partial_\mu \pi \partial^\mu \pi \right]
+
a g^2 F_\pi^2 \, \mbox{tr} \left[ \rho_\mu \rho^\mu \right]
+ 2 i \left( \frac{1}{2} a g \right) \, \mbox{tr} 
\left[ \rho^\mu \left[ \partial_\mu \pi \,,\, \pi \right] \right]
\nonumber\\
&&
{}- 2 \left( a g F_\pi^2 \right)
\,\mbox{tr} \left[ \rho_\mu {\cal V}^\mu \right] 
{}+ 2i \left( 1 - \frac{a}{2} \right) \mbox{tr} 
\left[ 
  {\cal V}^\mu \left[ \partial_\mu \pi \,,\, \pi \right] 
\right]
+ \cdots
\ ,
\label{Lag:expand}
\end{eqnarray}
where
the vector meson field $\rho_\mu$ is introduced by
\begin{equation}
V_\mu = g \rho_\mu \ ,
\end{equation}
and 
vector external gauge field ${\cal V}_\mu$ is defined 
as~\footnote{%
  Note that the photon field $A_\mu$ for $N_f = 3$ is 
  embedded into ${\cal V}_\mu$ as
  \begin{displaymath}
   {\cal V}_\mu = eA_\mu Q,\quad
    Q = \left(\begin{array}{ccc}
        2/3 &      &  \\
            & -1/3 &  \\
            &      & -1/3
        \end{array}\right)\ ,
  \end{displaymath}
   with $e$ being the electromagnetic coupling constant.
}
\begin{equation}
{\cal V}_\mu \equiv \frac{1}{2} \left(
  {\cal R}_\mu + {\cal L}_\mu
\right)
\ .
\end{equation}
{}From the expansion in Eq.~(\ref{Lag:expand})
we find the following expressions for
the mass of vector meson $M_\rho$,
the $\rho\pi\pi$ coupling $g_{\rho\pi\pi}$,
the $\rho$-$\gamma$ mixing strength $g_\rho$ and
the direct $\gamma\pi\pi$ coupling $g_{\gamma\pi\pi}$:
\begin{eqnarray}
 {M_\rho}^2 &=& ag^2{F_\pi}^2, \\
 g_{\rho\pi\pi} &=& \frac{1}{2}ag, \\
 g_\rho &=& ag{F_\pi}^2, \\
 g_{\gamma\pi\pi} &=& e\bigl( 1-\frac{1}{2}a \bigr).
\end{eqnarray}
By taking the parameter choice $a=2$ at tree level,
the HLS reproduces the following 
phenomenological facts~\cite{BKUYY}:
 \begin{enumerate}
  \item $g_{\rho\pi\pi}=g$ \quad
        (universality of the $\rho$ coupling)~\cite{Sakurai};
  \item ${M_\rho}^2=2g_{\rho\pi\pi}^2{F_\pi}^2$ \quad
        (the KSRF relation, version II)~\cite{KSRF};
  \item $g_{\gamma\pi\pi}=0$ \quad
        (vector dominance of the electromagnetic form factor
         of the pion)~\cite{Sakurai}.
 \end{enumerate}
Moreover, the KSRF relation (version I)~\cite{KSRF}
is predicted as a low-energy theorem of the HLS~\cite{BKY:NPB}:
 \begin{equation}
  g_\rho = 2{F_\pi}^2 g_{\rho\pi\pi},
 \end{equation}
which is independent of the parameter $a$.
This relation was first proved at tree level~\cite{BKY:PT} and
at one-loop level~\cite{HY:PLB} and
then at any loop order~\cite{HKY}.

\subsubsection{Chiral perturbation theory with HLS}

In the HLS theory it is possible to perform the derivative
expansion systematically owing to the gauge invariance
~\cite{HY:PRep,Georgi,Tanabashi}.
In this ChPT with  HLS the
vector meson mass is considered as small
compared with the chiral symmetry breaking scale 
$\Lambda_\chi$, by assigning ${\cal O}(p)$ to 
the HLS gauge coupling~\cite{Georgi,Tanabashi}: 
\begin{equation}
 g \sim {\cal O}(p).
\end{equation}
According to the entire list shown in Ref.~\cite{Tanabashi},
there are 35 counter terms at ${\cal O}(p^4)$ for general
$N_f$.
However, only three terms are relevant 
when we consider two-point functions
at chiral limit:
 \begin{equation}
  {\cal{L}}_{(4)} = z_1\mbox{tr}\bigl[ \hat{\cal{V}}_{\mu\nu}
                       \hat{\cal{V}}^{\mu\nu} \bigr] +
                    z_2\mbox{tr}\bigl[ \hat{\cal{A}}_{\mu\nu}
                       \hat{\cal{A}}^{\mu\nu} \bigr] +
                    z_3\mbox{tr}\bigl[ \hat{\cal{V}}_{\mu\nu}
                       V^{\mu\nu} \bigr], \label{eq:L(4)}
 \end{equation}
where
 \begin{eqnarray}
  \hat{\cal{A}}_{\mu\nu}=\frac{1}{2}
                         \bigl[ \xi_R{\cal{R}}_{\mu\nu}\xi_R^{\dagger}-
                                \xi_L{\cal{L}}_{\mu\nu}\xi_L^{\dagger}
                         \bigr]\ ,
  \label{def A mn}\\
  \hat{\cal{V}}_{\mu\nu}=\frac{1}{2}
                         \bigl[ \xi_R{\cal{R}}_{\mu\nu}\xi_R^{\dagger}+
                                \xi_L{\cal{L}}_{\mu\nu}\xi_L^{\dagger}
                         \bigr]\ ,
  \label{def V mn}
 \end{eqnarray}
with ${\cal{R}}_{\mu\nu}\ \mbox{and}\ {\cal{L}}_{\mu\nu}$ being
the field strengths of ${\cal{R}}_{\mu}\ \mbox{and}\ {\cal{L}}_{\mu}$.

There is a limit that the vector meson mass is small, 
in large $N_f$ QCD as shown in Ref.~\cite{HY:PRep}.
Then it can be justified to perform the derivative expansion by 
$m_\rho^2 / \Lambda_\chi^2$ under such an extreme condition.
We obtain physical quantities in the real world with $N_f = 2$ or $3$
by extrapolating the results in large $N_f$.
Numerically the expansion parameter is not very small, 
$m_\rho^2 / \Lambda_\chi^2 \sim 0.5$.
However,
the Wilsonian matching at $T=0$ with $N_f = 3$,
through which the parameters of the HLS Lagrangian are determined 
by the underlying QCD at the matching scale $\Lambda$
(see chapter~\ref{ch:WMFT}),
was shown to
give several predictions in remarkable agreement with experiments
~\cite{HY:WM,HY:PRep},
where the validity of the ChPT with HLS is essential.
This strongly suggests that the extrapolation mentioned above is 
valid even numerically.

As discussed in Ref.~\cite{HY:PRep}, 
$\Lambda \ll \Lambda_\chi$ can be
justified in the $N_c$ limit of QCD~\footnote{
We consider that
the scale at which the HLS theory breaks down is the same order
as $\Lambda_\chi$ since the chiral symmetry is non-linearly
realized.
}:
In this limit,
$F_\pi^2(\Lambda)$ scales as $N_c$,
which implies that $\Lambda_\chi$ becomes large in the large $N_c$
limit.
On the other hand,
the meson masses do not scale with $N_c$,
so that we can introduce the $\Lambda$ which has no large $N_c$
scaling property.
Thus the quadratic divergent at $n$-th loop order
is suppressed by $[ \Lambda^2 / \Lambda_\chi^2 ]^n \sim [ 1/ N_c ]^n$.
As a result,
we can perform the loop expansion with quadratic divergences
in the large $N_c$ limit, and extrapolate the results to the real QCD
with $N_c = 3$.

In the following, 
we will apply the ChPT with HLS combined with the Wilsonian matching
to finite temperature.  There the expansion parameter 
is $T/F_\pi(\Lambda)$ instead of $T/F_\pi(0)$ used in the 
standard ChPT.  Since $F_\pi(\Lambda) > F_\pi(0)$, 
the present formalism can be applied in the higher
temperature region than the standard ChPT.


\section{Two-Point Functions in Background Field Gauge}
\label{sec:TPFBFG}

In the present approach
hadronic thermal effects are included
by calculating pseudoscalar and vector meson loop contributions.
In this section we show details of the 
calculation of the hadronic thermal corrections 
as well as the quantum corrections
to the two-point functions  
in the background field gauge.

\subsection{Background field gauge}

In this subsection we briefly review the background field gauge
following 
Ref.~\cite{Tanabashi,HY:WM,HY:PRep}.
We introduce the background fields $\overline{\xi}_{L,R}$ as
 \begin{equation}
  \xi_{L,R}=\check{\xi}_{L,R}\overline{\xi}_{L,R},
 \end{equation}
where $\check{\xi}_{L,R}$ are the quantum fields.
It is convenient to write
\begin{eqnarray}
&&
\check{\xi}_{\rm L} = 
\check{\xi}_{\rm S} \cdot
\check{\xi}_{\rm P}^\dag \ ,
\qquad
\check{\xi}_{\rm R} = 
\check{\xi}_{\rm S} \cdot
\check{\xi}_{\rm P} \ ,
\nonumber\\
&& \quad
\check{\xi}_{\rm P} =
\exp \left[ i\, \check{\pi}^a T_a /F_\pi\right] \ ,
\quad
\check{\xi}_{\rm S} =
\exp \left[ i \, \check{\sigma}^a T_a/F_\sigma \right] \ ,
\end{eqnarray}
with $\check{\pi}$ and $\check{\sigma}$ being the
quantum fields corresponding to the pseudoscalar NG boson $\pi$ 
and the would-be NG boson $\sigma$. 
The background gauge field $\overline{V}_\mu$ for
the HLS gauge boson is introduced as
 \begin{equation}
  V_\mu = \overline{V}_\mu + g\check{\rho}_\mu ,
 \end{equation}
where $\check{\rho}_\mu$ is the quantum field.
It is convenient to introduce the following fields corresponding
to $\hat{\alpha}_{\perp \mu}\ \mbox{and}\ 
\hat{\alpha}_{\parallel \mu} + V_\mu$:
 \begin{eqnarray}
  \overline{\cal{A}}_\mu
   = \frac{1}{2i}\bigl[ \partial_\mu\overline{\xi}_R \cdot
                        \overline{\xi}_R^{\dagger} -
                        \partial_\mu\overline{\xi}_L \cdot
                        \overline{\xi}_L^{\dagger}
                 \bigr] +
     \frac{1}{2}\bigl[ \overline{\xi}_R{\cal{R}}_\mu
                       \overline{\xi}_R^{\dagger} -
                       \overline{\xi}_L{\cal{L}}_\mu
                       \overline{\xi}_L^{\dagger}
                 \bigr] \ , \nonumber\\
  \overline{\cal{V}}_\mu
   = \frac{1}{2i}\bigl[ \partial_\mu\overline{\xi}_R \cdot
                        \overline{\xi}_R^{\dagger} +
                        \partial_\mu\overline{\xi}_L \cdot
                        \overline{\xi}_L^{\dagger}
                 \bigr] +
     \frac{1}{2}\bigl[ \overline{\xi}_R{\cal{R}}_\mu
                       \overline{\xi}_R^{\dagger} +
                       \overline{\xi}_L{\cal{L}}_\mu
                       \overline{\xi}_L^{\dagger}
                 \bigr] \ .
 \end{eqnarray}
The field strengths of $\overline{\cal A}_\mu$ and
$\overline{\cal V}_\mu$ are defined as
\begin{eqnarray}
 \overline{\cal A}_{\mu\nu}
  &=& \partial_\mu \overline{\cal A}_\nu -
     \partial_\nu \overline{\cal A}_\mu -
     i \Bigl[ \overline{\cal V}_\mu,\overline{\cal A}_\nu \Bigr]-
     i \Bigl[ \overline{\cal A}_\mu,\overline{\cal V}_\nu \Bigr]
\ ,\nonumber\\
 \overline{\cal V}_{\mu\nu}
  &=& \partial_\mu \overline{\cal V}_\nu -
     \partial_\nu \overline{\cal V}_\mu -
     i \Bigl[ \overline{\cal V}_\mu,\overline{\cal V}_\nu \Bigr]-
     i \Bigl[ \overline{\cal A}_\mu,\overline{\cal A}_\nu \Bigr]
\ .
\end{eqnarray}
Note that both $\overline{\cal A}_{\mu\nu}$ and 
$\overline{\cal V}_{\mu\nu}$ do not include any derivatives of the
background fields $\overline{\xi}_{\rm R}$ and 
$\overline{\xi}_{\rm L}$:
\begin{eqnarray}
\overline{\cal V}_{\mu\nu} &=&
  \frac{1}{2}
  \left[
    \overline{\xi}_{\rm R} {\cal R}_{\mu\nu}
      \overline{\xi}_{\rm R}^\dag
    + \overline{\xi}_{\rm L} {\cal L}_{\mu\nu}
      \overline{\xi}_{\rm L}^\dag
  \right]
\ ,
\nonumber\\
\overline{\cal A}_{\mu\nu} &=&
  \frac{1}{2}
  \left[
    \overline{\xi}_{\rm R} {\cal R}_{\mu\nu}
      \overline{\xi}_{\rm R}^\dag
    - \overline{\xi}_{\rm L} {\cal L}_{\mu\nu}
      \overline{\xi}_{\rm L}^\dag
  \right]
\ .
\end{eqnarray}
Then $\overline{\cal A}_{\mu\nu}$ and
$\overline{\cal V}_{\mu\nu}$
correspond to $\hat{A}_{\mu\nu}$ and $\hat{V}_{\mu\nu}$
in Eqs.~(\ref{def A mn}) and (\ref{def V mn}), respectively.

In the background field gauge
the quantum fields as well as the background
fields $\overline{\xi}_{\rm R,L}$ transform homogeneously
under the background gauge transformation, while the background gauge
field $\overline{V}_\mu$ transforms inhomogeneously:
\begin{eqnarray}
 \overline{\xi}_{\rm R,L} &\rightarrow& 
  h(x) \cdot \overline{\xi}_{\rm R,L} \cdot g_{\rm R,L}^\dag \ ,
\nonumber\\
 \overline{V}_\mu &\rightarrow& 
  h(x) \cdot \overline{V}_\mu \cdot h^\dag(x)
  + i h(x) \cdot \partial_\mu h^\dag(x) \ ,
\nonumber\\
 \check{\xi}_{\rm L,R} &\to& 
  h(x) \cdot \check{\xi}_{\rm L,R} \cdot h^{\dagger}(x)\ , 
\nonumber\\
 \check{\pi} &\rightarrow& 
  h(x) \cdot \check{\pi} \cdot h^\dag(x) \ ,
\nonumber\\
 \check{\sigma} &\rightarrow& 
  h(x) \cdot \check{\sigma} \cdot h^\dag(x) \ ,
\nonumber\\
 \check{\rho}_\mu &\rightarrow& 
  h(x) \cdot \check{\rho}_\mu \cdot h^\dag(x) \ .
\end{eqnarray}
Then
the expansion of the Lagrangian in terms of the quantum field
manifestly keeps the HLS of the background field
$\overline{V}_\mu$~\cite{Tanabashi}
and thus
the gauge invariance of the result is manifest.
We fix the background field gauge in 
't\,Hooft-Feynman gauge as~\cite{Tanabashi}
\begin{equation}
 {\cal L}_{\rm GF} = -\mbox{tr}\Bigl[ \bigl( \overline{D}^\mu
                   \check{\rho}_\mu + g F_\sigma \check{\sigma}
                  \bigr)^2 \Bigr],
\end{equation}
where $\overline{D}^\mu$ denotes the covariant derivative
with respect to the background fields:
\begin{equation}
 \overline{D}^\mu \check{\rho}_\nu = \partial^\mu \check{\rho}_\nu
     {} - i\bigl[ \overline{V}^\mu, \check{\rho}_\nu \bigr].
\end{equation}
The FP ghost associated with ${\cal L}_{GF}$ is
given by~\cite{Tanabashi}
\begin{equation}
 {\cal L}_{\rm FP} = 2i\,\mbox{tr}\Bigl[ \overline{C} \bigl(
                  \overline{D}^\mu \overline{D}_\mu +
                  (g F_\sigma)^2 \bigr) C \Bigr] 
  + \cdots \ ,
\end{equation}
where ellipses stand for the terms including at least three
quantum fields.

Now the ${\cal O}(p^2)$ Lagrangian, ${\cal L}_{(2)}
+ {\cal L}_{\rm GF} + {\cal L}_{\rm FP}$, is expanded in terms
of the quantum fields, $\check{\pi}$, $\check{\sigma}$, $\check{\rho}$,
$C$ and $\overline{C}$.
The terms which do not include the quantum fields are nothing but the
original ${\cal O}(p^2)$ Lagrangian with the fields replaced by the
corresponding background fields.
The terms which are of first order in the quantum fields 
lead to the equations of motions for the background 
fields:~\cite{HY:WM,HY:PRep}
\begin{eqnarray}
&&
\overline{D}_\mu \overline{\cal A}^\mu =
- i \left( a- 1 \right)
\left[
  \overline{\cal V}_\mu - \overline{V}_\mu \,,\,
  \overline{\cal A}^\mu
\right]
+ {\cal O}(p^4)
\ ,
\label{EOM Npi B}
\\
&&
\overline{D}_\mu 
\left( \overline{\cal V}^\mu - \overline{V}^\mu \right)
= {\cal O}(p^4)
\ ,
\label{EOM Nsig B}
\\
&&
\overline{D}_\nu \overline{V}^{\nu\mu}
=
g^2 F_\sigma^2 
\left( \overline{\cal V}^\mu - \overline{V}^\mu \right)
+ {\cal O}(p^4)
\ .
\label{EOM Nvec B}
\end{eqnarray}
The quantum correction as well as the hadronic thermal corrections
are calculated from the terms of quadratic order in the quantum
fields.
The explicit forms of the terms as well as
the Feynman rules obtained from them
are shown in Ref.~\cite{HY:PRep}.

\subsection{Two-point functions at $T=0$}
\label{ssec:TPFT0}

In this subsection, we calculate
the quantum corrections
from pseudoscalar and vector mesons 
to the two-point functions of $\overline{\cal A}_\mu$,
$\overline{\cal V}_\mu$ and $\overline{V}_\mu$ at zero 
temperature.
In the present analysis, as was done in Ref.~\cite{HKRS:SUS}
(see also subsection~\ref{ssec:WMCTf}),
we neglect possible Lorenz symmetry violating effects caused by the
intrinsic temperature dependences of the bare parameters, and use
the bare HLS Lagrangian with Lorenz invariance even at non-zero
temperature.  Then the results obtained in this subsection are
identified with  the quantum corrections from pseudoscalar 
and vector mesons at non-zero temperature
by taking the intrinsic
temperature dependences of the parameters into account.~\footnote{%
  See next subsection for dividing the quantum corrections from the
  hadronic thermal corrections.
}
In the following,
we write the two-point functions of 
$\overline{\cal A}_\mu$-$\overline{\cal A}_\nu$,
$\overline{\cal V}_\mu$-$\overline{\cal V}_\nu$,
$\overline{V}_\mu$-$\overline{V}_\nu$ and
$\overline{V}_\mu$-$\overline{\cal V}_\nu$
as $\Pi_\perp^{\mu\nu}$, $\Pi_\parallel^{\mu\nu}$,
$\Pi_V^{\mu\nu}$ and
$\Pi_{V\parallel}^{\mu\nu}$, respectively.

In the present analysis, it is important to include the quadratic
divergences to obtain the RGEs in the Wilsonian sense.
Here, following 
Ref.~\cite{HY:conformal,HY:WM,HY:PRep}, we adopt the
dimensional regularization and identify the quadratic divergences with
the presence of poles of ultraviolet origin at $n=2$~\cite{Veltman}.
This can be done by the following replacement in the Feynman
integrals:
\begin{equation}
\int \frac{d^n k}{i (2\pi)^n} \frac{1}{-k^2} \rightarrow 
\frac{\Lambda^2} {(4\pi)^2} \ ,
\qquad
\int \frac{d^n k}{i (2\pi)^n} 
\frac{k_\mu k_\nu}{\left[-k^2\right]^2} \rightarrow 
- \frac{\Lambda^2} {2(4\pi)^2} g_{\mu\nu} \ .
\label{quad repl}
\end{equation}
On the other hand, 
the logarithmic divergence is identified with the pole at 
$n=4$:
\begin{equation}
\frac{1}{\bar{\epsilon}} + 1 \rightarrow
\ln \Lambda^2
\ ,
\label{logrepl:2}
\end{equation}
where
\begin{equation}
\frac{1}{\bar{\epsilon}} \equiv
\frac{2}{4 - n } - \gamma_E + \ln (4\pi)
\ ,
\end{equation}
with $\gamma_E$ being the Euler constant.

\begin{figure}
 \begin{center}
  \includegraphics[width = 10cm]{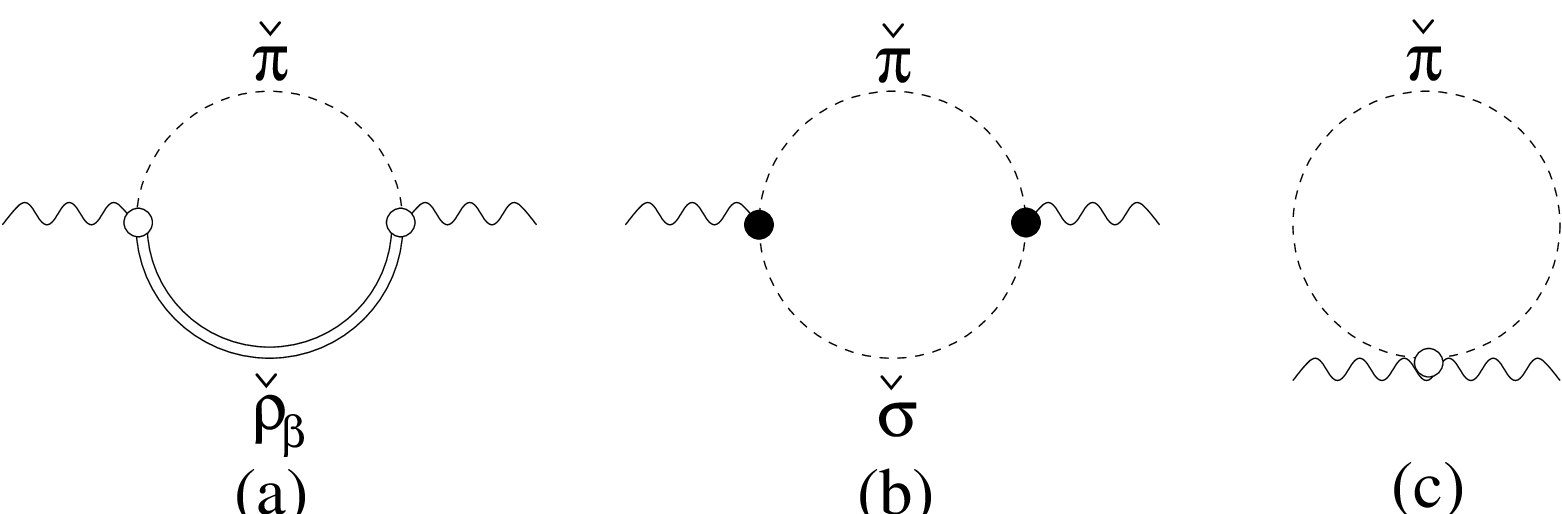}
 \end{center}
 \caption{Diagrams for contributions to $\Pi_\perp^{\mu\nu}$
         at one-loop level. The circle
 $(\circ)$ denotes
         the momentum-independent vertex and the dot
      $(\bullet)$ denotes the momentum-dependent vertex.}
 \label{fig:AAdiagrams}
\end{figure}

Let us start from the one-loop corrections to
the two-point function of 
$\overline{\cal A}_\mu$-$\overline{\cal A}_\nu$
denoted by $\Pi_\perp^{\mu\nu}$.
We show the diagrams for contributions to
$\Pi_\perp^{\mu\nu}$
at one-loop level in Fig.~\ref{fig:AAdiagrams}.
We obtain these contributions as~\footnote{%
  Here we put the superscript $\mbox{(vac)}$ in the functions in the
  right-hand-side to distinguish them from the functions expressing
  the hadronic thermal corrections [see next subsection].
}
\begin{eqnarray}
 \Pi_\perp^{(a)\mu\nu}(p)&=&
  -N_f a{M_\rho}^2 g^{\mu\nu} B_0^{\rm(vac)}(p;M_\rho ,0)\ ,
                                            \label{eq:AAa}\\
 \Pi_\perp^{(b)\mu\nu}(p)&=&
  N_f \frac{a}{4}B^{{\rm(vac)}\mu\nu}(p;M_\rho ,0)\ ,
                                           \label{eq:AAb}\\
 \Pi_\perp^{(c)\mu\nu}(p)&=&
  N_f (a-1)g^{\mu\nu}A_0^{\rm(vac)}(0)\ , \label{eq:AAc}
\end{eqnarray}
where the functions $B_0^{\rm(vac)}$, $B^{{\rm(vac)}\mu\nu}$ and
$A_0^{\rm(vac)}$ are defined
by the following integrals~\cite{HY:PRep}:
\begin{eqnarray}
  A_0^{\rm(vac)}(M)
   &=& \int \frac{d^n k}{i(2\pi)^4}
      \frac{1}{M^2 - k^2}\ , 
\label{A0vac def}
\\
  B_0^{\rm(vac)}(p;M_1,M_2)
   &=& \int \frac{d^n k}{i(2\pi)^4}
      \frac{1}{[M_1^2-k^2][M_2^2-(k-p)^2]}\ , 
\label{B0vac def}
\\
  B^{{\rm(vac)}\mu\nu}(p;M_1,M_2)
   &=& \int \frac{d^n k}{i(2\pi)^4}
      \frac{(2k-p)^\mu (2k-p)^\nu }{[M_1^2-k^2][M_2^2-(k-p)^2]}\ .
\label{Bmnvac def}
\end{eqnarray}

\begin{figure}
 \begin{center}
  \includegraphics[width = 13cm]{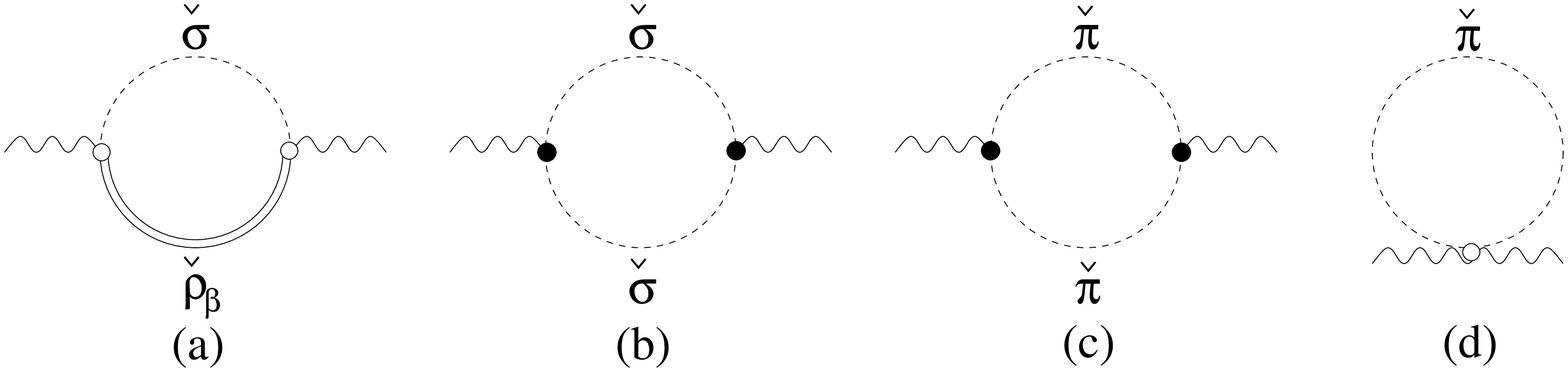}
 \end{center}
 \caption{Diagrams for contributions to
  $\Pi_\parallel^{\mu\nu}$ at one-loop level.}
 \label{fig:VVdiagrams}
\end{figure}

Next we calculate the one-loop corrections to 
the $\overline{\cal V}_\mu$-$\overline{\cal V}_\nu$
two-point function denoted by
$\Pi_\parallel^{\mu\nu}$.
We show the diagrams for contributions to 
$\Pi_\parallel^{\mu\nu}$
at one-loop level in Fig.~\ref{fig:VVdiagrams}.
We obtain these contributions as
\begin{eqnarray}
 \Pi_\parallel^{(a)\mu\nu}(p)&=&
  -N_f {M_\rho}^2 g^{\mu\nu}
   B_0^{\rm(vac)}(p;M_\rho ,M_\rho ), \label{eq:VVa} \\
 \Pi_\parallel^{(b)\mu\nu}(p)&=&
  \frac{1}{8}N_f B^{{\rm(vac)}\mu\nu}(p;M_\rho ,M_\rho ),
                                            \label{eq:VVb} \\
 \Pi_\parallel^{(c)\mu\nu}(p)&=&
  \frac{(2-a)^2}{8}N_f B^{{\rm(vac)}\mu\nu}(p;0,0),
                                             \label{eq:VVc} \\
 \Pi_\parallel^{(d)\mu\nu}(p)&=&
  -(a-1)N_f g^{\mu\nu}A_0^{\rm(vac)}(0). \label{eq:VVd}
 \end{eqnarray}

\begin{figure}
 \begin{center}
  \includegraphics[width = 13cm]{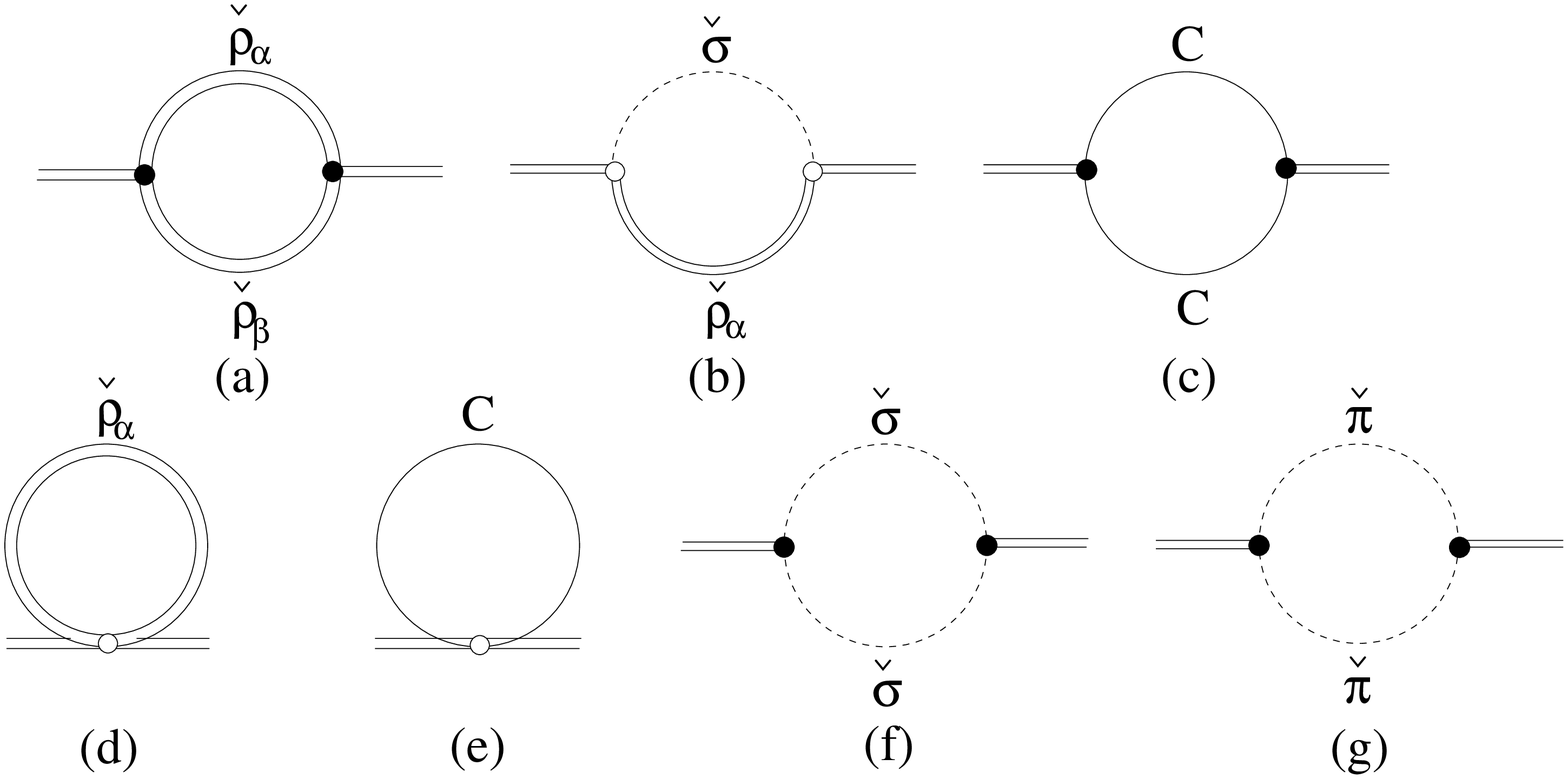}
 \end{center}
 \caption{Diagrams for contributions to $\Pi_V^{\mu\nu}$
         at one-loop level.}
 \label{fig:RRdiagrams}
\end{figure}
The one-loop corrections to $\Pi_V^{\mu\nu}$, 
the diagrams for contributions to which are shown in 
Fig.~\ref{fig:RRdiagrams},
are evaluated as
 \begin{eqnarray}
  \Pi_V^{(a)\mu\nu}(p)&=&
       \frac{n}{2}N_f 
        B^{\rm{(vac)}\mu\nu}(p;M_\rho ,M_\rho )+ 
           4N_f (g^{\mu\nu}p^2 - p^\mu p^\nu)
           B_0^{\rm{(vac)}}(p;M_\rho ,M_\rho ),
                                             \label{eq:RRa} \\
  \Pi_V^{(b)\mu\nu}(p)&=&
       -N_f {M_\rho}^2 g^{\mu\nu}
           B_0^{\rm{(vac)}}(p;M_\rho ,M_\rho ),
                                             \label{eq:RRb} \\
  \Pi_V^{(c)\mu\nu}(p)&=&
       -N_f B^{\rm{(vac)}\mu\nu}(p;M_\rho ,M_\rho ),
                                             \label{eq:RRc} \\
  \Pi_V^{(d)\mu\nu}(p)&=&
       nN_f g^{\mu\nu}A_0^{\rm{(vac)}}(M_\rho ),
                                             \label{eq:RRd} \\
  \Pi_V^{(e)\mu\nu}(p)&=&
       -2N_f g^{\mu\nu}A_0^{\rm{(vac)}}(M_\rho),
                                             \label{eq:RRe} \\
  \Pi_V^{(f)\mu\nu}(p)&=&
       \frac{1}{8}N_f 
        B^{\rm{(vac)}\mu\nu}(p;M_\rho ,M_\rho ),
                                             \label{eq:RRf} \\
  \Pi_V^{(g)\mu\nu}(p)&=&
       \frac{a^2}{8}N_f B^{\rm{(vac)}\mu\nu}(p;0,0),
                                             \label{eq:RRg}
 \end{eqnarray}
where $n$ denotes the dimension of the spacetime.

\begin{figure}
 \begin{center}
  \includegraphics[width = 12cm]{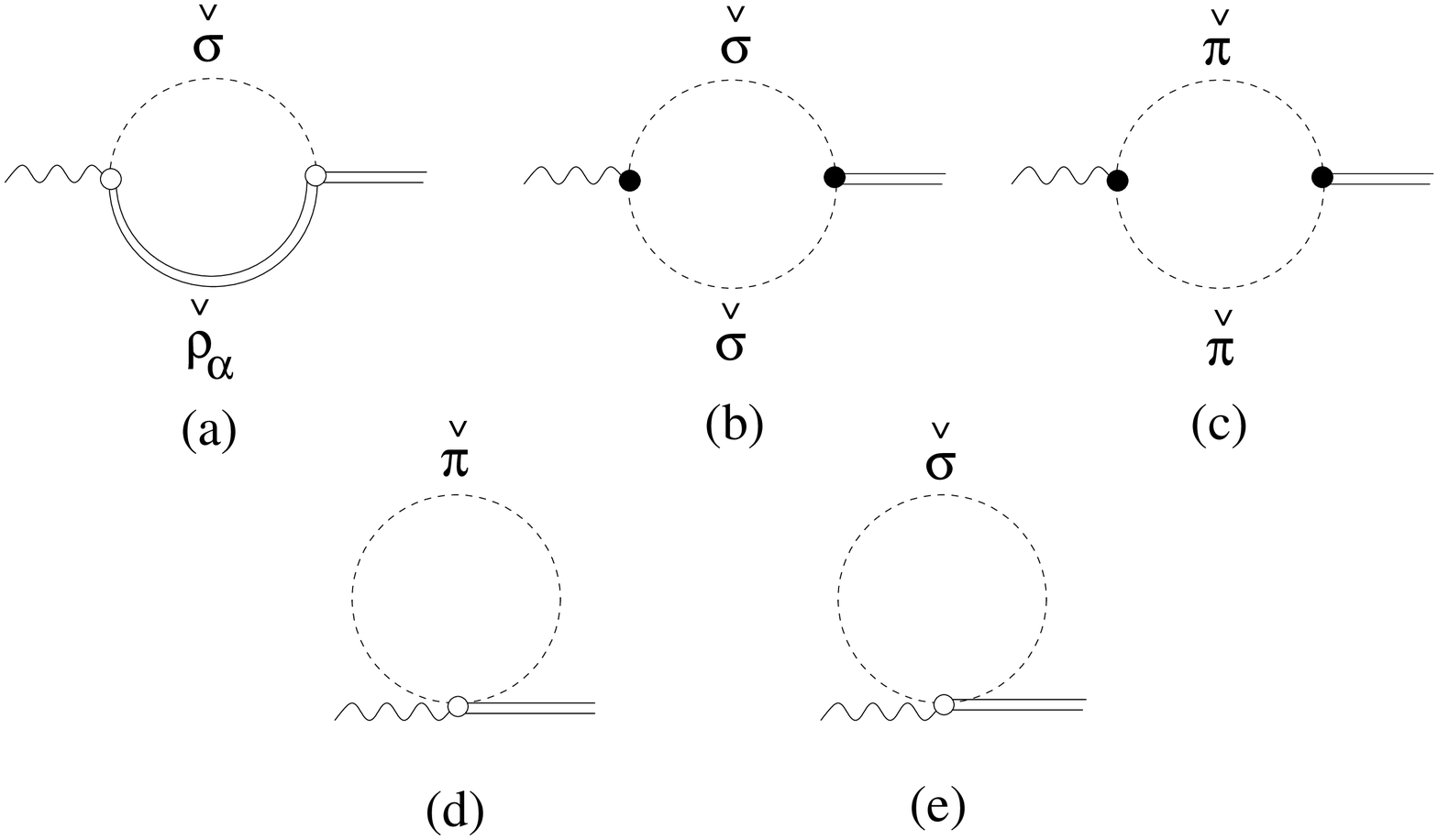}
 \end{center}
 \caption{Diagrams for contributions to
  $\Pi_{V\parallel}^{\mu\nu}$ 
         at one-loop level.}
 \label{fig:VRdiagrams}
\end{figure}

We also show the one-loop diagrams for contributions to
$\Pi_{V\parallel}^{\mu\nu}$ in Fig.~\ref{fig:VRdiagrams}.
These are given by
\begin{eqnarray}
 \Pi_{V\parallel}^{(a)\mu\nu}(p)
  &=& N_f M_\rho^2 g^{\mu\nu} B_0^{\rm{(vac)}}(p;M_\rho,M_\rho), 
       \label{eq:RVa}\\
 \Pi_{V\parallel}^{(b)\mu\nu}(p)
  &=& \frac{1}{8}N_f B^{\rm{(vac)}\mu\nu}(p;M_\rho,M_\rho), \label{eq:RVb}\\
 \Pi_{V\parallel}^{(c)\mu\nu}(p)
  &=& \frac{a(2-a)}{8}N_f B^{\rm{(vac)}\mu\nu}(p;0,0), \label{eq:RVc}\\
 \Pi_{V\parallel}^{(d)\mu\nu}(p)
  &=& \frac{a}{2}N_f g^{\mu\nu} A_0^{\rm{(vac)}}(0), \label{eq:RVd}\\
 \Pi_{V\parallel}^{(e)\mu\nu}(p)
  &=& \frac{1}{2}N_f g^{\mu\nu} A_0^{\rm{(vac)}}(M_\rho). \label{eq:RVe}
\end{eqnarray}

At tree level
the two-point functions of $\overline{\cal A}_\mu$,
$\overline{\cal V}_\mu$ and $\overline{V}_\mu$ are
\begin{eqnarray}
 \Pi_\perp^{\rm{(tree)}\mu\nu}(p)
  &=& F_{\pi, \rm{bare}}^2 g^{\mu\nu} +
     2z_{2, \rm{bare}}(p^2 g^{\mu\nu} - p^\mu p^\nu)\ , \nonumber\\
 \Pi_\parallel^{\rm{(tree)}\mu\nu}(p)
  &=& F_{\sigma, \rm{bare}}^2 g^{\mu\nu} +
     2z_{1, \rm{bare}}(p^2 g^{\mu\nu} - p^\mu p^\nu)\ , \nonumber\\
 \Pi_V^{\rm{(tree)}\mu\nu}(p)
  &=& F_{\sigma, \rm{bare}}^2 g^{\mu\nu} -
     \frac{1}{g_{\rm{bare}}^2}
     (p^2 g^{\mu\nu} - p^\mu p^\nu)
\ ,\nonumber\\
 \Pi_{V\parallel}^{\rm{(tree)}\mu\nu}(p)
  &=& {}- F_{\sigma, \rm{bare}}^2 g^{\mu\nu} +
     z_{3, \rm{bare}}(p^2 g^{\mu\nu} - p^\mu p^\nu)
\ .
\label{TP tree}
\end{eqnarray}
Thus the one-loop contributions to
$\Pi_\perp^{\mu\nu}$
give the quantum corrections to $F_\pi ^2$ and $z_2$.
Similarly,
each of the one-loop contributions to
$\Pi_\parallel^{\mu\nu}$, $\Pi_V^{\mu\nu}$ and
$\Pi_{V\parallel}^{\mu\nu}$ includes the quantum corrections to 
two parameters shown above.
For distinguishing the quantum corrections to two parameters included
in the two-point function,
it is convenient to decompose each two-point function as
\begin{equation}
 \Pi^{\mu\nu}(p)=\Pi^S(p)g^{\mu\nu} +
                 \Pi^{LT}(p)(g^{\mu\nu}p^2 - p^\mu p^\nu).
\label{decomp T0}
\end{equation}
Using the above decomposition, we identify 
$\Pi_\perp^{S\mbox{\scriptsize(1-loop)}}(p^2)$ with the quantum
correction to $F_\pi^2$, 
$\Pi_\perp^{LT\mbox{\scriptsize(1-loop)}}(p^2)$ with that to $z_2$,
and so on.
It should be noticed that
the following relation is satisfied~\cite{HY:WM,HY:PRep}:
\begin{equation}
\Pi_V^S(p^2) = \Pi_\parallel^S(p^2) =
- \Pi_{V\parallel}^S(p^2) \ .
\label{Pi V S equal}
\end{equation}
Then the quantum correction to $F_\sigma^2$ can be extracted from
any of $\Pi_\parallel^{\mu\nu}$,
$\Pi_V^{\mu\nu}$ and $\Pi_{V\parallel}^{\mu\nu}$.
The divergent contributions 
coming from the diagrams shown in 
Figs.~\ref{fig:AAdiagrams}-\ref{fig:VRdiagrams}
are listed in Refs.~\cite{HY:WM,HY:PRep}.
We list them in Appendix~\ref{app:QC} for convenience.

\subsection{Two-point functions at $T \neq 0$}
\label{ssec:TFf}

Let us consider the loop corrections to the
two-point functions at non-zero temperature.
For this purpose it is convenient to introduce the following
functions:
\begin{eqnarray}
A_0(M^2;T) &\equiv&
T \sum_{n=-\infty}^{\infty}
\int \frac{d^3k}{(2\pi)^3}
\frac{1}{M^2-k^2}
\ ,
\label{def:A0 2}
\\
B_0(p_0,\bar{p};M_1,M_2;T) &\equiv&
T \sum_{n=-\infty}^{\infty}
\int \frac{d^3k}{(2\pi)^3}
\frac{1}{ [M_1^2-k^2] [M_2^2-(k-p)^2] }
\ ,
\label{def:B0 2}
\\
B^{\mu\nu}(p_0,\bar{p};M_1,M_2;T) &\equiv&
T \sum_{n=-\infty}^{\infty}
\int \frac{d^3k}{(2\pi)^3}
\frac{\left(2k-p\right)^\mu \left(2k-p\right)^\nu}{%
 [M_1^2-k^2] [M_2^2-(k-p)^2] }
\ ,
\label{def:Bmunu 2}
\end{eqnarray}
where $\bar{p} = | \vec{p} |$
and the $0$th component of the loop momentum is taken
as $k^0 = i 2 n \pi T$,
while that of the external momentum is taken as 
$p^0 = i 2 n^\prime \pi T$ [$n, n^\prime$: integer].
Using the standard formula (see, e.g., Ref.~\cite{Kapusta}),
these functions are divided
into two parts as
\begin{eqnarray}
A_0(M;T) &=&
A_{0}^{\rm(vac)}(M)
+ \bar{A}_{0}(M;T) 
\ ,
\label{def:A0 11}
\\
B_{0}(p_0,\bar{p};M_1,M_2;T) &=&
B_{0}^{\rm(vac)}(p;M_1,M_2)
+ \bar{B}_{0}(p_0,\bar{p};M_1,M_2;T)
\ ,
\label{def:B0 11}
\\
B^{\mu\nu}(p_0,\bar{p};M_1,M_2;T) &=&
B^{{\rm(vac)}\mu\nu}(p;M_1,M_2)
+ \bar{B}^{\mu\nu}(p_0,\bar{p};M_1,M_2;T) 
\ ,
\label{def:Bmunu 11}
\end{eqnarray}
where
$A_{0}^{\rm(vac)}$, $B_{0}^{\rm(vac)}$
and $B^{{\rm(vac)}\mu\nu}$
express the quantum corrections given in 
Eqs.~(\ref{A0vac def})--(\ref{Bmnvac def}),
and 
$\bar{A}_{0}$, 
$\bar{B}_{0}$ and
$\bar{B}^{\mu\nu}$ 
the hadronic thermal corrections.
We summarize the explicit forms of the functions
$\bar{A}_{0}$, 
$\bar{B}_{0}$ and
$\bar{B}^{\mu\nu}$ in various limits relevant to the present
analysis in Appendix~\ref{app:LIFT}.
Note that 
the $0$th component of the momentum
$p_0$ in the right-hand-sides 
of the above expressions
is analytically continued to the Minkowsky variable:
$p_0$ is understood as $p_0 + i\epsilon$
($\epsilon \rightarrow +0$) for the retarded function and 
$p_0 - i \epsilon$ for the advanced function.
Then,
$A_{0}^{\rm(vac)}$, $B_{0}^{\rm(vac)}$
and $B^{{\rm(vac)}\mu\nu}$ have no explicit temperature dependence,
while they have intrinsic temperature dependence which is introduced
through the Wilsonian matching as we will see later.

Now, 
the contributions to the two-point functions at
non-zero temperature
are obtained by replacing,
in the expressions listed in the previous subsection,
the functions $A_0^{\rm(vac)}$, 
$B_0^{\rm(vac)}$ and 
$B^{{\rm(vac)}\mu\nu}$ defined in 
Eqs.~(\ref{A0vac def})--(\ref{Bmnvac def})
with $A_{0}$, $B_{0}$ and 
$B^{\mu\nu}$ defined in 
Eqs.~(\ref{def:A0 2})--(\ref{def:Bmunu 2}).

At non-zero temperature
there are four independent polarization tensors,
which we choose as
defined in Appendix~\ref{app:PTFT}.
We decompose 
the two-point functions
$\Pi_\perp^{\mu\nu}$, $\Pi_V^{\mu\nu}$ 
$\Pi_{V\parallel}^{\mu\nu}$ and $\Pi_\parallel^{\mu\nu}$ 
as
\begin{eqnarray}
 \Pi^{\mu\nu}
&=&
u^\mu u^\nu \Pi^t +
   (g^{\mu\nu}-u^\mu u^\nu)\Pi^s +
   P_L^{\mu\nu}\Pi^L + P_T^{\mu\nu}\Pi^T \ ,
\label{Pi decomp}
\end{eqnarray}
where $u^\mu=(1,\vec{0})$
and $P_L^{\mu\nu}$ and $P_T^{\mu\nu}$ are defined in
Eq.~(\ref{A.1}).
Similarly, we decompose the function $B^{\mu\nu}$
as
\begin{equation}
 B^{\mu\nu}
 =u^\mu u^\nu B^t +
   (g^{\mu\nu}-u^\mu u^\nu) B^s +
   P_L^{\mu\nu} B^L + P_T^{\mu\nu} B^T
\ .
\label{Bmn decomp}
\end{equation}
Furthermore, similarly to the division of the functions
into one part for expressing the quantum correction
and another for the hadronic thermal correction done in 
Eqs.~(\ref{def:A0 11})--(\ref{def:Bmunu 11}),
we divide the two-point functions into two parts as
 \begin{equation}
    \Pi^{\mu \nu}(p_0,\bar{p};T) =
           \Pi^{\rm{(vac)}\mu \nu}(p_0,\bar{p}) +
  \overline{\Pi}^{\mu \nu}(p_0,\bar{p};T)
 \ .
      \label{eq:TPart}
 \end{equation}
Accordingly each of four components of the two-point functions
defined in Eqs.~(\ref{Pi decomp})
is divided into the one part for
the quantum correction and another for the hadronic thermal
correction.
We note that
all the divergences are included in zero temperature part
$\Pi^{\rm(vac)}$.

{}From the contributions in Eqs.~(\ref{eq:AAa})-(\ref{eq:AAc}),
the temperature dependent parts are obtained as
\begin{eqnarray}
 \overline{\Pi}_\perp^t(p_0,\bar{p};T)
  &=& -N_f a M_\rho^2 \overline{B}_{0}(p_0,\bar{p};M_\rho,0;T) +
     N_f \frac{a}{4}\overline{B}^t(p_0,\bar{p};M_\rho,0;T) \nonumber\\ 
  &&{}+ N_f (a-1) \overline{A}_{0}(0;T),
\label{Pi perp t}
\\
 \overline{\Pi}_\perp^s(p_0,\bar{p};T)
  &=& -N_f a M_\rho^2 \overline{B}_{0}(p_0,\bar{p};M_\rho,0;T) +
     N_f \frac{a}{4}\overline{B}^s(p_0,\bar{p};M_\rho,0;T) \nonumber\\
  &&{}+ N_f (a-1) \overline{A}_{0}(0;T), 
\label{Pi perp s}
\\
 \overline{\Pi}_\perp^L(p_0,\bar{p};T)
   &=& 
  N_f \frac{a}{4}\overline{B}^L(p_0,\bar{p};M_\rho,0;T),
\label{Pi perp L}
\\
 \overline{\Pi}_\perp^T(p_0,\bar{p};T)
   &=& N_f \frac{a}{4}\overline{B}^T(p_0,\bar{p};M_\rho,0;T).
\end{eqnarray}

As we will see later in subsection~\ref{ssec:VMM-lowT},
we define 
the vector meson mass as the pole
of longitudinal or transverse component 
of the vector meson propagator.
{}From the contributions in 
Eqs.~(\ref{eq:RRa})-(\ref{eq:RRg}),
we obtain 
the hadronic thermal corrections to 
$\overline{\Pi}_V^{\mu\nu}$
as
\begin{eqnarray}
 \overline{\Pi}_V^t(p_0,\bar{p};T)
 &=& 2N_f\overline{A}_{0}(M_\rho;T) 
    - M_\rho^2 N_f
    \overline{B}_{0}(p_0,\bar{p};M_\rho,M_\rho;T) \nonumber\\
 &&{}+\frac{a^2}{8}N_f
       \overline{B}^t(p_0,\bar{p};0,0;T) 
   {}+\frac{9}{8}N_f
       \overline{B}^t(p_0,\bar{p};M_\rho,M_\rho;T)\ ,
\label{PiRt}
\\
 \overline{\Pi}_V^s(p_0,\bar{p};T)
 &=& 2N_f\overline{A}_{0}(M_\rho;T) 
    - M_\rho^2 N_f
    \overline{B}_{0}(p_0,\bar{p};M_\rho,M_\rho;T) \nonumber\\
 &&{}+\frac{a^2}{8}N_f
       \overline{B}^s(p_0,\bar{p};0,0;T) 
   {}+\frac{9}{8}N_f
       \overline{B}^s(p_0,\bar{p};M_\rho,M_\rho;T)\ ,
\label{PiRs}
\\
 \overline{\Pi}_V^L(p_0,\bar{p};T)
 &=& -4p^2 N_f \overline{B}_{0}(p_0,\bar{p};M_\rho,M_\rho;T)\nonumber\\
 &&{}+\frac{a^2}{8}N_f
       \overline{B}^L(p_0,\bar{p};0,0;T) 
   {}+\frac{9}{8}N_f
       \overline{B}^L(p_0,\bar{p};M_\rho,M_\rho;T),
\label{PiRL}
\\
 \overline{\Pi}_V^T(p_0,\bar{p};T)
 &=& -4p^2 N_f \overline{B}_{0}(p_0,\bar{p};M_\rho,M_\rho;T)\nonumber\\
 &&{}+\frac{a^2}{8}N_f
       \overline{B}^T(p_0,\bar{p};0,0;T) 
   {}+\frac{9}{8}N_f
       \overline{B}^T(p_0,\bar{p};M_\rho,M_\rho;T).
\label{PiRT}
\end{eqnarray}
Another two-point function associated with the vector current
correlator 
is $\Pi_{V\parallel}^{\mu\nu}$.
{}From the corrections in
Eqs.~(\ref{eq:RVa})-(\ref{eq:RVe}), 
we get the temperature dependent parts as
\begin{eqnarray}
 \overline{\Pi}_{V\parallel}^t(p_0,\bar{p};T)
 &=& \frac{a}{2}N_f \overline{A}_{0}(0;T)
   {}+ \frac{1}{2}N_f \overline{A}_{0}(M_\rho;T) 
{}+N_f M_\rho^2 
  \overline{B}_{0}(p_0,\bar{p};M_\rho,M_\rho;T)\nonumber\\
 &&{}+\frac{1}{8}N_f \overline{B}^t(p_0,\bar{p};M_\rho,M_\rho;T)
   {}+\frac{a(2-a)}{8}N_f 
       \overline{B}^t(p_0,\bar{p};0,0;T) ,
\label{PiVt}
\\
 \overline{\Pi}_{V\parallel}^s(p_0,\bar{p};T)
 &=& \frac{a}{2}N_f \overline{A}_{0}(0;T)
   {}+ \frac{1}{2}N_f \overline{A}_{0}(M_\rho;T) 
   {}+N_f M_\rho^2 
      \overline{B}_{0}(p_0,\bar{p};M_\rho,M_\rho;T)\nonumber\\
 &&{}+\frac{1}{8}N_f \overline{B}^s(p_0,\bar{p};M_\rho,M_\rho;T)
   {}+\frac{a(2-a)}{8}N_f \overline{B}^s(p_0,\bar{p};0,0;T),
\label{PiVs}
\\
 \overline{\Pi}_{V\parallel}^L(p_0,\bar{p};T)
 &=& \frac{1}{8}N_f \overline{B}^L(p_0,\bar{p};M_\rho,M_\rho;T)
   {}+\frac{a(2-a)}{8}N_f\overline{B}^L(p_0,\bar{p};0,0;T),
\\
 \overline{\Pi}_{V\parallel}^T(p_0,\bar{p};T)
 &=& \frac{1}{8}N_f \overline{B}^T(p_0,\bar{p};M_\rho,M_\rho;T)
   {}+\frac{a(2-a)}{8}N_f \overline{B}^T(p_0,\bar{p};0,0;T).
\end{eqnarray}

In subsection~\ref{ssec:PP-lowT}, we will define the parameter $a$
from the direct $\gamma\pi\pi$ coupling
using the two-point function 
$\Pi_\parallel^{\mu\nu}$.
{}From the corrections in
Eqs.~(\ref{eq:VVa})-(\ref{eq:VVd}),
we get the temperature dependent parts as
\begin{eqnarray}
 \overline{\Pi}_\parallel^t(p_0,\bar{p};T) 
 &=&-(a-1)N_f \overline{A}_{0}(0;T)
   {}-N_f M_\rho^2 
     \overline{B}_{0}(p_0,\bar{p};M_\rho,M_\rho;T)\nonumber\\
 &&{}+\frac{1}{8}N_f \overline{B}^t(p_0,\bar{p};M_\rho,M_\rho;T)
   {}+\frac{(2-a)^2}{8}N_f \overline{B}^t(p_0,\bar{p};0,0;T) \ , 
\label{PiVRt}
\\
 \overline{\Pi}_\parallel^s(p_0,\bar{p};T) 
 &=&-(a-1)N_f \overline{A}_{0}(0;T)
   {}-N_f M_\rho^2 
    \overline{B}_{0}(p_0,\bar{p};M_\rho,M_\rho;T)\nonumber\\
 &&{}+\frac{1}{8}N_f \overline{B}^s(p_0,\bar{p};M_\rho,M_\rho;T)
   {}+\frac{(2-a)^2}{8}N_f \overline{B}^s(p_0,\bar{p};0,0;T)\ ,
\label{PiVRs}
\\
 \overline{\Pi}_\parallel^L(p_0,\bar{p};T)
 &=& \frac{1}{8}N_f \overline{B}^L(p_0,\bar{p};M_\rho,M_\rho;T)
   {}+\frac{(2-a)^2}{8}N_f\overline{B}^L(p_0,\bar{p};0,0;T),
\\
 \overline{\Pi}_\parallel^T(p_0,\bar{p};T)
 &=& \frac{1}{8}N_f \overline{B}^T(p_0,\bar{p};M_\rho,M_\rho;T)
   {}+\frac{(2-a)^2}{8}N_f \overline{B}^T(p_0,\bar{p};0,0;T).
\end{eqnarray}

At the end of this subsection, we note that,
using the relations shown in Eq.~(\ref{B.8}),
we obtain
\begin{eqnarray}
&& \overline{\Pi}_V^t =
- \overline{\Pi}_{V\parallel}^t = \overline{\Pi}_{\parallel}^t
\nonumber\\
&& \qquad
  = \overline{\Pi}_V^s =
-\overline{\Pi}_{V\parallel}^s = \overline{\Pi}_{\parallel}^s
\nonumber\\
&& \qquad
 = - N_f \frac{1}{4} 
  \left[
    \overline{A}_{0}(M_\rho;T) 
     + \overline{A}_{0}(0;T)
  \right]
  - N_f M_\rho^2 \overline{B}_{0}(p_0,\bar{p};M_\rho,M_\rho;T)
\ .
\label{Pi bar V ts}
\end{eqnarray}
Since the quantum corrections to the
corresponding components are identical
as we have shown in Eq.~(\ref{Pi V S equal}),
the above relation implies that 
the components
$\Pi_V^t$, $\Pi_{V\parallel}^t$, $\Pi_\parallel^t$, 
$\Pi_V^s$, $\Pi_{V\parallel}^s$ and $\Pi_\parallel^t$ 
agree:
\begin{equation}
\Pi_V^t = 
- \Pi_{V\parallel}^t = \Pi_{\parallel}^t
  = \Pi_V^s =
-\Pi_{V\parallel}^s = \Pi_{\parallel}^s
\ .
\label{Pi V ts}
\end{equation}
This relation is important to prove the conservation of the
vector current correlator as we will show in 
subsection~\ref{ssec:VCC}.~\footnote{
  Actually, for the conservation of the vector current correlator,
  $\Pi_V^t = -\Pi_{V\parallel}^t = \Pi_{\parallel}^t$ and
  $\Pi_V^s = -\Pi_{V\parallel}^s = \Pi_{\parallel}^s$ are enough,
  and $\Pi_V^t$ and $\Pi_V^s$ can be generally different.
  }


\section{Current Correlators}
\label{sec:CC}

In this section, following Ref.~\cite{HKRS:SUS},
we review how to 
construct the axial-vector and vector current
correlators from the two-point functions calculated in the previous
section.
The correlators are defined by
\begin{eqnarray}
G_A^{\mu\nu}(p_0=i\omega_n,\vec{p};T) \delta_{ab}
=
\int_0^{1/T} d \tau \int d^3\vec{x}
e^{-i(\vec{p}\cdot\vec{x}+\omega_n\tau)}
\left\langle
  J_{5a}^\mu(\tau,\vec{x}) J_{5b}^\nu(0,\vec{0})
\right\rangle_\beta
\ ,
\nonumber\\
G_V^{\mu\nu}(p_0=i\omega_n,\vec{p};T) \delta_{ab}
=
\int_0^{1/T} d \tau \int d^3\vec{x}
e^{-i(\vec{p}\cdot\vec{x}+\omega_n\tau)}
\left\langle
  J_a^\mu(\tau,\vec{x}) J_b^\nu(0,\vec{0})
\right\rangle_\beta
\ ,
\end{eqnarray}
where $J_{5a}^\mu$ and $J_a^\mu$ are, respectively, the
axial-vector and vector currents, $\omega_n=2n\pi T$ is the
Matsubara frequency, $(a,b)=1,\ldots,N_f^2-1$ denotes the flavor
index and $\langle ~\rangle_\beta$ the thermal average. The
correlators for Minkowski momentum are obtained by the analytic
continuation of $p_0$.

\subsection{Axial-vector current correlator}

Let us consider the axial-vector current correlator $G_A^{\mu\nu}$.
For constructing it, we first parameterize the background field
$\bar{\xi}_{\rm R}$ and
$\bar{\xi}_{\rm L}$.
It is convenient to take the unitary gauge of the background HLS.
Then, 
$\bar{\xi}_{\rm R}$ and
$\bar{\xi}_{\rm L}$ are
parameterized 
\begin{eqnarray}
\bar{\xi}_{\rm L} = e^{-\bar{\phi}} \ ,
\quad
\bar{\xi}_{\rm R} = e^{\bar{\phi}} \ ,
\quad
\bar{\phi} = \bar{\phi}_a T_a \ ,
\end{eqnarray}
where $\bar{\phi}$ is the interpolating background
field corresponding to the pion field.
In terms of $\bar{\phi}$, the background 
$\overline{\cal A}_\mu$ is expanded as
\begin{equation}
\overline{\cal A}_\mu
= {\cal A}_\mu + i \partial_\mu \bar{\phi} + \cdots \ ,
\label{A bar exp}
\end{equation}
where ellipses denote the terms including two or more fields
and the axial-vector external gauge field ${\cal A}_\mu$ is defined
as
\begin{equation}
{\cal A}_\mu \equiv \frac{1}{2} \left(
  {\cal R}_\mu - {\cal L}_\mu
\right)
\ .
\end{equation}

There are two contributions to the axial-vector current
correlator
$G_A^{\mu\nu}$:
one from the $\bar{\phi}$-exchange diagram and another from the 
one-particle-irreducible (1PI) diagram
of ${\cal A}_\mu$-${\cal A}_\nu$ interaction.
By adding these two contributions, 
$G_A^{\mu\nu}$ is expressed as
\begin{equation}
 G_A^{\mu\nu}
  =\frac{p_\alpha p_\beta \Pi_\perp^{\mu\alpha}\Pi_\perp^{\nu\beta}}
        {-p_{\bar{\mu}}p_{\bar{\nu}}\Pi_\perp^{\bar{\mu}\bar{\nu}}}
   {}+ \Pi_\perp^{\mu\nu}
\ .
\end{equation}
Substituting 
the decomposition of $\Pi_\perp^{\mu\nu}$ 
in Eq.~(\ref{Pi perp decomp}) into the above
$G_A^{\mu\nu}$, we obtain
\begin{equation}
 G_A^{\mu\nu}=P_L^{\mu\nu}G_A^L + P_T^{\mu\nu}G_A^T,
\label{eq:G_A}
\end{equation}
where
\begin{eqnarray}
 G_A^L &=& \frac{p^2 \Pi_\perp^t \Pi_\perp^s}
               {-\bigl[ p_0^2 \Pi_\perp^t -
                 \bar{p}^2 \Pi_\perp^s \bigr]}
          {}+ \Pi_\perp^L, \nonumber\\
 G_A^T &=& -\Pi_\perp^s + \Pi_\perp^T.
\label{eq:G_A^L,T}
\end{eqnarray}
{}From this form we can easily see that
the current conservation $p_\mu G_A^{\mu\nu}=0$ is satisfied,
and the pion couples only to the longitudinal part $G_A^L$.

\subsection{Vector current correlator}
\label{ssec:VCC}

In the background field gauge
the vector meson inverse propagator is
related to 
the two-point function of the background field
$\overline{V}_\mu$, $\Pi_V^{\mu\nu}$ 
as
\begin{equation}
 i\bigl( D^{-1} \bigr)^{\mu \nu}
  =  \Pi_V^{\mu\nu}
  = u^\mu u^\nu \Pi_V^t + 
    (g^{\mu\nu}-u^\mu u^\nu)\Pi_V^s +
    P_L^{\mu\nu}\Pi_V^L + P_T^{\mu\nu}\Pi_V^T
\ .
       \label{eq:InvProp}
\end{equation}
It is convenient to decompose the full propagator in a same way:
\begin{equation}
  -iD^{\mu\nu}
  = u^\mu u^\nu D_V^t +
    (g^{\mu\nu} - u^\mu u^\nu)D_V^s +
    P_L^{\mu\nu}D_V^L + P_T^{\mu\nu}D_V^T
\ .
\label{eq:rho-propagator}
\end{equation}
After some algebra, these four components are expressed as
\begin{eqnarray}
 D_V^t
 &=& \frac{p^2\bigl( \Pi_V^L - \Pi_V^s \bigr)}
         {p_0^2 \Pi_V^t \bigl( \Pi_V^L - \Pi_V^s \bigr)-
          \bar{p}^2 \Pi_V^s \bigl( \Pi_V^L - \Pi_V^t \bigr)},\nonumber\\
 D_V^s
 &=& \frac{p^2\bigl( \Pi_V^L - \Pi_V^t \bigr)}
         {p_0^2 \Pi_V^t \bigl( \Pi_V^L - \Pi_V^s \bigr)-
          \bar{p}^2 \Pi_V^s \bigl( \Pi_V^L - \Pi_V^t \bigr)},\nonumber\\
 D_V^L
 &=& \frac{p^2 \Pi_V^L}
         {p_0^2 \Pi_V^t \bigl( \Pi_V^L - \Pi_V^s \bigr)-
          \bar{p}^2 \Pi_V^s \bigl( \Pi_V^L - \Pi_V^t \bigr)},\nonumber\\
 D_V^T
 &=& D_V^s + \frac{1}{\Pi_V^T - \Pi_V^s},
\end{eqnarray}
where we define $\bar{p}=|\vec{p}|$.
By using the above vector meson propagator
$-iD^{\mu\nu}$
and the two-point functions $\Pi_\parallel^{\mu\nu}$
and $\Pi_{V\parallel}^{\mu\nu}$, 
the vector current correlator $G_V^{\mu\nu}$,
which consists of the contributions 
from the vector meson exchange
and 1PI diagrams,
is obtained as
\begin{equation}
 G_V^{\mu\nu} = \Pi_{V\parallel}^{\mu\alpha}iD_{\alpha\beta}
                \Pi_{V\parallel}^{\beta\nu} + \Pi_\parallel^{\mu\nu}
\ .
\label{eq:G_V-def}
\end{equation}
Substituting 
Eq.~(\ref{eq:rho-propagator}) together with 
Eqs.~(\ref{Pi decomp}) into Eq.~(\ref{eq:G_V-def}),
we obtain $G_V^{\mu\nu}$ as
\begin{equation}
 G_V^{\mu\nu}
 = u^\mu u^\nu G_V^t +
    (g^{\mu\nu} - u^\mu u^\nu)G_V^s +
    P_L^{\mu\nu}G_V^L + P_T^{\mu\nu}G_V^T,
\end{equation}
where each component is expressed as
\begin{eqnarray}
 G_V^t
 &=& \frac{D_V^L}{\Pi_V^L}
   \Bigl[ \frac{\bar{p}^2}{p^2}\Pi_{V\parallel}^L
    \bigl\{ \Pi_V^s \Pi_{V\parallel}^t - \Pi_V^t \Pi_{V\parallel}^s
    \bigr\} \nonumber\\ 
   &&{}- \frac{\Pi_{V\parallel}^t}{p^2}
    \bigl\{ -p_0^2 \Pi_{V\parallel}^t
       \bigl( \Pi_V^s - \Pi_V^L \bigr) +
       p^2 \bigl( \Pi_{V\parallel}^t \Pi_V^s - \Pi_{V\parallel}\Pi_V^L \bigr)
    \bigr\}
   \Bigr] + \Pi_\parallel^t, \nonumber\\
 G_V^s
 &=& \frac{D_V^L}{\Pi_V^L}
   \Bigl[ \frac{p_0^2}{p^2}\Pi_{V\parallel}^L
    \bigl\{ \Pi_V^s \Pi_{V\parallel}^t - \Pi_V^t \Pi_{V\parallel}^s
    \bigr\} \nonumber\\ 
  &&{}- \frac{\Pi_{V\parallel}^s}{p^2}
    \bigl\{ -p_0^2 \bigl( \Pi_V^t \Pi_{V\parallel}^s - 
            \Pi_{V\parallel}^t \Pi_V^L \bigr) +
            \bar{p}^2 \Pi_{V\parallel}^s \bigl(\Pi_V^t-\Pi_V^s \bigr)
    \bigr\}
   \Bigr] + \Pi_\parallel^s, \nonumber\\
 G_V^L
 &=& \frac{D_V^L}{\Pi_V^L}
   \Bigl[ -\Pi_V^L \Pi_{V\parallel}^t + 
          \Pi_{V\parallel}^L \bigl( \Pi_{V\parallel}^t \Pi_V^s +
          \Pi_{V\parallel}^s \Pi_V^t \bigr) \nonumber\\
  &&{}- \frac{1}{p^2}
     \bigl( p_0^2 \Pi_V^t - \bar{p}^2 \Pi_V^s \bigr)
     \bigl( \Pi_{V\parallel}^L \bigr)^2
   \Bigr] + \Pi_\parallel^L, \nonumber\\
 G_V^T
 &=& \frac{D_V^L}{\Pi_V^L}
   \Bigl[ -\frac{p_0^2}{p^2}\Pi_{V\parallel}^L
    \bigl\{ \Pi_V^t \Pi_{V\parallel}^s -
            \Pi_V^s \Pi_{V\parallel}^t
    \bigr\} \nonumber\\ 
  &&{}- \frac{\Pi_{V\parallel}^s}{p^2}
    \bigl\{ -p_0^2 \bigl( \Pi_V^t \Pi_{V\parallel}^s -
            \Pi_{V\parallel}^t \Pi_V^L \bigr) +
            \bar{p}^2 \Pi_{V\parallel}^s
            \bigl( \Pi_V^t - \Pi_V^L \bigr)
    \bigr\} 
   \Bigr]\nonumber\\
  &&{}+ \frac{\bigl( \Pi_{V\parallel}^s - \Pi_{V\parallel}^T \bigr)^2}
           {\Pi_V^s - \Pi_V^T} + \Pi_\parallel^T.
\end{eqnarray}
One might worry that the above form does not satisfy the current
conservation $p_\mu G_V^{\mu\nu} =0$.
However, 
since the following equalities are satisfied [see Eq.~(\ref{Pi V ts})],
\begin{eqnarray}
 \Pi_V^t &=& -\Pi_{V\parallel}^t = \Pi_\parallel^t, \nonumber\\
 \Pi_V^s &=& -\Pi_{V\parallel}^s = \Pi_\parallel^s,
\end{eqnarray}
each component of the above $G_V^{\mu\nu}$ is rewritten as
\begin{eqnarray}
 G_V^t &=& G_V^s = 0, \nonumber\\
 G_V^L
 &=& -\frac{D_V^L}{\Pi_V^L}
    \Bigl[ \Pi_V^t \Pi_V^s \bigl( \Pi_V^L + 2\Pi_{V\parallel}^L \bigr)+
           \frac{p_0^2 \Pi_V^t - \bar{p}^2 \Pi_V^s}{p^2}
           \bigl( \Pi_{V\parallel}^L \bigr)^2
    \Bigr] + \Pi_\parallel^L, \nonumber\\
 G_V^T
 &=& \frac{\Pi_V^s \bigl( \Pi_V^T + 2\Pi_{V\parallel}^T \bigr)+
          \bigl( \Pi_{V\parallel}^T \bigr)^2}
         {\Pi_V^s - \Pi_V^T} + \Pi_\parallel^T.
\end{eqnarray}
Now we can easily see that
the current conservation $p_\mu G_V^{\mu\nu} = 0$ is satisfied
since $p_\mu P_L^{\mu\nu} = p_\mu P_T^{\mu\nu} = 0$.
In the present analysis we further have $\Pi_V^t = \Pi_V^s$
as can be seen from Eq.~(\ref{Pi V ts}).
Finally we obtain the following $G_V^{\mu\nu}$:
\begin{equation}
 G_V^{\mu\nu}=P_L^{\mu\nu}G_V^L + P_T^{\mu\nu}G_V^T,
\label{eq:G_V}
\end{equation}
where
\begin{eqnarray}
 G_V^L
 &=& \frac{\Pi_V^s \bigl( \Pi_V^L + 2\Pi_{V\parallel}^L \bigr)}
         {\Pi_V^s - \Pi_V^L} + \Pi_\parallel^L, \nonumber\\
 G_V^T
 &=& \frac{\Pi_V^s \bigl( \Pi_V^T + 2\Pi_{V\parallel}^T \bigr)}
         {\Pi_V^s - \Pi_V^T} + \Pi_\parallel^T.
\label{GV LT components}
\end{eqnarray}
Here we have dropped the terms 
$\bigl( \Pi_{V\parallel}^L \bigr)^2$ and 
$\bigl( \Pi_{V\parallel}^T \bigr)^2$
since they are of higher order.


\section{Thermal Properties of $\pi$-$\rho$ Parameters}
\label{sec:TPpi-rhoP}

In this section,
we briefly summarize the temperature dependences of
several parameters of pions and vector mesons 
in hot medium
following Refs.~\cite{HS:VM,HKRS:SUS,HS:VVD}.

\subsection{Vector meson mass}
\label{ssec:VMM-lowT}

In this subsection,
we first define the vector meson pole 
masses of both the longitudinal and transverse modes
at non-zero temperature from the vector current correlator 
in the background field gauge and show
the explicit forms of the hadronic thermal corrections
from the
vector and pseudoscalar meson loop.

Let us define pole masses of longitudinal and transverse modes of
the vector meson 
from the poles of longitudinal and transverse components 
of the vector current correlator in the  rest frame,
which are given by Eqs.~(\ref{eq:G_V}) and (\ref{GV LT components}).
Then, the pole masses are obtained as the solutions of the following
on-shell conditions:
\begin{eqnarray}
&&
0 = 
\mbox{Re}
\left[
  \Pi_V^s(p_0=m_{\rho}^L(T),0;T) - 
  \Pi_V^L(p_0=m_{\rho}^L(T),0;T)
\right]
\ ,
\nonumber\\
&&
0 = 
\mbox{Re}
\left[
  \Pi_V^s(p_0=m_{\rho}^T(T),0;T) - 
  \Pi_V^T(p_0=m_{\rho}^T(T),0;T)
\right]
\ ,
\label{rho on shell cond}
\end{eqnarray}
where $m_\rho^L(T)$ and $m_\rho^T(T)$ denote the pole masses of
the longitudinal and transverse modes, respectively.
As we have calculated in section~\ref{sec:TPFBFG},
$\Pi_V^s(p_0,\bar{p};T)$,
$\Pi_V^L(p_0,\bar{p};T)$ and
$\Pi_V^T(p_0,\bar{p};T)$ 
in the HLS at one-loop level
are expressed as
\begin{eqnarray}
\Pi_V^s(p_0,\bar{p};T) 
&=&
F_\sigma^2(M_\rho) + \widetilde{\Pi}_V^S(p^2)
+ \overline{\Pi}_V^{s} (p_0,\bar{p};T)
\ ,
\nonumber\\
\Pi_V^L(p_0,\bar{p};T) 
&=&
\frac{p^2}{g^2(M_\rho)} + \widetilde{\Pi}_V^{LT}(p^2)
+ \overline{\Pi}_V^{L} (p_0,\bar{p};T)
\ ,
\nonumber\\
\Pi_V^T(p_0,\bar{p};T) 
&=&
\frac{p^2}{g^2(M_\rho)} + \widetilde{\Pi}_V^{LT}(p^2)
+ \overline{\Pi}_V^{T} (p_0,\bar{p};T)
\ ,
\label{Pi V s L T forms}
\end{eqnarray}
where the explicit forms of 
the finite renormalization effects
$\widetilde{\Pi}_V^S(p^2)$ and
$\widetilde{\Pi}_V^{LT}(p^2)$ 
are given in Eqs.~(\ref{C.2}) and (\ref{C.3}),
and those of the hadronic thermal effects
$\overline{\Pi}_V^{s} (p_0,\bar{p};T)$,
$\overline{\Pi}_V^{L} (p_0,\bar{p};T)$ and
$\overline{\Pi}_V^{T} (p_0,\bar{p};T)$ are
given in Eqs.~(\ref{Pi bar V ts}), (\ref{PiRL}) and (\ref{PiRT}).
Substituting Eq.~(\ref{Pi V s L T forms}) into
Eq.~(\ref{rho on shell cond}),
we obtain
\begin{eqnarray}
\left[m_\rho^L(T)\right]^2 
&=&
M_\rho^2 
{}+
g^2(M_\rho) \,
\Biggl[
  \mbox{Re}\,\widetilde{\Pi}_V^S( p_0^2 )
  + \mbox{Re}\,\overline{\Pi}_V^{s} (p_0,0;T)
\nonumber\\
&& \qquad\qquad\qquad
  {}- \mbox{Re}\,\widetilde{\Pi}_V^{T}(p_0^2)
  - \mbox{Re}\,\overline{\Pi}_V^{L} (p_0,0;T)
\Biggr]_{p_0 = m_\rho^L(T)}
\ ,
\nonumber\\
\left[m_\rho^T(T)\right]^2 
&=&
M_\rho^2 
{}+
g^2(M_\rho) \,
\Biggl[
  \mbox{Re}\,\widetilde{\Pi}_V^S( p_0^2 )
  + \mbox{Re}\,\overline{\Pi}_V^{s} (p_0,0;T)
\nonumber\\
&& \qquad\qquad\qquad
  {}- \mbox{Re}\,\widetilde{\Pi}_V^{T}(p_0^2)
  - \mbox{Re}\,\overline{\Pi}_V^{T} (p_0,0;T)
\Biggr]_{p_0 = m_{\rho}^T(T)}
\ .
\end{eqnarray}
We can replace $m_\rho^L(T)$ and $m_\rho^T(T)$ with
$M_\rho$ in the hadronic
thermal effects as well as in 
the finite renormalization effect,
since the difference is of higher order.
Then, 
noting that 
$\mbox{Re}\,\widetilde{\Pi}_V^S(p^2=M_\rho^2) =
\mbox{Re}\,\widetilde{\Pi}_V^{LT}(p^2=M_\rho^2) = 0$
as shown in Eq.~(\ref{zero FRE}),
we obtain
\begin{eqnarray}
\left[m_{\rho}^L(T)\right]^2
&=&
M_\rho^2 -
g^2(M_\rho) \, \mbox{Re}
\left[
  \overline{\Pi}_V^L(M_\rho;0;T) 
  - \overline{\Pi}_V^s (M_\rho;0;T)
\right]
\ ,
\label{eq:OnShellL}
\\
\left[m_{\rho}^T(T)\right]^2
&=& M_\rho^2 -
g^2(M_\rho) \, \mbox{Re}
\left[
  \overline{\Pi}_V^T(M_\rho;0;T) 
  - \overline{\Pi}_V^s (M_\rho;0;T)
\right]
\ .
\label{eq:OnSellT}
\end{eqnarray}

Let us calculate the explicit form of the pole mass using the
expression of $\overline{\Pi}_V^s$,
$\overline{\Pi}_V^L$ and $\overline{\Pi}_V^T$ obtained in 
section~\ref{sec:TPFBFG}.
Here we note that 
Eqs.~(\ref{BL rest}) and (\ref{BT rest}) imply that 
$\overline{B}^L-\overline{B}^s$ agrees with 
$\overline{B}^T-\overline{B}^s$ in the rest frame. 
Then in the rest frame $\overline{\Pi}_V^L-\overline{\Pi}_V^s$
agrees with $\overline{\Pi}_V^T-\overline{\Pi}_V^s$.
Thus the longitudinal pole mass
is the same as the transverse one:
\begin{equation}
m_\rho^L(T) = m_\rho^T(T)\equiv m_\rho(T) \ .
\end{equation}
By using the low momentum limits of the functions
shown in Eqs.~(\ref{B0B rest}) and (\ref{BL rest}),
$\overline{\Pi}^{L}_V-\overline{\Pi}^{s}_V$ in the rest frame is
expressed as
\begin{eqnarray}
 &&\overline{\Pi}^{L}_V(p_0,\bar{p}=0;T)-
 \overline{\Pi}^{s}_V(p_0,\bar{p}=0;T) \nonumber\\
 &&= N_f 
   \Bigl[ \frac{a^2}{24}\tilde{G}_{2}(p_0;T) - 
           \tilde{J}_{1}^2(M_\rho;T) +
          \bigl( \frac{M_\rho^2}{4}-p_0^2 \bigr)
                  \tilde{F}_{3}^2(p_0;M_\rho;T) +
          \frac{3}{8}\tilde{F}_{3}^4(p_0;M_\rho;T)
   \Bigr], 
\end{eqnarray}
where the functions $\tilde{I}_n$, $\tilde{J}^n_m$, $\tilde{F}^n_m$
and $\tilde{G}^n_m$
are defined in Appendix~\ref{app:Functions}.
Substituting the above expression
into Eq.~(\ref{eq:OnShellL}) and using the relation
\begin{eqnarray}
 -\frac{1}{3M_\rho^2}
 \tilde{F}_{3}^4(M_\rho;M_\rho;T)
 &=& \frac{1}{4} \tilde{F}_{3}^2(M_\rho;M_\rho;T) 
   - \frac{1}{3M_\rho^2} \tilde{J}_{1}^2(M_\rho;T)\ ,
\end{eqnarray}
we find that the vector meson pole mass is expressed as
\begin{eqnarray}
  m^2_\rho (T) 
&=& {M_\rho}^2
    +N_f\,g^2
    \Biggl[- \frac{a^2}{12}\tilde{G}_{2}(M_\rho;T)
      + \frac{4}{5} \tilde{J}^2_{1}(M_\rho;T)
      + \frac{33}{16} M_\rho^2 \tilde{F}^2_{3}(M_\rho;M_\rho;T)
    \Biggr].
\label{eq:Mass} 
 \end{eqnarray}
The contribution in this expression
agrees with the result 
in Ref.~\cite{HS:VM} which is derived from the 
hadronic thermal correction calculated
in the Landau gauge 
in Ref.~\cite{HShibata} by
taking the on-shell condition (\ref{eq:OnShellL}).~\footnote{%
  The functions used in this paper are related to the ones used
  in Ref.~\cite{HS:VM} 
  as $\tilde{J}_{1}^2 = \frac{1}{2\pi^2}J_1^2$, 
  $\tilde{G}_{2}=\frac{1}{2\pi^2}\bar{G}_2$, and so on.
  }

We study the behavior of the pole mass in the low temperature
region, $T \ll M_\rho$.
In this region the functions
$F^n_{m}$ and $J^n_{m}$ are suppressed by $e^{-M_\rho/T}$, 
and thus give negligible contributions.
Since $G_{2} \approx -\frac{{\pi}^4}{15}
                         \frac{T^4}{{M_\rho}^2}$,
the vector meson pole mass increases as $T^4$
in the low temperature region,
dominated by the $\pi$-loop effects:
 \begin{equation}
  {m_\rho}^2(T)\approx {M_\rho}^2 +
                 \frac{N_f {\pi}^2 a}{360 {F_\pi}^2}T^4
  \qquad \mbox{for} \quad T \ll M_\rho.
  \label{m rho small T}
 \end{equation}
Note that the lack of $T^2$-term in the above expression is consistent
with the result by the current algebra 
analysis~\cite{Dey-Eletsky-Ioffe}.

\subsection{Pion parameters}
\label{ssec:PP-lowT}

First we study the on-shell structure of the pion. 
For this, we look at the pole of
the longitudinal component $G_A^L$ in Eq.~(\ref{eq:G_A^L,T}). 
Since both $\Pi_\perp^t$ and $\Pi_\perp^s$ have imaginary parts, 
we choose to determine the pion energy $E$ from the real part 
by solving the dispersion formula
\begin{eqnarray}
  0
&=&
  \left[
    p_0^2 \, \mbox{Re} \Pi^{t}_\perp (p_0,\bar{p};T)
    - \bar{p}^2 \, \mbox{Re} \Pi^{s}_\perp (p_0,\bar{p};T)
  \right]_{p_0=E}
\ ,
\label{pi on shell cond}
\end{eqnarray}
where $\bar{p}\equiv\vert\vec{p}\vert$. 
As remarked in section~\ref{sec:TPFBFG}, 
in HLS at one-loop level, $\Pi^{t}_\perp (p_0,\bar{p};T)$ and 
$\Pi^{s}_\perp (p_0,\bar{p};T)$ are of the form
\begin{eqnarray}
\Pi^{t}_\perp (p_0,\bar{p};T)
&=&
F_\pi^2(0) + \widetilde{\Pi}_\perp^S(p^2) +
\bar{\Pi}^{t}_\perp (p_0,\bar{p};T)
\ ,
\nonumber\\
\Pi^{s}_\perp (p_0,\bar{p};T)
&=&
F_\pi^2(0) + \widetilde{\Pi}_\perp^S(p^2) +
\bar{\Pi}^{s}_\perp (p_0,\bar{p};T)
\ ,
\label{Pi t s forms}
\end{eqnarray}
where $\widetilde{\Pi}_\perp^S(p^2)$ is the finite renormalization
contribution, and $\bar{\Pi}^{t}_\perp (p_0,\bar{p};T) $ and
$\bar{\Pi}^{s}_\perp (p_0,\bar{p};T) $ are the hadronic thermal
contributions. Substituting Eq.~(\ref{Pi t s forms}) into
Eq.~(\ref{pi on shell cond}), we obtain
\begin{eqnarray}
&&0 =
\left( E^2 - \bar{p}^2 \right)
\left[
  F_\pi^2(0) +
  \mbox{Re} \,\widetilde{\Pi}_\perp^S( p^2=E^2 - \bar{p}^2 )
\right]
\nonumber\\
&&\qquad\qquad\qquad\qquad\qquad\quad
{}+
  E^2
  \mbox{Re} \,\bar{\Pi}^{t}_\perp (E,\bar{p};T)
  - \bar{p}^2
  \mbox{Re} \, \bar{\Pi}^{s}_\perp (E,\bar{p};T)
\ .
\end{eqnarray}
The pion velocity $v_\pi(\bar{p}) \equiv E / \bar{p}$ is then obtained
by solving
\begin{eqnarray}
v_\pi^2(\bar{p})
&=&
\frac{
  F_\pi^2(0) +
  \mbox{Re} \, \bar{\Pi}^{s}_\perp (\bar{p},\bar{p};T)
}{
  F_\pi^2(0) +
  \mbox{Re} \, \bar{\Pi}^{t}_\perp (\bar{p},\bar{p};T)
}
\ .
\label{v2: form}
\end{eqnarray}
Here we replaced $E$ by $\bar{p}$ in the hadronic thermal terms
$\bar{\Pi}_\perp^t (E,\vec{p})$ and $\bar{\Pi}_\perp^s
(E,\vec{p})$ as well as in the finite renormalization contribution
$\widetilde{\Pi}_\perp^S( p^2 = E^2 - \bar{p}^2)$, since the
difference is of higher order. [Note that $
\widetilde{\Pi}_\perp^S( p^2=0 )= 0$.]

Next we determine the wave function renormalization of the pion
field, which relates the background field $\bar{\phi}$ to the pion
field $\bar{\pi}$ in the momentum space as
 \begin{equation}
\bar{\phi} = \bar{\pi}/\widetilde{F}(\bar{p};T).
 \end{equation}
We follow the analysis in Ref.~\cite{MOW} to obtain
\begin{equation}
\widetilde{F}^2(\bar{p};T) =
\mbox{Re} \Pi_\perp^t(E,\bar{p};T)
= F_\pi^2(0) +
  \mbox{Re} \, \bar{\Pi}^{t}_\perp (\bar{p},\bar{p};T)
\ .
\label{Ftil def}
\end{equation}
Using this wave function renormalization and the velocity in
Eq.~(\ref{v2: form}), we can rewrite the longitudinal part of the
axial-vector current correlator as
\begin{eqnarray}
G_A^L(p_0,\vec{p})
&=&
\frac{ p^2
  \Pi_\perp^t(p_0,\bar{p};T) \Pi_\perp^s(p_0,\bar{p};T)
  / \widetilde{F}^2(\bar{p};T)
}{
  - \left[
      p_0^2 - v_\pi^2(\bar{p}) \bar{p}^2 + \Pi_\pi(p_0,\bar{p};T)
  \right]
}
+ \Pi_\perp^L(p_0,\bar{p};T)
\ ,
\label{GAL 2}
\end{eqnarray}
where the pion self energy $\Pi_\pi(p_0,\bar{p};T)$ is given by
\begin{eqnarray}
&&
\Pi_\pi(p_0,\bar{p};T)
=
\frac{1}{
  \mbox{Re} \, \Pi^{t}_\perp (E,\bar{p};T)
}
\nonumber\\
&& \quad
\times
\Biggl[
  p_0^2
  \left\{
    \Pi^{t}_\perp (p_0,\bar{p};T)
    -
    \mbox{Re} \, \Pi^{t}_\perp (E,\bar{p};T)
  \right\}
  -
  \bar{p}^2
  \left\{
    \Pi^{s}_\perp (p_0,\bar{p};T)
    -
    \mbox{Re} \, \Pi^{s}_\perp (E,\bar{p};T)
  \right\}
\Biggr]
\ .
\end{eqnarray}

Let us now define the pion decay constant. A natural procedure is
to define the pion decay constant from the pole residue of the
axial-vector current correlator. From Eq.~(\ref{GAL 2}), the pion
decay constant is given by
\begin{eqnarray}
f_\pi^2(\bar{p};T)
&=&
\frac{
 \Pi_\perp^t(E,\bar{p};T)
 \Pi_\perp^s(E,\bar{p};T)
}{ \widetilde{F}^2(\bar{p};T) }
\nonumber\\
&=&
\frac{
 \left[ F_\pi^2(0) + \bar{\Pi}^{t}_\perp (\bar{p},\bar{p};T) \right]
 \left[ F_\pi^2(0) + \bar{\Pi}^{s}_\perp (\bar{p},\bar{p};T) \right]
}{ \widetilde{F}^2(\bar{p};T) }
\ .
\label{fpi2 def}
\end{eqnarray}
We now address how $f_\pi^2(\bar{p};T)$ is related to the temporal
and spatial components of the pion decay constant introduced in
Ref.~\cite{PT:96}. Following their notation, let $f_\pi^t$ denote
the decay constant associated with the temporal component of the
axial-vector current and $f_\pi^s$ the one with the spatial
component. In the present analysis, they can be read off from the
coupling of the $\bar{\pi}$ field to the axial-vector external
field ${\cal A}_\mu$:
\begin{eqnarray}
  f_\pi^t(\bar{p};T)
&\equiv&
  \frac{
    \Pi^{t}_\perp (E,\bar{p};T)
  }{\widetilde{F}(\bar{p};T)}
=
  \frac{ F_\pi^2(0) + \bar{\Pi}^{t}_\perp (\bar{p},\bar{p};T) }
    {\widetilde{F}(\bar{p};T)}
\ ,
\label{fpit def}
\\
  f_\pi^s(\bar{p};T)
&\equiv&
  \frac{
    \Pi^{s}_\perp (\tilde{E},\bar{p};T)
  }{\widetilde{F}(\bar{p};T)}
=
  \frac{ F_\pi^2(0) + \bar{\Pi}^{s}_\perp (\bar{p},\bar{p};T) }
    {\widetilde{F}(\bar{p};T)}
\ .
\label{fpis def}
\end{eqnarray}
Comparing Eqs.~(\ref{fpit def}) and (\ref{fpis def}) with
Eqs.~(\ref{v2: form}), (\ref{Ftil def}) and (\ref{fpi2 def}), we
have~\cite{PT:96,MOW}
\begin{eqnarray}
\widetilde{F}(\bar{p};T)
&=& \mbox{Re} \, f_\pi^t(\bar{p};T)
\ ,
\label{wave-func-ren-const}
\\
f_\pi^2(\bar{p};T) &=&
  f_\pi^t(\bar{p};T) f_\pi^s(\bar{p};T)
\ .
\label{v2 rel}
\end{eqnarray}
By using the above decay constants, the pion velocity at one-loop level
is expressed as~\cite{PT:96,HKRS:SUS}
~\footnote{
 This form of the pion velocity looks slightly different from 
 Eq.~(\ref{v2: form}).
 However, Eq.~(\ref{fpts rels}) is actually equivalent to Eq.~(\ref{v2: form})
 at one-loop order, and more convenient to study the temperature
 dependence of the pion velocity in the low temperature region.
}
\begin{equation}
v_\pi^2(\bar{p};T) =
 1 + \frac{\tilde{F}(\bar{p};T)}{F_\pi^2}
  \Bigl[ \mbox{Re}f_\pi^s (\bar{p};T) - \mbox{Re}f_\pi^t (\bar{p};T) 
  \Bigr].
\label{fpts rels}
\end{equation}

Substituting Eqs.~(\ref{A0BM}) and (\ref{A0B0})
into Eqs.~(\ref{Pi perp t}) and (\ref{Pi perp s}),
we obtain 
$\overline{\Pi}_\perp^t$ and $\overline{\Pi}_\perp^s$ for the on-shell pion as
\begin{eqnarray}
 &&\overline{\Pi}^{t}_\perp(p_0 = \bar{p}+i\epsilon,\bar{p};T)
 = N_f(a-1) \tilde{I}_{2}(T) \nonumber\\ 
 &&\qquad{}+ \frac{N_f}{2}a
   \left[ \frac{1}{2}
   \overline{B}^t(\bar{p}+i\epsilon,\bar{p};M_\rho,0;T)
 {}-2M_\rho^2 
   \overline{B}_{0}(\bar{p}+i\epsilon,\bar{p};M_\rho,0;T)
   \right]\ , \label{Pit A on-shell}\\
 &&\overline{\Pi}^{s}_\perp(p_0=\bar{p}+i\epsilon,\bar{p};T)
 = N_f(a-1) \tilde{I}_{2}(T) \nonumber\\ 
 &&\qquad{}+ \frac{N_f}{2}a
  \left[ 
    \frac{1}{2}
      \overline{B}^s(\bar{p}+i\epsilon,\bar{p};M_\rho,0;T)
    {}-2M_\rho^2 
      \overline{B}_{0}(\bar{p}+i\epsilon,\bar{p};M_\rho,0;T)
  \right]\ ,
\label{Pits A on-shell}
\end{eqnarray}
where we put $\epsilon \rightarrow +0$ to make the analytic
continuation for the frequency $p_0=i 2\pi nT$ to the Minkowski
variable.
We show the explicit forms of the functions
$\overline{B}^t$, $\overline{B}^s$ and
$\overline{B}_{0}$ in Eqs.~(\ref{B0 A on-shell}),
(\ref{Bt A on-shell}) and (\ref{Bs A on-shell}).

In general the pion velocity in medium
does not agree with the value at $T=0$
due to the interaction with the heat bath.
Below $T_c$,
since the temporal component does not agree with the spatial one
due to the thermal vector meson effect,
$\overline{\Pi}_\perp^t \neq \overline{\Pi}_\perp^s$,
the pion velocity $v_\pi(\bar{p};T)$ is not the speed of light.
As we will see below, in the framework of HLS the pion velocity receives
a change from the $\rho$-loop effect for $0 < T < T_c$.
When we take the low temperature limit ($T \ll M_\rho$)
and the soft pion limit ($\bar{p} \ll M_\rho$ and $\bar{p} \ll T$),
the real parts of
$\overline{\Pi}_\perp^t$ and $\overline{\Pi}_\perp^s$ are approximated as
\begin{eqnarray}
 \mbox{Re}\overline{\Pi}_\perp^t (p_0 = \bar{p}+i\epsilon,\bar{p};T)
  &\simeq&
   {}-N_f \tilde{I}_{2}(T) + N_f \frac{a}{M_\rho^2}\tilde{I}_{4}(T)
   {}- N_f\,a \sqrt{\frac{M_\rho}{8\pi^3}}\,e^{-M_\rho/T}T^{3/2},
\label{t-low-T}\\
 \mbox{Re}\overline{\Pi}_\perp^s (p_0 = \bar{p}+i\epsilon,\bar{p};T)
  &\simeq&
   {}-N_f \tilde{I}_{2}(T) - N_f \frac{a}{M_\rho^2}\tilde{I}_{4}(T)
   {}+ N_f\,a \sqrt{\frac{M_\rho}{8\pi^3}}\,e^{-M_\rho/T}T^{3/2}. 
\label{s-low-T}
\end{eqnarray}
By using Eqs.~(\ref{t-low-T}) and (\ref{s-low-T}) and neglecting the
terms proportional to the suppression factor $e^{-M_\rho / T}$,
the pion velocity is expressed as
\begin{eqnarray}
 v_\pi^2 (\bar{p};T)
 &\simeq& 1 - N_f \frac{2a}{F_\pi^2 M_\rho^2}\tilde{I}_4 (T)
\nonumber\\
 &=& 1 - \frac{N_f}{15}\,a \pi^2 \frac{T^4}{F_\pi^2 M_\rho^2}
 \, <  1\,.
\end{eqnarray}
This shows that the pion velocity is smaller than the speed of light
already at one-loop level in the HLS due to the $\rho$-loop effect.
This is different from the result obtained in the ordinary
ChPT including only the pion at one-loop [see for example, 
Ref.~\cite{PT:96} and references therein].
Generally, the longitudinal $\rho$ contribution to $\overline{\Pi}_\perp^t$
expressed by $\overline{B}^t$ in Eq.~(\ref{Pit A on-shell}) differs from
that to $\overline{\Pi}_\perp^s$ by $\overline{B}^s$ in Eq.~(\ref{Pits A
on-shell}), which implies that, below the critical temperature,
there always exists a deviation of the pion velocity from
the speed of light due to the longitudinal $\rho$-loop effect:
\begin{equation}
 v_\pi^2 (\bar{p};T) < 1 \qquad \mbox{for} \quad 0 < T < T_c\,.
\label{v_pi<1}
\end{equation}

Next, we study the temporal and spatial pion decay constants in hot matter
defined by Eqs.~(\ref{fpit def}) and (\ref{fpis def}).
The imaginary parts of $\overline{\Pi}_\perp^t$ and $\overline{\Pi}_\perp^s$
in the low temperature region
are given by
\begin{eqnarray}
 \mbox{Im}\overline{\Pi}_\perp^t (p_0 = \bar{p}+i\epsilon,\bar{p};M_\rho,0;T)
 &\stackrel{\bar{p}\ll T}{\simeq}&
  \frac{N_f}{4}\,a\,\mbox{Im}\overline{B}^t (p_0 = \bar{p}+i\epsilon,\bar{p};
   M_\rho,0;T) \nonumber\\
 &\stackrel{T \ll M_\rho}{\simeq}&
  \frac{N_f}{8\pi}\,a\,M_\rho^2 e^{-M_\rho/T}, \\
 \mbox{Im}\overline{\Pi}_\perp^s (p_0 = \bar{p}+i\epsilon,\bar{p};M_\rho,0;T)
 &\stackrel{\bar{p}\ll T}{\simeq}&
  \frac{N_f}{4}\,a\,\mbox{Im}\overline{B}^s (p_0 = \bar{p}+i\epsilon,\bar{p};
   M_\rho,0;T) \nonumber\\
 &\stackrel{T \ll M_\rho}{\simeq}&
  {}- \frac{N_f}{8\pi}\,a\,M_\rho^2 e^{-M_\rho/T}. 
\end{eqnarray}
Thus the contributions from the imaginary parts 
$\mbox{Im}\overline{\Pi}_\perp^{t,s}$ are small because of the suppression
factor $e^{-M_\rho/T}$.
From Eqs.~(\ref{t-low-T}) and (\ref{s-low-T}) with
$\tilde{I}_{2}(T)=T^2/12$ and $\tilde{I}_{4}(T)=\pi^2T^4/30$,
we obtain the following results for the real parts of 
$f_\pi^t$ and $f_\pi^s$:
\begin{eqnarray}
 \bigl[ \mbox{Re}f_\pi^t \bigr] \tilde{F}
 &\simeq& F_\pi^2 - N_f \Biggl[ \,\frac{T^2}{12} - 
       \frac{a\,\pi^2}{30 M_\rho^2}T^4 + 
       a \sqrt{\frac{M_\rho}{8\pi^3}}\,e^{-M_\rho/T}T^{3/2} \,\Biggr],
\nonumber\\
 \bigl[ \mbox{Re}f_\pi^s \bigr] \tilde{F}
 &\simeq& F_\pi^2 - N_f \Biggl[ \,\frac{T^2}{12} + 
       \frac{a\,\pi^2}{30 M_\rho^2}T^4 -
       a \sqrt{\frac{M_\rho}{8\pi^3}}\,e^{-M_\rho/T}T^{3/2} \,\Biggr].
\label{fpi-ts low-T}
\end{eqnarray}
We note that $\tilde{F}=\mbox{Re}f_\pi^t$ as shown in 
Eq.~(\ref{wave-func-ren-const}).
Then, neglecting the terms suppressed by $e^{-M_\rho / T}$, 
we obtain the difference between $(f_\pi^t)^2$ and $f_\pi^t f_\pi^s$ as
\begin{equation}
 (f_\pi^t)^2 - f_\pi^t f_\pi^s \,
 \simeq \, \frac{N_f}{15}\,a \pi^2 \frac{T^4}{M_\rho^2} \, > 0.
\label{fpit fpits diff}
\end{equation}
This implies that the contribution from the $\rho$-loop 
(the second and third terms in the brackets)
generates a difference between the temporal and spatial pion decay constants
in the low temperature region.
Similarly, at general temperature below $T_c$,
there exists a difference between $f_\pi^t \tilde{F}$ and $f_\pi^s \tilde{F}$
due to the $\rho$-loop effects: 
$\bigl[\mbox{Re}f_\pi^t \bigr] \tilde{F} > 
 \bigl[\mbox{Re}f_\pi^s \bigr] \tilde{F}$.
Since we can always take $\tilde{F}$ to be positive,
we find that $\mbox{Re}f_\pi^t$ is larger than 
$\mbox{Re}f_\pi^s$:
\begin{equation}
 \mbox{Re}f_\pi^t (\bar{p};T) > \mbox{Re}f_\pi^s (\bar{p};T) 
 \qquad \mbox{for} \quad 0 < T < T_c\,.
\label{t>s}
\end{equation}
This result is consistent with Eq.~(\ref{v_pi<1})
because $v_\pi^2 - 1 = (\tilde{F}/F_\pi^2)
[\mbox{Re}f_\pi^s - \mbox{Re}f_\pi^t]$ by definition.
The difference between 
$(f_\pi^t)^2$ and $f_\pi^t f_\pi^s$ shown in 
Eq.~(\ref{fpit fpits diff})
is tiny, and the hadronic thermal corrections to them are
dominated by the first term ($T^2/12$) in the bracket 
in Eq.~(\ref{fpi-ts low-T}).
Then, in the very low temperature region,
the above expressions are further reduced to
\begin{equation}
 f_\pi^2 
  = \bigl( f_\pi^t \bigr)\bigl( f_\pi^s \bigr)
  \sim \bigl( f_\pi^t \bigr)^2
   \sim  {F_\pi}^2 - \frac{N_f}{12}T^2, 
                 \label{eq:f_piGL}
\end{equation}
which is consistent with the result obtained in Ref.~\cite{GL}.


\subsection{Vector meson dominance}
\label{ssec:VDHM}

As shown in section~\ref{sec:HLS},
the HLS theory can reproduce the vector dominance (VD)
of the electromagnetic form factor of pion,
which is phenomenologically successful in mesonic system.
In this subsection,
we study the validity of the VD in hot matter.

\begin{figure}
 \begin{center}
  \includegraphics[width = 12cm]{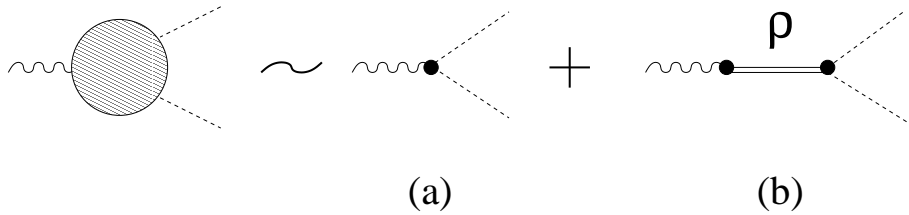}
 \end{center}
 \caption{Leading contributions to the $\gamma\pi\pi$ interaction.}
 \label{fig:VD}
\end{figure}

We first study the direct $\gamma\pi\pi$ interaction at zero
temperature. 
We expand the Lagrangian (\ref{eq:L(2)}) in terms of 
the $\pi$ field with taking the unitary gauge of the 
HLS ($\sigma = 0$) to obtain
\begin{eqnarray}
{\cal{L}}_{(2)} 
&=&
\mbox{tr} \left[ \partial_\mu \pi \partial^\mu \pi \right]
+
a g^2 F_\pi^2 \, \mbox{tr} \left[ \rho_\mu \rho^\mu \right]
+ 2 i \left( \frac{1}{2} a g \right) \, \mbox{tr} 
\left[ \rho^\mu \left[ \partial_\mu \pi \,,\, \pi \right] \right]
\nonumber\\
&&
{}- 2 \left( a g F_\pi^2 \right)
\,\mbox{tr} \left[ \rho_\mu {\cal V}^\mu \right] 
{}+ 2i \left( 1 - \frac{a}{2} \right) \mbox{tr} 
\left[ 
  {\cal V}^\mu \left[ \partial_\mu \pi \,,\, \pi \right] 
\right]
+ \cdots
\ ,
\label{Lag:expand-2}
\end{eqnarray}
where
the vector meson field $\rho_\mu$ is introduced by
\begin{equation}
V_\mu = g \rho_\mu \ ,
\end{equation}
and 
vector external gauge field ${\cal V}_\mu$ is defined as
\begin{equation}
{\cal V}_\mu \equiv \frac{1}{2} \left(
  {\cal R}_\mu + {\cal L}_\mu
\right)
\ .
\end{equation}
At the leading order of the derivative expansion in the HLS,
there are two contributions shown in Fig.~\ref{fig:VD},
i.e., the direct $\gamma\pi\pi$ interaction [Fig.~\ref{fig:VD} (a)]
and the $\rho$-mediated interaction [Fig.~\ref{fig:VD} (b)].
The form of the direct $\gamma\pi\pi$ interaction is easily read from
Eq.~(\ref{Lag:expand-2}) as
\begin{equation}
 \Gamma_{\gamma\pi\pi{\rm(tree)}}^{{\rm (a)} \mu} 
  = e(q - k)^\mu \Bigl( 1-\frac{a}{2} \Bigr)\ ,
            \label{eq:gpp}
\end{equation}
where 
$e$ is the electromagnetic coupling constant and
$q$ and $k$ denote outgoing momenta of the pions. 
This shows that,
for the parameter choice $a=2$,
the direct $\gamma\pi\pi$ coupling vanishes
(the VD of the electromagnetic form factor of the pion).

At the next order there exist quantum corrections.
In the background field gauge adopted in the present analysis the
background fields $\overline{\cal A}_\mu$ and 
$\overline{\cal V}_\mu$ include the 
photon field $A_\mu$ and the background pion field $\bar{\pi}$ as
\begin{eqnarray}
\overline{\cal A}_\mu &=&
  \frac{1}{F_\pi(0)}\partial_\mu \bar{\pi} 
  + \frac{i\,e}{F_\pi(0)} A_\mu \left[ Q\,,\, \bar{\pi} \right]
  + \cdots \ ,
\nonumber\\
\overline{\cal V}_\mu &=&
  e \, Q A_\mu 
  - \frac{i}{2F_\pi^2(0)}
    \left[ \partial_\mu \bar{\pi} \,,\, \bar{\pi} \right]
  + \cdots \ ,
\label{expand V A}
\end{eqnarray}
where 
$F_\pi(0)$ is the on-shell pion decay constant
(residue of the pion pole) in order to identify the field $\bar{\pi}$ with the
on-shell pion field, and
the charge matrix $Q$ is given by
\begin{equation}
Q = 
\left(\begin{array}{ccc}
2/3 &      &  \\
    & -1/3 &  \\
    &      & -1/3
\end{array}\right)
\ .
\label{charge}
\end{equation}
The direct $\gamma\pi\pi$ interaction including 
the next order correction is determined from the 
two-point functions of
$\overline{\cal V}_\mu$-$\overline{\cal V}_\nu$ and 
$\overline{\cal A}_\mu$-$\overline{\cal A}_\nu$
and three-point function of 
$\overline{\cal V}_\mu$-$\overline{\cal A}_\alpha$-$%
\overline{\cal A}_\beta$.
We can easily show that contribution from the
three-point function vanishes in the low energy limit as follows:
Let $\Gamma_{\overline{\cal V}\overline{\cal A}\overline{\cal A}}
  ^{\mu\alpha\beta}$
denote the 
$\overline{\cal V}_\mu$-$\overline{\cal A}_\alpha$-$%
\overline{\cal A}_\beta$
three-point function.
Then the direct $\gamma\pi\pi$ coupling 
derived from this three-point function
is proportional to 
$q_{\alpha} k_{\beta} 
 \Gamma_{\overline{\cal V}\overline{\cal A}\overline{\cal A}}
  ^{\mu\alpha\beta}$.
Since the legs $\alpha$ and $\beta$ of
$\Gamma_{\overline{\cal V}\overline{\cal A}\overline{\cal A}}
  ^{\mu\alpha\beta}$ are carried by $q$ or $k$, 
$q_{\alpha} k_{\beta} 
 \Gamma_{\overline{\cal V}\overline{\cal A}\overline{\cal A}}
  ^{\mu\alpha\beta}$ is generally proportional to two of
$q^2$, $k^2$ and $q\cdot k$.
Since the loop integral does not generate any massless poles,
this implies that
$q_{\alpha} k_{\beta} 
 \Gamma_{\overline{\cal V}\overline{\cal A}\overline{\cal A}}
  ^{\mu\alpha\beta}$
vanishes in the low energy limit
$q^2 = k^2 = q \cdot k = 0$.
Thus, the direct $\gamma\pi\pi$ interaction in the low energy limit
can be read from the two-point functions of
$\overline{\cal V}_\mu$-$\overline{\cal V}_\nu$ and 
$\overline{\cal A}_\mu$-$\overline{\cal A}_\nu$ as
\begin{equation}
\Gamma_{\gamma\pi\pi}^\mu
=  e \frac{1}{F_\pi^2(0)}
  \left[ 
    q_\nu \Pi_\perp^{\mu\nu}(q) - k_\nu \Pi_\perp^{\mu\nu}(k)
    - \frac{1}{2} (q - k)_\nu \Pi_\parallel^{\mu\nu}(p)
  \right]
\ ,
\label{gpp T0 0}
\end{equation}
where $p_\nu = (q+k)_\nu$ is the photon momentum.
Substituting the decomposition of the two-point function given in
Eq.~(\ref{decomp T0}) and taking the low-energy limit
$q^2=k^2=p^2=0$, we obtain
\begin{equation}
\Gamma_{\gamma\pi\pi}^\mu
=  e (q-k)^\mu 
  \left[ 
    1
    - \frac{1}{2} \frac{\Pi_\parallel^{{\rm (vac)}S}(p^2=0)}{F_\pi^2(0)}
  \right]
\ ,
\label{gpp T0}
\end{equation}
where we used $\Pi_\perp^{{\rm (vac)}S}(q^2=0) = F_\pi^2(0)$.
Comparing the above expression with the one in Eq.~(\ref{eq:gpp}),
we define the parameter $a(0)$ at one-loop level as
\begin{equation}
 a(0) = \frac{\Pi_\parallel^{{\rm (vac)}S} (p^2=0)}
          {F_\pi^2(0)}\ . \label{eq:Defa}
\end{equation}
We note that, in Ref.~\cite{HY:VM},
$a(0)$ is defined by the ratio $F_\sigma^2(M_\rho)/F_\pi^2(0)$
by neglecting the finite renormalization effect
which depends on the details of the renormalization condition.
While
the above $a(0)$ in Eq.~(\ref{eq:Defa}) 
defined from the direct $\gamma\pi\pi$ interaction
is equivalent to the one 
used in Ref.~\cite{HKRS:SUS}.
In the present renormalization condition (\ref{Fs ren cond}),
$\Pi_\parallel^{{\rm (vac)}S}(p^2=0)$
is given by
\begin{equation}
\Pi_\parallel^{{\rm (vac)}S} (p^2=0) 
=
F_\sigma^2(M_\rho) +
\frac{N_f}{(4\pi)^2} M_\rho^2
\bigl( 2 - \sqrt{3}\tan^{-1}\sqrt{3} \bigr)
\ .
\end{equation}
By adding this, the parameter $a(0)$ is expressed as
\begin{equation}
  a(0) =
  \frac{F_\sigma^2(M_\rho)}{ F_\pi^2(0) }
  {}+\frac{N_f}{(4\pi)^2}\frac{M_\rho^2}{F_\pi^2(0)}
     \bigl( 2 - \sqrt{3}\tan^{-1}\sqrt{3} \bigr) 
\ .
\label{a0 expression}
\end{equation}
Using $M_\rho=771.1\,\mbox{MeV}$,
$F_\pi(\mu = 0)=86.4\,\mbox{MeV}$ estimated in the chiral 
limit~\cite{Gas:84,Gas:85a,Gas:85b}~\footnote{
 In Ref.~\cite{HY:PRep}, it was assumed that 
 the scale dependent $F_\pi (\mu)$
 agrees with the scale-independent parameter $F_\pi$ 
 in the ChPT at $\mu = 0$ to obtain 
 $F_\pi(\mu=0)=86.4\pm9.7\,$MeV.
 Here, we simply
 use this value to get a rough estimate of the value of the
 parameter $a(0)$ as done in Ref.~\cite{HY:PRep}.
} and
$F_\sigma^2(M_\rho)/F_\pi^2(0) = 2.03$ obtained
through the Wilsonian matching 
for $\Lambda_{\rm QCD}=400\,\mbox{MeV}$ and 
the matching scale $\Lambda=1.1\,\mbox{GeV}$
in Ref.~\cite{HY:PRep},
we estimate the value of $a(0)$ at zero temperature as
\begin{equation}
 a(0) \simeq 2.31 \ .
\label{a0 val}
\end{equation}
This implies that the VD is well satisfied at $T=0$ even though
the value of the parameter $a$ at the scale $M_\rho$ is 
close to one~\cite{Harada:2001rf}.

Now, let us study
the direct $\gamma\pi\pi$ coupling in hot matter.
In general,
the electric mode and the magnetic mode of the photon
couple to the pions differently in hot matter,
so that there are two parameters as extensions of the parameter $a$.
Similarly to the one obtained at $T=0$ in Eq.~(\ref{gpp T0 0}),
in the low energy limit
the direct $\gamma\pi\pi$ interaction derived from
$\overline{\cal A}_\mu$-$\overline{\cal A}_\nu$ and
$\overline{\cal V}_\mu$-$\overline{\cal V}_\nu$ two-point
functions is expressed as
\begin{eqnarray}
&&
\Gamma_{\gamma\pi\pi}^{\mu}(p;q,k)
\nonumber\\
&& \quad
=
\frac{1}{\widetilde{F}(\bar{q};T)\widetilde{F}(\bar{k};T)}
\Biggl[
  q_\nu \, 
  \Pi_\perp^{\mu\nu}(q_0,\bar{q};T)
  -
  k_\nu \, 
  \Pi_\perp^{\mu\nu}(k_0,\bar{k};T)
\nonumber\\
&& \qquad\qquad\qquad\qquad\qquad
  {}- \frac{1}{2} (q-k)_\nu \,
    \Pi_\parallel^{\mu\nu}(p_0,\bar{p};T)
\Biggr]
\ ,
\label{direct g pi pi 0}
\end{eqnarray}
where $q$ and $k$ denote outgoing
momenta of the pions and $p = (q+k)$ is the photon momentum.
$\widetilde{F}$ is the wave function renormalization of 
the background 
$\bar{\pi}$ field given in Eq.~(\ref{wave-func-ren-const}).
Note that each pion is on its mass shell, so that
$q_0 = v_\pi(\bar{q}) \bar{q}$ and 
$k_0 = v_\pi(\bar{k}) \bar{k}$.
To define extensions of the parameter $a$, 
we consider the soft limit of the photon:
$p_0 \to 0$ and $\bar{p} \to 0$.~\footnote{%
  Note that we take the $p_0 \rightarrow 0$ limit first and then take
  the $\bar{p}\rightarrow0$ limit for definiteness.
  This is natural since the form factor in the vacuum
  is defined for space-like momentum of the photon.
  }
Then the pion momenta become
\begin{equation}
 q_0 = - k_0 \ , \quad  \bar{q}=- \bar{k} \ .
\end{equation}
Note that while
only two components $\Pi_\perp^t$ and $\Pi_\perp^s$ appear
in $q_\nu \, \Pi_\perp^{\mu\nu}$
or $k_\nu \, \Pi_\perp^{\mu\nu}$,
$(q-k)_\nu \, \Pi_\parallel^{\mu\nu}$
includes all four components $\Pi_\parallel^t$,
$\Pi_\parallel^s$, $\Pi_\parallel^L$ and $\Pi_\parallel^T$.
Since the tree part of $\Pi_\parallel^L$ and $\Pi_\parallel^T$ is
$-2z_2\,p^2$ which vanishes at $p^2=0$, 
it is natural to use
only $\Pi_\parallel^t$ and $\Pi_\parallel^s$ to define the extensions
of the parameter $a$.
By including these two parts only, the temporal and
the spatial components of 
$\Gamma_{\gamma\pi\pi}^\mu$ are given by
\begin{eqnarray}
 \Gamma_{\gamma\pi\pi}^0(0;q,-q)
 &=& 
  \frac{2q_0}{\widetilde{F}^2(\bar{q};T)}
   \Bigl[ \Pi_\perp^t(q_0,\bar{q};T)-
                \frac{1}{2}\Pi_\parallel^t(0,0;T)
         \Bigr], \nonumber\\
 \Gamma_{\gamma\pi\pi}^i(0;q,-q)
 &=& 
  \frac{-2q_i}{\widetilde{F}^2(\bar{q};T)}
    \Bigl[ \Pi_\perp^s(q_0,\bar{q};T)-
                \frac{1}{2}\Pi_\parallel^s(0,0;T)
         \Bigr].
\end{eqnarray}
Thus we define $a^t(T)$ and $a^s(T)$ as
\begin{eqnarray}
 a^t(\bar{q};T)
 &=& \frac{\Pi_\parallel^t(0,0;T)}
         {\Pi_\perp^t(q_0,\bar{q};T)} \ , \nonumber\\
 a^s(\bar{q};T)
 &=& \frac{\Pi_\parallel^s(0,0;T)}
         {\Pi_\perp^s(q_0,\bar{q};T)} \ .
\end{eqnarray}
Here we should stress again that the pion momentum $q_\mu$ is
on mass-shell: $q_0 = v_\pi(\bar{q}) \bar{q}$.

In the HLS at one-loop level the above
$a^t(\bar{q};T)$ and $a^s(\bar{q};T)$
are expressed as
\begin{eqnarray}
 a^t (\bar{q};T)
 &=& a(0)\Biggl[ 1 + \frac{\overline{\Pi}_\parallel^t (0,0;T) -
     a(0)\overline{\Pi}_\perp^t (\bar{q},\bar{q};T)}
     {a(0) F_\pi^2(0;T)} \Biggr], \label{at expression}
\\
 a^s (\bar{q};T)
 &=& a(0)\Biggl[ 1 + \frac{\overline{\Pi}_\parallel^s (0,0;T) -
     a(0)\overline{\Pi}_\perp^s (\bar{q},\bar{q};T)}
     {a(0) F_\pi^2(0;T)} \Biggr], \label{as expression}
\end{eqnarray}
where $a(0)$ is defined in Eq.~(\ref{eq:Defa}).
{}From Eq.~(\ref{Pi bar V ts}) together with
Eq.~(\ref{B0B_static}) we obtain 
$\overline{\Pi }^{t}_\parallel$ and $\overline{\Pi }^{s}_\parallel$
in Eqs.~(\ref{at expression}) and (\ref{as expression}) as
\begin{eqnarray}
 \overline{\Pi }^{t}_\parallel(0,0;T)
 &=& \overline{\Pi }^{s}_\parallel(0,0;T) \nonumber\\
 &=& - \frac{N_f}{4}
   \Biggl[ 
     a^2 \tilde{I}_{2}(T) 
     - \tilde{J}_{1}^2(M_\rho;T) 
     + 2 \tilde{J}_{-1}^0(M_\rho;T) 
   \Biggr].
\label{PiVts zero limits}
\end{eqnarray}

Let us study the temperature dependence of the parameters
$a^t(\bar{q};T)$ and $a^s(\bar{q};T)$ in the low temperature
region.
At low temperature $T \ll M_\rho$
the functions $\tilde{J}_{1}^2(M_\rho;T)$ and
$\tilde{J}_{-1}^0(M_\rho;T)$ are suppressed by 
$e^{-M_\rho/T}$.
Then the dominant contribution is given by 
$\tilde{I}_{2}(T)=T^2/12$.
Combining this with Eq.~(\ref{t-low-T}) and (\ref{s-low-T}), we obtain
\begin{equation}
 a^t \simeq a^s \simeq
         a(0) \left[
         1 + \frac{N_f}{12} \left( 1 - \frac{a^2}{4 a(0)} \right)
         \frac{T^2}{F_\pi^2(0;T)} \right],
  \label{at as form low T}
 \end{equation}
where $a$ is the parameter renormalized at the scale $\mu=M_\rho$,
while $a(0)$ is defined in Eq.~(\ref{eq:Defa}).
We expect that the temperature dependences of the parameters are small 
in the low temperature region, so that we use the values of parameters at
$T=0$ to estimate the hadronic thermal corrections to $a^{t,s}(T)$.
By using $F_\pi(0)=86.4\,\mbox{MeV}$,
$a(0) \simeq 2.31$ given in Eq.~(\ref{a0 val}) and
$a(M_\rho) = 1.38$ 
obtained through the Wilsonian matching for
$(\Lambda_{\rm QCD}\,,\,\Lambda) = (0.4\,,\,1.1)\,\mbox{GeV}$
and $N_f = 3$~\cite{HY:PRep},
$a^t$ and $a^s$ in Eq.~(\ref{at as form low T}) are evaluated as
\begin{equation}
a^t \simeq a^s \simeq
a(0) \left[
  1 + 0.066 \left( \frac{T}{50\,\mbox{MeV}} \right)^2 
\right]
\ .
\end{equation}
This implies that the parameters $a^t$ and $a^s$ increase with
temperature in the low temperature region.
However, since the correction is small, we conclude that the
vector dominance is well satisfied in the low temperature region.


\section{Lorentz Non-invariance at Bare Level}
\label{sec:LNBL}

As stressed in section~\ref{sec:AHDM},
the Lorentz non-invariance appears in the bare HLS theory
as a result of including the intrinsic temperature dependence.
Once the temperature dependence of the bare parameters is determined 
through the matching with QCD mentioned above,
the parameters 
appearing in the hadronic corrections pick up the intrinsic 
thermal effects through the RGEs.
Then the HLS Lagrangian in hot
and/or dense matter is generically Lorentz non-invariant. 
Its explicit form was presented in Ref.~\cite{HKR:VM}. 
In this section, we show the HLS Lagrangian at leading order
including the effects of
Lorentz non-invariance.

Two matrix valued
variables $\xi_L(x)$ and $\xi_R(x)$
are now parameterized as
\footnote{
 The wave function renormalization constant of the pion field
 is given by the temporal component of the pion decay constant
 ~\cite{MOW}.
 Thus we normalize $\pi$ and $\sigma$ by $F_\pi^t$ and $F_\sigma^t$
 respectively.
}
\begin{equation}
\qquad
  \xi_{L,R}(x)=e^{i\sigma (x)/{F_\sigma^t}}
     e^{\mp i\pi (x)/{F_\pi^t}},
\end{equation}
where $F_\pi^t$ and $F_\sigma^t$ denote
the temporal components of the decay constant of
$\pi$ and $\sigma$, respectively.
The leading order Lagrangian with
Lorentz non-invariance can be written as~\cite{HKR:VM}
\begin{eqnarray}
&&
{\cal L}
=
\biggl[
  (F_{\pi}^t)^2 u_\mu u_\nu
  {}+ F_{\pi}^t F_{\pi}^s
    \left( g_{\mu\nu} - u_\mu u_\nu \right)
\biggr]
\mbox{tr}
\left[
  \hat{\alpha}_\perp^\mu \hat{\alpha}_\perp^\nu
\right] \nonumber\\
&&\quad{}+
\biggl[
  (F_{\sigma}^t)^2 u_\mu u_\nu
  {}+  F_{\sigma}^t F_{\sigma}^s
    \left( g_{\mu\nu} - u_\mu u_\nu \right)
\biggr]
\mbox{tr}
\left[
  \hat{\alpha}_\parallel^\mu \hat{\alpha}_\parallel^\nu
\right]
\nonumber\\
&&\quad
{} +
\Biggl[
  - \frac{1}{ g_{L}^2 } \, u_\mu u_\alpha g_{\nu\beta}
  {}- \frac{1}{ 2 g_{T}^2 }
  \left(
    g_{\mu\alpha} g_{\nu\beta}
   - 2 u_\mu u_\alpha g_{\nu\beta}
  \right)
\Biggr]
\mbox{tr}
\left[ V^{\mu\nu} V^{\alpha\beta} \right]
\ .
\label{Lag}
\end{eqnarray}
Here $F_{\pi}^s$ denote the spatial pion decay constant and similarly
$F_{\sigma}^s$ for the $\sigma$. The rest frame
of the medium is specified by $u^\mu = (1,\vec{0})$ and
$V_{\mu\nu}$ is the field strength of $V_\mu$. $g_{L}$ and $g_{T}$
correspond in medium to the HLS gauge coupling $g$. The parametric
$\pi$ and $\sigma$ velocities are defined by~\cite{PT:96}
\begin{equation}
\qquad
 V_\pi^2 = {F_\pi^s}/{F_\pi^t}, \qquad
 V_\sigma^2 = {F_\sigma^s}/{F_\sigma^t}.
\end{equation}

The axial-vector and vector current correlators at bare level
are constructed in terms of bare parameters and
are divided into the longitudinal and transverse components:
\begin{equation}
\qquad
 G_{A,V}^{\mu\nu} = P_L^{\mu\nu}G_{A,V}^L + P_T^{\mu\nu}G_{A,V}^T,
\end{equation}
where $P_{L,T}^{\mu\nu}$ are the longitudinal and transverse 
projection operators, respectively.
The axial-vector current correlator 
in the HLS around the matching scale $\Lambda$ is well described 
by the following forms with the bare parameters~\cite{HKR:VM,HKRS:SUS}:
\begin{eqnarray}
 &&
 G_{A{\rm(HLS)}}^L(p_0,\bar{p})
 =
 \frac{ p^2 F_{\pi,{\rm bare}}^t F_{\pi,_{\rm bare}}^s }{
    - [ p_0^2 - V_{\pi,{\rm bare}}^2 \bar{p}^2 ] }
 -2p^2z^L_{2,\rm bare},
\label{gal} \nonumber\\
 &&
 G_{A{\rm(HLS)}}^T(p_0,\bar{p})
 =
 -F_{\pi,\rm bare}^tF_{\pi,\rm bare}^s
 {} - 2 \left(
    p_0^2 z_{2,{\rm bare}}^L  - \bar{p}^2 z_{2,{\rm bare}}^T
  \right)
\ ,
\label{gat}
\end{eqnarray}
where $z_{2,{\rm bare}}^{L}$ and $z_{2,{\rm bare}}^{T}$ are
the coefficients of the higher order terms,
and $V_{\pi,{\rm bare}}$ is the bare pion velocity related to
$F_{\pi,{\rm bare}}^t$ and $F_{\pi,{\rm bare}}^s$ as
\begin{equation}
\qquad
V_{\pi,{\rm bare}}^2
= \frac{F_{\pi, {\rm bare}}^s}{F_{\pi, {\rm bare}}^t}
\ .
\end{equation}
Similarly, two components of 
the vector current correlator have the following forms:
\begin{eqnarray}
&&G_{V{\rm(HLS)}}^L(p_0,\bar{p})
\nonumber\\
&&=
\frac{
  p^2 \, F_{\sigma,{\rm bare}}^t F_{\sigma,{\rm bare}}^s
  \left( 1 - 2 g_{L,{\rm bare}}^2 z_{3,{\rm bare}}^L \right)}
 { -
  \left[
    p_0^2 - V_{\sigma,{\rm bare}}^2 \bar{p}^2
    - M_{\rho,{\rm bare}}^2
  \right] } 
{}- 2 p^2 z_{1,{\rm bare}}^L
+ {\cal O}(p^4)
\ ,
\nonumber\\
&&G_{V{\rm(HLS)}}^T(p_0,\bar{p})
\nonumber\\
&&=
\frac{
  F_{\sigma,{\rm bare}}^t F_{\sigma,{\rm bare}}^s}
 { - \left[
    p_0^2 - V_{T,{\rm bare}}^2 \bar{p}^2 - M_{\rho,{\rm bare}}^2
  \right] }
\nonumber\\
&&\qquad\qquad
 \times
  \Bigl[
    p_0^2 \left(1 - 2 g_{L,{\rm bare}}^2 z_{3,{\rm bare}}^L\right)
   {} - V_{T,{\rm bare}}^2 \bar{p}^2
    \left(1 - 2 g_{T,{\rm bare}}^2 z_{3,{\rm bare}}^T\right)
  \Bigr]
\nonumber\\
&&\quad
{}- 2 \left( p_0^2 z_{1,{\rm bare}}^L - \bar{p}^2 z_{1,{\rm bare}}^T \right)
+ {\cal O}(p^4)
\ ,
\end{eqnarray}
where $z_{1,2,{\rm bare}}^L$ and $z_{1,2,{\rm bare}}^T$ denote
the coefficients of the higher order terms.
In the above expressions, the bare vector meson mass in the rest frame, 
$M_{\rho,{\rm bare}}$, is
\begin{equation}
\qquad
 M_{\rho,{\rm bare}}^2
 \equiv g_{L,{\rm bare}}^2 F_{\sigma,{\rm bare}}^t
        F_{\sigma,{\rm bare}}^s \ .
\end{equation}
We define the bare parameters $a^t_{\rm bare}$ and $a^s_{\rm bare}$ as
\begin{equation}
\quad
  a^t_{\rm bare}
  = \Biggl( \frac{F_{\sigma,{\rm bare}}^t}
    {F_{\pi,{\rm bare}}^t} \Biggr)^2, \quad
  a^s_{\rm bare}
  = \Biggl( \frac{F_{\sigma,{\rm bare}}^s}
    {F_{\pi,{\rm bare}}^s} \Biggr)^2,
\end{equation}
and the bare $\sigma$ and transverse $\rho$ velocities as
\begin{equation}
\quad
 V_{\sigma,{\rm bare}}^2
 = \frac{F_{\sigma,{\rm bare}}^s}{F_{\sigma,{\rm bare}}^t}, \quad
 V_{T,{\rm bare}}^2
 = \frac{g_{L,{\rm bare}}^2}{g_{T,{\rm bare}}^2}.
\end{equation}

\newpage

\chapter{Wilsonian Matching at Finite Temperature}
\label{ch:WMFT}

In order to fix full temperature dependences of physical quantities,
parameters of the EFT should be determined through the matching to QCD.
In this chapter, we perform the matching in the Wilsonian sense 
discussed in section~\ref{sec:MWS}.
The bare parameters of HLS theory at zero temperature were
originally determined in Ref.~\cite{HY:WM},
where they matched the bare HLS theory to the operator product
expansion (OPE).
The Wilsonian matching well describes the real world
(for details, see Ref.~\cite{HY:PRep}).
Applying this scheme to QCD in hot/dense matter,
we obtain the bare parameters in terms of the OPE variables
like expectation value of an operator.
The intrinsic thermal effects of bare parameters are caused by
the temperature dependences of such condensations at a matching scale,
which are evaluated in the {\it thermal} vacuum.

In this chapter,
we briefly review the Wilsonian matching proposed at zero temperature.
Next we extend the Wilsonian matching
to the version at non-zero temperature in order to incorporate
the intrinsic thermal effect into the bare parameters of the HLS
Lagrangian following 
Refs.~\cite{HS:VM,HKRS:SUS,HS:VVD,Sasaki:2003qj,HKRS:PV}.
There we also discuss the effect of Lorentz symmetry violation
at bare level.

\section{Wilsonian Matching Conditions at $T=0$}
\label{ssec:WMCT0}

The Wilsonian matching proposed in Ref.~\cite{HY:WM}
is done by matching
the axial-vector and vector current correlators derived from the
HLS with those by the operator product expansion (OPE) in
QCD at the matching scale $\Lambda$.~\footnote{%
  For the validity of the expansion in the HLS the
  matching scale $\Lambda$ must be smaller than the chiral symmetry
  breaking scale $\Lambda_\chi$
  as we stressed in chapter~\ref{ch:EFTRG}.
}
The axial-vector and vector current correlators in the OPE
up until ${\cal O}(1/Q^6)$
at $T=0$ are expressed as~\cite{SVZ}
\begin{eqnarray}
G_A^{\rm(QCD)}(Q^2) &=& \frac{1}{8\pi^2}
\left( \frac{N_c}{3} \right)
\Biggl[
  - \left(
      1 +  \frac{3(N_c^2-1)}{8N_c}\, \frac{\alpha_s}{\pi}
  \right) \ln \frac{Q^2}{\mu^2}
\nonumber\\
&& \qquad
  {}+ \frac{\pi^2}{N_c}
    \frac{
      \left\langle
        \frac{\alpha_s}{\pi} G_{\mu\nu} G^{\mu\nu}
      \right\rangle
    }{ Q^4 }
  {}+ \frac{\pi^3}{N_c} \frac{96(N_c^2-1)}{N_c^2}
    \left( \frac{1}{2} + \frac{1}{3N_c} \right)
    \frac{\alpha_s \left\langle \bar{q} q \right\rangle^2}{Q^6}
\Biggr]
\ ,
\label{Pi A OPE}
\\
G_V^{\rm(QCD)}(Q^2) &=& \frac{1}{8\pi^2}
\left( \frac{N_c}{3} \right)
\Biggl[
  - \left(
      1 +  \frac{3(N_c^2-1)}{8N_c}\, \frac{\alpha_s}{\pi}
  \right) \ln \frac{Q^2}{\mu^2}
\nonumber\\
&& \qquad
  {}+ \frac{\pi^2}{N_c}
    \frac{
      \left\langle
        \frac{\alpha_s}{\pi} G_{\mu\nu} G^{\mu\nu}
      \right\rangle
    }{ Q^4 }
  {}- \frac{\pi^3}{N_c} \frac{96(N_c^2-1)}{N_c^2}
    \left( \frac{1}{2} - \frac{1}{3N_c} \right)
    \frac{\alpha_s \left\langle \bar{q} q \right\rangle^2}{Q^6}
\Biggr]
\ ,
\label{Pi V OPE}
\end{eqnarray}
where $\mu$ is the renormalization scale of QCD
and we
wrote the $N_c$-dependences explicitly
(see, e.g., Ref.~\cite{Bardeen-Zakharov}).
In the HLS the same correlators are
well described by the tree contributions with including
${\cal O}(p^4)$ terms
when the momentum is around the matching scale, $Q^2 \sim \Lambda^2$:
\begin{eqnarray}
G_A^{\rm(HLS)}(Q^2) &=&
\frac{F_\pi^2(\Lambda)}{Q^2} - 2 z_2(\Lambda)
\ ,
\label{Pi A HLS}
\\
G_V^{\rm(HLS)}(Q^2) &=&
\frac{
  F_\sigma^2(\Lambda)
}{
  M_\rho^2(\Lambda) + Q^2
}
\left[ 1 - 2 g^2(\Lambda) z_3(\Lambda) \right]
- 2 z_1(\Lambda)
\ ,
\label{Pi V HLS}
\end{eqnarray}
where we defined the bare $\rho$ mass $M_\rho(\Lambda)$ as
\begin{equation}
M_\rho^2(\Lambda) \equiv g^2(\Lambda) F_\sigma^2(\Lambda)
\ .
\label{on-shell cond 5}
\end{equation}

We require that current correlators in the HLS
in Eqs.~(\ref{Pi A HLS}) and (\ref{Pi V HLS})
can be matched with those in QCD in
Eqs.~(\ref{Pi A OPE}) and (\ref{Pi V OPE}).
Of course,
this matching cannot be made for any value of $Q^2$,
since the $Q^2$-dependences of the current correlators
in the HLS are completely
different from those in the OPE:
In the HLS the derivative expansion (in {\it positive} power of $Q$)
is used, and the expressions for
the current correlators are valid in the low energy region.
The OPE, on the other hand, is an asymptotic expansion
(in {\it negative} power of $Q$), and it is
valid in the high energy region.
Since we calculate the current correlators in the HLS including the
first non-leading order [${\cal O}(p^4)$], we expect that we can match
the correlators with those in the OPE up until the first derivative
~\footnote{
 If there exists an overlapping area around a scale $\widetilde{\Lambda}$,
 we can require the matching condition at that $\widetilde{\Lambda}$.
 In fact, the Wilsonian matching at $T=0$ in three flavor QCD was shown
 to give several predictions in remarkable agreement with experiments
 ~\cite{HY:WM,HY:PRep}.
 This strongly suggests that there exists such an overlapping region.
 As discussed in Ref.~\cite{HY:PRep}, 
 $\Lambda \ll \Lambda_{\rm HLS}$ can be
 justified in the large $N_c$ limit, where $\Lambda_{\rm HLS}$ denotes
 the scale at which the HLS theory breaks down
 (see also section~\ref{sec:HLS}).
 We obtain the matching conditions in $N_c=3$ by extrapolating 
 the conditions in large $N_c$.
 As we mentioned above, the success of the Wilsonian matching at $T=0$
 with taking the matching scale as $\Lambda = 1.1\,\mbox{GeV}$
 shows that this extrapolation is valid.  
}.
Then we obtain the following Wilsonian matching
conditions~\cite{HY:WM,HY:PRep}~\footnote{%
  One might think that there appear corrections from $\rho$ and/or
  $\pi$ loops in the left-hand-sides of Eqs.~(\ref{match A}) and
  (\ref{match V}).
  However, such corrections are of higher order in the present
  counting scheme, and thus we neglect them here
  at $Q^2 \sim {\Lambda}^2$.
  In the low-energy scale
  we incorporate the loop effects into the correlators.
  }
\begin{eqnarray}
&&
   \frac{F^2_\pi (\Lambda)}{{\Lambda}^2}
  = \frac{1}{8{\pi}^2} \left( \frac{N_c}{3} \right)
  \Biggl[
    1 +
    \frac{3(N_c^2-1)}{8N_c}\, \frac{\alpha_s}{\pi}
    + \frac{2\pi^2}{N_c}
      \frac{
        \left\langle
          \frac{\alpha_s}{\pi} G_{\mu\nu} G^{\mu\nu}
        \right\rangle
      }{ \Lambda^4 }
\nonumber\\
&& \qquad\qquad\qquad\qquad
    {}+ \frac{288\pi(N_c^2-1)}{N_c^3}
      \left( \frac{1}{2} + \frac{1}{3N_c} \right)
      \frac{\alpha_s \left\langle \bar{q} q \right\rangle^2}
           {\Lambda^6}
  \Biggr]
\ ,
\label{match A}
\\
&&
   \frac{F^2_\sigma (\Lambda)}{{\Lambda}^2}
        \frac{{\Lambda}^4[1 - 2g^2(\Lambda)z_3(\Lambda)]}
         {({M_\rho}^2(\Lambda) + {\Lambda}^2)^2}
\nonumber\\
&& \qquad
= \frac{1}{8\pi^2} \left( \frac{N_c}{3} \right)
  \Biggl[
    1 +
    \frac{3(N_c^2-1)}{8N_c}\, \frac{\alpha_s}{\pi}
    + \frac{2\pi^2}{N_c}
      \frac{
        \left\langle
          \frac{\alpha_s}{\pi} G_{\mu\nu} G^{\mu\nu}
        \right\rangle
      }{ \Lambda^4 }
\nonumber\\
&& \qquad\qquad\qquad\qquad
    {}- \frac{288\pi(N_c^2-1)}{N_c^3}
      \left( \frac{1}{2} - \frac{1}{3N_c} \right)
      \frac{\alpha_s \left\langle \bar{q} q \right\rangle^2}
           {\Lambda^6}
  \Biggr]
\ ,
\label{match V}
\\
&&
   \frac{F^2_\pi (\Lambda)}{{\Lambda}^2} -
            \frac{F^2_\sigma (\Lambda)[1 -
              2g^2(\Lambda)z_3(\Lambda)]}
             {{M_\rho}^2(\Lambda) + {\Lambda}^2} -
            2[z_2(\Lambda) - z_1(\Lambda)]
\nonumber\\
&& \qquad
  =  \frac{4\pi(N_c^2-1)}{N_c^2}
  \frac{\alpha_s \left\langle \bar{q} q \right\rangle^2}{\Lambda^6}
\ .
\label{match z}
\end{eqnarray}
The above three equations (\ref{match z}), (\ref{match A}) and
(\ref{match V}) are the Wilsonian matching conditions
proposed in Ref.~\cite{HY:WM}.
They determine several bare parameters of the HLS without much
ambiguity.  Especially, the first condition (\ref{match A})
determines the ratio $F_\pi(\Lambda)/\Lambda$ directly from QCD.


\section{Wilsonian Matching Conditions at $T \neq 0$}
\label{ssec:WMCTf}

Next we consider the extension of the matching conditions at $T = 0$ 
to the analysis in hot matter.
We present the matching conditions to determine
the bare pion decay constants including the effect of Lorentz symmetry
breaking at the bare level which is caused by the intrinsic thermal effect.

\subsubsection{Case neglecting Lorentz non-invariance}

Before going to study the matching conditions taking into account
the possible Lorentz non-invariance,
we consider a naive extension of the matching to the one in hot matter:
Strictly speaking,
inclusion of intrinsic effects generates Lorentz non-invariance
in bare theory.
Then we should include
the Lorentz non-scalar operators such as
$\bar{q}\gamma_\mu D_\nu q$ into 
the form of the current correlators derived from the OPE~\cite{HKL},
which leads to a difference between the temporal and spatial
bare pion decay constants.
However, we neglect the contributions from these operators
since they give a small correction compared with 
the main term $1 + \frac{\alpha_s}{\pi}$
as discussed in Ref.~\cite{HS:VM}.
This implies that the Lorentz symmetry breaking effect in
the bare pion decay constant is small, 
$F_{\pi,\rm{bare}}^t \simeq F_{\pi,\rm{bare}}^s$~\cite{HKRS:SUS}.
In fact, we will see below that the difference between them is caused
by an existence of a  higher spin operator
in the OPE side of the Wilsonian matching condition.
Thus to a good approximation we determine the pion decay
constant at non-zero temperature through the matching 
condition at zero temperature, putting possible 
temperature dependences into the gluonic 
and quark condensates~\cite{HS:VM, HKRS:SUS}:
\begin{equation}
 \frac{F^2_\pi (\Lambda ;T)}{{\Lambda}^2} 
  = \frac{1}{8{\pi}^2}\Bigl[ 1 + \frac{\alpha _s}{\pi} +
     \frac{2{\pi}^2}{3}\frac{\langle \frac{\alpha _s}{\pi}
      G_{\mu \nu}G^{\mu \nu} \rangle_T }{{\Lambda}^4} +
     {\pi}^3 \frac{1408}{27}\frac{\alpha _s{\langle \bar{q}q
      \rangle }^2_T}{{\Lambda}^6} \Bigr]
\ .
\label{eq:WMC A}
\end{equation}
Through this condition
the temperature dependences of the quark and gluonic condensates
determine the intrinsic temperature dependences 
of the bare parameter $F_\pi(\Lambda;T)$,
which is then converted into 
those of the on-shell parameter $F_\pi(\mu=0;T)$ 
through the Wilsonian RGEs.

\subsubsection{Case taking into account Lorentz non-invariance}

Now we study the Wilsonian matching conditions for the bare
pion decay constants without Lorentz invariance.
{}From the bare Lagrangian with replacement of the parameters
by the bare ones in Eq.~(\ref{Lag}), 
the current correlator at the matching
scale is constructed as Eq.~(\ref{gat}):
\begin{eqnarray}
&& G_{A{\rm(HLS)}}^L(q_0,\bar{q}) =
  \frac{ F_{\pi,{\rm bare}}^t F_{\pi,{\rm bare}}^s }{
   - [ q_0^2 - V_{\pi,{\rm bare}}^2 \bar{q}^2 ] }
 - 2 z_{2,{\rm bare}}^L
\ ,
\nonumber\\
&&
G_{A{\rm(HLS)}}^T(q_0,\bar{q})
=
-\frac{F_{\pi,\rm bare}^tF_{\pi,\rm bare}^s}{q^2}
  - 2 \frac{ q_0^2 z_{2,{\rm bare}}^L - \bar{q}^2 z_{2,{\rm bare}}^T }
      { q^2 }
\ ,
\label{galt}
\end{eqnarray}
where
$V_{\pi,{\rm bare}}=F_{\pi,{\rm bare}}^s / F_{\pi,{\rm bare}}^t$
is the bare pion velocity.
To perform the matching, we regard $G_A^{L,T}$ as functions of 
$-q^2$ and $\bar{q}^2$ instead of $q_0$ and $\bar{q}$,
and expand $G_A^{L,T}$ in
a Taylor series around $\bar{q}=|\vec{q}|=0$ in $\bar{q}^2/(-q^2)$ as
follows:
\begin{eqnarray}
 G_A^L(-q^2,\bar{q}^2)
 = G_A^{L(0)}(-q^2) + G_A^{L(1)}(-q^2)\bar{q}^2 + \cdots,
\nonumber\\
 G_A^T(-q^2,\bar{q}^2)
 = G_A^{T(0)}(-q^2) + G_A^{T(1)}(-q^2)\bar{q}^2 + \cdots.
\end{eqnarray}
In the following, we determine the bare pion velocity $V_{\pi,{\rm
bare}}$ from $G_A^{L(0)}$ and $G_A^{L(1)}$ via the matching.

Expanding $G_A^{{\rm (HLS)}L}$ in Eq.~(\ref{galt})
in terms of $\bar{q}^2/(-q^2)$, we obtain
\begin{eqnarray}
 &&
 G_A^{{\rm (HLS)}L(0)}(-q^2)
  = \frac{F_{\pi,{\rm bare}}^t F_{\pi,{\rm bare}}^s}{-q^2}
    {}- 2z_{2,{\rm bare}}^L~,
\label{HLS-L(0)}
\\
 &&
 G_A^{{\rm (HLS)}L(1)}(-q^2)
  = \frac{F_{\pi,{\rm bare}}^t F_{\pi,{\rm bare}}^s
    (1 - V_{\pi,{\rm bare}}^2)}{(-q^2)^2}.
\label{HLS-L(1)}
\end{eqnarray}
On the other hand, the correlator $G_A^{\mu\nu}$ in the QCD sector
to be given in OPE is more involved. Our strategy goes as follows.
Since the effect of Lorentz non-invariance in medium has been more
extensively studied in dense matter, we first examine the form of
the relevant correlator in dense matter following
Refs.~\cite{FLK, LM}.
The current correlator $\tilde{G}^{\mu\nu}$ constructed from
the current defined by
\begin{equation}
 J_\mu^{(q)} = \bar{q}\gamma_\mu q, \quad
 \mbox{or} \quad
 J_{5\mu}^{(q)} = \bar{q}\gamma_5\gamma_\mu q,
\end{equation}
is given by
\begin{eqnarray}
 &&
 {\tilde G}^{\mu\nu}(q_0,\bar{q})
 = (q^\mu q^\nu - g^{\mu\nu}q^2)
   \Biggl[ -c_0 \ln |Q^2| + \sum_n \frac{1}{Q^n}A^{n,n} \Biggr]
\nonumber\\
 &&\qquad{}+
   \sum_{\tau = 2}\sum_{k = 1}
   \bigl[ -g^{\mu\nu}q^{\mu_1}q^{\mu_2} + g^{\mu\mu_1}q^\nu q^{\mu_2}
    + q^\mu q^{\mu_1}g^{\nu\mu_2} + g^{\mu\mu_1}g^{\nu\mu_2}Q^2
   \bigr] \nonumber\\
 &&\qquad\qquad \times
   q^{\mu_3}\cdots q^{\mu_{2k}}
   \frac{2^{2k}}{Q^{4k + \tau - 2}}
   A_{\mu_1 \cdots \mu_{2k}}^{2k + \tau, \tau} \nonumber\\
 &&\qquad{}+
  \sum_{\tau = 2}\sum_{k=1}\Bigl[ g^{\mu\nu} - \frac{q^\mu q^\nu}{q^2}
   \Bigr] q^{\mu_1}\cdots q^{\mu_{2k}}
   \frac{2^{2k}}{Q^{4k + \tau -2}}
   C_{\mu_1 \cdots \mu_{2k}}^{2k + \tau, \tau},
\label{acc-ope}
\end{eqnarray}
where $Q^2 = -q^2$. $\tau = d - s$ denotes the twist, and $s = 2k$
is the number of spin indices of the operator of dimension $d$.
Here
$A^{n,n}$ represents the contribution from the Lorentz invariant
operators such as
$A^{4,4} = \frac{1}{6}
  \left\langle \frac{\alpha_s}{\pi} G^2 \right\rangle_\rho$.
$A_{\mu_1 \cdots \mu_{2k}}^{2k + \tau, \tau}$  and $C_{\mu_1
\cdots \mu_{2k}}^{2k + \tau, \tau}$ are the residual Wilson
coefficient times matrix element of dimension $d$ and twist
$\tau$;
e.g., $A_{\mu_1\mu_2\mu_3\mu_4}^{6,2}$ is given by
\begin{equation}
A_{\mu_1\mu_2\mu_3\mu_4}^{6,2} = i
\left\langle
  {\mathcal ST}
  \left(
    \bar{q} \gamma_{\mu_1} D_{\mu_2}D_{\mu_3}D_{\mu_4} q
  \right)
\right\rangle_\rho
\ ,
\end{equation}
where we have introduced the symbol ${\mathcal ST}$ which makes the
operators symmetric and traceless with respect to the Lorentz indices.
The general tensor structure of the matrix element of
$A_{\mu_1 \cdots \mu_{2k}}^{2k + \tau, \tau}$ is given in
Ref.~\cite{HKL}. For $k=2$, it takes the following form:
\begin{eqnarray}
&&A_{\alpha\beta\lambda\sigma}=\Bigl[ p_\alpha p_\beta p_\lambda
p_\sigma
    {}- \frac{p^2}{8}\bigl( p_\alpha p_\beta g_{\lambda\sigma}
     {}+ p_\alpha p_\lambda g_{\beta\sigma}
     {}+ p_\alpha p_\sigma g_{\lambda\beta}
     {}+ p_\beta p_\lambda g_{\alpha\sigma}
\nonumber\\
 &&\qquad\qquad
     {}+ p_\beta p_\sigma g_{\alpha\lambda}
     {}+ p_\lambda p_\sigma g_{\alpha\beta} \bigr)
    {}+ \frac{p^4}{48}\bigl( g_{\alpha\beta}g_{\lambda\sigma}
     {}+ g_{\alpha\lambda}g_{\beta\sigma}
     {}+ g_{\alpha\sigma}g_{\beta\lambda} \bigr) \Bigr]A_4
\end{eqnarray}
For $\tau=2$ with arbitrary $k$, we have ~\cite{FLK}:
\begin{eqnarray}
A_{2k}^{2k+2,2}&=&C_{2,2k}^q
A_{2k}^q+C_{2,2k}^{\rm G} A_{2k}^{\rm G}
\nonumber\\
C_{2k}^{2k+2,2}&=&C_{L,2k}^q A_{2k}^q+C_{L,2k}^{\rm G} A_{2k}^{\rm
G}~,
 \end{eqnarray}
where $C_{2,2k}^q= 1+{\mathcal O}(\alpha_s)$, $C_{L,2k}^{q, {\rm
G}}\sim {\mathcal O}(\alpha_s)$ and $C_{2,2k}^{\rm G}\sim
{\mathcal O}(\alpha_s)$ (with the superscripts $q$ and $G$
standing respectively for quark and gluon) are the Wilson
coefficients in the OPE ~\cite{FLK}. The quantities $ A_{n}^q$ and
$ A_{n}^{\rm G}$ are defined by 
\begin{eqnarray}
A_n^q
(\mu)&=&2\int_0^1dx~x^{n-1}[q(x,\mu) + \bar q (x,\mu)]
\nonumber\\
A_n^{\rm
G} (\mu)&=&2\int_0^1dx~x^{n-1} G (x,\mu)~, 
\end{eqnarray} 
where $q(x,\mu)$
and $G (x,\mu)$ are quark and gluon distribution functions
respectively. We observe that (\ref{acc-ope}) consists of three
classes of terms: One is independent of the background, i.e.,
density in this case, the second consists of scalar operators with
various condensates $\langle{\mathcal O}\rangle_\rho$ and the third is
made up of non-scalar operators whose matrix elements in dense
matter could not be simply expressed in terms of various
condensates $\langle{\mathcal O}\rangle_\rho$.

It is clear that Eq.~(\ref{acc-ope}) is a general expression that
can be applied equally well to heat-bath systems. Thus we can
simply transcribe (\ref{acc-ope}) to the temperature case by
replacing the condensates $\langle{\mathcal O}\rangle_\rho$ by
$\langle{\mathcal O }\rangle_T$ and the quantities $A_{\mu_1 \cdots
\mu_{2k}}^{2k + \tau, \tau}$ and $C_{\mu_1 \cdots \mu_{2k}}^{2k +
\tau, \tau}$ by the corresponding quantities in heat bath. 
(We show the matching condition on the bare pion velocity 
at finite density in Appendix~\ref{app:CCOPE}.)
The higher the twist of operators becomes,
the more these operators are suppressed since the dimensions of such
operators become higher and the power of $1/Q^2$ appear.
Thus in the following, 
we restrict ourselves to contributions from the twist 2
$(\tau = 2)$ operators. 
Then the temperature-dependent correlator
can be written as
\begin{eqnarray}
 &&
 G_A^{\mu\nu}(q_0,\bar{q})
 = (q^\mu q^\nu - g^{\mu\nu}q^2)\frac{-1}{4}
   \Biggl[ \frac{1}{2\pi^2}
   \Biggl( 1+\frac{\alpha_s}{\pi} \Biggr) \ln \Biggl( \frac{Q^2}{\mu^2}
   \Biggr) + \frac{1}{6Q^4} \Big\langle \frac{\alpha_s}{\pi} G^2
   \Big\rangle_T \nonumber\\
 &&\qquad\qquad{}- \frac{2\pi\alpha_s}{Q^6} \Big\langle
   \Bigl( \bar{u}\gamma_\mu \gamma_5 \lambda^a u -
   \bar{d}\gamma_\mu \gamma_5 \lambda^a d \Bigr)^2 \Big\rangle_T
\nonumber\\
 &&\qquad\qquad{}- \frac{4\pi\alpha_s}{9Q^6}
   \Big\langle
   \Bigl( \bar{u}\gamma_\mu\lambda^a u + \bar{d}\gamma_\mu\lambda^a d
   \Bigr)\sum_q^{u,d,s}\bar{q}\gamma^\mu\lambda^a q \Big\rangle_T
   \Biggr] \nonumber\\
 &&\qquad{}+
[-g^{\mu\nu}q^{\mu_1}q^{\mu_2} + g^{\mu\mu_1}q^\nu q^{\mu_2}
    + q^\mu q^{\mu_1}g^{\nu\mu_2} + g^{\mu\mu_1}g^{\nu\mu_2}Q^2]
\nonumber\\
&&\qquad\qquad{}\times [\frac{4}{Q^4}A_{\mu_1\mu_2}^{4,2}
+\frac{16}{Q^8}q^{\mu_3}q^{\mu_4}A_{\mu_1\mu_2\mu_3\mu_4}^{6,2}] \
,\label{ope-t} 
\end{eqnarray}
where $G_A^{\mu\nu}$ is constructed from the
axial-vector current associated with the iso-triplet channel
defined by
\begin{eqnarray}
 J_{5\mu}
 &=& \frac{1}{2}(\bar{u}\gamma_5\gamma_\mu u -
     \bar{d}\gamma_5\gamma_\mu d),
\end{eqnarray}
and we keep terms only up to the order of $1/Q^8$ for $ A_{\mu_1
\cdots \mu_{2k}}^{2k+2, 2}$. The $\lambda^a$ denote the $SU(3)$
color matrices normalized as $\mbox{tr}[\lambda^a \lambda^b ] =
2\delta^{ab}$. Here we have dropped the terms with $ C_{\mu_1
\cdots \mu_{2k}}^{2k+2, 2} $ in the non-scalar operators since
they are of higher order in both $1/(Q^2)^n$  and $\alpha_s$
compared to the terms in the first line of Eq. (\ref{ope-t}). The
temperature dependence of $A_{\mu_1\mu_2}^{4,2}$ and
$A_{\mu_1\mu_2\mu_3\mu_4}^{6,2}$, implicit in Eq.~(\ref{ope-t}),
will be specified below.

In order to effectuate the Wilsonian matching, we should in
principle evaluate the condensates and temperature-dependent
matrix elements of the non-scalar operators in Eq.(\ref{ope-t}) at
the given scale $\Lambda$ and temperature $T$ in terms of QCD
variables only. This can presumably be done on lattice. However no
complete information is as yet available from model-independent
QCD calculations. We are therefore compelled to resort to indirect
methods and we adopt here an approach borrowed from QCD sum-rule
calculations.

Let us first evaluate the quantities that figure in
Eq.(\ref{ope-t}) at low temperature. In low temperature regime,
only the pions are expected to be thermally excited. In the dilute
pion-gas approximation, $\langle {\mathcal O}\rangle_T$ is evaluated as
\begin{eqnarray}
\langle {\mathcal O}\rangle_T\simeq \langle {\mathcal O}\rangle_0 +
\sum_{a=1}^3\int\frac{d^3p}{2\epsilon(2\pi)^3}\langle\pi^a(\vec p)|
{\mathcal
  O}|\pi^a(\vec p)\rangle n_B(\epsilon/T),\label{dpa}
\end{eqnarray} 
where $\epsilon=\sqrt{\bar p^2+m_\pi^2}$ and $n_B$ is the
Bose-Einstein distribution.
As an
example, we consider the operator of $(\tau,s)=(2,4)$ that
 contributes to both $G_A^{L(0)}$ and $G_A^{L(1)}$.
Noting that $G_A^{L}(q_0,\bar q) =G_{A00}/\bar q^2$, we evaluate
 $G_{A00}(q_0,\bar q)$.
\begin{eqnarray}
G_{{\rm A}00}^{(\tau=2,s=4)}(q_0,\bar q)
&=& \frac{3}{4}\int\frac{d^3p}{2\epsilon(2\pi)^3}
\frac{16}{Q^8}[-q^{\alpha}q^{\beta} + g^{0\alpha}q^0 q^{\beta}
    + q^0 q^{\alpha}g^{0\beta} + g^{0\alpha}g^{0\beta}Q^2]
\nonumber\\
&&\qquad\times q^\lambda q^\sigma A_{\alpha\beta\lambda\sigma}^{6,2
 (\pi)}
n_B(\epsilon/T),\label{Ga00} 
\end{eqnarray}
where
$A_{\alpha\beta\lambda\sigma}^{6,2 (\pi)}$ is given
by~\footnote{%
 For the general
 tensor structure of the matrix elements with
 a polarized spin-one target, say, along the beam direction in
 scattering process,
 see Ref. \cite{hjm}.
}
\begin{eqnarray}
&&A_{\alpha\beta\lambda\sigma}^{6,2 (\pi)}=\Bigl[ p_\alpha p_\beta p_\lambda p_\sigma
    {}- \frac{p^2}{8}\bigl( p_\alpha p_\beta g_{\lambda\sigma}
     {}+ p_\alpha p_\lambda g_{\beta\sigma}
     {}+ p_\alpha p_\sigma g_{\lambda\beta}
     {}+ p_\beta p_\lambda g_{\alpha\sigma}
\nonumber\\
 &&\qquad\qquad
     {}+ p_\beta p_\sigma g_{\alpha\lambda}
     {}+ p_\lambda p_\sigma g_{\alpha\beta} \bigr)
    {}+ \frac{p^4}{48}\bigl( g_{\alpha\beta}g_{\lambda\sigma}
     {}+ g_{\alpha\lambda}g_{\beta\sigma}
     {}+ g_{\alpha\sigma}g_{\beta\lambda} \bigr) \Bigr]A_4^{\pi}
\ ,\label{A4T}
\end{eqnarray}
where $A_4^\pi$ carries the temperature dependence.
Taking $m_\pi^2=0$, we see that the terms with $p^2$ and
 $p^4$ in Eq. (\ref{A4T}) are zero.

{}From Eqs. (\ref{ope-t}), (\ref{dpa}) and (\ref{Ga00}), we obtain
\begin{eqnarray}
 G_A^{{\rm (OPE)}L(0)}(-q^2)
  &=& \frac{-1}{3}g^{\mu\nu}G_{A,\mu\nu}^{{\rm (OPE)}(0)}
\nonumber\\
  &=& \frac{-1}{4}
   \Biggl[ \frac{1}{2\pi^2}
   \Biggl( 1+\frac{\alpha_s}{\pi} \Biggr) \ln \Biggl( \frac{Q^2}{\mu^2}
   \Biggr) + \frac{1}{6Q^4} \Big\langle \frac{\alpha_s}{\pi} G^2
   \Big\rangle_T
\nonumber\\
 &&{}- \frac{2\pi\alpha_s}{Q^6} \Big\langle
   \Bigl( \bar{u}\gamma_\mu \gamma_5 \lambda^a u -
   \bar{d}\gamma_\mu \gamma_5 \lambda^a d \Bigr)^2 \Big\rangle_T
\nonumber\\
 &&{}- \frac{4\pi\alpha_s}{9Q^6}
   \Big\langle
   \Bigl( \bar{u}\gamma_\mu\lambda^a u + \bar{d}\gamma_\mu\lambda^a d
   \Bigr)\sum_q^{u,d,s}\bar{q}\gamma^\mu\lambda^a q \Big\rangle_T
   \Biggr]
\nonumber\\
&&{}+ \frac{\pi^2}{30}\frac{T^4}{Q^4}A_2^{\pi (u+d)}
  {}- \frac{16\pi^4}{63}\frac{T^6}{Q^6}A_4^{\pi (u+d)}.
\label{OPE-L(0)}
\end{eqnarray}
  $G_A^{{\rm (OPE)}L(1)}$ takes the following form
\begin{eqnarray}
 G_A^{{\rm (OPE)}L(1)}
  = \frac{32}{105}\pi^4\frac{T^6}{Q^8} A_4^{\pi (u+d)} .\label{L(1)exp}
 \end{eqnarray}

We now proceed to estimate the pion velocity by matching to
QCD.

First we consider the matching between $G_A^{{\rm (HLS)}L(0)}$ and
$G_A^{{\rm (OPE)}L(0)}$. {}From Eqs.~(\ref{HLS-L(0)}) and
(\ref{OPE-L(0)}), we obtain
\begin{eqnarray}
 (-q^2)\frac{d}{d(-q^2)}G_A^{{\rm (HLS)}L(0)}
 &=& - \frac{F_{\pi,{\rm bare}}^t F_{\pi,{\rm bare}}^s}{Q^2},
\nonumber\\
 (-q^2)\frac{d}{d(-q^2)}G_A^{{\rm (OPE)}L(0)}
 &=& \frac{-1}{8\pi^2}\Biggl[ \Bigl( 1 + \frac{\alpha_s}{\pi} \Bigr) +
   \frac{2\pi^2}{3}\frac{\big\langle \frac{\alpha_s}{\pi}G^2
   \big\rangle_T }{Q^4}
   {}+ \pi^3 \frac{1408}{27}\frac{\alpha_s
    \langle \bar{q}q \rangle_T^2}{Q^6} \Biggr]
\nonumber\\
 &&{}- \frac{\pi^2}{15}\frac{T^4}{Q^4}A_2^{\pi (u+d)}
   {}+ \frac{16\pi^4}{21}\frac{T^6}{Q^6}A_4^{\pi (u+d)}.
\end{eqnarray}
Matching them at $Q^2 = \Lambda^2$, we obtain
\begin{eqnarray}
 \frac{F_{\pi,{\rm bare}}^t F_{\pi,{\rm bare}}^s}{\Lambda^2}
 &=& \frac{1}{8\pi^2}\Biggl[ \Bigl( 1 + \frac{\alpha_s}{\pi} \Bigr) +
   \frac{2\pi^2}{3}\frac{\big\langle \frac{\alpha_s}{\pi}G^2
   \big\rangle_T }{\Lambda^4}
   {}+ \pi^3 \frac{1408}{27}\frac{\alpha_s
    \langle \bar{q}q \rangle_T^2}{\Lambda^6}  \Biggr]
\nonumber\\
 &&{}+ \frac{\pi^2}{15}\frac{T^4}{\Lambda^4}A_2^{\pi (u+d)}
   {}- \frac{16\pi^4}{21}\frac{T^6}{\Lambda^6}A_4^{\pi (u+d)}
\nonumber\\
&\equiv& G_0.
\label{ts}
\end{eqnarray}
Next we consider the matching between $G_A^{{\rm (HLS)}L(1)}$ and
$G_A^{{\rm (OPE)}L(1)}$.
{}From Eqs.~(\ref{HLS-L(1)}) and (\ref{L(1)exp}), we have
\begin{equation}
 \frac{F_{\pi,{\rm bare}}^t F_{\pi,{\rm bare}}^s
 (1 - V_{\pi,{\rm bare}}^2)}{\Lambda^2}
  = \frac{32}{105}\pi^4\frac{T^6}{\Lambda^6} A_4^{\pi (u+d)} .
\label{1-v^2}
\end{equation}
Noting that the right-hand-side of this expression is positive, we
verify that
\begin{equation}
 V_{\pi,{\rm bare}} < 1
\end{equation}
which is consistent with the causality.

The bare pion velocity can be obtained by dividing
Eq.~(\ref{1-v^2}) with Eq.~(\ref{ts}). What we obtain is the
deviation from the speed of light:
\begin{equation}
 \delta_{\rm bare}\equiv 1 - V_{\pi,{\rm bare}}^2
 = \frac{1}{G_0}
   \frac{32}{105}\pi^4\frac{T^6}{\Lambda^6} A_4^{\pi (u+d)} .
\label{deviation-rho}
\end{equation}
This should be valid at low temperature. We note that the Lorentz
non-invariance does not appear when we consider the operator with
$s=2$, and that the operator with $s=4$ generates the Lorentz
non-invariance. This is consistent with the fact that $G_A^L$
including up to the operator with $s=2$ is expressed as the
function of only $Q^2$ ~\cite{Mallik:1997kj}.
Equation~(\ref{deviation-rho}) implies that the intrinsic
temperature dependence starts from the ${\cal O}(T^6)$
contribution. On the other hand, the hadronic thermal correction
to the pion velocity starts from the ${\cal O}(T^4)$.
[There are ${\mathcal O}(T^2)$ corrections to $[f_\pi^t]^2$ and
$[f_\pi^t f_\pi^s]$, but they are canceled with each other in the
pion velocity. See subsection~\ref{ssec:PP-lowT}.] 
Thus the hadronic thermal effect is dominant in
low temperature region.

\newpage

\chapter{Vector Manifestation in Hot Matter}
\label{ch:VMHM}

The vector manifestation (VM) is a new pattern for 
Wigner realization of chiral symmetry,
in sharp contrast to the standard scenario of 
chiral symmetry restoration.
In order to clarify the difference between the standard scenario 
and the VM, we consider the chiral representations of the low-lying mesons.

In the broken phase, the eigenstate of the chiral representation under 
$SU(3)_L \times SU(3)_R$ does not generally agree with
the mass eigenstate due to the existence of the Nambu-Goldstone bosons:
There exists a representation mixing.
Then the scalar, pseudoscalar, vector and axial-vector mesons
belong to the following representations~\cite{Gilman:1967qs,Weinberg:hw}:
\begin{eqnarray}
 |s \rangle 
   &=& |(3,3^*) \oplus (3^*,3) \rangle, \nonumber\\
 |\pi \rangle 
   &=& |(3,3^*) \oplus (3^*,3) \rangle \sin\psi +
       |(1,8) \oplus (8,1) \rangle \cos\psi, \nonumber\\
 |\rho \rangle 
   &=& |(1,8) \oplus (8,1) \rangle, \nonumber\\
 |A_1 \rangle 
   &=& |(3,3^*) \oplus (3^*,3) \rangle \cos\psi -
       |(1,8) \oplus (8,1) \rangle \sin\psi,
\label{rep.mixing}
\end{eqnarray}
where $\psi$ denotes the mixing angle 
and is determined as $\psi \simeq 45\,{}^\circ$.

Now we consider the chiral symmetry restoration, 
where it is expected that the above representation mixing is dissolved.
{}From Eq.~(\ref{rep.mixing}), one can easily see that 
there are two possibilities for pattern of chiral symmetry restoration.
One possible pattern is the case where $\cos\psi \to 0$ 
when we approach the critical point.
In this case, the pion belongs to $|(3,3^*) \oplus (3^*,3)\rangle$
and becomes the chiral partner of the scalar meson:
\begin{eqnarray}
 |\pi \rangle 
   &=& |(3,3^*) \oplus (3^*,3) \rangle, \nonumber\\
 |s \rangle 
   &=& |(3,3^*) \oplus (3^*,3) \rangle
 \qquad \mbox{for}\quad \cos\psi \to 0.
\end{eqnarray}
The vector and axial-vector mesons are in the same multiplet
$|(1,8) \oplus (8,1) \rangle$.
This is the standard scenario of chiral symmetry restoration.

Another possibility is the case where $\sin\psi \to 0$ 
when we approach the critical point~\cite{HY:VM}.
In this case, the pion belongs to pure $|(1,8) \oplus (8,1)\rangle$
and so its chiral partner is the vector meson:
\begin{eqnarray}
 |\pi \rangle 
   &=& |(1,8) \oplus (8,1) \rangle, \nonumber\\
 |\rho \rangle 
   &=& |(1,8) \oplus (8,1) \rangle
 \qquad \mbox{for}\quad \sin\psi \to 0.
\end{eqnarray}
The scalar meson joins with the axial-vector meson 
in the same representation $|(3,3^*) \oplus (3^*,3) \rangle$.
This is nothing but the VM of chiral symmetry.

In order to formulate the VM, 
we need a theory including both pions and vector mesons.
One of such theories is the one based on the hidden local symmetry (HLS).
In the following sections, we will show how the VM is formulated
at the critical point in the framework of the HLS theory.
We also study the predictions of the VM in hot matter.


\section{Conditions for Bare Parameters}
\label{sec:CBP}

In this section,
we summarize the conditions for the bare parameters obtained
in Ref.~\cite{HS:VM} through the Wilsonian matching at the critical
temperature.

\subsection{Case with Lorentz invariance}
\label{ssec:CLI}

We consider the Wilsonian matching near the
chiral symmetry restoration point.
Here we assume that the order of the chiral phase transition is
second or weakly first order,
and thus the quark condensate becomes zero
continuously for $T \to T_c$.
First, note that
the Wilsonian matching condition~(\ref{eq:WMC A}) 
provides
\begin{equation}
  \frac{F^2_\pi (\Lambda ;T_c)}{{\Lambda}^2} 
  = \frac{1}{8{\pi}^2}\Bigl[
                            1 + \frac{\alpha _s}{\pi} +
                             \frac{2{\pi}^2}{3}
                            \frac{\langle \frac{\alpha _s}{\pi}
                            G_{\mu \nu}G^{\mu \nu} \rangle_{T_c} }
                             {{\Lambda}^4}
                 \Bigr]
 \neq 0 
\ ,
\label{eq:WMC A Tc}
\end{equation}
which implies that the matching with QCD dictates
\begin{equation}
F^2_\pi (\Lambda ;T_c) \neq 0 
\label{Fp2 Lam Tc}
\end{equation}
even at the critical temperature where the on-shell pion decay constant 
vanishes by adding the quantum corrections through
the RGE including the quadratic divergence~\cite{HY:WM}
and hadronic thermal corrections,
as we will show in section~\ref{sec:PDCPV}.
As was shown in Ref.~\cite{HKR:VM} for the VM in dense matter,
the Lorentz non-invariant version of
the VM conditions for the bare parameters are obtained 
by the requirement of the equality between the axial-vector
and vector current correlators in the HLS,
which is valid also in hot matter (see next subsection):
\begin{eqnarray}
 && a_{\rm{bare}}^t \equiv
  \Biggl( \frac{F_{\sigma,\rm{bare}}^t}{F_{\pi,\rm{bare}}^t} \Biggr)^2
  \stackrel{T \to T_c}{\to} 1, \quad
  a_{\rm{bare}}^s \equiv
  \Biggl( \frac{F_{\sigma,\rm{bare}}^s}{F_{\pi,\rm{bare}}^s} \Biggr)^2
  \stackrel{T \to T_c}{\to} 1, \label{EVM a}\\
 && g_{T,\rm{bare}} \stackrel{T \to T_c}{\to} 0, \quad
    g_{L,\rm{bare}} \stackrel{T \to T_c}{\to} 0, \label{EVM g}
\end{eqnarray}
where $a^t_{\rm{bare}}, a^s_{\rm{bare}}, g_{T,\rm{bare}}$ and 
$g_{L,\rm{bare}}$ are the extensions of the parameters
$a_{\rm{bare}}$ and $g_{\rm{bare}}$ in the bare Lagrangian
with the Lorentz symmetry breaking effect included as in Appendix A
of Ref.~\cite{HKR:VM}.

When we use the conditions for the parameters $a^{t,s}$ 
in Eq.~(\ref{EVM a})
and the above result that the Lorentz symmetry violation 
between the bare pion decay constants 
$F_{\pi,\rm{bare}}^{t,s}$ is small, 
we can easily show
that the Lorentz symmetry breaking effect between
the temporal and spatial bare sigma decay constants is also small,
$F_{\sigma,\rm{bare}}^t \simeq F_{\sigma,\rm{bare}}^s$.
While we cannot determine the ratio $g_{L,\rm{bare}}/g_{T,\rm{bare}}$
through the Wilsonian matching
since the transverse mode of vector meson decouples near
the critical temperature.~\footnote{
 In Ref.~\cite{HKR:VM}, the analysis including the Lorentz
 non-invariance at bare HLS theory was carried out.
 Due to the equality between axial-vector and vector current
 correlators, $(g_{L,{\rm bare}}, g_{T,{\rm bare}}) \to (0,0)$ is
 satisfied when we approach the critical point.
 This implies that at the bare level the longitudinal mode becomes the real
 NG boson and couples to the vector current correlator, 
 while the transverse mode decouples.
 Furthermore $g_L \to 0$ is a fixed point for the RGE~\cite{Sasaki:2003qj}.
 Thus in any energy scale the transverse mode decouples from the vector
 current correlator.
 For details, see next subsection.
}
However this implies that the transverse mode is irrelevant
for the quantities studied in this thesis.
Therefore in the present analysis, we set
$g_{L,\rm{bare}}=g_{T,\rm{bare}}$ for simplicity and
use the Lorentz invariant Lagrangian at bare level.
In the low temperature region, the intrinsic temperature dependences
are negligible, so that we also use the 
Lorentz invariant Lagrangian at bare level
as in the analysis by the ordinary chiral
Lagrangian in Ref.~\cite{GL}.

At the critical temperature,
the axial-vector and vector current correlators
derived in the OPE
agree with each other for any value of $Q^2$.
Thus we require that
these current correlators in the HLS are
equal at the critical temperature
for any value of $Q^2\ \mbox{around}\ {\Lambda}^2$.
As we discussed above, we start from the Lorentz invariant
bare Lagrangian even in hot matter, and then
the axial-vector current correlator $G_A^{\rm{(HLS)}}$ and 
the vector current correlator $G_V^{\rm{(HLS)}}$
are expressed by the same forms
as those at zero temperature with the bare parameters
having the intrinsic temperature dependences:
\begin{eqnarray}
 G^{\rm{(HLS)}}_A (Q^2;T) 
  &=& \frac{F^2_\pi (\Lambda;T)}{Q^2} -
      2z_2(\Lambda;T), \nonumber\\
 G^{\rm{(HLS)}}_V (Q^2;T) 
  &=& \frac{F^2_\sigma (\Lambda;T)[1 - 2g^2(\Lambda;T)z_3(\Lambda;T)]}
           {{M_\rho}^2(\Lambda;T) + Q^2} - 2z_1(\Lambda;T).
  \label{correlator HLS at zero-T}
  \end{eqnarray}
By taking account of the fact 
$F^2_\pi (\Lambda ;T_c) \neq 0$ derived from
the Wilsonian matching condition 
given in Eq.~(\ref{eq:WMC A Tc}),
the requirement 
$G_A^{(\rm{HLS})}=G_V^{(\rm{HLS})}$ is satisfied
only if the following conditions are met~\cite{HS:VM}: 
\begin{eqnarray}
 g(\Lambda ;T) &\stackrel{T \to T_c}{\to}& 0, \label{eq:VMg}\\
 a(\Lambda ;T) = F_\sigma^2(\Lambda;T)/F_\pi^2(\Lambda;T)
               &\stackrel{T \to T_c}{\to}& 1, \label{eq:VMa}\\
 z_1(\Lambda ;T) - z_2(\Lambda ;T)
               &\stackrel{T \to T_c}{\to}& 0. \label{eq:VMz}
\end{eqnarray}
Note that the intrinsic thermal effects act on the parameters 
in such a way that they become the values 
of Eqs.~(\ref{eq:VMg})-(\ref{eq:VMz}).

Through the Wilsonian matching at non-zero temperature mentioned above,
the parameters appearing in the hadronic thermal
corrections calculated in section~\ref{sec:TPFBFG}
have the intrinsic temperature dependences:
$F_\pi$, $a$ and $g$ appearing there should be regarded as
\begin{eqnarray}
F_\pi &\equiv& F_\pi(\mu=0;T) \ ,
\nonumber\\
a &\equiv& a\left(\mu=M_\rho(T);T\right) \ ,
\nonumber\\
g &\equiv& g\left(\mu=M_\rho(T);T\right) \ ,
\end{eqnarray}
where $M_\rho$ is determined from the on-shell condition:
  \begin{equation}
   M_\rho^2 \equiv M_\rho^2(T) = 
   a(\mu=M_\rho ;T)g^2(\mu=M_\rho ;T)F_\pi^2(\mu=M_\rho ;T)\ .
                                  \label{eq:Mdef}
  \end{equation}
{}From the RGEs for $g\ \mbox{and}\ a$ in Eqs.(\ref{eq:RGEg})
and (\ref{eq:RGEa}), 
we find that $(g,a)=(0,1)$ is the fixed point.
Therefore, Eqs.~(\ref{eq:VMg}) and (\ref{eq:VMa}) imply that $g$ and $a$
at the on-shell of the vector meson take the same values:
\begin{eqnarray}
&&
 a\,(\mu = M_\rho(T);T) 
\ \mathop{\longrightarrow}^{T\rightarrow T_c}\  1\ , 
\nonumber\\
&&
 g\,(\mu = M_\rho(T);T) 
\ \mathop{\longrightarrow}^{T\rightarrow T_c}\  0
\ ,
\label{VM a g}
\end{eqnarray}
where the parametric vector meson mass $M_\rho(T)$ is determined from
the condition (\ref{eq:Mdef}).
The above conditions with Eq.~(\ref{eq:Mdef}) imply that
$M_\rho(T)$ also vanishes:
\begin{equation}
M_\rho(T) 
\ \mathop{\longrightarrow}^{T\rightarrow T_c}\  0 \ .
\label{VM Mrho}
\end{equation}

\subsection{Case without Lorentz invariance}
\label{sec:VMC}

In this subsection, 
we start from the Lagrangian with Lorentz
non-invariance, and requiring that the axial-vector current correlator be
equal to the vector current correlator at the critical point, we
obtain the conditions for the bare parameters.
Then we show that the conditions are satisfied in any energy scale,
which is protected by the fixed point of the RGEs.

First 
we show the Lorentz non-invariant version of the conditions satisfied 
at the critical temperature for the bare parameters,
following Ref.~\cite{HKR:VM} where the conditions
for the current correlators
with the bare parameters in dense matter were presented.
We consider the matching near the critical temperature.
At the chiral phase transition point, the axial-vector and vector
current correlators must agree with each other: $G_{A{\rm(HLS)}}^L
= G_{V{\rm(HLS)}}^L$ and $G_{A{\rm(HLS)}}^T = G_{V{\rm(HLS)}}^T$.
These equalities are satisfied for
any values of $p_0$ and $\bar{p}$ around the matching scale
only if the following conditions are met:
\begin{eqnarray}
&&
  a_{\rm bare}^t \to 1, \quad
  a_{\rm bare}^s \to 1,
\nonumber\\
&&
  g_{L,{\rm bare}} \to 0, \quad
  g_{T,{\rm bare}} \to 0  \quad \mbox{for}\,\,T \to T_c.
\end{eqnarray}
This implies that at bare level the longitudinal mode of the vector 
meson becomes the real NG boson and couples to the vector current correlator, 
while the transverse mode decouples.

Next, 
we show that the conditions for the bare parameters
for $T \to T_c$ are satisfied in any energy scale and that
this is protected by the fixed point of the RGEs.

It was shown that the HLS gauge coupling $g=0$ is a fixed point of the RGE
for $g$ at one-loop level~\cite{HY:PLB,HY:WM}.
The existence of the fixed point $g=0$ is guaranteed by
the gauge invariance.
This is easily understood from the fact that the gauge field is
normalized as $V_\mu = g \rho_\mu$.
In the present case without Lorentz symmetry,
the gauge field is normalized by $g_L$ as $V_\mu = g_L \rho_\mu$
and thus $g_L = 0$ becomes a fixed point of the RGE for $g_L$.

\begin{figure}
 \begin{center}
  \includegraphics[width = 10cm]{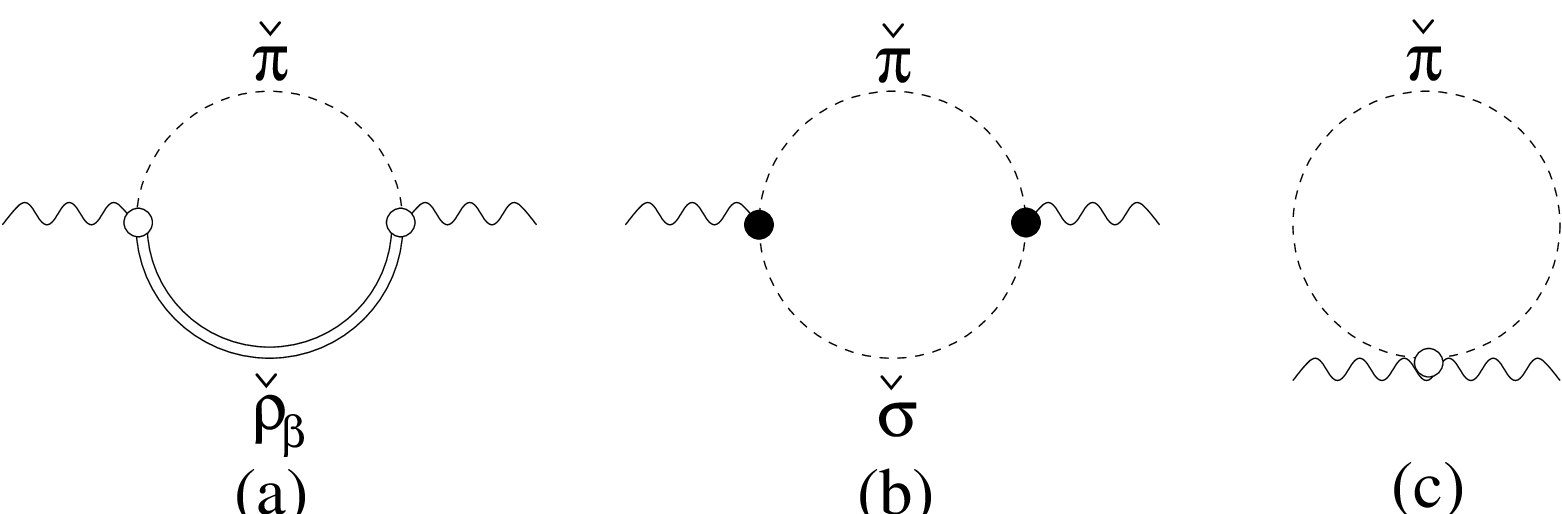}
 \end{center}
 \caption{Diagrams for contributions to $\Pi_\perp^{\mu\nu}$
         at one-loop level.}
 \label{fig:AA}
\end{figure}

\begin{figure}
 \begin{center}
  \includegraphics[width = 13cm]{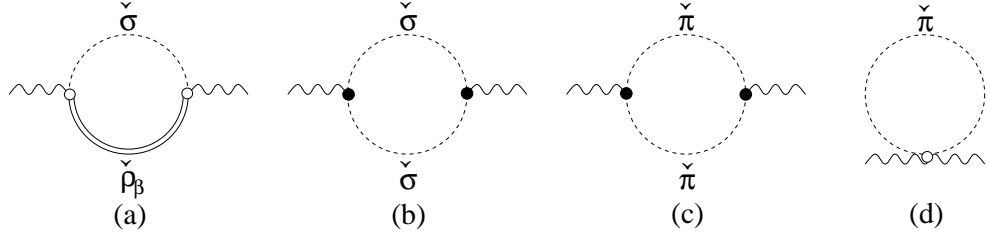}
 \end{center}
 \caption{Diagrams for contributions to
  $\Pi_\parallel^{\mu\nu}$ at one-loop level.}
 \label{fig:VV}
\end{figure}

Provided that $g_L = 0$ is a fixed point,
we can show that $a^t = a^s = 1$ is also a fixed point 
of the coupled RGEs for $a^t$ and $a^s$ as follows:
We start from the bare theory defined at a scale $\Lambda$
with $a_{\rm bare}^t = a_{\rm bare}^s = 1$ (and $g_L = 0$).
The parameters $a^t$ and $a^s$ at $(\Lambda - \delta\Lambda)$
are calculated by integrating out the modes in
[$\Lambda - \delta\Lambda, \Lambda$].
They are obtained from the two-point functions of
${\cal A}_\mu$ and ${\cal V}_\mu$,
denoted by $\Pi_\perp^{\mu\nu}$ and $\Pi_\parallel^{\mu\nu}$.
We decompose these functions into
\begin{eqnarray}
&&
 \Pi_{\perp,\parallel}^{\mu\nu}
  =u^\mu u^\nu \Pi_{\perp,\parallel}^t +
   (g^{\mu\nu}-u^\mu u^\nu)\Pi_{\perp,\parallel}^s 
{}+
   P_L^{\mu\nu}\Pi_{\perp,\parallel}^L + 
   P_T^{\mu\nu}\Pi_{\perp,\parallel}^T,
\label{Pi perp decomp}
\end{eqnarray}
where $u^\mu u^\nu$, $(g^{\mu\nu}-u^\mu u^\nu)$, $ P_L^{\mu\nu}$
and $P_T^{\mu\nu}$ denote the temporal, spatial, longitudinal and
transverse projection operators, respectively.
The parameters $a^t$ and $a^s$ are defined by
$ a^t = {\Pi_\parallel^t}/{\Pi_\perp^t}\,,
  a^s = {\Pi_\parallel^s}/{\Pi_\perp^s}$~\cite{HS:VVD}.
These expressions are further reduced to
\begin{eqnarray}
 &&
 a^t(\Lambda - \delta\Lambda) = a_{\rm bare}^t
 {}+ \frac{1}{\bigl( F_{\pi,{\rm bare}}^t \bigr)^2}
   \Bigl[ \Pi_\parallel^t (\Lambda;\Lambda - \delta\Lambda)
    {}- a_{\rm bare}^t \Pi_\perp^t (\Lambda;\Lambda - \delta\Lambda)
   \Bigr],
\nonumber\\
 &&
 a^s(\Lambda - \delta\Lambda) = a_{\rm bare}^s
 {}+ \frac{1}{F_{\pi,{\rm bare}}^t F_{\pi,{\rm bare}}^s}
   \Bigl[ \Pi_\parallel^s (\Lambda;\Lambda - \delta\Lambda)
    {}- a_{\rm bare}^s \Pi_\perp^s (\Lambda;\Lambda - \delta\Lambda)
   \Bigr],
\label{at as}
\end{eqnarray}
where $\Pi_{\perp,\parallel}^{t,s}(\Lambda;\Lambda - \delta\Lambda)$
denotes the quantum correction obtained by integrating the modes out 
between $[\Lambda - \delta\Lambda, \Lambda ]$.
We show the diagrams for contributions to $\Pi_\perp^{\mu\nu}$
and $\Pi_\parallel^{\mu\nu}$ at one-loop level
in Figs.~\ref{fig:AA} and~\ref{fig:VV}.
The contributions (a) in Fig.~\ref{fig:AA} and (a) in
Fig.~\ref{fig:VV} are proportional to $g_{L,{\rm bare}}^2$. The
contributions (c) in Fig.~\ref{fig:AA} and (d) in
Fig.~\ref{fig:VV} are proportional to $(a_{\rm bare}^t-1)$. 
Taking $g_{L,{\rm bare}} = 0$ and $a_{\rm bare}^t = a_{\rm bare}^s =1$, 
these contributions vanish. 
We note that $\sigma$ (i.e., longitudinal vector meson) is massless and
the chiral partner of pion at the critical temperature. 
Then the contributions (b) and (c) in Fig.~\ref{fig:VV} have a
symmetry factor $1/2$ respectively and are obviously equal to the
contribution (b) in Fig.~\ref{fig:AA}, i.e.,
$\Pi_\perp^{(b)\mu\nu} = \Pi_\parallel^{(b)+(c)\mu\nu}$. 
Thus from Eq.~(\ref{at as}), we obtain
\begin{eqnarray}
 &&
 a^t(\Lambda - \delta\Lambda)
 = a_{\rm bare}^t = 1,
\nonumber\\
 &&
 a^s(\Lambda - \delta\Lambda)
 = a_{\rm bare}^s = 1.
\end{eqnarray}
This implies that $a^t$ and $a^s$ are not renormalized at the scale
$(\Lambda - \delta\Lambda)$. Similarly, we include the corrections
below the scale $(\Lambda - \delta\Lambda)$ in turn, and find that
$a^t$ and $a^s$ do not receive the quantum corrections. Eventually
we conclude that $a^t = a^s =1$ is a fixed point of the RGEs for
$a^t$ and $a^s$.

{}From the above, we find that $(g_L, a^t, a^s) = (0,1,1)$ is 
a fixed point of the combined RGEs for $g_L, a^t$ and $a^s$. 
Thus the VM condition is given by
\begin{eqnarray}
\qquad
 & g_L \to 0,&  \nonumber\\
 & a^t \to 1, \quad
    a^s \to 1&  \quad \mbox{for}\,\,T \to T_c.
\label{evm}
\end{eqnarray}
The vector meson mass is never generated at the critical temperature
since the quantum correction to $M_\rho^2$ is proportional to $g_L^2$.
Because of $g_L \to 0$,
the transverse vector meson at the critical point, in any energy scale, 
decouples from the vector current correlator.
The VM condition for $a^t$ and $a^s$ leads to the equality between
the $\pi$ and $\sigma$ (i.e., longitudinal vector meson) velocities:
\begin{eqnarray}
\qquad
 \bigl( V_\pi / V_\sigma \bigr)^4
 &=& \bigl( F_\pi^s F_\sigma^t / F_\sigma^s F_\pi^t \bigr)^2
\nonumber\\
 &=& a^t / a^s 
 \stackrel{T \to T_c}{\to} 1.
\label{vp=vs}
\end{eqnarray}
This is easily understood from a point of view of the VM
since the longitudinal vector meson becomes the chiral partner
of pion.
We note that this condition
$V_\sigma = V_\pi$ holds independently of the value of the bare
pion velocity which is to be determined through the Wilsonian matching.


\section{Vector Meson Mass}
\label{sec:VMM}

In this section,
we briefly review that the vector manifestation (VM) in hot matter
can be formulated following Ref.~\cite{HS:VVD}.
Including the intrinsic temperature dependences of the
parameters near the critical temperature
determined in the previous section, 
we show that the vector meson mass vanishes at the critical temperature.

Let us study the vector meson pole mass near the critical temperature.
As shown in section~\ref{sec:CBP}, 
the parametric vector meson mass $M_\rho$ vanishes 
at the critical temperature, which is driven by the intrinsic effects.
Then, near the critical temperature we should take 
$M_\rho \ll T$ 
in Eq.~(\ref{eq:Mass})
instead of $T \ll M_\rho$ which was taken to reach the 
expression in Eq.~(\ref{m rho small T}) in the low temperature region.
Thus, by noting that
\begin{eqnarray}
 \tilde{G}_{2}(M_\rho ;T)
   &\stackrel{M_\rho \to 0}{\to}&\tilde{I}_{2}(T), \nonumber\\
 \tilde{J}^2_{1}(M_\rho ;T)
   &\stackrel{M_\rho \to 0}{\to}&\tilde{I}_{2}(T), \nonumber\\
 {M_\rho}^2\tilde{F}^2_{3}(M_\rho ;M_\rho ;T)
   &\stackrel{M_\rho \to 0}{\to}& 0, 
\end{eqnarray}
the pole mass of the vector meson
at $T \lesssim T_c$ becomes
\begin{eqnarray}
 {m_\rho}^2(T) 
  &=& {M_\rho}^2 + 
      N_f\,g^2 \frac{15-a^2}{12}\tilde{I}_{2}(T) 
\nonumber\\
  &=& {M_\rho}^2 +
      N_f\,g^2 \frac{15 - a^2}{144}T^2.
\label{eq:Mass2}
\end{eqnarray}
In the vicinity of $a \simeq 1$
the hadronic thermal effect gives a positive correction,
and then the vector meson pole mass is actually larger than the
parametric mass $M_\rho$.
However, the intrinsic temperature dependences of the parameters
obtained in section~\ref{sec:CBP} lead to 
$g \to 0$ and $M_\rho \rightarrow0$ for $T \to T_c$.
Then, from  Eq.~(\ref{eq:Mass2}) we conclude that
the pole mass of the vector meson $m_\rho$
also vanishes at the critical temperature:
  \begin{equation}
   m_\rho (T) \to 0 \quad \mbox{for} \ T \rightarrow T_c \ .
  \end{equation}
This implies that the VM is formulated 
at the critical temperature in the framework of the HLS theory,
which is consistent with the picture shown in
Refs.~\cite{BR,Brown-Rho:96,Brown:2001jh,Brown:2001nh}.


\section{Pion Decay Constants}
\label{sec:PDCPV}

In this section, 
we show how the two decay constants, $f_\pi^t$ and $f_\pi^s$, 
vanish at the critical temperature following Ref.~\cite{HKRS:SUS}.

As we discussed around Eq.~(\ref{VM a g}),
at the critical temperature the intrinsic thermal effects lead to
$(g,a) \to (0,1)$ which is a fixed point of the coupled RGEs.
Then the RGE for $F_\pi$ becomes
        \begin{equation}
         \mu \frac{d{F_\pi}^2}{d\mu} = \frac{N_f}{(4\pi)^2}{\mu}^2.
        \end{equation}
This RGE is easily solved and
the relation between $F_\pi (\Lambda ;T_c)\ \mbox{and}\ 
F_\pi (0;T_c)$ is given by
        \begin{equation}
         F^2_\pi (0;T_c) = F^2_\pi (\Lambda ;T_c) -
                           \frac{N_f}{2(4\pi)^2}{\Lambda}^2.
              \label{eq:F_pi0}
        \end{equation}
Finally we obtain the pion decay constants as follows:
\begin{eqnarray}
 f_\pi^t \tilde{F} = F^2_\pi (0;T) + 
        \overline{\Pi}_\perp^t(\bar{p},\bar{p};T), \nonumber\\
 f_\pi^s \tilde{F} = F^2_\pi (0;T) + 
        \overline{\Pi}_\perp^s(\bar{p},\bar{p};T), \label{eq:f_pi}
\end{eqnarray}
where the second terms are the hadronic thermal effects.
In the VM limit $(g,a) \to (0,1)$ the temperature dependent parts become
\begin{equation}
  \overline{\Pi}_\perp^t(\bar{p},\bar{p};T) \stackrel{T \to T_c}{\to}
  {} - \frac{N_f}{24}{T_c}^2, \quad
  \overline{\Pi}_\perp^s(\bar{p},\bar{p};T) \stackrel{T \to T_c}{\to}
  {} - \frac{N_f}{24}{T_c}^2. \label{eq:hadronicT_c}
\end{equation}
{}From Eqs.~(\ref{v2 rel}), (\ref{eq:f_pi}) and (\ref{eq:hadronicT_c}), 
the order parameter $f_\pi$ becomes
\begin{equation}
 f_\pi^2(\bar{p};T) \stackrel{T \to T_c}{\to}
 f_\pi^2(\bar{p};T_c) = 
 F_\pi^2(0;T_c) - \frac{N_f}{24}T_c^2.
\label{fpi critical}
\end{equation}
Since the order parameter $f_\pi$ vanishes at the critical temperature,
this implies
\begin{equation}
 F_\pi^2(0;T) \stackrel{T \to T_c}{\to}
  F_\pi^2(0;T_c) = \frac{N_f}{24}T_c^2. \label{eq:intrinsic}
\end{equation}
Thus we obtain
\begin{equation}
 (f_\pi^t)^2 \stackrel{T \to T_c}{\to} 0, \quad
 f_\pi^t f_\pi^s \stackrel{T \to T_c}{\to} 0.
\end{equation}
{}From Eq.~(\ref{t>s}) or equivalently Eq.~(\ref{v_pi<1}),
the above results imply that the temporal and spatial 
pion decay constants vanish simultaneously 
at the critical temperature~\cite{HKRS:SUS}:
\begin{equation}
 f_\pi^t(T_c) = f_\pi^s(T_c) = 0.
\end{equation}
Comparing Eq.~(\ref{fpi critical}) with the expression 
in the low temperature region in Eq.~(\ref{eq:f_piGL})
where the vector meson is decoupled,
we find that the coefficient of ${T_c}^2$ is different
by the factor $\frac{1}{2}$.
This is the contribution from the $\sigma$-loop 
(longitudinal $\rho$-loop).
In the low temperature region
the $\pi$-loop effects give the dominant contributions
to $f_\pi (T)$ and the $\rho$-loop effects are negligible.
However {\it by the vector manifestation
the $\rho$ contributions become essentially equal
to the one of $\pi$}, 
are then incorporated
into $f_\pi (T)$ near the critical temperature.


\section{Predictions of the Vector Manifestation}
\label{sec:PVM}

In this section,
we summarize the predictions of the VM in hot matter
studied in Refs.~\cite{HS:VM,HKRS:SUS,HS:VVD,Sasaki:2003qj,HKRS:PV}.

\subsection{Critical temperature}
\label{ssec:Tc}

In this subsection we estimate the value of the critical temperature
$T_c$
where the order parameter $f_\pi^2$ vanishes.

We first determine $T_c$
by naively extending
the expression (\ref{eq:f_piGL}) to the higher temperature region
to get $ T_c^{(\rm{hadron})} = 180\,\mbox{MeV}$ for $N_f = 3$.
However this naive extension is inconsistent with 
the chiral restoration in QCD
since the axial vector and vector current correlators do not agree with
each other at that temperature.
As is stressed in Ref.~\cite{HS:VM},
the disagreement between two correlators is cured by including
the intrinsic thermal effect.
As can be seen from Eq.~(\ref{eq:WMC A}),
the intrinsic temperature dependence of the parameter $F_\pi$
is determined from  
${\langle \frac{\alpha _s}{\pi}G_{\mu \nu}G^{\mu \nu}
\rangle}_T$ and  ${\langle \bar{q}q \rangle}_T $,
and gives only a small contribution compared with the main term
$1 + \frac{\alpha_s}{\pi}$.
However it is important that 
the parameters in the hadronic corrections
have the intrinsic temperature dependences as 
$(a,g) \to (1,0)$ for $T \to T_c$,
which carry the information of QCD. 
Actually the inclusion of the intrinsic thermal effects provides
the formula (\ref{fpi critical}) for
the pion decay constant at the critical temperature,
in which the second term has an extra factor of $1/2$ compared with
the one in Eq.~(\ref{eq:f_piGL}).

In Ref.~\cite{HS:VM} we determined the critical temperature
from $f_\pi (T_c) = 0$ and estimated the value which is dependent on
the matching scale $\Lambda$:
{}From Eq.~(\ref{eq:intrinsic}) 
we obtain
  \begin{equation}
   T_c = \sqrt{\frac{24}{N_f}}F_\pi (0;T_c).
  \end{equation}
Using Eqs.(\ref{eq:WMC A Tc}) and (\ref{eq:F_pi0}) we get~\cite{HS:VM}
  \begin{equation}
   \frac{T_c}{\Lambda} = \sqrt{\frac{3}{N_f {\pi}^2}}
         \Bigl[ 1 + \frac{\alpha _s}{\pi} +
                \frac{2{\pi}^2}{3}
                   \frac{\langle \frac{\alpha _s}{\pi}
                     G_{\mu \nu}G^{\mu \nu} \rangle_{T_c}}{{\Lambda}^4} -
         \frac{N_f}{4} \Bigr]^{\frac{1}{2}}. \label{eq:T_c}
  \end{equation}
We would like to stress that the critical temperature is expressed
in terms of the parameters appearing in the OPE.
We evaluate the critical temperature for $N_f=3$.
The gluonic condensate at $T_c$ is about half of the value
at $T=0$~\cite{Miller,Brown:2001nh}
and we use 
$\langle \frac{\alpha_s}{\pi}G_{\mu\nu}G^{\mu\nu} \rangle_{T_c}
= 0.006\,\mbox{GeV}^4$.
We show the predicted values of $T_c$ for several choices of
$\Lambda_{\rm{QCD}}$ and $\Lambda$ in Table \ref{fig:T_c}.
\begin{table}
 \begin{center}
  \begin{tabular}{|c||c|c|c|c|c|c|c|c|}
    \hline
    $\Lambda_{\rm{QCD}}$ & \multicolumn{3}{c}{} & 0.30 &
                           \multicolumn{3}{c}{} & 0.35 \\
    \hline
    $\Lambda$ & 0.8 & 0.9 & 1.0 & 1.1 &
                0.8 & 0.9 & 1.0 & 1.1 \\
    \hline
    $T_c$ & 0.20 & 0.20 & 0.22 & 0.23 &
            0.20 & 0.21 & 0.22 & 0.24 \\
    \hline
    \hline
    $\Lambda_{\rm{QCD}}$ & \multicolumn{3}{c}{} & 0.40 &
                           \multicolumn{3}{c}{} & 0.45 \\
    \hline
    $\Lambda$ & 0.8 & 0.9 & 1.0 & 1.1 &
                0.8 & 0.9 & 1.0 & 1.1 \\
    \hline
    $T_c$ & 0.21 & 0.22 & 0.23 & 0.25 &
            0.22 & 0.23 & 0.24 & 0.25 \\
    \hline
  \end{tabular}
 \end{center}
 \caption[The critical temperature]
          {Values of the critical temperature for several choices of
           $\Lambda_{\rm{QCD}}\ \mbox{and}\ \Lambda$.
           The units of $\Lambda_{\rm{QCD}}, \Lambda \ \mbox{and}\ 
           T_c$ are GeV.}
 \label{fig:T_c}
\end{table}
Note that the Wilsonian matching describes
the experimental results very well
for $\Lambda_{QCD}=0.4\,\mbox{GeV}$ and $\Lambda=1.1\,\mbox{GeV}$
at $T=0$~\cite{HY:PRep}.
At non-zero temperature, however,
the matching scale $\Lambda$ may be dependent 
on temperature.
The smallest $\Lambda$ in Table~{\ref{fig:T_c}} is determined 
by requiring 
$(2\pi^2/3) \,\langle \frac{\alpha_s}{\pi}G_{\mu\nu}G^{\mu\nu} 
\rangle/ \Lambda^4\,< 0.1$.

We should note that the above values may be changed when we adopt
a different way to estimate the matrix elements of operators 
in the OPE side,
e.g., an estimation with the dilute pion gas approximation~\cite{HKL}
or that by the lattice QCD calculation~\cite{lattice}. 
Even when we choose one way to estimate the matrix elements in the OPE, 
some temperature effects are supposed to be left out 
due to the truncation of the OPE to neglect higher order operators,
inclusion of which will cause a small change of the above values
of the critical temperature.
The important point is that as a result of the Wilsonian matching,
$T_c$ is obtained as in Eq.~(\ref{eq:T_c}) 
in terms of the quark and gluonic condensates, not hadronic degrees of
freedom.   It is expected that the value of $T_c$ may become smaller
than that obtained in this paper by including the higher order
corrections.

\subsection{Axial-vector and vector charge susceptibilities}
\label{sec:SUS}

In this subsection, we address the issue of what the relevant degrees
of freedom can be at the chiral transition induced by high
temperature and their possible implications on observables in
heavy-ion physics. In doing this, we focus on the vector and
axial-vector susceptibilities very near the critical temperature $T_c$.
The issue of what happens at high density is discussed in
\cite{LPRV}.

Susceptibility is related with the fluctuation of 
a conserved quantity.
The statistical expectation value of a conserved operator ${\cal O}$
is given by
\begin{equation}
 \langle {\cal O} \rangle
 = \frac{\mbox{Tr}\bigl[ {\cal O} e^{-({\cal H}-\mu{\cal O})/T} \bigr]}
        {Z},
\end{equation}
where ${\cal H}$ denotes the Hamiltonian of system and
$\mu$ is the chemical potential associated with ${\cal O}$
and we define the partition function $Z$ as
\begin{equation}
 Z = \mbox{Tr}\bigl[ e^{-({\cal H}-\mu{\cal O})/T} \bigr].
\end{equation}
The mean square deviation of ${\cal O}$ is
\begin{eqnarray}
 (\delta {\cal O})^2
 &\equiv& \langle {\cal O}^2 \rangle - \langle {\cal O} \rangle^2
 \nonumber\\
 &=& T \frac{\partial{\langle {\cal O} \rangle}}
           {\partial \mu}.
\end{eqnarray}
Then we define the susceptibility of ${\cal O}$ as follows:
\begin{eqnarray}
 \chi (T)
 &\equiv& \frac{\partial{\langle {\cal O} \rangle}}
              {\partial \mu} \big{|}_{\mu = 0} 
 \nonumber\\
 &=& \int_0^{1/T}d\tau \int d^3\vec{x} \langle {\cal O}(\tau,\vec{x})
     {\cal O}(0,\vec{0}) \rangle .
\end{eqnarray}

Consider the vector isospin susceptibility (VSUS) $\chi_V$
and the axial-vector isospin
susceptibility (ASUS) $\chi_A$
defined in terms of the vector charge density $V_0^a (x)$ and the
axial-vector charge density $A_0^a (x)$ by the Euclidean
correlators:
 \begin{eqnarray}
\delta^{ab}\chi_V&=&
\int^{1/T}_0 d\tau\int d^3\vec{x}\langle V_0^a
(\tau, \vec{x}) V_0^b (0,\vec{0})\rangle_\beta,\\
\delta^{ab}\chi_A&=& \int^{1/T}_0 d\tau\int d^3\vec{x}\langle A_0^a
(\tau, \vec{x}) A_0^b (0,\vec{0})\rangle_\beta
 \end{eqnarray}
where $\langle ~\rangle_\beta$ denotes thermal average and
 \begin{equation}
V_0^a\equiv \bar{\psi}\gamma^0\frac{\tau^a}{2}\psi, \ \
A_0^a\equiv \bar{\psi}\gamma^0\gamma^5\frac{\tau^a}{2}\psi
 \end{equation}
with the quark field $\psi$ and the $\tau^a$ Pauli matrix the
generator of the flavor $SU(2)$.

\begin{figure}
 \begin{center}
  \includegraphics[width = 12cm]{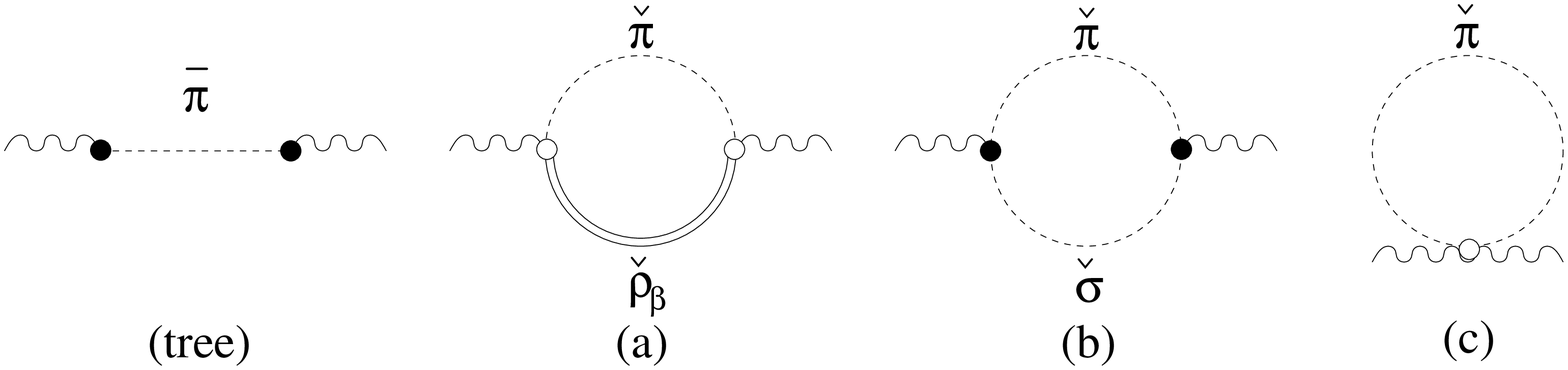}
 \end{center}
 \caption{Diagrams for contributions to $\chi_A$.}
 \label{fig:ASUS}
\end{figure}

\begin{figure}
 \begin{center}
  \includegraphics[width = 15cm]{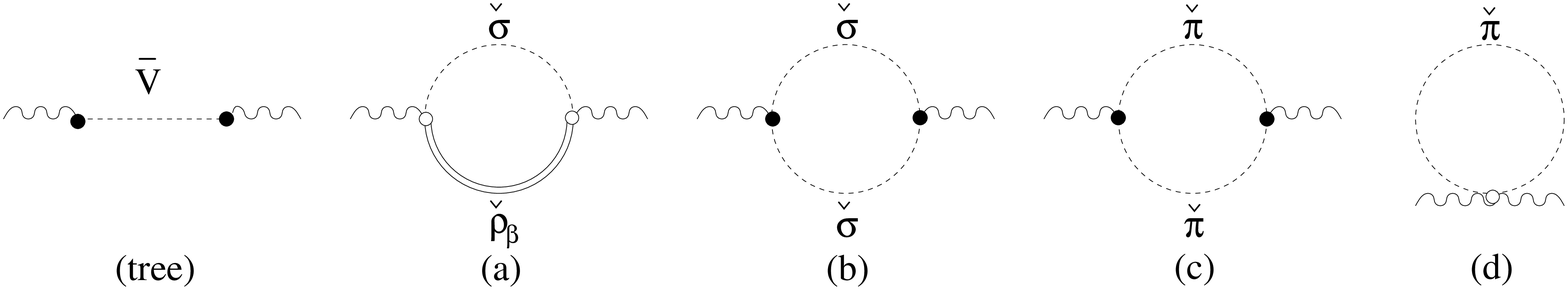}
 \end{center}
 \caption{Diagrams for contributions to $\chi_V$.}
 \label{fig:VSUS}
\end{figure}

The axial-vector susceptibility
$\chi_A(T)$ and the vector susceptibility $\chi_V(T)$ for
non-singlet currents~\footnote{We will confine ourselves to
non-singlet (that is, isovector) susceptibilities, so we won't
specify the isospin structure from here on.} are given by the
$00$-component of the axial-vector and vector current correlators
in the static--low-momentum limit:
\begin{eqnarray}
&&
\chi_A(T)  = 2 N_f \,
  \lim_{\bar{p}\rightarrow0}
  \lim_{p_0\rightarrow0}
  \left[ G_A^{00}(p_0,\vec{p};T) \right]
\ ,
\nonumber\\
&&
\chi_V(T)
= 2 N_f \, \lim_{\bar{p}\rightarrow0}
  \lim_{p_0\rightarrow0}
  \left[ G_V^{00}(p_0,\vec{p};T) \right]
\ ,
\label{def chiV}
\end{eqnarray}
where we have included the normalization factor of $2 N_f$. Using
the current correlators given in Eqs.~(\ref{eq:G_A}) 
and (\ref{eq:G_V}) 
and noting that $ \lim_{p_0\rightarrow0} P_L^{00} =
\lim_{p_0\rightarrow0} \bar{p}^2/p^2 = - 1 $, we can express
$\chi_A(T)$ and $\chi_V(T)$ as
\begin{eqnarray}
&&
\chi_A(T)  = - 2 N_f \,\lim_{\bar{p}\rightarrow0}
\lim_{p_0\rightarrow0}
\left[
  \Pi_\perp^L(p_0,\vec{p};T) - \Pi_\perp^t(p_0,\vec{p};T)
\right]
\ ,
\nonumber\\
&&
\chi_V(T)  = - 2 N_f \,\lim_{\bar{p}\rightarrow0}
\lim_{p_0\rightarrow0}
\left[
  \frac{
    \Pi_V^t \left( \Pi_V^L + 2 \Pi_{V\parallel}^L \right)
  }{
    \Pi_V^t - \Pi_V^L
  }
  + \Pi_{\parallel}^L
\right]
\ ,
\end{eqnarray}
where for simplicity of notation, we have suppressed the argument
$(p_0,\vec{p};T)$ in the right-hand-side of the expression for $\chi_V(T)$. 
In HLS theory at one-loop level, the susceptibilities read
\begin{eqnarray}
&&
\chi_A(T)  = 2 N_f \left[
  F_\pi^2(0)
  + \lim_{\bar{p}\rightarrow0}
  \lim_{p_0\rightarrow0}
  \left\{
    \bar{\Pi}_\perp^t(p_0,\vec{p};T) -
    \bar{\Pi}_\perp^L(p_0,\vec{p};T)
  \right\}
\right]
\ ,
\nonumber\\
&&
\chi_V(T)  =
- 2 N_f \,\lim_{\bar{p}\rightarrow0}
\lim_{p_0\rightarrow0}
\left[
  \frac{
    \left( a(0) F_\pi^2(0) + \bar{\Pi}_V^t \right)
    \left( \bar{\Pi}_V^L + 2 \bar{\Pi}_{V\parallel}^L \right)
  }{
    a(0) F_\pi^2(0) + \bar{\Pi}_V^t
    - \bar{\Pi}_V^L
  }
  + \Pi_{\parallel}^L
\right]
\ ,
\label{chiV}
\end{eqnarray}
where the parameter $a(0)$ is defined by (see section~\ref{sec:PaVVD})
\begin{eqnarray}
a(0) &=&
\frac{ \Pi_V^{{\rm(vac)}t}(p_0=0,\bar{p}=0) }{ F_\pi^2(0) }
= \frac{ \Pi_V^{{\rm(vac)}s}(p_0=0,\bar{p}=0) }{ F_\pi^2(0) }
\ .
\end{eqnarray}
We show the tree and one-loop contributions to $\chi_A$ and $\chi_V$
in Figs.~\ref{fig:ASUS} and~\ref{fig:VSUS}
It follows from the static--low-momentum limit of
$(\bar{\Pi}_\perp^t - \bar{\Pi}_\perp^L)$ given in Eq.~(\ref{PiA
tmL SL}) that the axial-vector susceptibility $\chi_A(T)$ takes
the form
\begin{eqnarray}
\chi_A(T) &=& 2N_f
\Biggl[
  F_\pi^2(0)
  - N_f \widetilde{J}_{1}^2(0;T)
  + N_f a\, \widetilde{J}_{1}^2(M_\rho;T)
\nonumber\\
&& \qquad
  {}- N_f \frac{a}{M_\rho^2}
  \left\{
    \widetilde{J}_{-1}^2(M_\rho;T)
    - \widetilde{J}_{-1}^2(0;T)
  \right\}
\Biggr]
\ .
\label{chiA}
\end{eqnarray}
Near the critical temperature ($T\rightarrow T_c$), we have
$M_\rho \rightarrow 0$, $a\rightarrow1$ due to the intrinsic
temperature dependence in the VM in hot
matter~\cite{HS:VM}.
Furthermore, from Eq.~(\ref{eq:intrinsic}), we see that the
parameter $F_\pi^2(0)$ approaches
$\frac{N_f}{24} T_c^2$ for $T\rightarrow T_c$. Substituting these
conditions into Eq.~(\ref{chiA}) and noting that
\begin{equation}
\lim_{M_\rho\rightarrow0}
\left[
  {}- \frac{1}{M_\rho^2}
  \left\{
    \widetilde{J}_{-1}^2(M_\rho;T)
    - \widetilde{J}_{-1}^2(0;T)
  \right\}
\right]
= \frac{1}{2} \widetilde{J}_{1}^2(0;T) = \frac{1}{24} T^2 \ ,
\end{equation}
we obtain
\begin{equation}
\chi_A(T_c) = \frac{N_f^2}{6} T_c^2
\ .
\label{axial SUS}
\end{equation}

To obtain the vector susceptibility near the critical temperature,
we first consider $ a(0) F_\pi^2(0) + \bar{\Pi}_V^t$ appearing in the
numerator of the first term in the right-hand-side of
Eq.~(\ref{chiV}). Using Eq.~(\ref{Pir t SL}), we get for the
static--low-momentum limit of $ a(0) F_\pi^2(0) + \bar{\Pi}_V^t$ as
\begin{eqnarray}
&&
\lim_{\bar{p}\rightarrow0}
\lim_{p_0\rightarrow0}
\left[ a(0) F_\pi^2(0) + \bar{\Pi}_V^t(p_0,\bar{p};T) \right]
\nonumber\\
&& \qquad
=
a(0) F_\pi^2(0)
- \frac{N_f}{4} \left[
  2 \widetilde{J}_{-1}^0(M_\rho;T)
  - \widetilde{J}_{1}^2(M_\rho;T)
  + a^2 \, \widetilde{J}_{1}^2(0;T)
\right]
\ .
\label{aF Pit SL}
\end{eqnarray}
{}From Eq.~(\ref{a0 expression}) we can see that $a(0) \rightarrow 1$ as
$T\rightarrow T_c$ since $F_\sigma^2(M_\rho) \rightarrow
F_\pi^2(0)$ and $M_\rho \rightarrow 0$. Furthermore,
$F_\pi^2(0)\rightarrow \frac{N_f}{24}T_c^2$ as we have shown in
Eq.~(\ref{eq:intrinsic}). Then, the first term of Eq.~(\ref{aF Pit SL})
approaches $\frac{N_f}{24}T_c^2$. The second term, on the other
hand, approaches $- \frac{N_f}{24}T_c^2$ as $M_\rho \rightarrow0$
and $a \rightarrow1$ for $T\rightarrow T_c$. Thus, we have
\begin{equation}
\lim_{\bar{p}\rightarrow0}
\lim_{p_0\rightarrow0}
\left[ a(0) F_\pi^2(0) + \bar{\Pi}_V^t(p_0,\bar{p};T) \right]
\mathop{\longrightarrow}_{T \rightarrow T_c}
0 \ .
\label{aF Pit 0}
\end{equation}
This implies that
only the second term $\Pi_\parallel^L$
in the right-hand-side of Eq.~(\ref{chiV}) contributes
to the vector susceptibility
near the critical temperature.
Thus, taking $M_\rho \rightarrow 0$ and $a\rightarrow1$,
in Eq.~(\ref{Piv L SL}),
we obtain
\begin{equation}
\chi_V(T_c) = \frac{N_f^2}{6} T_c^2
\ ,
\label{vector SUS}
\end{equation}
which agrees with the axial-vector susceptibility in
Eq.~(\ref{axial SUS}). This is a prediction, not an input
condition, of the theory. For $N_f=2$, we have
\begin{equation}
\chi_A(T_c) =
\chi_V(T_c) = \frac{2}{3} T_c^2
\ .
\label{SUS value}
\end{equation}
The result $\chi_V(T_c) = \frac{2}{3} T_c^2$ 
is consistent with the lattice result as
interpreted in \cite{Brown-Rho:96}. It is interesting to note that
the RPA result obtained in \cite{Brown-Rho:96} in NJL model in
terms of a quasi-quark-quasi-antiquark bubble is reproduced
quantitatively by the one-loop graphs in HLS with the VM.

One might think that the VSUS diverges at $T_c$ because of
the massless $\rho$ meson pole.
However the resultant VSUS is finite, 
and it will imply that the screening mass of $\rho$ meson is also finite.

It should be noticed that the pion pole effect does not
contribute to the ASUS in Eq.~(\ref{axial SUS})
since the pion decay constant
$f_\pi^t$ vanishes at the critical temperature 
as we have shown in section~\ref{sec:PDCPV}, and that
the contribution to the ASUS
comes from the
non-pole contribution expressed in Fig.~\ref{fig:ASUS}.
In three diagrams, the third diagram in Fig.~\ref{fig:ASUS}(c)
is proportional to $(1-a)$ and then it vanishes at the
critical temperature due to the VM.
Similarly, since the transverse $\rho$ decouples
at the critical point in the VM~\cite{HY:VM,HS:VM},
the first diagram in Fig.~\ref{fig:ASUS}(a) does not contribute.
Thus,
the above result for the ASUS in Eq.~(\ref{axial SUS})
comes from only the contribution
generated via the (longitudinal) vector
meson plus pion loop [see Fig.~\ref{fig:ASUS}(b)].
Similarly,
the vector meson pole effect to the VSUS
vanishes at the critical
temperature as shown in Eq.~(\ref{aF Pit 0}),
and the contribution
to the VSUS is generated via the pion loop
[Fig.~\ref{fig:VSUS}(c)]
and
the longitudinal vector meson loop
[Fig.~\ref{fig:VSUS}(b)].~\footnote{%
  Note that the contribution from Fig.~\ref{fig:VSUS}(a) vanishes since
  the transverse $\rho$ decouples and that the one from
  Fig.~\ref{fig:VSUS}(d) also vanishes since it is proportional
  to $(1-a)$.
} Since the longitudinal vector meson becomes massless, degenerate
with the pion as the chiral partner in the VM, loop contribution
to the ASUS becomes identical to that to the VSUS. Thus, the
massless vector meson predicted by the VM fixed point plays an
essential role to obtain the above equality between the ASUS and
the VSUS.

In the present analysis, our aim is
to show the qualitative structure of the ASUS and the VSUS
in the VM,
i.e., {\it the equality between them is predicted by the VM}.
In order to compare our qualitative results with the lattice result,
we need to
go beyond the one-loop approximation.
We note here that there is a
result from a hard thermal loop calculation which gives $\chi_V
(T_c)\approx 1.3 T_c^2$~\cite{CMT}. However this result cannot be
compared to ours for two reasons. First we need to sum higher
loops in our formalism which may be done in random phase
approximation as in \cite{Kunihiro:91}. Second, the perturbative
QCD with a hard thermal loop approximation may not be valid in the
temperature regime we are considering. Even at $T\gg T_c$, the
situation is not clear as pointed out in Ref.~\cite{blaizot}.

\subsection{Violation of vector dominance}
\label{sec:PaVVD}

In this subsection,
we shed some light on the validity of the vector dominance (VD)
of electromagnetic form factor of the pion 
near the critical temperature.
In several analyses such as the one on the dilepton spectra 
in hot matter carried out
in Ref.~\cite{Rapp-Wambach:00}, the VD is assumed to be held even in
the high temperature region.
There are several analyses resulting in the dropping mass consistent
with the VD, as shown in Refs.~\cite{GomezNicola:2002tn,Dobado:2002xf}.
On the other hand,
the analysis done in Ref.~\cite{Pisarski} shows that 
the thermal vector meson mass goes up if the VD holds.
Thus, it is interesting to study what the VM predicts on the VD.
In the present analysis we
present a new prediction of the VM in hot matter
on the direct photon-$\pi$-$\pi$ coupling which measures the validity
of the VD of the electromagnetic form factor of the
pion. 
We find that {\it the VM predicts a large violation of the VD at
the critical temperature}.
This indicates that the assumption of 
the VD may need to be weakened, at least in some amounts,
for consistently including the
effect of the dropping mass of the vector meson.

In Ref.~\cite{Harada:2001rf} 
it has been shown that VD is 
accidentally satisfied in $N_f=3$ QCD at zero temperature and zero
density, and that it is largely 
violated in large $N_f$ QCD when the VM occurs.
At non-zero temperature there exists the hadronic thermal correction
to the parameters.
Thus it is nontrivial whether or not the
VD is realized in hot matter,
especially near the critical temperature.
Here we will show that the intrinsic temperature dependences
of the parameters
of the HLS Lagrangian play essential roles, and then
the VD is largely violated near the critical temperature.

Let us study the validity of the VD near the critical temperature.
As we have shown in section~\ref{sec:PDCPV},
near the critical temperature
$\Pi_\perp^t$ and $\Pi_\perp^s$
in Eqs.~(\ref{at expression}) and (\ref{as expression})
approach the following expressions:
\begin{eqnarray}
 \overline{\Pi}_\perp^t (\bar{q},\bar{q};T)
  \stackrel{T \to T_c}{\to} {}- \frac{N_f}{24}T^2 , \nonumber\\
 \overline{\Pi}_\perp^s (\bar{q},\bar{q};T)
  \stackrel{T \to T_c}{\to} {}- \frac{N_f}{24}T^2 .
\label{VM limits Pit Pis}
\end{eqnarray}
On the other hand, the functions
$\Pi_\parallel^{t}$ and
$\Pi_\parallel^{s}$ in Eq.~(\ref{PiVts zero limits})
in the limit of $M_\rho/T \rightarrow0$ and $a\rightarrow1$
become
\begin{equation}
 \bar{\Pi }^{t}_\parallel(0,0;T)
 = \bar{\Pi }^{s}_\parallel(0,0;T)
 \, \rightarrow \, - \frac{N_f}{2} \tilde{I}_{2}(T) 
 = - \frac{N_f}{24} T^2 \ .
\label{VM limit PiVts}
\end{equation}
Furthermore, from Eq.~(\ref{a0 expression}), the parameter
$a(0)$ approaches $1$ in the limit of $M_\rho\rightarrow0$ and
$F_\sigma^2(M_\rho)/F_\pi^2(0)\rightarrow1$:
\begin{equation}
a(0) \rightarrow 1 \ .
\label{VM limit a0}
\end{equation}
{}From the above limits in Eqs.~(\ref{VM limits Pit Pis}),
(\ref{VM limit PiVts})
and (\ref{VM limit a0}), the numerators of 
$a^t(\bar{q};T)$ and $a^s(\bar{q};T)$ 
in Eqs.~(\ref{at expression}) and (\ref{as expression})
behave as
\begin{eqnarray}
 \overline{\Pi}_\parallel^t (0,0;T) - a(0)\overline{\Pi}_\perp^t
 (\bar{q},\bar{q};T) \to 0, \nonumber\\
 \overline{\Pi}_\parallel^s (0,0;T) - a(0)\overline{\Pi}_\perp^s
 (\bar{q},\bar{q};T) \to 0.
\end{eqnarray}
Thus 
we obtain
 \begin{equation}
  a^t(\bar{q};T), a^s(\bar{q};T) 
   \stackrel{T \to T_c}{\to} 1 \ ,
 \end{equation}
namely, the vector dominance is largely
violated near the critical temperature.

\subsection{Pion velocity}
\label{sec:PV}

In this subsection, 
we focus on the pion velocity at the critical temperature
and study the quantum and hadronic thermal effects
based on the VM.
The pion velocity is one of the important quantities 
since it is a dynamical object, 
which controls
the pion propagation in medium through the dispersion relation.

\subsubsection{Non-renormalization theorem on pion velocity at $T_c$}

As we mentioned in chapter~\ref{ch:EFTRG},
the intrinsic temperature dependence generates the effect of Lorentz
symmetry breaking at bare level.
Then how does the Lorentz non-invariance at bare level influence
physical quantities?
In order to make it clear, in this section
we study the pion decay constants and the pion velocity 
near the critical temperature.

Following subsection~\ref{ssec:PP-lowT},
we define the on-shell of the pion from the pole
of the longitudinal component $G_A^L$
of the axial-vector current correlator.
This pole structure is expressed
by temporal and spatial components of the two-point function 
$\Pi_\perp^{\mu\nu}$. 
The temporal and spatial pion decay constants are expressed as follows
~\cite{HKRS:SUS,HS:VVD}:
\begin{eqnarray}
 \qquad
 &\bigl( f_\pi^t (\bar{p};T) \bigr)^2 
  = \Pi_\perp^t(V_\pi\bar{p},\bar{p};T),&
\nonumber\\
 & f_\pi^t (\bar{p};T) f_\pi^s (\bar{p};T) 
  = \Pi_\perp^s(V_\pi\bar{p},\bar{p};T),&
\end{eqnarray}
where the on-shell condition $p_0 \to V_\pi\bar{p}$ was taken.
We divide the two-point function $\Pi_\perp^{\mu\nu}$ into two parts, 
zero temperature (vacuum) and non-zero temperature parts, as $
\Pi_\perp^{\mu\nu} = \Pi_\perp^{(\rm vac)\mu\nu} +
   \bar{\Pi}_\perp^{\mu\nu}$.
The quantum correction is included in the vacuum part
$\Pi_\perp^{(\rm vac)\mu\nu}$, and the hadronic thermal correction
is in $\bar{\Pi}_\perp^{\mu\nu}$. In the present
perturbative analysis, we obtain the pion velocity
as~\cite{HS:VVD}
\begin{eqnarray}
 v_\pi^2 (\bar{p};T)
 &=& \frac{f_\pi^s (\bar{p};T)}
          {f_\pi^t (\bar{p};T)}
\nonumber\\
 &=& V_\pi^2 + \widetilde{\Pi}_\perp (V_\pi\bar{p},\bar{p})
 {}+ \frac{\bar{\Pi}_\perp^s (V_\pi\bar{p},\bar{p};T) -
    V_\pi^2\bar{\Pi}_\perp^t (V_\pi\bar{p},\bar{p};T)}
   {\bigl( F_\pi^t \bigr)^2}\,,
\label{v_pi-def}
\end{eqnarray}
where $\widetilde{\Pi}_\perp (V_\pi\bar{p},\bar{p})$ denotes
a possible finite renormalization effect.
Note that the renormalization condition on $V_\pi$ is
determined as $\widetilde{\Pi}_\perp (V_\pi\bar{p},\bar{p})|_{\bar{p}=0}=0$.

In the following, we study the quantum and hadronic corrections to
the pion velocity for $T \to T_c$, on the assumption
of the VM conditions~(\ref{evm}). As we defined above, the
two-point function associated with the pion velocity
$v_\pi(\bar{p};T)$ is $\Pi_\perp^{\mu\nu}(p_0,\bar{p};T)$.
The diagrams contributing to $\Pi_\perp^{\mu\nu}$ are shown 
in Fig.~\ref{fig:AA}.
As mentioned in subsection~\ref{sec:VMC}, 
diagram (a) is proportional to $g_L$ and
diagram (c) has the factor $(a^t - 1)$.
Then these contributions vanish at the critical point. 
We consider the contribution from diagram (b) only.

\begin{flushleft}
\underline{\bf {\it Quantum correction at $T_c$}}
\end{flushleft}

First we evaluate the quantum correction to the vacuum part
$\Pi_\perp^{(\rm vac)(b)\mu\nu}$.
This is expressed as
\begin{eqnarray}
 \Pi_\perp^{\rm{(vac)}(b)\mu\nu}(p_0,\bar{p})
 = N_f \int\frac{d^n k}{i(2\pi)^n}
   \frac{\Gamma^\mu(k;p) \Gamma^\nu(-k;-p)}
    {[-k_0^2 + V_\pi^2 \bar{k}^2]
     [M_\rho^2 - (k_0 - p_0)^2 + V_\sigma^2 |\vec{k} - \vec{p}|^2]},
\label{Pi-AA-vac-b}
\nonumber\\
\end{eqnarray}
where $\Gamma^\mu$ denotes the $\bar{\cal A}\check{\pi}\check{\sigma}$
vertex as
\begin{eqnarray}
 \Gamma^\mu(k;p)
  = \frac{i}{2}\sqrt{a^t}\,g_{\bar{\mu}}^\mu
    [ u^{\bar{\mu}}u^{\bar{\nu}}
{}+
    V_\sigma^2 (g^{\bar{\mu}\bar{\nu}}-u^{\bar{\mu}}u^{\bar{\nu}}) ]
    (2k-p)_{\bar{\nu}}.
\label{vertex}
\end{eqnarray}
We note that the spatial component of this vertex $\Gamma^i$ has
an extra-factor $V_\sigma^2$ as compared with the temporal one. In
the present analysis it is important to include the quadratic
divergences to obtain the RGEs in the Wilsonian sense
as we have already seen in chapter~\ref{ch:WMFT} and~\ref{ch:VMHM}.  
In this section, when we evaluate four dimensional integral,
we first integrate over $k_0$ from $-\infty$ to $\infty$.
Then we carry out the integral over three-dimensional momentum
$\vec{k}$
with three-dimensional cutoff $\Lambda_3$.
In order to be consistent with ordinary regularization in
four dimension~\cite{HY:conformal,HY:WM,HY:PRep},
we use the following replacement associated with 
quadratic divergence:
\begin{eqnarray}
&&\qquad\qquad
 \Lambda_3  \, \to \, \frac{1}{\sqrt{2}}\Lambda_4
            = \frac{1}{\sqrt{2}}\Lambda,
\\
&& \int\frac{d^{n-1} \bar{k}}{(2\pi)^{n-1}}\frac{1}{\bar{k}}
  \to \frac{\Lambda^2}{8\pi^2}, 
\qquad
 \int\frac{d^{n-1} \bar{k}}{(2\pi)^{n-1}}
   \frac{\bar{k}^i \bar{k}^j}{\bar{k}^3}
  \to -\delta^{ij}\frac{\Lambda^2}{8\pi^2}.
\end{eqnarray}
When we make these replacement,
the present method of integral preserves chiral symmetry.

\begin{figure}
 \begin{center}
  \includegraphics[width = 12cm]{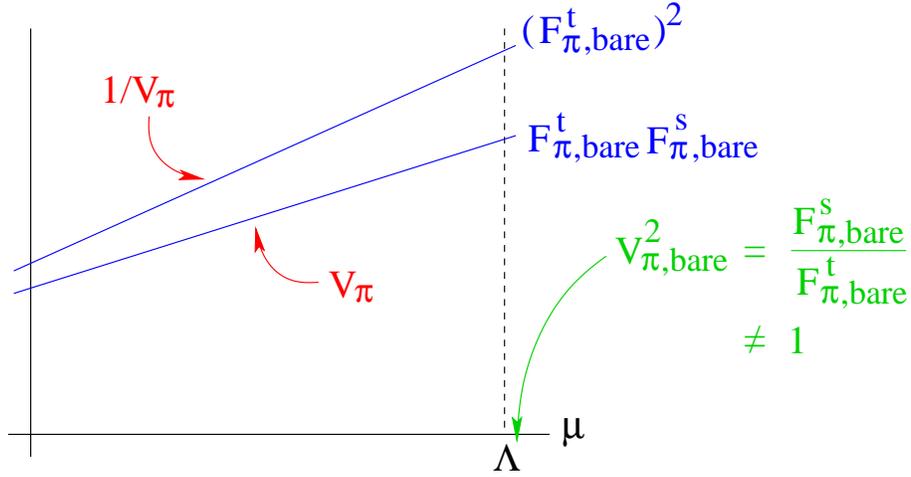}
 \end{center}
 \caption{Quadratic running of $(F_\pi^t)^2$ and 
  $F_\pi^t F_\pi^s$ at $T_c$.}
 \label{fig:Fpi-ts-RGE}
\end{figure}

As shown in Appendix A, 
$\Pi_\perp^{{\rm (vac)}t}$ and $\Pi_\perp^{{\rm (vac)}s}$ are 
independent of the external momentum.
Accordingly, the finite renormalization effect $\widetilde{\Pi}_\perp$ 
is also independent of the external momentum and then vanishes:
\begin{equation}
\qquad
 \widetilde{\Pi}_\perp (V_\pi \bar{p},\bar{p}) = 0.
\label{fre}
\end{equation}
Thus in the following,
we take the external momentum as zero.
In that case, the temporal and spatial components of
$\Pi_\perp^{{\rm (vac)}\mu\nu}$ are expressed as
$\Pi_\perp^{\rm (vac)t} = \Pi_\perp^{\rm (vac)00}$ and
$\Pi_\perp^{\rm (vac)s} = - (\delta^{ij}/3)\Pi_\perp^{{\rm (vac)}ij}$.
Taking the VM limit ($M_\rho \to 0$ and $V_\sigma \to V_\pi$),
these components become
\begin{eqnarray}
 \lim_{\rm VM}
 \Pi_\perp^{\rm{(vac)}00}(p_0 = \bar{p} = 0)
&=& \frac{N_f}{4}\int\frac{d k_0 d^{n-1}\bar{k}}{i(2\pi)^n}
   \frac{4k_0^2}{[-k_0^2 + V_\pi^2 \bar{k}^2]^2} \nonumber\\
&=& - \frac{N_f}{4}\int\frac{d^{n-1}\bar{k}}{(2\pi)^{n-1}}
     \frac{1}{V_\pi \bar{k}} \nonumber\\
&=& -\frac{N_f}{4}\frac{1}{V_\pi}\frac{\Lambda^2}{8\pi^2},
\nonumber\\
 \lim_{\rm VM}
 \Pi_\perp^{\rm{(vac)}ij}(p_0 = \bar{p} = 0)
&=& - \frac{N_f}{4} (V_\pi^2)^2
   \int\frac{d k_0 d^{n-1}\bar{k}}{i(2\pi)^n}
   \frac{4\bar{k}^i \bar{k}^j}{[-k_0^2 + V_\pi^2 \bar{k}^2]^2}
\nonumber\\
&=& - \frac{N_f}{4}\, V_\pi^4
   \int\frac{d^{n-1}\bar{k}}{(2\pi)^{n-1}}
   \frac{\bar{k}^i \bar{k}^j}{(V_\pi \bar{k})^3} \nonumber\\
&=& \frac{N_f}{4}\,V_\pi\,\delta^{ij}\,\frac{\Lambda^2}{8\pi^2}.
\end{eqnarray}
Thus we obtain the temporal and spatial parts as
\begin{eqnarray}
&&
 \lim_{\rm VM}
 \Pi_\perp^{\rm{(vac)}t}(p_0 = \bar{p} = 0) 
 = -\frac{N_f}{4}\frac{1}{V_\pi}\frac{\Lambda^2}{8\pi^2},
\nonumber\\
&&
 \lim_{\rm VM}
 \Pi_\perp^{\rm{(vac)}s}(p_0 = \bar{p} = 0) 
 = -\frac{N_f}{4}\,V_\pi\,\frac{\Lambda^2}{8\pi^2}.
\label{qc}
\end{eqnarray}
These quadratic divergences are renormalized by
$(F_{\pi,{\rm bare}}^t)^2$ and
$F_{\pi,{\rm bare}}^t F_{\pi,{\rm bare}}^s$, respectively.
Then RGEs for the parameters $(F_\pi^t)^2$ and
$F_\pi^t F_\pi^s$ are expressed as
\begin{eqnarray}
&&
 \mu \frac{d\bigl( F_\pi^t \bigr)^2}{d\mu}
 = \frac{N_f}{(4\pi)^2}\frac{1}{V_\pi}\mu^2,
\label{rge-tt}\\
&&
 \mu \frac{d\bigl( F_\pi^t F_\pi^s \bigr)}{d\mu}
 = \frac{N_f}{(4\pi)^2}V_\pi\,\mu^2.
\label{rge-ts}
\end{eqnarray}
Both $(F_\pi^t)^2$ and $F_\pi^t F_\pi^s$ scale following the quadratic
running $\mu^2$.
However, the coefficient of $\mu^2$ in the RGE for $(F_\pi^t)^2$
is different from that for $F_\pi^t F_\pi^s$ 
(see Fig.~\ref{fig:Fpi-ts-RGE}).

When we use these RGEs, the scale dependence of the parametric 
pion velocity is
\begin{eqnarray}
 \mu \frac{dV_\pi^2}{d\mu}
 &=& \mu \frac{d\bigl( F_\pi^t F_\pi^s / (F_\pi^t)^2 \bigr)}{d\mu}
 \nonumber\\
 &=& \frac{1}{\bigl(F_\pi^t\bigr)^4}\frac{N_f}{(4\pi)^2}
     \Bigl[ V_\pi \bigl(F_\pi^t\bigr)^2 - F_\pi^t F_\pi^s
      \frac{1}{V_\pi} \Bigr]\mu^2 \nonumber\\
 &=& 0\,. \label{v-scale}
\end{eqnarray}
This implies that {\it the parametric pion velocity at the critical
temperature does not scale.}
In other words, the Lorentz non-invariance at bare level is not
enhanced through the RGEs as long as we consider the pion velocity.
As we noted below Eq.~(\ref{vertex}),
the factor $V_\sigma^2$ is in the spatial component of the vertex
$\Gamma^\mu$.
If $V_\sigma$ were not equal to $V_\pi$,
the coefficients of running in the right-hand-side of
Eqs.~(\ref{rge-tt}) and (\ref{rge-ts}) would change.
However, since the VM conditions do guarantee $V_\sigma = V_\pi$,
the quadratic running caused from $\Lambda^2$ in $(F_\pi^t)^2$
and $F_\pi^t F_\pi^s$ are exactly canceled in the second line of
Eq.~(\ref{v-scale}).

\begin{flushleft}
\underline{\bf {\it Hadronic thermal correction at $T_c$}}
\end{flushleft}

Next we study the hadronic thermal correction to the pion velocity 
at the critical temperature. The temporal and spatial parts of the
hadronic thermal correction $\bar{\Pi}_\perp^{\mu\nu}$ contribute
to the pion velocity, which have the same structure as those of
the quantum correction $\Pi_\perp^{(\rm vac)\mu\nu}$, except for a
Bose-Einstein distribution function. Thus by the replacement of
$\Lambda^2 / (4\pi)^2$ with $T^2 / 12$ in $\Pi_\perp^{\rm
(vac)t,s}$, hadronic corrections to the temporal and spatial parts
of $\bar{\Pi}_\perp^{\mu\nu}$ are obtained as follows:
\begin{eqnarray}
&&
 \lim_{\rm VM}
 \bar{\Pi}_\perp^{t}(p_0,\bar{p};T)
 = -\frac{N_f}{24}\frac{1}{V_\pi}T_c^2,
\nonumber\\
&&
 \lim_{\rm VM}
 \bar{\Pi}_\perp^{s}(p_0,\bar{p};T)
 = -\frac{N_f}{24}\,V_\pi\,T_c^2.
\label{hc}
\end{eqnarray}

Substituting Eq.~(\ref{hc}) into Eq.~(\ref{v_pi-def})
with $\widetilde{\Pi}_\perp = 0$ as shown in Eq.~(\ref{fre}),
we obtain the physical pion velocity in the VM as
\begin{eqnarray}
 v_{\pi}^2(\bar{p};T)
&\stackrel{T \to T_c}{\to}&
    V_\pi^2
   {}+ \frac{\bar{\Pi}_\perp^s (V_\pi\bar{p},\bar{p};T_c) -
    V_\pi^2\bar{\Pi}_\perp^t (V_\pi\bar{p},\bar{p};T_c)}
   {\bigl( F_\pi^t \bigr)^2}
\nonumber\\
&\,=\,&
   V_\pi^2.
\label{velocity}
\end{eqnarray}
Since the parametric pion velocity in the VM does not scale
with energy [see Eq.~(\ref{v-scale})],
$V_\pi$ in the above expression is equivalent
to the bare pion velocity:
\begin{equation}
\qquad
 v_\pi(\bar{p};T) = V_{\pi,{\rm bare}}(T)
 \qquad \mbox{for}\, T \to T_c.
\label{phys=bare}
\end{equation}
This implies that {\it the pion velocity in the limit $T \to T_c$
receives neither hadronic nor quantum corrections due
to the protection by the VM.}

In order to clarify the reason of this non-renormalization property,
let us recall the fact that only the diagram (b) 
in Fig.~\ref{fig:AA} contributes to the pion velocity 
at the critical temperature.
Away from the critical temperature, the contribution of the massive 
$\sigma$ (i.e., the longitudinal mode of massive vector meson) is 
suppressed owing to the Boltzmann factor $\exp [-M_\rho / T]$, 
and then only the pion loop contributes to the pion velocity.
Then there exists the ${\cal O}(T^4)$ correction to the pion velocity
~\cite{HS:VVD} (see subsection~\ref{ssec:PP-lowT}).
Near the critical temperature, on the other hand,
$\sigma$ becomes massless due to the VM
since $\sigma$ (i.e., the longitudinal vector meson) becomes 
the chiral partner of the pion.
Thus the absence of the hadronic corrections in the pion velocity
at the critical temperature is due to 
the exact cancellation between the contribution of pion and that 
of its chiral partner $\sigma$.
Similarly the quantum correction generated from the pion loop 
is exactly canceled by that from the $\sigma$ loop.

\begin{flushleft}
\underline{\bf {\it Non-renormalization property}}
\end{flushleft}

Now we consider the meaning of our result~(\ref{phys=bare}). Based
on the point of view that the bare HLS theory is defined from QCD,
we presented the VM conditions realizing the chiral symmetry in
QCD consistently, i.e, $(g_L,a^t,a^s) \to (0,1,1)$ for $T \to
T_c$. This is the fixed point of the RGEs for the parameters
$g_L, a^t$ and $a^s$.  
Although both pion decay constants $(F_\pi^t)^2$ 
and $F_\pi^t F_\pi^s$ scale following the quadratic running, 
$(F_\pi^t)^2$ and $F_\pi^t F_\pi^s$ show a different running 
since the coefficient of $\mu^2$ in Eq.~(\ref{rge-tt}) is
different from that in Eq.~(\ref{rge-ts}).
Nevertheless in the
pion velocity at the critical temperature, the quadratic running
in $(F_\pi^t)^2$ is exactly canceled by that in $F_\pi^t F_\pi^s$
[see second line of Eq.~(\ref{v-scale})]. There it was crucial for
intricate cancellation of the quadratic running that the velocity
of $\sigma$ (i.e., longitudinal vector meson) is equal to its
chiral partner, i.e., $V_\sigma \to V_\pi$ for $T \to T_c$. Note
that this is not an extra condition but a consequence from the VM
conditions for $a^t$ and $a^s$; we started simply from the VM
conditions alone and found that $V_\pi$ does not receive quantum
corrections at the restoration point. 
As we showed in Eq.~(\ref{hc}), the
hadronic correction to $(F_\pi^t)^2$ is different from that to
$F_\pi^t F_\pi^s$. 
In the pion velocity, however,  
the hadronic correction from $(F_\pi^t)^2$ is exactly
cancelled by that from $F_\pi^t F_\pi^s$ [see second line of
Eq.~(\ref{velocity})]. The VM conditions guarantee these exact
cancellations of the quantum and hadronic corrections. 
This implies that $(g_L,a^t,a^s,V_\pi) = (0,1,1,\mbox{any})$ forms
a fixed line for four RGEs of $g_L, a^t, a^s$ and $V_\pi$.
When one point on this fixed line is selected through the matching
procedure as done in Ref.~\cite{HKRS:PV},
namely the value of $V_{\pi,{\rm bare}}$ is fixed,
the present result implies that the point does not move in a subspace
of the parameters. 
This is likely the manifestation of a new fixed point 
in the Lorentz non-invariant formulation of the VM. 
Approaching the restoration point of chiral symmetry, 
the physical pion velocity itself would flow into the fixed point.

We should distinguish the consequences within HLS/VM from those
beyond HLS/VM.
Clearly the determination of the definite value of
the bare pion velocity is done outside HLS/VM.
On the other hand, our main result~(\ref{phys=bare}) holds
independently of the value of the bare pion velocity itself.
Applying this result to the case
where one starts from the bare HLS theory with Lorentz invariance,
i.e., $V_{\pi,{\rm bare}}=1$,
one finds that the pion velocity at $T_c$ becomes the speed of light since
$v_\pi = V_{\pi,{\rm bare}} = 1$,
as obtained in Ref.~\cite{HKRS:SUS}.

As a consequence of the relation~(\ref{phys=bare}),
we can determine the temporal and spatial pion decay constants
at the critical temperature when we take the bare pion velocity
as finite.
In the following, we study these decay constants
and discuss their determinations based on Eq.~(\ref{phys=bare}).
Using Eq.~(\ref{v-scale}), we solve the RGEs~(\ref{rge-tt})
and (\ref{rge-ts}) and
obtain a relation between two parametric pion decay constants as
$F_\pi^t(0;T_c)F_\pi^s(0;T_c) = V_\pi^2 \bigl( F_\pi^t(0;T) \bigr)^2$.
From this and (\ref{hc}),
the temporal and spatial pion decay constants with the quantum and
hadronic corrections are obtained as
\begin{eqnarray}
\quad
 \bigl( f_\pi^t \bigr)^2
 &=& \Bigl( F_\pi^t(0;T_c)\Bigr)^2 -
\frac{N_f}{24}\frac{1}{V_\pi}T_c^2,
\nonumber\\
 f_\pi^t\,f_\pi^s
 &=& F_\pi^t(0;T_c)F_\pi^s(0;T_c) - \frac{N_f}{24}V_\pi\,T_c^2
\nonumber\\
 &=& V_\pi^2 \bigl( f_\pi^t \bigr)^2.
\end{eqnarray}
Since the order parameter $(f_\pi^t f_\pi^s)$ vanishes as expected
at the critical temperature, we find that $ f_\pi^t f_\pi^s =
V_\pi^2 \bigl( f_\pi^t \bigr)^2 = 0\,$. Multiplying both side by
$v_\pi^2 = V_\pi^2$, the above expression is reduced to
\begin{equation}
\qquad
 (f_\pi^s)^2 = V_\pi^4 (f_\pi^t)^2 = 0.
\label{s-t}
\end{equation}
Now, the spatial pion decay constant vanishes at the critical
temperature, $ f_\pi^s (T_c) = 0\,$. In the case of a vanishing
pion velocity, $f_\pi^t$ can be finite at the restoration point.
On the other hand, when $V_\pi$ is finite, Eq.~(\ref{s-t}) leads
to $f_\pi^t(T_c)=0$. Thus we find that both temporal and spatial
pion decay constants vanish simultaneously at the critical
temperature when the bare pion velocity is determined as finite.

In order to know the value of the (bare) pion velocity, we need to
specify a method that determines the bare parameters. 

As we will show below,
the analysis performed
on the basis of a Wilsonian matching gives the finie 
bare pion velocity at the critical temperature, i.e.,
$V_{\pi,{\rm bare}} \neq 0$.
Thus, by combining
Eq.~(\ref{phys=bare}) with estimation of $V_{\pi,{\rm bare}}$,
the value of the physical pion velocity $v_\pi(T)$ at
the critical temperature is obtained to be finite~\cite{HKRS:PV}.

\subsubsection{Estimation}

Now we evaluate the physical pion velocity at
the critical temperature $T_c$ starting from the Lorentz
non-invariant bare Lagrangian. According to the
non-renormalization theorem, 
the $bare$ velocity so
calculated should correspond to the $physical$ pion velocity $at$
the chiral transition point. Now, in the VM, bare parameters are
determined by matching the HLS to QCD at the matching scale
$\Lambda$ and at temperature $T=T_c$.

We begin with a summary of the pion
velocity found in
 the HLS/VM theory with Lorentz invariance
~\cite{HKRS:SUS, Sasaki:2003qj, HS:VVD}.
The pion velocity is given
by Eq.~(\ref{fpts rels}):
\begin{equation}
v_\pi^2(\bar{p};T) =
 1 + \frac{\tilde{F}(\bar{p};T)}{F_\pi^2}
  \Bigl[ \mbox{Re}f_\pi^s (\bar{p};T) - \mbox{Re}f_\pi^t (\bar{p};T) 
  \Bigr].
\end{equation}
The VM dictates that if one ignores Lorentz non-invariance in the
bare Lagrangian in medium, the pion velocity approaches the speed
of light as $T\rightarrow T_c$~\cite{HKRS:SUS, HS:VVD}.

In the following, we extend the matching condition valid at low
temperature, Eq.~(\ref{deviation-rho}), to near the critical
temperature, and determine the bare pion velocity at $T_c$. 
We should in principle evaluate
the matrix elements in terms of QCD variables only in order for
performing the Wilsonian matching, which is as yet unavailable from
model-independent QCD calculations.  Therefore, we make an
estimation by extending the dilute gas approximation adopted in
the QCD sum-rule analysis in the low-temperature region to the
critical temperature with including all the light degrees of
freedom expected in the VM. In the HLS/VM theory, both the
longitudinal and transverse $\rho$ mesons become massless at the
critical temperature since the HLS gauge coupling constant $g$
vanishes. At the critical point, the longitudinal $\rho$ meson
which becomes the NG boson $\sigma$ couples to the vector current
whereas the transverse $\rho$ mesons decouple from the theory
because of the vanishing $g$. Thus we assume that thermal
fluctuations of the system are dominated near $T_c$ not only by
the pions but also by the longitudinal $\rho$ mesons. In
evaluating the thermal matrix elements of the non-scalar operators
in the OPE, we extend the thermal pion gas approximation employed
in Ref.~\cite{HKL} to the longitudinal $\rho$ mesons that figure
in our approach~\cite{HKRS:PV}. 
This is feasible since at the critical
temperature, we expect the equality $A_4^\rho(T_c) = A_4^\pi(T_c)$
to hold as the massless $\rho$ meson is the chiral partner of the
pion in the VM~\footnote{
 We observe from Refs.~\cite{best, SMRS} that
 $A_4^{\pi (u+d)}(\mu=2.4 ~{\rm GeV})\sim  A_4^{\rho
  (u+d)}(\mu=2.4~{\rm GeV})$ even at zero temperature.
}. It should be noted that, although we use the dilute gas
approximation, the treatment here is already beyond the
low-temperature approximation adopted in Eq.~(\ref{dpa}) because
the contribution from $\rho$ meson is negligible in the
low-temperature region. Since we treat the pion as a massless
particle in the present analysis, it is reasonable to take
$A_4^\pi(T) \simeq A_4^\pi(T=0)$. We therefore use
\begin{equation}
 A_4^\rho(T) \simeq A_4^{\pi}(T) \simeq A_4^\pi(T=0)
 \quad \mbox{for}\quad T \simeq T_c.
\label{matrix Tc}
\end{equation}

The properties of the scalar operators giving rise to the
condensates are fairly well understood at chiral restoration. We
know that the quark condensate must be zero at the critical
temperature. Furthermore the value of the gluon condensate at the
phase transition is known from lattice calculations to be roughly
half of the one in the free space~\cite{Miller}. We therefore can
use in what follows the following values at $T=T_c$:
 \begin{eqnarray}
 \langle\bar q q\rangle_T= 0,~~~\langle
\frac{\alpha_s}{\pi} G^2\rangle_T\sim 0.006 {\rm GeV}^4~.
\label{Tcondensates}
 \end{eqnarray}
Including the contributions from both pions and massless $\rho$
mesons, Eq.~(\ref{L(1)exp}) can be expressed as
 \begin{eqnarray}
 G_A^{{\rm (OPE)}L(1)}
  = \frac{32}{105}\pi^4\frac{T^6}{Q^8}\Biggl[ A_4^{\pi (u+d)}+  A_4^{\rho (u+d)} \Biggr].\label{L(1)-pi-rho-exp}
 \end{eqnarray}
Therefore from Eq.~(\ref{deviation-rho}), we obtain the deviation
$\delta_{\rm bare}$ as
\begin{equation}
 \delta_{\rm bare} = 1 - V_{\pi,{\rm bare}}^2
 = \frac{1}{G_0}
   \frac{32}{105}\pi^4\frac{T^6}{\Lambda^6}\Biggl[ A_4^{\pi (u+d)}+  A_4^{\rho (u+d)} \Biggr].
\label{deviation-pi-rho}
\end{equation}
This is the matching condition to be used for determining the
value of the bare pion velocity near the critical temperature.

To make a rough estimate of $\delta_{\rm bare}$, we use $ A_4^{\pi
(u+d)}(\mu=1~{\rm GeV}) =0.255$~\cite{HKL}. This value is arrived
at by following Appendix B of ~\cite{HKL}. $A_n^{\pi (q)}$ is
defined by
 \begin{eqnarray}
\langle\pi| \bar q\gamma_{\mu_1} D_{\mu_2}...
D_{\mu_ n}q (\mu)|\pi\rangle =(-i)^{n-1} (p_{\mu_1}...p_{\mu_n} -{\rm
traces})A_n^{\pi (q)} (\mu)~, 
\end{eqnarray} 
where 
\begin{eqnarray}
A_n^{\pi (q)}(\mu)=
2\int_0^1 dx x^{n-1}[q(x,\mu)+(-1)^n \bar q (x,\mu)]~.
 \end{eqnarray}
For any charge state of the pion ($\pi^0, \pi^+, \pi^-$),
$A_n^{\pi (u+d)}$ ($n=2,4$) can be written in terms of the $n$th
moment of valence quark distribution $V_n^\pi (\mu)$ and sea quark
distribution $S_n^\pi (\mu)$~\cite{HKL},
 \begin{eqnarray}
A_n^{\pi
(u+d)}(\mu)=4V_n^\pi (\mu) +8S_n^\pi (\mu)~, 
\end{eqnarray}
where 
\begin{eqnarray}
V_n^\pi
&=&\int_0^1 dx x^{n-1} v^\pi (x,\mu),
\nonumber\\
S_n^\pi &=&\int_0^1 dx
x^{n-1} s^\pi (x,\mu).
 \end{eqnarray}
Simple parameterizations of the valence distribution $ v^\pi
(x,\mu)$ and the sea distribution $s^\pi (x,\mu)$ can be found in
Ref. \cite{GRV}-- see Ref.~\cite{GRS} for the updated results --
where the parton distributions in the pions are determined through
the $\pi$-N Drell-Yan and direct photon production processes. With
the leading-order parton distribution functions given in
\cite{GRV}, we obtain $A_4^{\pi (u+d)}=0.255 ~{\rm at}~\mu=1 ~{\rm
GeV}$~\cite{HKL}. For the purpose of comparison with the lattice
QCD result~\cite{best}, we need to calculate the value at $
\mu=2.4 ~{\rm GeV}$; it comes out to be $A_4^{\pi (u+d)}=0.18$.
The value $A_4^{\pi (u+d)}=0.18$ is slightly bigger than the $\sim
0.13$ calculated by the lattice QCD~\cite{best}, while it is a bit
smaller than the $\sim 0.22$ of Ref.~\cite{SMRS}. Note that the
value $\sim  0.18$ agrees with the one determined by lattice QCD
~\cite{best} within quoted errors. Using $A_2^{\pi (u+d)}=0.972$
and $A_4^{\pi (u+d)}=0.255$ and for the range of matching scale
$(\Lambda = 0.8 - 1.1\, \mbox{GeV})$, that of QCD scale
$(\Lambda_{QCD} = 0.30 - 0.45\, \mbox{GeV})$ and critical
temperature $(T_c = 0.15 - 0.20\, \mbox{GeV})$, we get
\begin{equation}
 \delta_{\rm bare}(T_c) = 0.0061 - 0.29\,,
\end{equation}
where the $\Lambda$ dependence of $A_{2,4}^{\pi (u+d)}$ is
ignored as it is expected to be suppressed by more than
$1/\Lambda^6$. Thus we find the $bare$ pion velocity to be close
to the speed of light:
\begin{equation}
 V_{\pi,{\rm bare}}(T_c) = 0.83 - 0.99\,.
\end{equation}
Thanks to the non-renormalization theorem, 
i.e., $v_\pi (T_c)=V_{\pi, {\rm bare}}(T_c)$, 
we arrive at the physical pion velocity at chiral
restoration~\cite{HKRS:PV}:
\begin{equation}
 v_{\pi}(T_c) = 0.83 - 0.99\,.
\end{equation}

This is in contrast to the result obtained from the chiral theory
~\cite{SS},
where the relevant degree of freedom near $T_c$ is only the pion.
Their result is that the pion velocity becomes zero for $T \to T_c$.
Therefore by experiment and lattice analysis,
we may be able to distinguish which theory is appropriate
to describe the chiral phase transition.


\section{Critical Behavior of Vector Meson Mass}
\label{sec:SPVMM}

We showed that the vector meson pole mass $m_\rho(T)$ goes
to zero in the limit $T \to T_c$ and the chiral symmetry is restored
as the VM in the framework of the HLS theory.
In this section, 
we study how $m_\rho (T)$ falls in the limit $T \to T_c$.

We expand the vector current correlator $G_V^{\rm (HLS)}$ 
at the matching scale around $T_c$ taking the limits 
$g(\Lambda;T) \ll 1$, $a(\Lambda;T) - 1 \ll 1$ and
$M_\rho (\Lambda;T)/\Lambda \ll 1$.
Then the quantity $G_A - G_V$ in the HLS sector is obtained as
\begin{eqnarray}
&&
 G_A^{\rm (HLS)}(\Lambda;T) - G_V^{\rm (HLS)}(\Lambda;T)
\nonumber\\
&&\qquad\qquad\qquad
 \sim g^2(\Lambda;T)
      \Biggl( \frac{F_\pi^2(\Lambda;T)}{\Lambda^2} \Biggr)^2
 {}-  \Bigl( a(\Lambda;T) -1 \Bigr)
      \frac{F_\pi^2 (\Lambda;T)}{\Lambda^2}\,,
\label{scaling-HLS}
\end{eqnarray}
where we neglected the higher order coefficients.
The same quantity using the OPE has the following form:
\begin{equation}
 G_A^{\rm (OPE)}(\Lambda;T) - G_V^{\rm (OPE)}(\Lambda;T)
 = \frac{4\pi (N_c^2 - 1)}{N_c^2}
   \frac{\alpha_s \langle \bar{q}q \rangle^2_T}{\Lambda^6}\,.
\label{scaling-OPE}
\end{equation}
Here we assume that all the terms of Eq.~(\ref{scaling-HLS})
have same scaling behaviors near $T_c$. 
{}From Eqs.~(\ref{scaling-HLS}) and (\ref{scaling-OPE}),
we obtain the scaling behavior on the gauge coupling constant as
\begin{equation}
 g(\Lambda;T) \sim \langle \bar{q}q \rangle_T\,.
\end{equation}
As shown in section~\ref{sec:VMM},
the vector meson pole mass $m_\rho(T)$ near $T_c$ is
\begin{equation}
 m_\rho^2(T) \simeq g^2(T)\Bigl( a(T)F_\pi^2 + \delta T^2 \Bigr)\,,
\end{equation}
where $\delta T^2$ denotes the hadronic thermal correction
and gives a positive contribution [see Eq.~(\ref{eq:Mass2})].
Thus $m_\rho(T)$ vanishes as $\langle \bar{q}q \rangle_T$:
\begin{equation}
 m_\rho(T) \sim \langle \bar{q}q \rangle_T\,.
\end{equation}
If $\langle \bar{q} q \rangle_T^2$ falls as $f_\pi^2(T)$ near $T_c$,
then the $\rho$ pole mass $m_\rho^2(T)$ vanishes as $f_\pi^2(T)$.
In such a case the scaling property in the VM may be consistent 
with the Brown-Rho scaling 
$m_\rho(T)/m_\rho(0) \sim f_\pi(T)/f_\pi(0)$~\cite{BR}.

\newpage

\chapter{Chiral Doubling of Heavy-Light Mesons}
\label{ch:CDHLM}

Based on the manifestation of chiral symmetry \`a la linear sigma
model, it was predicted a decade
ago~\cite{Nowak-Rho-Zahed:93,Bardeen-Hill:94} that the mass
splitting $\Delta M$ between the ${\cal M} (0^-, 1^-)$ and
$\tilde{\cal M}(0^+,1^+)$ mesons where ${\cal M}$ denotes a
heavy-light meson consisting of heavy quark $Q$ and light
antiquark $\bar{q}$ should be of the size of the constituent quark
mass. Recently, BaBar~\cite{BABAR}, CLEO~\cite{CLEO} and
subsequently the Belle collaboration~\cite{Belle}, discovered new
$D$ mesons with $Q=c$, $c$ being charm quark, which most likely
have spin-party $0^+$ and $1^+$ and the mass difference to the
$D(0^-, 1^-)$ is in fair agreement with the prediction of
\cite{Nowak-Rho-Zahed:93,Bardeen-Hill:94}. 
In Ref.~\cite{Nowak-Rho-Zahed:03}, it was proposed that the
splitting of ${\cal M}$ and $\tilde{\cal M}$ mesons could carry direct
information on the property of chiral symmetry at some critical
density or temperature at which the symmetry is
restored~\footnote{In what follows, unless otherwise specified, we
will refer to heavy-light mesons generically as $D$ but the
arguments should apply better to heavier-quark mesons. Numerical
estimates will however be made solely for the (open charm) $D$
mesons.}. 
In this chapter, we pick up this idea and make a first step
in consolidating the proposal of Ref.~\cite{Nowak-Rho-Zahed:03}
following Ref.~\cite{HRS:CD}.
In doing this, we shall take the reverse direction: Instead of
starting with a Lagrangian defined in the chiral-symmetry broken
phase and then driving the system to the chiral symmetry
restoration point by an external disturbance, we will start from
an assumed structure of chiral symmetry at its restoration point
and then make a prediction as to what happens to the splitting in
the broken phase. We find that the splitting is directly
proportional to the light-quark condensate and comes out to be of
the size of the constituent quark mass consistent with the
prediction of Refs.~\cite{Nowak-Rho-Zahed:93,Bardeen-Hill:94}. 
We shall associate this result as giving a link between the assumed
structure of the chiral restoration point and the broken phase.

Our procedure is following:
We assume that the heavy-light hadrons are described by a VM-fixed
point theory at the chiral
restoration point generically denoted $C_\chi$
(critical temperature $T_c$~\cite{HS:VM}
or density $n_c$~\cite{HKR:VM}
or number of flavors $N_f^c$~\cite{HY:VM})
and by introducing the simplest form of
the VM breaking terms, we compute the mass splitting of the chiral
doublers in matter-free space in terms of the quantities that
figure in the QCD correlators~\footnote{
  Introducing vector mesons in the light-quark sector
  of heavy-light mesons was considered in
  Ref.~\cite{Nowak-Rho-Zahed:93} but without
  the matching to QCD and hence without the VM fixed point.
}.


\section{Heavy Quark Symmetry}
\label{sec:HQS}

Like the approximate light quark flavor symmetry (chiral symmetry)
in the light quark sector of QCD,
the heavy quark symmetry is also approximate in the situation
where the heavy quark mass $m_Q$ is much larger than 
$\Lambda_{\rm QCD}$, $\Lambda_{\rm QCD}/m_Q \ll 1$.
The heavy quark symmetry arises in the limit of QCD
where the heavy quark, charm and bottom, masses $m_Q$ are
taken to be infinity.
In this limit, dynamics of the heavy quark $Q$ is independent
of its mass and spin~\cite{Isgur:1989vq}.
In this section,
we give a brief review of the heavy quark symmetry and
the multiplet of hadrons containing a single heavy quark.

In the limit where the heavy quark mass goes to infinity, 
$m_Q \to \infty$, the four velocity of the heavy quark $v_\mu$
is fixed.
The size of the meson is of order $1/\Lambda_{\rm QCD}$,
and then the typical momentum scale of the light degrees of freedom
is of order $\Lambda_{\rm QCD}$.
We decompose the four momentum of the heavy quark into
\begin{equation}
 p_\mu = m_Q v_\mu + k_\mu \,,
\label{residual} 
\end{equation}
where $k_\mu$ is called the residual momentum and is of order 
$\Lambda_{\rm QCD}$.
The propagator of the heavy quark is given by
\begin{equation}
 \frac{- (p_\mu \gamma^\mu + m_Q)}{p^2 - m_Q^2 + i\epsilon}\,.
\end{equation}
Substituting Eq.~(\ref{residual}) into the propagator,
we obtain the leading form as follows:
\begin{equation}
 \frac{-1}{v \cdot k + i\epsilon}\,,
\end{equation}
where we moved the projection operator $(v_\mu \gamma^\mu + 1)/2$
taking out of the large component.
We should note that it is independent of the heavy quark mass
$m_Q$ and the gamma matrices are completely disappeared.
Similarly the vertices are also independent of $m_Q$ and gamma matrices.
The heavy quark is sitting with infinite mass $m_Q \to \infty$,
and thus the spin of the heavy quark does not flip at all
by the interaction of the light degrees of freedom 
which is of order $\Lambda_{\rm QCD}$.
Thus the absence of gamma matrices indicates that the spin of
heavy quark is conserved.
Hence if there are $N_f$ heavy quarks with the same four velocity,
the effective heavy quark theory has a $SU(2N_f)$ spin-flavor 
symmetry~\cite{Isgur:1989vq}.

Let us consider the heavy-light meson multiplet.
For the hadrons containing a single heavy quark $Q$,
there is a $SU(2)$ spin symmetry between $Q$ with up-spin, 
$Q(\uparrow)$ and $Q$ with down-spin, $Q(\downarrow)$.
Thus the hadrons with the spin combined $\uparrow$ and 
$\downarrow$ with the spin of the light degrees of freedom $s_{\rm light}$
are the spin partners each other.
In general for each $s_{\rm light}$,
there is a degenerate doublet with the total spin $s_+$ and $s_-$:
\begin{equation}
 s_{\pm} = s_{\rm light} \pm \frac{1}{2}\,.
\end{equation}
Hadron in the ground state has $s_{\rm light}=1/2$ and
negative parity and then the hadron with $J^P = 0^-$ is the spin partner
of the one with $J^P = 1^-$.


\section{An Effective Field Theory}

In this section we give our reasoning that leads to the Lagrangian
that defines our approach.
Here we construct the Lagrangian using the approximate chiral
$SU(3)_{\rm L}\times SU(3)_{\rm R}$
symmetry in the light-quark sector and the
heavy quark symmetry in the heavy-quark sector.
We will start
from the Lagrangian given at the vector manifestation
(VM) fixed point.
We first describe
how to construct the fixed point Lagrangian based on the VM.
Then,
we account for
the effect of spontaneous chiral symmetry breaking by adding a
$bare$ parameter for the mass splitting in the heavy sector and
including the deviation of the HLS parameters from the values
at the VM fixed point.
The explicit form of the Lagrangian so
constructed is shown in subsection~\ref{ssec:Lag}.

\subsection{The fixed point Lagrangian}
\label{ssec:FPL}

Let us consider the VM at the point at which chiral symmetry
is restored (in the chiral limit). At the VM at its fixed point
characterized by $(g,a)=(0,1)$, the two 1-forms become
\begin{eqnarray}
  {\alpha}_{\parallel\mu} =
  \frac{1}{2i} \left(
    {\partial}_\mu\xi_{\rm R} \cdot \xi_{\rm R}^\dag +
    {\partial}_\mu \xi_{\rm L} \cdot \xi_{\rm L}^\dag
  \right)
\ ,
\nonumber\\
  {\alpha}_{\perp\mu} =
  \frac{1}{2i} \left(
    {\partial}_\mu \xi_{\rm R} \cdot \xi_{\rm R}^\dag -
    {\partial}_\mu \xi_{\rm L} \cdot \xi_{\rm L}^\dag
  \right)
\ .
\end{eqnarray}
Note that the above ${\alpha}_{\parallel\mu}$ and
${\alpha}_{\perp\mu}$ do not
contain the HLS gauge field since the gauge coupling $g$ vanishes
at the VM fixed point. It is convenient to define the (L,R)
1-forms:
\begin{eqnarray}
{\alpha}_{R\mu} &=&
  {\alpha}_{\parallel\mu} + {\alpha}_{\perp\mu}
  = \frac{1}{i} \partial_\mu \xi_{\rm R} \cdot \xi^\dag_{\rm R}
\ ,
\nonumber\\
{\alpha}_{L\mu} &=&
  {\alpha}_{\parallel\mu} - {\alpha}_{\perp\mu}
  = \frac{1}{i} \partial_\mu \xi_{\rm L} \cdot \xi^\dag_{\rm L}
\ ,
\label{def:1forms}
\end{eqnarray}
which can be regarded as belonging to the chiral representation
$(1,8)$ and $(8,1)$, respectively, transforming under chiral
$SU(3)_{\rm L} \times SU(3)_{\rm R}$ as
\begin{eqnarray}
{\alpha}_{R\mu} \rightarrow
g_{\rm R} {\alpha}_{R\mu} g_{\rm R}^\dag
\ ,
\nonumber\\
{\alpha}_{L\mu} \rightarrow
g_{\rm L} {\alpha}_{L\mu} g_{\rm L}^\dag
\ .
\end{eqnarray}
By using these 1-forms, the HLS Lagrangian at the VM fixed point
can be written as~\cite{HY:PRep}
\begin{eqnarray}
{\mathcal L}_{\rm light}^{\ast}
=
\frac{1}{2} F_\pi^2 \,\mbox{tr}\left[
  {\alpha}_{R\mu} {\alpha}_{R}^\mu
\right]
+
\frac{1}{2} F_\pi^2 \,\mbox{tr}\left[
  {\alpha}_{L\mu} {\alpha}_{L}^\mu
\right]
\ ,
\label{light fixed Lag}
\end{eqnarray}
where the $*$ affixed to the
Lagrangian denotes that it is a fixed-point Lagrangian,
and $F_\pi$ denotes the bare pion decay constant.
Note that the physical pion decay
constant $f_\pi$ vanishes at the VM fixed point by the quadratic
divergence although the bare one is non-zero~\cite{HY:PRep}. It
should be stressed that the above fixed point Lagrangian is
approached only as a limit of chiral symmetry
restoration~\cite{HY:PRep}.

Next we consider the fixed-point Lagrangian of the heavy meson
sector at the chiral restoration point identified with the VM
fixed point. Let us introduce two heavy-meson fields
${\mathcal H}_R$ and
${\mathcal H}_L$ transforming under chiral
$\mbox{SU}(3)_{\rm R} \times\mbox{SU}(3)_{\rm L}$ as
\begin{equation}
{\mathcal H}_R \rightarrow {\mathcal H}_R \, g_{\rm R}^\dag \ ,
\quad
{\mathcal H}_L \rightarrow {\mathcal H}_L \, g_{\rm L}^\dag \ .
\end{equation}
By using these fields together with
the light-meson 1-forms $\alpha_{L,R}^\mu$,
the fixed point Lagrangian of the heavy
mesons
is expressed as~\footnote{
  We assign the right chirality to ${\mathcal H}_R$, and the left
  chirality to ${\mathcal H}_L$. Then the interaction term has the
  right and left projection operators. Note that the insertion of
  $(1\pm\gamma_5)$
  to kinetic and mass termes does not cause any difference.
}
\begin{eqnarray}
{\mathcal L}_{\rm heavy}^{\ast}
&=&
- \mbox{tr} \left[
  {\mathcal H}_{R} i v_\mu \partial^\mu \bar{\mathcal H}_{R}
\right]
- \mbox{tr} \left[
  {\mathcal H}_{L} i v_\mu \partial^\mu \bar{\mathcal H}_{L}
\right]
\nonumber\\
&&
{} + m_0 \,\mbox{tr} \left[
  {\mathcal H}_R \bar{\mathcal H}_R
  + {\mathcal H}_L \bar{\mathcal H}_L
\right]
\nonumber\\
&&
{} + 2k \,\mbox{tr}\left[
  {\mathcal H}_R {\alpha}_{R\mu} \gamma^\mu \frac{1+\gamma_5}{2}
  \bar{\mathcal H}_R
  +
  {\mathcal H}_L {\alpha}_{L\mu} \gamma^\mu \frac{1-\gamma_5}{2}
  \bar{\mathcal H}_L
\right]
\ ,
\label{heavy fixed Lag}
\end{eqnarray}
where $v_\mu$ is the velocity of heavy meson, $m_0$ represents the
mass generated by the interaction between heavy quark and the
``pion cloud'' surrounding the heavy quark, and $k$ is a real
constant to be determined.

\subsection{Effects of spontaneous chiral symmetry breaking}
\label{ssec:TESCSB}

Next we consider what happens in the broken phase of chiral
symmetry.
In the real world at low temperature and low density, chiral
symmetry is spontaneously broken by the non-vanishing quark
condensate. In the scenario of chiral-symmetry manifestation \`a
la linear sigma model, the effect of spontaneous chiral symmetry
breaking is expressed by the vacuum expectation value
(VEV) of the scalar fields.
In the VM,
on the other hand, it is signaled by the HLS
Lagrangian departing from the VM fixed point:
There the gauge
coupling constant $g \neq 0$~\footnote{
  Actually, near the chiral restoration point, the
  Wilsonian matching between HLS and QCD dictates~\cite{HY:PRep}
  that (in the chiral limit) the HLS gauge coupling be proportional
  to the quark condensate:
  $g \sim \langle \bar{q} q \rangle$.
  See section~\ref{sec:SPVMM}.
}
and we have the kinetic term of the
HLS gauge bosons ${\mathcal L}_{\rho {\rm kin}} = - \frac{1}{2}
\mbox{tr} \left[
  \rho_{\mu\nu} \rho^{\mu\nu}
\right]$.
The derivatives in the HLS 1-forms become the covariant
derivatives and then $\alpha_{L\mu}$ and $\alpha_{R\mu}$ are
covariantized:
\begin{eqnarray}
 &\partial_\mu \to D_\mu = \partial_\mu - ig\rho_\mu,&
\nonumber\\
 &\alpha_{R\mu}
  \to \hat{\alpha}_{R\mu} = \alpha_{R\mu} - g\rho_\mu,&
\nonumber\\
 &\alpha_{L\mu}
  \to \hat{\alpha}_{L\mu} = \alpha_{L\mu} - g\rho_\mu.&
\label{covariant}
\end{eqnarray}
These 1-forms transform as $\hat{\alpha}_{R(L)\mu} \to h\,
\hat{\alpha}_{R(L)\mu}h^\dagger$ with
$h \in [SU(3)_{\rm V}]_{\rm local}$
as shown in Eq.~(\ref{transform}).

Although $a=1$ at the VM fixed point, generally $a \neq 1$ in the
broken phase. We therefore expect to have a term of the form
$\frac{1}{2}(a-1)F_\pi^2 \mbox{tr}[\hat{\alpha}_{L\mu}
\hat{\alpha}_R^\mu]$. Thus the Lagrangian for the light mesons
takes the following form:
\begin{eqnarray}
 {\cal L}_{\rm light}
 &=& \frac{a+1}{4}F_\pi^2 \mbox{tr}
  [ \hat{\alpha}_{R\mu}\hat{\alpha}_R^\mu
    {}+ \hat{\alpha}_{L\mu}\hat{\alpha}_L^\mu ]
\nonumber\\
 &&\qquad
  {}+ \frac{a-1}{2}F_\pi^2 \mbox{tr}
   [ \hat{\alpha}_{R\mu}\hat{\alpha}_L^\mu ]
  {}+ {\cal L}_{\rho{\rm kin}}.
\end{eqnarray}
By using $\hat{\alpha}_{\parallel\mu}$ and $\hat{\alpha}_{\perp\mu}$
given in Eq.~(\ref{alpha}), this Lagrangian is rewritten as
\begin{equation}
{\cal L}_{\rm light} =
  F_\pi^2 \mbox{tr}[ \hat{\alpha}_{\perp\mu}\hat{\alpha}_\perp^\mu ]
 {}+ F_\sigma^2 \mbox{tr}[ \hat{\alpha}_{\parallel\mu}
     \hat{\alpha}_\parallel^\mu ]
 {}+ {\cal L}_{\rho {\rm kin}},
\end{equation}
which is nothing but the general HLS Lagrangian.

We next consider the spontaneous breaking of chiral symmetry in
the heavy-meson sector. One of the most important effects of the
symmetry breaking is to generate the mass splitting between the
odd parity multiplet and the even parity
multiplet~\cite{Nowak-Rho-Zahed:93}. This effect can be
represented by the Lagrangian of the form:
\begin{eqnarray}
{\mathcal L}_{\chi{\rm{SB}}}
= \frac{1}{2}
\Delta M \,\mbox{tr}\left[
  {\mathcal H}_{L} \bar{{\mathcal H}}_R +
  {\mathcal H}_R \bar{{\mathcal H}}_L
\right]
\ ,
\label{bare Delta M}
\end{eqnarray}
where ${\mathcal H}_{R(L)}$ transforms under the HLS as
${\mathcal H}_{R(L)} \rightarrow {\mathcal H}_{R(L)}\,h^\dag$.
Here
$\Delta M$ is the $bare$ parameter corresponding to the mass
splitting between the two multiplets. An important point of our
work is that the bare $\Delta M$ can be determined by matching the
EFT with QCD as we will show in
section~\ref{sec:DMD}:
The
matching actually shows that $\Delta M$ is proportional to the
quark condensate:
\begin{equation}
\Delta M \sim \langle \bar{q} q \rangle.
\label{bare mass diff 0}
\end{equation}

\subsection{Lagrangian in parity eigenfields}
\label{ssec:Lag}

In order to compute the mass splitting between ${\cal M}$ and
$\tilde{\cal M}$, it is convenient to go to the corresponding
fields in parity eigenstate, $H$ (odd-parity) and $G$
(even-parity) as defined, e.g., in Ref.~\cite{Nowak-Rho-Zahed:03};
\begin{eqnarray}
{\mathcal H}_R &=&
  \frac{1}{\sqrt{2}} \left[ G - i H \gamma_5 \right] \ ,
\nonumber\\
{\mathcal H}_L &=&
  \frac{1}{\sqrt{2}} \left[ G + i H \gamma_5 \right] \ .
\end{eqnarray}
Here, the pseudoscalar meson $P$ and the vector meson $P_\mu^{\ast}$
are included in the $H$ field as
\begin{eqnarray}
H &=& \frac{1+ v_\mu \gamma^\mu }{2} \left[
  i \gamma_5 P + \gamma^\mu P_\mu^{\ast}
\right]
\ ,
\label{H}
\end{eqnarray}
and the scalar meson $Q^{\ast}$ and the axial-vector meson
$Q_\mu$ are in $G$ as
\begin{eqnarray}
G &=& \frac{1+ v_\mu \gamma^\mu }{2} \left[
  Q^\ast - i \gamma^\mu \gamma_5 Q_\mu
\right]
\ .
\label{G}
\end{eqnarray}
In terms of the $H$ and $G$ fields, the heavy-meson Lagrangian
{\it off} the VM fixed point is of the form
\begin{equation}
 {\cal L}_{\rm heavy} = {\cal L}_{\rm kin} + {\cal L}_{\rm int}
 \ ,
\end{equation}
with
\begin{eqnarray}
{\mathcal L}_{\rm kin} &=&
 \mbox{tr} \left[ H \,
   \left( i v_\mu D^\mu - M_H \right)
 \bar{H} \right]
 -
 \mbox{tr} \left[ G \,
   \left( i v_\mu D^\mu - M_G \right)
 \bar{G} \right]
\ ,
\\
{\cal L}_{\rm int}
 &=& k\, \Biggl[ \mbox{tr}[ H \gamma_\mu \gamma_5
     \hat{\alpha}_\perp^\mu \bar{H}]
 {}- \mbox{tr}[ H v_\mu
     \hat{\alpha}_\parallel^\mu \bar{H}]
\nonumber\\
 &&\qquad
 {}+ \mbox{tr}[ G \gamma_\mu \gamma_5
     \hat{\alpha}_\perp^\mu \bar{G}]
 {}+ \mbox{tr}[ G v_\mu
     \hat{\alpha}_\parallel^\mu \bar{G}]
\nonumber\\
 &&\qquad
 {}- i\mbox{tr}[ G \hat{\alpha}_{\perp\mu} \gamma^\mu
        \gamma_5 \bar{H}]
 {}+ i\mbox{tr}[ H \hat{\alpha}_{\perp\mu} \gamma^\mu
        \gamma_5 \bar{G}]
\nonumber\\
 &&\qquad
 {}- i\mbox{tr}[ G \hat{\alpha}_{\parallel\mu}
      \gamma^\mu \bar{H}]
 {}+ i\mbox{tr}[ H \hat{\alpha}_{\parallel\mu}
      \gamma^\mu \bar{G}] \Biggr],
\label{Lag:component}
\end{eqnarray}
where
the covariant derivatives acting on $\bar{H}$ and $\bar{G}$
are defined as
\begin{equation}
D_\mu \bar{H} = \left( \partial_\mu - i g \rho_\mu \right) \bar{H}
\ , \quad
D_\mu \bar{G} = \left( \partial_\mu - i g \rho_\mu \right) \bar{G}
\ .
\end{equation}
In the above expression, $M_H$ and $M_G$ denote the masses of
the parity-odd multiplet $H$ and the parity-even multiplet $G$,
respectively.
They are related to $m_0$ and $\Delta M$ as
\begin{eqnarray}
M_H &=& - m_0 - \frac{1}{2} \Delta M \ , \nonumber\\
M_G &=& - m_0 + \frac{1}{2} \Delta M \ .
\end{eqnarray}
The mass splitting between $G$ and $H$ is therefore given by
\begin{equation}
M_G - M_H = \Delta M \ .
\end{equation}


\section{Matching to the Operator Product Expansion}
\label{sec:MOPE}

The bare parameter $\Delta M_{{\rm bare}}$ which carries
information on QCD should be determined by matching the EFT
correltators to QCD ones. We are concerned with the pseudoscalar
correlator $G_P$ and the scalar correlator $G_S$. In the EFT
sector, the correlators at the matching scale are of the
form~\footnote{Here and in the rest of this chapter, the heavy meson
is denoted $D$ with the open charm heavy meson in mind. However
the arguments (except for the numerical values) are generic for
all heavy mesons ${\cal M}$.}
\begin{eqnarray}
 &&
 G_P(Q^2) = \frac{F_D^2 M_D^4}{M_D^2 + Q^2},
\nonumber\\
 &&
 G_S(Q^2) = \frac{F_{\tilde{D}}^2 M_{\tilde{D}}^4}
  {M_{\tilde{D}}^2 + Q^2},
\end{eqnarray}
where $F_D$ ($F_{\tilde{D}}$) denotes the $D$-meson
($\tilde{D}$-meson) decay constant and the space-like momentum
$Q^2 = (M_{D} + \Lambda_M)^2$ with $\Lambda_M$ being the matching scale.
We note that the heavy quark limit $M_D \to \infty$
should be taken with $\Lambda_M$ kept fixed
since $\Lambda_M$ must be smaller than the chiral symmetry breaking
scale
characterized by $\Lambda_\chi \sim 4\pi f_\pi$.
Then, $Q^2$ should be regarded as $Q^2 \simeq M_D^2$
in the present framework based on the chiral and heavy quark symmetries.
If we ignore the difference between $F_D$ and
$F_{\tilde{D}}$ which can be justified by the QCD sum rule
analysis~\cite{Narison:2003td}, then we get
 \begin{equation}
\Delta_{SP}(Q^2) \equiv G_S(Q^2) - G_P(Q^2)\simeq \frac{3 F_D^2
M_D^3}{M_D^2 + Q^2}
   \Delta M_D.
\label{diff-eft}
 \end{equation}
In the QCD sector, the correlators $G_S$ and $G_P$ are given by
the operator product expansion (OPE) as~\cite{Narison:1988ep}
\begin{eqnarray}
  G_S(Q^2) &=&
  \left. G(Q^2) \right\vert_{\rm pert}\nonumber\\
   &+&
  \frac{m_H^2}{m_H^2 + Q^2}
  \Biggl[ {}- m_H \langle \bar{q}q \rangle +
   \frac{\alpha_s}{12\pi}\langle G^{\mu\nu}G_{\mu\nu} \rangle \Biggr],
\nonumber\\
  G_P(Q^2) &=&
  \left. G(Q^2) \right\vert_{\rm pert}\nonumber\\
   &+&
  \frac{m_H^2}{m_H^2 + Q^2}
  \Biggl[ m_H \langle \bar{q}q \rangle +
   \frac{\alpha_s}{12\pi}\langle G^{\mu\nu}G_{\mu\nu} \rangle \Biggr],
\end{eqnarray}
where $m_H$ is the heavy-quark mass. To the accuracy we are aiming
at, the OPE can be truncated at ${\mathcal O}(1/Q^2)$. The
explicit expression for the perturbative contribution $\left.
G(Q^2) \right\vert_{\rm pert}$ is available in the literature but
we do not need it since it drops out in the difference. {}From
these correlators, the $\Delta_{SP}$ becomes
\begin{equation}
 \Delta_{SP}(Q^2) = - \frac{2 m_H^3}{m_H^2 + Q^2}
  \langle \bar{q}q \rangle.
\label{diff-ope}
\end{equation}
Equating Eq.~(\ref{diff-eft}) to Eq.~(\ref{diff-ope}) and
neglecting the difference $(m_H-M_D)$, we obtain the following
matching condition:
\begin{equation}
 3 F_D^2 \, \Delta M \simeq - 2 \langle \bar{q}q \rangle.
\label{bare mass diff}
\end{equation}
Thus at the matching scale, the splitting is
\begin{equation}
 \Delta M_{\rm bare} \simeq
  -\frac{2}{3}\frac{\langle \bar{q}q \rangle}{F_D^2}.
 \label{bare diff}
\end{equation}
As announced, the $bare$ splitting is indeed proportional to the
light-quark condensate. The quantum corrections do not change the
dependence on the quark condensate [see section~\ref{sec:QC}].


\section{Quantum Corrections and RGE}
\label{sec:QC}

Given the bare Lagrangian whose parameters are fixed at the
matching scale $\Lambda_M$, the next step is to decimate the
theory \`a la Wilson to the scale at which $\Delta M$ is measured.
This amounts to calculating quantum corrections to the mass
difference $\Delta M$ in the framework of the present EFT.

\begin{figure}
\begin{center}
 \includegraphics
  [width = 6cm]
  {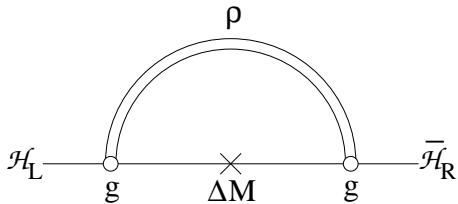}
\end{center}
\caption{Diagram contributing to the mass difference.}
\label{fig:mass}
\end{figure}

This calculation
turns out to be surprisingly simple for $a\approx 1$. If one sets
$a=1$ which is the approximation we are adopting here, $\alpha_L$
does not mix with $\alpha_R$ in the light sector, and then $\alpha_L$
couples to only ${\cal H}_L$ and $\alpha_R$ to only ${\cal H}_R$. As a
result ${\cal H}_{L(R)}$ cannot connect to ${\cal H}_{R(L)}$ by the
exchange of $\alpha_L$ or $\alpha_R$.
Only the $\rho$-loop links between the fields with different
chiralities as shown in Fig.~\ref{fig:mass}. We have verified this
approximation to be reliable since corrections to
the result with $a=1$ come only
at higher loop orders [see the next paragraph].
The diagram shown in Fig.~\ref{fig:mass} contributes to
the two-point function as
\begin{eqnarray}
 \biggl. \Pi_{LR} \biggr\vert_{\rm div}
 = - \frac{1}{2} \Delta M \,
   {\mathcal C}_2(N_f) \,
   \frac{g^2}{2\pi^2}\Bigl( 1 - 2k - k^2 \Bigr)
   \ln\Lambda,
\label{quantum correction}
\end{eqnarray}
where ${\mathcal C}_2(N_f)$ is the second Casimir defined by
$(T_a)_{ij}(T_a)_{jl}= {\mathcal C}_2(N_f)\delta_{il}$ with $i, j$
and $l$ denoting the flavor indices of the light quarks. This
divergence is renormalized by the bare contribution of the form
$\Pi_{LR,{\rm bare}} = \frac{1}{2}\Delta M_{\rm bare}$. Thus the
renormalization-group equation (RGE) takes the form
\begin{equation}
 \mu\frac{d\,\Delta M}{d\mu}
 =  {\mathcal C}_2(N_f)\,
   \frac{g^2}{2\pi^2}\Bigl( 1 - 2k - k^2 \Bigr)\Delta M.
\label{rge a=1}
\end{equation}
For an approximate estimate that we are interested in at this point,
it seems reasonable to ignore the scale dependence in $g$ and $k$.
Then the solution is simple:
\begin{equation}
 \Delta M = \Delta M_{{\rm bare}}
 \times C_{\rm quantum} \ ,
\label{mass diff}
\end{equation}
where we define $C_{\rm quantum}$ by
\begin{equation}
 C_{\rm quantum} =
 \exp\Bigl[ - {\mathcal C}_2(N_f)\,
   \frac{g^2}{2\pi^2}\Bigl( 1 - 2k - k^2 \Bigr)
   \ln\frac{\Lambda}{\mu} \Bigr]\ .
\label{Cquantum}
\end{equation}
This shows unequivocally that the mass
splitting is dictated by the ``bare'' splitting $\Delta M_{\rm
bare}$ proportional to $\langle\bar{q}q\rangle$ corrected by the quantum
effect $C_{\rm quantum}$.

Next we lift the condition $a=1$ made in the above analysis. For
this purpose, we compute the quantum effects to the masses of
$0^-$ $(P)$ and $0^+$ $(Q^\ast)$ $D$-mesons by calculating the
one-loop corrections to the two-point functions of $P$ and
$Q^\ast$ denoted by $\Pi_{PP}$ and $\Pi_{Q^\ast Q^\ast}$ [for the
explicit calculation, see Appendix~\ref{app:ECQC}]. We find that
amazingly, the resultant form of the quantum correction exactly
agrees with the previous one which was obtained by taking $a=1$.
To arrive at this result, it is essential that $P$ (or
$P^\ast_\mu$) be the chiral partner of $Q^\ast$ (or $Q_\mu$) as
follows: The loop diagrams shown in Fig.~\ref{fig:PP} and
Fig.~\ref{fig:QQ} in Appendix~\ref{app:ECQC} have power and
logarithmic divergences. However all the divergences of the
diagrams with pion loop are exactly canceled among themselves
since the internal (or external) particles are chiral partners. In
a similar way, the exact cancellation takes place in the diagrams
with $\sigma$ loop. Finally, the logarithmic divergence from the
$\rho$ loop does contribute to the mass difference. This shows
that the effect of spontaneous chiral symmetry breaking introduced
as the deviation of $a$ from 1 does not get transferred to the
heavy sector. Thus even in the case of $a \neq 1$, the bare mass
splitting is enhanced by only the vector meson loop, with the
pions not figuring in the quantum corrections at least at one-loop
order. Solving the RGE (\ref{rge}), which is exactly same as
Eq.~(\ref{rge a=1}), we obtain exactly the same mass splitting as
the one given in Eq.~(\ref{mass diff}).


\section{Mass Splitting}
\label{sec:DMD}

In this section we make a numerical estimation of the mass
splitting for the chiral doublers in the open charm system. (Here
$D$ denotes the open charm meson.) Since we are considering the
chiral limit, strictly speaking, a precise comparison with
experiments is not feasible particularly if the light quark is
strange, so what we obtain should be considered as
semi-quantitative at best. This caveat should be kept in mind in
what follows.

Determining the bare mass splitting from the matching condition
(\ref{bare diff}) requires the quark condensate at that scale and
the $D$-meson decay constant $F_D$. For the quark condensate, we
shall use the so-called ``standard value''~\cite{GL:mass}
$\langle\bar{q}q\rangle=-(225\pm 25\,\mbox{MeV})^3$ at $1$\,GeV.
Extrapolated to the scale $\Lambda_M=1.1$ GeV we shall adopt here,
this gives
 \begin{equation} \langle\bar{q}q\rangle_{\Lambda_M} =
-(228\pm25\,\mbox{MeV})^3.
 \end{equation}
Unfortunately this value is not firmly established, there being no
consensus on it. The values found in the literature vary widely,
even by a factor of $\sim 2$, some higher~\cite{BL} and some
lower~\cite{Moussallam}.
(We will study the dependence on the values of the mass splitting
of those of the quark condensate later.) 
Here we take the standard value
as a median~\footnote{
 It was shown in Ref.~\cite{Moussallam} that there is a
 strong $N_f$ dependence on the quark condensate and
 the value of the quark condensate for
 QCD with three massless quarks is smaller than
 the value used in estimating the value of the mass splitting
 in section~\ref{sec:DMD}.
 In the present analysis, we extract the value of the coupling
 constant $k$ from the experiment.  To be consistent, we need
 to use the quark condensate together with other parameters involved
 determined at the same scale from
 experimental and/or lattice data. This corresponds to the standard
value of the condensate we are using here.}.
As for the $D$-meson decay constant,  we take as a typical value
$F_D=0.205\pm 0.020\,\mbox{GeV}$ obtained from the QCD sum rule
analysis~\cite{Narison:2003td}. Plugging the above input values
into Eq.~(\ref{bare diff}) we obtain
\begin{equation}
\Delta M_{\rm bare} \simeq 0.19 \,\mbox{GeV} \ .
\end{equation}
By taking $\mu = m_\rho = 771\,\mbox{MeV}$,
$\Lambda = \Lambda_M = 1.1\,\mbox{GeV}$,
$g = g(m_\rho) = 6.27$ determined via the
Wilsonian matching for
$(\Lambda_M,\Lambda_{\rm QCD})=(1.1,0.4)\,\mbox{GeV}$
in Ref.~\cite{HY:PRep} and $k \simeq 0.59$
extracted from the $D^*\rightarrow D\pi$ decay
[see section~\ref{D*Dpi}] in Eq.~(\ref{Cquantum}),
we find for $N_f=3$
\begin{equation}
  C_{\rm quantum} = 1.6 \ .
\end{equation}
This is a sizable quantum correction involving only the vector
meson. If one takes into account the uncertainties involved in the
condensate and the decay constant, the quantum-corrected splitting
$\Delta M$ comes out to be
\begin{equation}
 \Delta M \approx 0.31\pm 0.12\,\mbox{GeV} \ .
\label{mass diff num}
\end{equation}
Despite the uncertainty involved, (\ref{mass diff num}) is a
pleasing result. It shows that the splitting is indeed of the size
of the constituent quark mass of a chiral quark $\Sigma \sim
m_p/3\sim 310$ MeV and is directly proportional to the quark
condensate.

We emphasized in the above analysis that there
is a great deal of uncertainty on the value of the quark
condensate at the relevant matching scale $\Lambda_M$. 
Here we list a few examples to show what sort of uncertainty
we are faced with.
We took $\left\langle \bar{q} q \right\rangle_{\rm 1\,GeV}
=-(225\pm25\,\mbox{MeV})^3$ in the above analysis 
as a ``standard value.'' 
For comparison, we shall take two other values quoted in
Ref.~\cite{BL} (without making any judgments on their validity).
Consider therefore
\begin{eqnarray}
\left\langle \bar{q} q \right\rangle_{\rm 1\,GeV}
&=& -(225\pm25\,\mbox{MeV})^3 \ ,
\nonumber\\
\left\langle \bar{q} q \right\rangle_{\rm 2\,GeV}
&=&
\left\{\begin{array}{l}
 -(273\pm19\,\mbox{MeV})^3 \ , \\
 -(316\pm24\,\mbox{MeV})^3 \ .
\end{array}\right.
\end{eqnarray}
Brought by RGE to the scale we are working at, $\Lambda_M=1.1$
GeV, and substituted into our formula for $\Delta M$, we get the
corresponding quantum corrected splitting
\begin{eqnarray}
\Delta M = \left\{\begin{array}{l}
 0.31\pm 0.12\,\mbox{GeV} \ , \\
 0.43\pm 0.12\,\mbox{GeV} \ , \\
 0.67\pm0.20\,\mbox{GeV} \ .
\end{array}\right.
\end{eqnarray}
This result clearly shows that the splitting cannot be pinned down
unless one has a confirmed quark condensate.

We should stress other several caveats associated 
with (\ref{mass diff num}).
Apart from the sensitivity to the quark condensate, if one
naively plugs in the matching scale $\Lambda_M$ into the RGE
solution, one finds the splitting is not insensitive as it should
be to the scale change. This is neither surprising nor too
disturbing since our RGE solution is obtained with the scale
dependence in both $g$ and $k$ ignored. In order to eliminate this
dependence on the matching scale, it will be necessary to solve
the RGE with the full scale dependence taken into account -- which
is at the moment beyond our scope here. The best we can do within
the scheme adopted is to pick the optimal $\Lambda_M$ determined
phenomenologically from elsewhere~\cite{HY:PRep} and this is what
we have done above.


\section{Hadronic Decay Modes}
\label{sec:HDM}

In this section we turn to the hadronic decay processes of the
$\tilde{D}$ mesons and make predictions of our scenario based on
the vector manifestation (VM) of chiral symmetry. Here we adopt
the notations $D_{u,d}$ and $\tilde{D}_{u,d}$ for the heavy
ground-state mesons and heavy excited mesons composed of
$c\bar{u}$ and $c\bar{d}$, and $D_{s}$ and $\tilde{D}_{s}$ for
those composed of $c\bar{s}$. The spin-parity quantum numbers will
be explicitly written as $D_{u,d}(0^-)$. For the heavy vector
meson, we follow the notation adopted by the Particle Data Group
(PDG)~\cite{Hagiwara:fs} and write $D_{u,d}^\ast(1^-)$ and
$D_{s}^\ast(1^-)$. Unless otherwise noted, the masses of the
ground-state heavy mesons will be denoted as $M_{D}$ and those of
the excited states as $M_{\tilde{D}}$.

\subsection{$D^\ast \rightarrow D + \pi$}
\label{D*Dpi}

Before studying the decay processes of the excited heavy mesons,
we first calculate the decay width of $D_{u,d}^\ast \rightarrow
D_{u,d} + \pi$ so as to determine the coupling constant $k$. The
decay widths of the $\pi^0$ and the $\pi^{\pm}$ modes are given by
\begin{eqnarray}
\Gamma(D_{u,d}^\ast(1^-) \rightarrow D_{u,d}(0^-) + \pi^0)
&=&
\frac{\bar{p}_\pi^3}{24\pi M_{D^\ast}^2 }
    \left( M_Q \, \frac{k}{F_\pi} \right)^2
\ ,
\nonumber\\
\Gamma(D^\ast_{u,d}(1^-) \rightarrow D_{u,d}(0^-) + \pi^{\pm})
&=&
\frac{\bar{p}_\pi^3}{12\pi M_{D^\ast}^2 }
    \left( M_Q \, \frac{k}{F_\pi} \right)^2
\ ,
\end{eqnarray}
where $\bar{p}_\pi \equiv |\vec{p}_\pi|$ denotes the
three-momentum of the
pion in the rest frame of the decaying particle $D^\ast_{u,d}(1^-)$,
and $M_Q$ the ``heavy quark mass'' introduced for
correctly normalizing the heavy meson field.
In the present analysis we use the following reduced mass
for definiteness:
\begin{equation}
 M_Q \equiv \frac{1}{4}\Bigl( M_{D(0^-)} + 3 M_{D^\ast(1^-)} \Bigr)
 = 1974 \, \mbox{MeV} \ .
\label{reduced mass}
\end{equation}

The total width is not determined for $D_u^\ast (1^-)$, although
the branching fractions for both the $\pi^0$ and the $\pi^+$ decay
modes are known experimentally. For $D_d^\ast (1^-)$ meson, on the
other hand, the total width is also determined.  Using the values
listed in PDG table~\cite{Hagiwara:fs}, the partial decay widths
are estimated to be
\begin{eqnarray}
&& \Gamma(D_{d}^\ast(1^-) \rightarrow D_{d}(0^-) + \pi^0)
  = 29.5 \pm 6.8 \,\mbox{keV}
\ ,
\nonumber\\
&& \Gamma(D_{d}^\ast(1^-) \rightarrow D_{u}(0^-) + \pi^+)
  = 65 \pm 15 \,\mbox{keV}
\ .
\end{eqnarray}
Here the $\pi^0$ mode will be used as an input to fix $k$. From
the experimental masses $M_{D_u^\ast (1^-)} = 2010.1$\,MeV,
$M_{D_d(0^-)} = 1869.4$\,MeV and $M_{\pi^0} = 134.9766$\,MeV
together with the value of the pion decay constant $F_\pi = 92.42
\pm 0.26$\,MeV, we obtain
\begin{equation}
k = 0.59 \pm 0.07 \ .
\end{equation}
Note that the error is mainly from that of the
$D_{d}^\ast(1^-) \rightarrow D_{d}(0^-) + \pi^0$ decay width.

In the following analysis, we shall use the central value of $k$
to make predictions for the decay widths of $\tilde{D}$ mesons.
Each prediction includes at least about 25\% error from the value
of $k$. For the masses of excited $D$ mesons, we use
$M_{\tilde{D_s}(0^+)}=2317$\,MeV determined by BaBar~\cite{BABAR},
$M_{\tilde{D_s}(1^+)}=2460$\,MeV by CLEO~\cite{CLEO} and
$(M_{\tilde{D}_{u,d}(0^+)},M_{\tilde{D}_{u,d}(1^+)})
=(2308,2427)$\,MeV by Belle~\cite{Belle}. Table~\ref{input}
summarizes the input parameters used in the present analysis.
\begin{table}
 \begin{center}
  \begin{tabular*}{16.8cm}{@{\extracolsep{\fill}}cccccc}
    \hline
    $D_{u,d}$ meson masses
     &  $M_{\tilde{D}_{u,d}(1^+)}$  &  $M_{\tilde{D}_{u,d}(0^+)}$
     &  $M_{D_{u,d}^\ast(1^-)}$     &  $M_{D_{u,d}(0^-)}$
     &  {}\\
    (MeV)
     &  2427                        &  2308
     &  2010                        &  1865
     &  {}\\
    \hline
    $D_s$ meson masses
     &  $M_{\tilde{D}_s(1^+)}$      &  $M_{\tilde{D}_s(0^+)}$
     &  $M_{D_s^\ast (1^-)}$        &  $M_{D_s(0^-)}$
     &  {}\\
    (MeV)
     &  2460                        &  2317
     &  2112                        &  1969
     &  {}\\
    \hline
    Light meson masses
     &  $M_\pi$          &  $M_\rho$         & $M_\eta$
     &  $M_\phi$          & {} \\
    (MeV)
     &  138.039          &  771.1            &  547.30
     &  1019.456         & {}\\
    \hline
    $\pi^0$-$\eta$ mixing
     &  $A_{11}$                         & $A_{21}$
     & $\Pi_{\pi^0\eta}\,(\mbox{MeV})^2$ &  $K_{\pi^0\eta}$       & {}\\
    {}
     &  0.71                             &  $-0.52$
     & $-4.25 \times 10^3$               & $-1.06 \times 10^{-2}$ & {}\\
    \hline
    $\phi$-$\rho$ mixing
     & $\Gamma_{\phi \to \pi^+\pi^-}\,$(MeV)
     & $\Gamma_{\rho \to \pi^+\pi^-}\,$(MeV)
     & {}                       & {}                     & {} \\
    {}
     & $3.11 \times 10^{-4}$  & 149.2
     & {}                     & {}                     & {}\\
    \hline
  \end{tabular*}
 \end{center}
 \caption{The values of input parameters.
  We use the values of $M_{\tilde{D}_s(0^+)}$~\cite{BABAR},
  $M_{\tilde{D}_s(1^+)}$~\cite{CLEO}
  and $M_{\tilde{D}_{u,d}(0^+,1^+)}$~\cite{Belle}.
  The $D$ mesons in the ground state, light mesons
  and decay widths $\Gamma (\phi, \rho)$ are
  the values listed by the PDG table~\cite{Hagiwara:fs}.
  As for the parameters associated with the $\pi^0$-$\eta$ mixing,
  we use the values given in
  Refs.~\cite{Schechter:1992iz,Harada:1995sj}.}
 \label{input}
\end{table}

\subsection{$\tilde{D} \to D + \pi$}

For the systems of $c\bar{u}$ and $c\bar{d}$,
the following decay processes of the $\tilde{D}_{u,d}$ meson
into the $D_{u,d}$ meson and one pion are allowed by the
spin and parity:
\begin{equation}
 \tilde{D}_{u,d}(0^+) \to D_{u,d}(0^-) + \pi
\qquad
 \tilde{D}_{u,d}(1^+) \to D_{u,d}^\ast (1^-) + \pi.
\end{equation}
Their partial decay widths are given by
\begin{eqnarray}
 && \Gamma (\tilde{D}_{u,d} \to D_{u,d} + \pi^{\pm})
   = \frac{\bar{p}_\pi}{4\pi}
   \Biggl( \frac{k}{F_\pi}\frac{M_Q}{M_{\tilde{D}}}E_\pi \Biggr)^2,
\nonumber\\
 && \Gamma (\tilde{D}_{u,d} \to D_{u,d} + \pi^0)
   = \frac{\bar{p}_\pi}{8\pi}
   \Biggl( \frac{k}{F_\pi}\frac{M_Q}{M_{\tilde{D}}}E_\pi \Biggr)^2,
\end{eqnarray}
where $E_\pi$ is the energy of the pion, and the reduced mass
$M_Q$ is defined in Eq.~(\ref{reduced mass}). With the input
parameters given in Table~\ref{input}, these decay widths come out
to be
\begin{eqnarray}
 && \Gamma (\tilde{D}_{u,d}(0^+) \to D_{u,d}(0^-) + \pi^0)
   = 74 \pm 18\,\mbox{MeV} \ ,
\nonumber\\
 && \Gamma (\tilde{D}_{u,d}(0^+) \to D_{u,d}(0^-) + \pi^{\pm})
   = 147 \pm 37 \,\mbox{MeV} \ ,
\nonumber\\
 && \Gamma (\tilde{D}_{u,d}(1^+) \to D_{u,d}^\ast (1^-) + \pi^0)
   = 57 \pm 14\,\mbox{MeV} \ ,
\nonumber\\
 && \Gamma (\tilde{D}_{u,d}(1^+) \to D_{u,d}^\ast (1^-) + \pi^{\pm})
   = 114 \pm 29 \,\mbox{MeV} \ ,
\end{eqnarray}
Here we have put the $25\%$ error from the value of $k$,
which we expect the dominant one.

For the system of $c\bar{s}$ there are two decay processes of the
$\tilde{D}_s$ meson into the $D_s$ meson and one pion:
\begin{equation}
 \tilde{D}_s(0^+) \to D_s(0^-) + \pi^0
\qquad
 \tilde{D}_s(1^+) \to D_s^\ast (1^-) + \pi^0.
\end{equation}
These processes violate the isospin invariance, and hence are
suppressed. In the present analysis we assume as in
Ref.~\cite{Bardeen-Eichten-Hill:03} that the isospin violation
occurs dominantly through the $\pi^0$-$\eta$ mixing. In other
words, we assume that the $\tilde{D}_s$ meson decays into the
$D_s$ meson and the virtual $\eta$ meson which mixes with
the $\pi^0$
through the $\pi^0$-$\eta$ mixing.
Then, the decay width is given by
\begin{equation}
 \Gamma (\tilde{D}_s \to D_s + \pi^0)
 = \frac{\bar{p}_\pi}{2\pi}
   \Biggl( \frac{k}{F_\pi} \frac{M_Q}{M_{\tilde{D}}} E_\pi
     \Delta_{\pi^0\eta} \Biggr)^2,
\end{equation}
where $\Delta_{\pi^0 \eta}$ denotes the $\pi^0$-$\eta$ mixing
and takes the following form~\cite{Schechter:1992iz,Harada:1995sj}:
\begin{equation}
 \Delta_{\pi^0 \eta}
 = - \frac{A_{11}A_{21}}{M_\eta^2 - M_{\pi^0}^2}
   \bigl( \Pi_{\pi^0 \eta} - K_{\pi^0 \eta}M_{\pi^0}^2 \bigr)
\end{equation}
with $\Pi_{\pi^0 \eta}$ and $K_{\pi^0 \eta}$ being the mass-type and
kinetic-type $\pi^0$-$\eta$ mixing, respectively.
$A_{11}$ and $A_{21}$ are the components of the $\eta$-$\eta^\prime$
mixing matrix in the two-mixing-angle scheme~\cite{Schechter:1992iz}.
By using the values listed in Table~\ref{input},
the $\pi^0$-$\eta$ mixing is estimated as
\begin{equation}
 \Delta_{\pi^0 \eta} = -5.3 \times 10^{-3}.
\end{equation}
{}From this value, the predicted decay widths are estimated as
\begin{eqnarray}
&& \Gamma(\tilde{D}_s(0^+) \to D_s(0^-) + \pi^0)
 \sim 4 \, \mbox{keV} \ ,
\nonumber\\
&& \Gamma(\tilde{D}_s(1^+) \to D_s^\ast (1^-) + \pi^0)
 \sim 4 \,\mbox{keV}
\ .
\end{eqnarray}

\subsection{$\tilde{D}(1^+) \rightarrow \tilde{D}(0^+) + \pi$}

With the masses of $\tilde{D}_{u,d}(1^+)$ and
$\tilde{D}_{u,d}(0^+)$ listed in Table~\ref{input}, the
intra-multiplet decay $\tilde{D}_{u,d}(1^+) \rightarrow
\tilde{D}_{u,d}(0^+) + \pi$ is not allowed kinematically. Since
the experimental errors for the masses are large~\footnote{
  The Belle collaboration~\cite{Abe:2003zm,Krokovny:2003rm,Belle:talk} gives
  $M_{\tilde{D}_{u,d}(1^+)} = 2427\pm26\pm20\pm17$\,MeV
  and $M_{\tilde{D}_{u,d}(0^+)} = 2308\pm17\pm15\pm28$\,MeV.
},
this decay mode
may still turn out to be possible. To show how large the possible
decay width is, we use $M_{\tilde{D}_{u,d}(0^+)} =
2272$\,\mbox{MeV} and $M_{\tilde{D}_{u,d}(1^+)} =
2464$\,\mbox{MeV} together with the formulas
\begin{eqnarray}
&&
\Gamma(\tilde{D}_{u,d}(1^+) \rightarrow \tilde{D}_{u,d}(0^+) + \pi^0 )
= \frac{\bar{p}_\pi^3}{24\pi}
    \left( \frac{M_Q}{M_{\tilde{D}(1^+)}} \,\frac{k}{F_\pi} \right)^2
\ ,
\nonumber\\
&&
\Gamma(\tilde{D}_{u,d}(1^+) \rightarrow \tilde{D}_{u,d}(0^+) + \pi^+ )
= \frac{\bar{p}_\pi^3}{12\pi}
    \left( \frac{M_Q}{M_{\tilde{D}(1^+)}} \,\frac{k}{F_\pi} \right)^2
\ .
\end{eqnarray}
The resultant decay widths are given by
\begin{eqnarray}
&&
\Gamma(\tilde{D}_{u,d}(1^+) \rightarrow \tilde{D}_{u,d}(0^+) + \pi^0 )
= 0.7 \pm 0.2\,\mbox{MeV}
\ ,
\nonumber\\
&&
\Gamma(\tilde{D}_{u,d}(1^+) \rightarrow \tilde{D}_{u,d}(0^+) + \pi^+ )
= 1.5 \pm 0.4\,\mbox{MeV}
\ ,
\end{eqnarray}
where we put $25\%$ error coming from the value of $k$.
We expect that this is the main source of the entire error.
They are smaller by the order of $10^{-2}$ than other one-pion modes
[see Table~\ref{predictions}].
This is caused by the suppression from the phase space.

With the present input values of $\tilde{D}$ masses, the process
$\tilde{D}_{s}(1^+) \rightarrow \tilde{D}_{s}(0^+) + \pi^0$ is
kinematically allowed. Similarly to the $\tilde{D}_s \to D_s +
\pi^0$ decay, we assume that this decay is dominated by the
process through the $\pi^0$-$\eta$ mixing. Then, the decay width
is estimated as
\begin{eqnarray}
 \Gamma (\tilde{D}_s(1^+) \to \tilde{D}_s(0^+) + \pi^0)
 &=& \frac{\bar{p}_\pi^3}{6\pi}
     \Biggl( \frac{k}{F_\pi}\frac{M_Q}{M_{\tilde{D}(1^+)}}
      \Delta_{\pi^0 \eta}
     \Biggr)^2
\nonumber\\
 &\sim& 2 \times 10^{-3}\,\mbox{keV}.
\end{eqnarray}
This is very tiny due to the isospin violation
and the phase-space suppression.

\subsection{$\tilde{D} \to D + 2\pi$}

There are several processes such as $\tilde{D} \rightarrow D +
\pi^{\pm}\pi^{\mp}$ to which the light scalar mesons could give
important contributions. In models based on the standard scenario
of the chiral symmetry restoration in the light quark sector, the
scalar-meson coupling to the heavy-quark system is related to the
pion coupling, enabling one to compute the decay width. In our
model based on the VM of the chiral symmetry restoration, on the
other hand, it is the coupling constant of the vector meson to the
heavy system that is related to the pion coupling constant: Here
coupling of the scalar meson is not directly connected, at least
in the present framework which contains no explicit scalar
fields~\footnote{Scalar excitations can of course be generated at
high loop level to assure unitarity or with the account of QCD
trace anomaly but we shall not attempt this extension in this
paper.}, to do that of the pion. So, while we cannot make firm
predictions to processes for which scalar mesons might contribute,
we can make definite predictions on certain decay widths for which
scalar mesons do not figure. If one ignores isospin violation, the
two-pion decay processes $\tilde{D}_{u,d} \rightarrow D_{u,d} +
\pi^{\pm} \pi^0$ receive no contributions from scalar mesons. We
give predictions for these processes below. As for the two-pion
decay modes of the $\tilde{D}_s$ meson, the scalar mesons could
give a contribution. To have an idea, we shall also compute the
vector-meson contribution to this process.

First, consider $\tilde{D}_{u,d}(0^+) \to D_{u,d}^\ast (1^-) +
\pi^{\pm}\pi^0$. In this process, there are two contributions:
\begin{eqnarray}
 \tilde{D}_{u,d}(0^+) &\to& D_{u,d}^\ast (1^-) + \pi^{\pm}\pi^0
 \qquad\qquad\qquad\qquad\mbox{(direct)}
\nonumber\\
  &\,\,& D_{u,d}^\ast (1^-) + (\rho^{\pm} \to \pi^{\pm}\pi^0)
  \qquad\qquad\mbox{($\rho$-mediation)}
\ .
\end{eqnarray}
The decay width is given by
\begin{eqnarray}
 &&\Gamma (\tilde{D}_{u,d}(0^+) \to D_{u,d}^\ast (1^-) + \pi^\pm\pi^0)
\nonumber\\
 &&\quad =
     \frac{M_Q^2}{64(2\pi)^3 M_{\tilde{D}}^3}
     \frac{k^2}{F_\pi^4}
     \int dm_{D\pi}^2 \int dm_{\pi\pi}^2
     |F_{\tilde{D}D}|^2
\nonumber\\
 &&\qquad\times
     \Biggl[ m_{\pi\pi}^2 - 4M_\pi^2 + \frac{1}{4M_D^2}
      \bigl( m_{\pi\pi}^2 - M_{\tilde{D}}^2 - M_D^2 - 2M_\pi^2
       {}+ 2m_{D\pi}^2 \bigr)^2
     \Biggr],
\end{eqnarray}
with $m_{D\pi}^2 = (p_D + p_\pi)^2$ and $m_{\pi\pi}^2 = (p_{1\pi}
+ p_{2\pi})^2$. The form factor $F_{\tilde{D}D}$ is taken to be of
the form
\begin{eqnarray}
  F_{\tilde{D}D}
  = 1 + \frac{M_\rho^2}{m_{\pi\pi}^2 - M_\rho^2}
  \ .
\end{eqnarray}
The first term of the form factor comes from the direct
contribution and the second from the $\rho$-mediation. Here we
have neglected the $\rho$ meson width in the propagator, since the
maximum value of $m_{\pi\pi}$ is about $300$\,MeV with the input
values listed in Table~\ref{input}. We can see that the form
factor $F_{\tilde{D}D}$ vanishes in the limit of $m_{\pi\pi} \to
0$, which is a consequence of chiral symmetry~\footnote{
  It should be stressed that this cancellation occurs
  because the vector meson is included consistently with
  chiral symmetry, and that it is $not$ a
  specific feature of the VM.  The chiral symmetry
  restoration based on the VM implies that the coupling
  constant of the vector meson to the heavy system is
  equal to that of the pion.
}. We note that $\left. m_{\pi\pi} \right\vert_{\rm max} \simeq
300$\,MeV makes this decay width strongly suppressed due to the
large cancellation between the direct and $\rho$-mediated
contributions. Furthermore, since $300$\,MeV is close to the
two-pion threshold, additional suppression comes from the phase
space. Due to these two types of suppressions the predicted decay
width is predicted to be very small, of the order of $10^{-2}$
keV.~\footnote{
  Note that the prediction on the decay width is very sensitive
  to the precise value of the mass of $\tilde{D}(0^+)$ meson.
}

Next we consider the process $\tilde{D}_{u,d}(1^+) \to
D_{u,d}^\ast (1^-) + \pi\pi$. Again there are two contributions,
direct and a $\rho$-mediated:
\begin{eqnarray}
 \tilde{D}_{u,d}(1^+) &\to& D_{u,d}^\ast (1^-) + \pi^{\pm}\pi^0
  \qquad\qquad\qquad\qquad\mbox{(direct)}
\nonumber\\
  &\,\,& D_{u,d}^\ast (1^-) + (\rho^{\pm} \to \pi^{\pm}\pi^0)
  \qquad\qquad\mbox{($\rho$-mediation)}
\end{eqnarray}
The resultant decay width is given by
\begin{eqnarray}
 &&\Gamma (\tilde{D}_{u,d}(1^+) \to D_{u,d}^\ast (1^-) + \pi^{\pm}\pi^0)
\nonumber\\
 &&\quad =
     \frac{M_Q^2}{96(2\pi)^3 M_{\tilde{D}}^3}
     \frac{k^2}{F_\pi^4}
     \int dm_{D\pi}^2 \int dm_{\pi\pi}^2
     |F_{\tilde{D}D}|^2
\nonumber\\
 &&\qquad\times
     \Biggl[ m_{\pi\pi}^2 - 4M_\pi^2 + \frac{1}{4M_{\tilde{D}}^2}
      \bigl( m_{\pi\pi}^2 - M_{\tilde{D}}^2 - M_D^2 - 2M_\pi^2
       {}+ 2m_{D\pi}^2 \bigr)^2
     \Biggr].
\end{eqnarray}
Similarly to $\Gamma (\tilde{D}_{u,d}(0^+) \to D_{u,d}^\ast
(1^-)+\pi^{\pm}\pi^0)$, the width is again suppressed due to the
large cancellation between the direct and $\rho$-mediated
contributions. The suppression from the phase space, on the other
hand, is not so large
since $\left. m_{\pi\pi} \right\vert_{\rm max} \simeq 420$\,MeV is
not so close to the two-pion threshold. The resulting decay width
is
\begin{equation}
\Gamma (\tilde{D}_{u,d}(1^+) \to D_{u,d}^\ast (1^-) + \pi^{\pm}\pi^0)
= 12 \pm 3 \, \mbox{keV} \ ,
\end{equation}
where the error comes from the value of $k$.

The decay width of the process
\begin{eqnarray}
 \tilde{D}_{u,d}(1^+) &\to& D_{u,d}(0^-) + \pi^{\pm}\pi^0
  \qquad\qquad\qquad\qquad\mbox{(direct)}
\nonumber\\
  &\,\,& D_{u,d}(0^-) + (\rho^{\pm} \to \pi^{\pm}\pi^0)
  \qquad\qquad\mbox{($\rho$-mediation)}
\end{eqnarray}
is given by
\begin{eqnarray}
 &&\Gamma (\tilde{D}_{u,d}(1^+) \to D_{u,d}(0^-) + \pi^\pm \pi^0)
\nonumber\\
 &&\quad =
     \frac{M_Q^2}{192(2\pi)^3 M_{\tilde{D}}^3}
     \frac{k^2}{F_\pi^4}
     \int dm_{D\pi}^2 \int dm_{\pi\pi}^2
     |F_{\tilde{D}D}|^2
\nonumber\\
 &&\qquad\times
     \Biggl[ m_{\pi\pi}^2 - 4M_\pi^2 + \frac{1}{4M_{\tilde{D}}^2}
      \bigl( m_{\pi\pi}^2 - M_{\tilde{D}}^2 - M_D^2 - 2M_\pi^2
       {}+ 2m_{D\pi}^2 \bigr)^2
     \Biggr].
\end{eqnarray}
In the present case, $m_{\pi\pi}|_{\rm max}\simeq 560\,\mbox{MeV}$
is much larger than the two-pion threshold and hence the width
becomes larger than other two-pion processes. We find
\begin{equation}
 \Gamma (\tilde{D}_{u,d}(1^+) \to D_{u,d}(0^-) + \pi^\pm \pi^0)
   = 310 \pm 80\,\mbox{keV},
\end{equation}
where the error comes from the value of $k$.

Finally we turn to the decay $\tilde{D}_s(1^+) \to D_s(0^-) +
\pi^+\pi^-$ which as mentioned could receive direct contributions
from scalar excitations. Since we have not incorporated scalar
degrees of freedom in the theory, we might not be able to make a
reliable estimate even if were to go to higher-loop orders. Just
to have an idea as to how important the vector meson contribution
can be, we calculate the decay width in which the $\tilde{D_s}$
meson decays into two pions through the $\phi$ meson. This isospin
violating decay can occur through the direct $\phi$-$\pi$-$\pi$
coupling and the $\phi$-$\rho$ mixing:
\begin{eqnarray}
 \tilde{D}_s(1^+) &\to& D_s(0^-) + (\phi \to \pi^+\pi^-)
  \qquad\qquad\qquad\quad\mbox{(direct)}
\nonumber\\
  &\,\,& D_s(0^-) + (\phi \to \rho^0 \to \pi^+\pi^-)
  \qquad\qquad\mbox{($\phi$-$\rho$ mixing)}
\end{eqnarray}
Since the main contribution to the $\phi \rightarrow \pi\pi$ is
expected to be given by the $\phi$-$\rho$ mixing, we shall neglect
the direct $\phi$-$\pi$-$\pi$-coupling contribution in the
following. Then the decay width is given by
\begin{eqnarray}
 &&\Gamma (\tilde{D}_s(1^+) \to D_s(0^-) + \pi^+\pi^-)
\nonumber\\
 &&\quad =
     \frac{M_Q^2}{192(2\pi)^3 M_{\tilde{D}}^3}
     \frac{k^2}{F_\pi^4}
     \int dm_{D\pi}^2 \int dm_{\pi\pi}^2
     \Biggl[ \frac{M_\rho^2 \Pi_{\phi\rho}}
      {(m_{\pi\pi}^2 - M_\phi^2)(m_{\pi\pi}^2 - M_\rho^2)}
     \Biggr]^2
\nonumber\\
 &&\qquad\times
     \Biggl[ m_{\pi\pi}^2 - 4M_\pi^2 + \frac{1}{4M_{\tilde{D}}^2}
      \bigl( m_{\pi\pi}^2 - M_{\tilde{D}}^2 - M_D^2 - 2M_\pi^2
       {}+ 2m_{D\pi}^2 \bigr)^2
     \Biggr],
\end{eqnarray}
where $\Pi_{\phi\rho}$ denotes the $\phi$-$\rho$ mixing
given by
\begin{equation}
 \Pi_{\phi\rho}^2
 = (M_\phi^2 - M_\rho^2)^2
   \Biggl( \frac{\bar{p}_\pi (\rho)}{\bar{p}_\pi (\phi)} \Biggr)^3
   \frac{M_\phi^2}{M_\rho^2}
   \frac{\Gamma (\phi \to \pi^+\pi^-)}{\Gamma (\rho \to \pi^+\pi^-)},
\end{equation}
with $\bar{p}_\pi (X)$ being the three-momentum of pion in the
rest frame of the decaying particle $X = \phi, \rho$. Using the
values listed in Table~\ref{input}, we have
\begin{equation}
\Pi_{\phi\rho} = 530\,(\mbox{MeV})^2
\end{equation}
so the decay width is predicted to be
\begin{equation}
\Gamma (\tilde{D}_s(1^+) \to D_s(0^-) + \pi^+\pi^-)
 \sim 2 \times 10^{-4} \,\mbox{keV} \ .
\end{equation}
The $\phi$-$\rho$ mixing is caused by the isospin violation, and
this process is highly suppressed. We conclude that should a
measured width come out to be substantially greater than what we
found here, it would mean that either scalars must figure
importantly or the VM is invalid in its present form.

\subsection{Summary of hadronic decay modes}

Our predictions of the decay widths are summarized in
Table~\ref{predictions}.
\begin{table}
 \begin{center}
  \begin{tabular*}{15cm}{@{\extracolsep{\fill}}cll}
    \hline
    Decaying particle       & Process
    & Width (MeV) \\
    \hline
    $\tilde{D}_{u,d}$       & $0^+ \to 0^- + \pi^0$
     & $(7.4 \pm 1.8) \times 10^1$\\
    {}                      & $0^+ \to 0^- + \pi^\pm$
     & $(1.5 \pm 0.4) \times 10^2$\\
    {}                      & $0^+ \to 1^- + \pi^\pm\pi^0$
     & $\sim 2 \times 10^{-5}$\\
    {}                      & $1^+ \to 1^- + \pi^0$
     & $(5.7 \pm 1.4) \times 10^1$\\
    {}                      & $1^+ \to 1^- + \pi^\pm$
     & $(1.1 \pm 0.3) \times 10^2$\\
    {}                      & $1^+ \to 0^- + \pi^\pm\pi^0$
     & $(3.1 \pm 0.8) \times 10^{-1}$\\
    {}                      & $1^+ \to 1^- + \pi^\pm\pi^0$
     & $(1.2 \pm 0.3) \times 10^{-2}$\\
    \hline
    $\tilde{D}_s$           & $0^+ \to 0^- + \pi^0$
     & $\sim 4 \times 10^{-3}$\\
    {}                      & $1^+ \to 0^+ + \pi^0$
     & $\sim 2 \times 10^{-6}$\\
    {}                      & $1^+ \to 1^- + \pi^0$
     & $\sim 4 \times 10^{-3}$\\
    {}                      & $1^+ \to 0^- + \pi^+\pi^-
       \quad (\mbox{through}\,\phi \to \rho^0 \to \pi^+\pi^-)$
     & $\sim 2 \times 10^{-7}$\\
    \hline
  \end{tabular*}
 \end{center}
 \caption{The predicted values of the hadronic decay processes.
  $25\%$ error comes from the value of $k$.
  ``$\sim$'' implies that the precise value also depends on
  parameters other than $k$, and we can give only the order estimation.}
 \label{predictions}
\end{table}
It should be stressed that the values obtained in this paper on
the one-pion reflect only that the $\tilde{D}$ meson is the chiral
partner of the $D$ meson. They are $not$ specific to the VM. We
therefore expect that as far as the one-pion processes are
concerned, there will be no essential differences between our
predictions and those in Ref.~\cite{Bardeen-Eichten-Hill:03}.
However, in the two-pion decay processes in which the scalar meson
does not mediate, our scenario based on the VM can make definite
predictions which might be distinguished from that based on the
standard picture. Especially for $\tilde{D}_{u,d}(1^+) \to
D_{u,d}(0^-) + \pi^{\pm}\pi^0$, we obtain a larger width than for
other two-pion modes. Although the predicted width is still small
-- perhaps too small to be detected experimentally, it is
important because of the following reason. In our approach, since
the excited heavy meson multiplets of $\tilde{D}(0^+)$ and
$\tilde{D}(1^+)$ denoted by $G$ are the chiral partners to the
ground-state multiplets denote by $H$, the $G$-$\bar{H}$-$\pi$
coupling is the same as the $H$-$\bar{H}$-$\pi$ coupling [see the
fifth and first terms of Eq.~(\ref{Lag:component})]. Thus the
width which is dependent on the strength of $k$ is a good probe to
test our scenario. The common $k$ is also essential for the ratio
of the widths of the two-pion modes to those of the one-pion
modes, which has no $k$ dependence. These are therefore are
definite predictions of our scenario. {}From the values listed in
Table~\ref{predictions}, we obtain
\begin{equation}
 \frac{\Gamma (\tilde{D}_{u,d}(1^+) \to D_{u,d}(0^-) + \pi^\pm\pi^0)}
      {\Gamma_{\pi + 2\pi}^{\rm (had)}}
 \simeq 2 \times 10^{-3},
\end{equation}
where $\Gamma_{\pi + 2\pi}^{\rm (had)}$ is the sum of the widths
of the one-pion and two-pion modes of the decaying
$\tilde{D}_{u,d}(1^+)$.

\newpage

\chapter{Summary and Discussions}
\label{ch:SD}

In this thesis,
we first showed the details of the hidden local symmetry (HLS) theory
at finite temperature.
Then in order to determine the parameters of the theory,
we presented the extension of the Wilsonian matching 
at zero temperature to the version in hot matter.
Next we showed the formulation of the vector manifestation (VM)
in hot matter using the HLS theory.
Based on the VM,
we gave several predictions which are testable by experiments
and lattice analysis.
In the following,
we summarize them and give discussions.


\subsubsection{HLS theory at finite temperature}

In chapter~\ref{ch:THLS},
we applied the HLS theory developed in Ref.~\cite{HY:PRep}
to QCD at finite temperature.
Both pions and vector mesons are introduced as 
the relevant degrees of freedom into the theory,
and thus the HLS theory has a wider range of the validity
than the ordinary chiral perturbation theory (ChPT)
including only pions.
Especially we can see it in studying the pion decay constants
which was carried out in section~\ref{sec:TPpi-rhoP}:
Two kinds of pion decay constants, temporal and spatial
components ($f_\pi^t$ and $f_\pi^s$), appear due to
the lack of Lorentz invariance in medium
and in general they do not agree with each other.
The splitting between $f_\pi^t$ and $f_\pi^s$ appears
at one-loop level in the HLS theory,
which is generated by the thermal loop contribution 
of the vector meson.
[The non-zero splitting also implies that the pion velocity
is not equal to the speed of light.]
On the other hand,
the splitting of the pion decay constant is not generated 
at one-loop level in the ordinary ChPT.

Further by using the Wilsonian matching studied 
in chapter~\ref{ch:WMFT},
the intrinsic temperature dependences of the bare parameters
are included.
We in fact determined the intrinsic thermal effects 
to the bare pion decay constants by performing the matching 
to the operator product expansion (OPE) in section~\ref{ssec:WMCTf}.

\subsubsection{Formulation of the VM in hot matter}

In chapter~\ref{ch:VMHM},
we first presented the formulation of the VM in hot matter
using the HLS theory combined with the Wilsonian matching.
Throughout this thesis, we assume that the chiral phase transition
is of second or weakly first order
so that the axial-vector and vector current correlators 
agree with each other at the critical temperature.
Thus the Wilsonian matching leads to the following conditions
on the parameters of the bare HLS Lagrangian:
\begin{eqnarray}
&&
 g_{\rm bare}(T) \stackrel{T \to T_c}{\to} 0
\qquad
 (\,\mbox{or}\quad M_{\rho,{\rm bare}}(T) \stackrel{T \to T_c}{\to} 0)\,,
\nonumber\\
&&
 a_{\rm bare}(T) = 
 \bigl( F_{\sigma,{\rm bare}}(T) / F_{\pi,{\rm bare}}(T) \bigr)^2
 \stackrel{T \to T_c}{\to} 1\,. \label{VMCs}
\end{eqnarray}
We should stress the following features:
(i) The above conditions for the bare parameters are
realized by the intrinsic thermal effect only.
(ii) $(g,a) = (0,1)$ is a fixed point of RGEs.
The agreement of both current correlators is satisfied 
only in the case that we take into account the intrinsic thermal
effects as well as the hadronic ones.

\begin{description}
\item[Vector meson mass :]
Near $T_c$, the vector meson mass has a positive thermal contribution,
which is proportional to the square of the HLS gauge coupling $g$,
as studied in section~\ref{sec:VMM}.
When we take the limit $T \to T_c$,
both the parametric mass $M_\rho$ and hadronic thermal correction 
go to zero following Eq.~(\ref{VMCs}).
Thus we showed that the vector meson mass vanishes:
\begin{eqnarray}
 m_\rho^2(T) \stackrel{T \lesssim T_c}{\to}
  g^2(T)\Bigl[ a(T)F_\pi^2(\mu = 0;T) + \delta T^2 \Bigr]
 \stackrel{T \to T_c}{\to} 0\,,
\nonumber
\end{eqnarray}
where $\delta T^2$ denotes the thermal contribution
and $F_\pi(\mu = 0;T)$ is the on-shell parametric pion decay
constant obtained in Eq.~(\ref{eq:intrinsic}) to be non-zero.
\item[Order parameter :]
If we write explicitly the quantum and hadronic thermal corrections
to the parameter $F_\pi$,
the pion decay constant at $T_c$ takes the following form
as studied in section~\ref{sec:PDCPV}:
\begin{eqnarray}
 f_\pi^2(T) \stackrel{T \to T_c}{\to}
  F_\pi^2(\Lambda;T_c) - \frac{N_f}{2(4\pi)^2}\Lambda^2
 {}- \frac{N_f}{24}T_c^2
 \,=\, 0\,.
\nonumber
\end{eqnarray}
This implies that
the order parameter goes to zero by the cancellation
among the non-zero bare pion decay constant,
the quantum correction and the leading contribution of matter effect
to $F_\pi^2$.
Here we should note that both the above quantum and 
hadronic contributions include the contribution of 
the massless vector meson
in the same footing as the pion.
\end{description}
The conditions for the bare parameters~(\ref{VMCs}),
which we call ``the VM conditions in hot matter'',
are essential for the VM to take place in hot matter.
Further even when we take into account the Lorentz non-invariance
in the bare HLS theory,
we obtained the similar conditions to Eq.~(\ref{VMCs}) and 
showed that the conditions are still non-renormalized:
There are no quantum corrections to the conditions.
In section~\ref{sec:SPVMM}
we studied the critical behavior of $m_\rho$ and 
showed that $m_\rho (T)$ falls as the quark condensate
in the limit $T \to T_c$:
\begin{eqnarray}
 m_\rho(T) \sim \langle \bar{q}q \rangle_T\,.
\nonumber
\end{eqnarray}

At present,
there are no clear lattice data for the $\rho$ pole mass 
in hot matter.
Our result here will be checked by 
lattice analyses in future.~\cite{lattice-vm}

\subsubsection{Predictions}

In section~\ref{sec:PVM},
we gave the following predictions of the VM in hot matter:
\begin{itemize}
\item
{\bf Critical temperature :}
{}From the order parameter $f_\pi$ including the intrinsic thermal effect
as well as the hadronic correction,
we obtained $T_c = 200\, \mbox{-}\, 250 \,\, \mbox{MeV}$
 for several choices of
the matching scale ($\Lambda = 0.8\, \mbox{-}\, 1.1$\,GeV)
and the scale of QCD 
($\Lambda_{\rm QCD}= 0.30 \, \mbox{-} 0.45$\,GeV).
\item
{\bf Axial-vector and vector charge susceptibilities (ASUS and VSUS):}
The masslss vector meson becomes a relevant degree
of freedom near $T_c$ and it contributes to the VSUS.
The equality between ASUS and VSUS is predicted by the VM.
\item
{\bf Violation of vector dominance :}
We showed that the vector dominance (VD) of the electromagnetic
form factor of the pion is largely violated near $T_c$.
This indicates that the assumption of 
the VD may need to be weakened, at least in some amounts,
for consistently including the
effect of the dropping mass of the vector meson.
\item
{\bf Pion velocity :}
We first showed that the pion velocity at $T_c$ never receive
any quantum and hadronic thermal corrections protected by the VM.
Using the non-renormalization property,
we next performed the Wilsonian matching to the OPE and then
obtained that the pion velocity is close to the speed of light.

This is in contrast to the result obtained from the chiral theory
~\cite{SS}, where the relevant degree of freedom near $T_c$ 
is only the pion.
Their result is that the pion velocity becomes zero for $T \to T_c$.
Therefore by experiment and lattice analysis,
we may be able to distinguish which theory is appropriate
to describe the chiral phase transition.
\end{itemize}

We note that the above estimated values may be changed
by including higher order effects.
On the other hand,
it is expected that the VM is governed by the fixed point 
and not changed by such higher order effects.

\subsubsection{Remnant of the VM in the real world}

In chapter~\ref{ch:CDHLM},
we studied the heavy meson system which consists of
one heavy quark and one light anti-quark,
which was 
motivated by the recent discovery of new $D$ mesons ($J^P = 0^+$ and $1^+$)
in Babar, CLEO and Belle~\cite{BABAR,CLEO,Belle}.
We showed that the mass splitting between new $D$ mesons 
(denoted by $\tilde{D}$ in the text) and 
the  existing $D$ mesons ($J^P = 0^-$ and $1^-$)
is directly proportional to the light quark condensate.
The estimated value was in good agreement with the experiments.
We also gave the predictions on the decay widths and the branching
ratios for the characteristic processes.
In the two-pion decay processes in which the scalar meson
does not mediate, our scenario gives definite predictions,
since the vector meson coupling to the heavy system is
equivalent to the pion coupling due to the VM.
Although the predicted values of widths are small,
we hope that they are clarified in future experiment.

One of the significant results of the analysis,
is that the vector meson plays an important role in
accounting for the splitting in the $D$ and $\tilde{D}$ mesons:
The {\it bare} mass splitting determined through the matching is
estimated as about $190$ MeV, too small to explain the observed
mass difference. 
However by including the quantum corrections
through the hadronic loop, the bare mass splitting is enhanced by
$\sim60\%$, where only the loop effect of the vector meson
contributes to the running of the mass splitting. 
The contributions from the pion loop are completely canceled among
themselves. 
This implies that the observed mass difference can not
be understood if one takes only the pion as the relevant degree of
freedom and that we need other degrees of freedom. 
In the VM, it is nothing but the vector meson. 
The situation here is much like in the calculation of pion velocity 
at the chiral restoration point: 
The pion velocity is zero if the pion is the only effective
degree of freedom but approaches 1 if the vector meson with the VM
is included~\cite{HKRS:SUS,HKRS:PV}.

In the construction of the EFT,
we assumed the VM as the chiral symmetry restoration of QCD
in the light sector, and
then we introduced the effects of the spontaneous chiral 
symmetry breaking by putting the mass term as soft breaking term.
In the analysis,
we assumed that 
the coefficients of the interaction such as $DD\pi$ and $DD\rho$
denoted by $k$ are universal,
which are the remnant of the VM.
Strictly speaking,
each interaction term may have its own
coefficient different from others.
However, we expect that
the effect of these interaction terms is suppressed by the factor
$1/\Lambda$ and as a result the contribution to $\Delta M$ is small
since the dimension of them is higher than that of the mass term.

Furthermore, the result is independent of the deviation of $a$ from
the fixed point value 1, at least at one-loop level. 
This implies that the deviation of
$a$ from 1 which reflects the effect of spontaneous chiral
symmetry breaking in the light quark sector does not get
transmitted to the heavy sector. 
This strongly suggests that the
deviation from $a=1$ involves physics that is not as primary as
the non-vanishing gauge coupling $g \neq 0$ in the description of
the broken phase: The deviation seems to be a ``secondary''
phenomenon, which is generated from $g \neq 0$ as expected in
Refs.~\cite{Georgi}.
 [Although $a=1$
  is the fixed point of the RGE at one-loop level, the deviation of
  $a$ from $1$ is generated by the finite renormalization part
  once we allow the deviation of the gauge coupling $g$
  from $0$ (see Appendix~\ref{app:QC}).]
In fact, even
when we start from the bare HLS theory with $g \neq 0$ and $a=1$, 
the physical quantities obtained through the Wilsonian matching are in
good agreement with experimental results as discussed in
~\cite{HY:PRep}. 
This observation supports the above argument.

The consistency of the physics taking $a_{\rm bare}=1$
and the universal coupling $k$
with the one in the real world remarkably indicates
that the VM is realized as the chiral symmetry restoration in QCD.


\vspace*{0.5cm}

Several comments are in order:

Although we concentrated on the hot matter calculation in this thesis,
the present approach can be applied to the general hot and/or dense
matter calculation.
As mentioned in chapter~\ref{ch:Intro},
the vector meson is one of the most important degrees of freedom
in hot and/or dense matter
since it will be lighter in medium than in the vacuum due to
the (partial) chiral symmetry restoration.
The HLS theory is an efficient implement to study hadron properties
in hot and/or dense matter as well as the chiral phase transition.

One might think that the VM is same as 
the Georgi's vector realization~\cite{Georgi},
in which the order parameter $f_\pi$ is non-zero 
although the chiral symmetry is unbroken.
However in the VM the order parameter certainly becomes zero
and the chiral symmetry is restored accompanied by massless vector meson.
Therefore the VM is consistent with the Ward-Takahashi identity
since it is the Wigner realization~\cite{Yamawaki}.

As shown in Ref.~\cite{HY:VM},
in the VM only the longitudinal $\rho$ couples to the vector current 
near the critical point, 
and the transverse $\rho$ is decoupled from it.
The $A_1$ in the VM is resolved and/or decoupled 
from the axial-vector current near $T_c$ 
since there is no contribution in the vector
current correlator to be matched with the axial-vector correlator.
We expect that the scalar meson is also resolved
and/or decoupled near $T_c$ since it in the VM is in the same
representation as the $A_1$ is in.
We also expect that excited mesons 
are also resolved and/or decoupled.

The VSUS was obtained to be finite, 
which will imply that the screening mass
of the vector meson is also finite at $T_c$.

In this thesis, we performed our analysis at the chiral limit.
We need to include the explicit chiral symmetry breaking
effect from the current quark masses when we apply the present
analysis to the real QCD.  
In such a case, we need the Wilsonian matching
conditions with including non-zero quark mass which have not yet
been established.
Here we expect that 
the qualitative structure obtained in the present analysis
will not be changed by the inclusion of
the current quark masses.

\newpage

\chapter*{Acknowledgment}
\addcontentsline{toc}{chapter}{Acknowledgment}

\begin{center}
\begin{minipage}{15cm}
\qquad
The author is sincerely grateful to Professor Masayasu Harada,
who is her adviser,
and Professor Mannque Rho for plenty of useful discussions
and continuous encouragements throughout her research.
She would like to thank Professor Koichi Yamawaki 
for valuable discussions and encouragement.
She would like to express her thanks to Doctor Youngman Kim,
who is a collaborator,
for many fruitful discussions and exchange of their opinions.

\qquad
She would like to express her sincere gratitude
for all the helps and encouragements she has received.
She would never complete this thesis without their helps.

\qquad
Support by the 21st Century COE
Program of Nagoya University provided by Japan Society for the
Promotion of Science (15COEG01) is much appreciated.
\end{minipage}
\end{center}

\newpage

\appendix

\setcounter{section}{0}
\renewcommand{\thesection}{\Alph{chapter}}
\setcounter{equation}{0}
\renewcommand{\theequation}{\Alph{chapter}.\arabic{equation}}


\chapter{Polarization Tensors at Finite Temperature}
\label{app:PTFT}

At non-zero temperature
there exist four independent polarization tensors,
$u^\mu u^\nu$, $(g^{\mu\nu}-u^\mu u^\nu)$, $P_L^{\mu\nu}$
and $P_T^{\mu\nu}$.
The rest frame of medium is shown by $u^\mu=(1,\vec{0})$.
$P_L^{\mu\nu}$ and $P_T^{\mu\nu}$ are given by~\cite{Kapusta}
  \begin{eqnarray}
   P_{T \mu \nu} &=& g_{\mu i}\Bigl( \delta _{ij} -
                     \frac{\vec{p_i}\vec{p_j}}{\bar{p}^2} \Bigr)
                     g_{j \nu}, \nonumber \\
   P_{L \mu \nu} &=& -\Bigl( g_{\mu \nu} - \frac{p_\mu p_\nu}{p^2} \Bigr)-
                     P_{T \mu \nu}, 
  \label{A.1}
  \end{eqnarray}
where we 
define $p^\mu = (p_0\,,\,\vec{p})$ and 
$\bar{p}=|\vec{p}|$.
They have the following properties:
\begin{eqnarray}
 &P_{L\mu}^\mu = -1,\, P_{T\mu}^\mu = -2,& \nonumber\\
 &P_{L\mu\alpha}P_L^{\alpha\nu} = -{P_{L\mu}}^\nu,\,
 P_{T\mu\alpha}P_T^{\alpha\nu} = -{P_{T\mu}}^\nu,& \nonumber\\
 &P_{L\mu\alpha}P_T^{\alpha\nu}  = 0.& 
\end{eqnarray}

Let us decompose a tensor $\Pi^{\mu\nu}(p_0,\bar{p})$ into
\begin{equation}
 \Pi^{\mu\nu}=u^\mu u^\nu \Pi^t +
              (g^{\mu\nu}-u^\mu u^\nu)\Pi^s +
              P_L^{\mu\nu}\Pi^L + P_T^{\mu\nu}\Pi^T. 
\end{equation}
Each component is obtained as
\begin{eqnarray}
 \Pi^t &=& \Pi^{00} + \frac{\vec{p}_i}{p_0}\Pi^{i0}, \nonumber\\
 \Pi^s &=& -\frac{\vec{p}_i}{\bar{p}}\Pi^{ij}\frac{\vec{p}_j}{\bar{p}} - 
          \frac{p_0 \vec{p}_i}{\bar{p}^2}\Pi^{i0}, \nonumber\\
 \Pi^L-\Pi^s &=& \frac{\vec{p}_i}{\bar{p}}\Pi^{ij}\frac{\vec{p}_j}{\bar{p}}+ 
                \frac{\vec{p}_i}{p_0}\Pi^{i0}, \nonumber\\
 \Pi^T-\Pi^s &=& \frac{1}{2}P_{T\mu\nu}\Pi^{\mu\nu}. 
\end{eqnarray}


\setcounter{equation}{0}
\chapter{Loop Integrals at Finite Temperature}
\label{app:LIFT}

In this appendix we list the explicit forms of the
functions appearing in the hadronic thermal corrections,
$\overline{A}_{0}$, $\overline{B}_{0}$ and
$\overline{B}^{\mu\nu}$ 
[see Eqs.~(\ref{def:A0 2})--(\ref{def:Bmunu 11}) for definitions]
in various limits relevant to the
present analysis.

The functions $\overline{A}_{0}$,
which are independent of
external momentum $p_\mu$,
is given by
 \begin{eqnarray}
  \overline{A}_{0}(M_\rho ;T)
     &=& \tilde{J}^2_{1}(M_\rho ;T)\ , 
  \label{A0BM} \\
  \overline{A}_{0}(0;T)
     &=& \tilde{I}_{2}(T)\ ,
  \label{A0B0}
 \end{eqnarray}
where $\tilde{J}^2_1$ and $\tilde{I}_2$ are 
defined 
in Eqs.~(\ref{I fun}) and (\ref{J fun}).

We list the relevant limits of the functions 
$\overline{B}_{0}(p_0,\bar{p};M_\rho,0;T)$ 
and $\overline{B}^{\mu\nu}(p_0,\bar{p};M_\rho,0;T)$ 
which appear in the two-point function
of $\overline{\cal A}_\mu$.
When 
the pion momentum is taken as its on-shell, 
the function $\overline{B}_{0}$ becomes
\begin{eqnarray}
 &&\overline{B}_{0}
 (p_0 = \bar{p}+i\epsilon, \bar{p};M_\rho,0;T) \nonumber\\
 &=& \int\,\frac{d^3 k}{(2\pi)^3}
   \Biggl[ \frac{-1}{2\omega_\rho}\frac{1}{e^{\omega_\rho/T}-1}
    \Biggl\{ \frac{1}{(\omega_\rho - \bar{p})^2 - (\omega_\pi^p)^2 -
                       2i\epsilon (\omega_\rho - \bar{p})} +
             \frac{1}{(\omega_\rho + \bar{p})^2 - (\omega_\pi^p)^2 +
                       2i\epsilon (\omega_\rho + \bar{p})}
    \Biggr\}\nonumber\\ 
  &&{}+
           \frac{-1}{2\omega_\pi^p}\frac{1}{e^{\omega_\pi^p/T}-1}
    \Biggl\{ \frac{1}{(\omega_\pi^p - \bar{p})^2 - \omega_\rho^2 -
                       2i\epsilon (\omega_\pi^p - \bar{p})} +
             \frac{1}{(\omega_\pi^p + \bar{p})^2 - \omega_\rho^2 +
                       2i\epsilon (\omega_\pi^p + \bar{p})}
    \Biggr\}
   \Biggr]\ , \nonumber\\
\label{B0 A on-shell}
\end{eqnarray}
where we put $\epsilon \rightarrow +0$ to make the analytic
continuation of the frequency $p_0=i2\pi n T$ to the Minkowski
variable, and 
we defined 
\begin{equation}
\omega_\rho = \sqrt{|\vec{k}|^2 + M_\rho^2} \ ,
\quad
\omega_\pi^p = |\vec{k}-\vec{p}| \ .
\label{def: omega-rho omega-pi}
\end{equation}
Two components $\overline{B}^t$ and $\overline{B}^s$ 
of $\overline{B}^{\mu\nu}$ 
in the same limit are given by
\begin{eqnarray}
 &&\overline{B}^t
  (p_0 = \bar{p}+i\epsilon,\bar{p};M_\rho,0;T) \nonumber\\
 &=& \int\,\frac{d^3 k}{(2\pi)^3}
   \Biggl[ \frac{-1}{2\omega_\rho}\frac{1}{e^{\omega_\rho/T}-1}
    \Biggl\{ \frac{(2\omega_\rho - \bar{p})^2}
                   {(\omega_\rho - \bar{p})^2 - (\omega_\pi^p)^2 -
                     2i\epsilon (\omega_\rho - \bar{p})} +
              \frac{(2\omega_\rho + \bar{p})^2}
                   {(\omega_\rho + \bar{p})^2 - (\omega_\pi^p)^2 +
                     2i\epsilon (\omega_\rho + \bar{p})} \nonumber\\
  &&{}-\frac{\vec{p}\cdot(2\vec{k}-\vec{p})}{\bar{p}}
             \Biggl( \frac{2\omega_\rho - \bar{p}}
                           {(\omega_\rho - \bar{p})^2 - (\omega_\pi^p)^2 -
                             2i\epsilon (\omega_\rho - \bar{p})} -
                      \frac{2\omega_\rho + \bar{p}}
                           {(\omega_\rho + \bar{p})^2 - (\omega_\pi^p)^2 +
                             2i\epsilon (\omega_\rho + \bar{p})}
              \Biggr)
    \Biggr\} \nonumber\\
  &&{}+\frac{-1}{2\omega_\pi^p}\frac{1}{e^{\omega_\pi^p/T}-1}
    \Biggl\{ \frac{(2\omega_\pi^p - \bar{p})^2}
                   {(\omega_\pi^p - \bar{p})^2 - \omega_\rho^2 -
                     2i\epsilon (\omega_\pi^p - \bar{p})} +
              \frac{(2\omega_\pi^p + \bar{p})^2}
                   {(\omega_\pi^p + \bar{p})^2 - \omega_\rho^2 +
                     2i\epsilon (\omega_\pi^p + \bar{p})} \nonumber\\
   &&{}+\frac{\vec{p}\cdot(2\vec{k}-\vec{p})}{\bar{p}}
             \Biggl( \frac{2\omega_\pi^p - \bar{p}}
                           {(\omega_\pi^p - \bar{p})^2 - \omega_\rho^2 -
                             2i\epsilon (\omega_\pi^p - \bar{p})} -
                      \frac{2\omega_\pi^p + \bar{p}}
                           {(\omega_\pi^p + \bar{p})^2 - \omega_\rho^2 +
                             2i\epsilon (\omega_\pi^p + \bar{p})}
              \Biggr)
    \Biggr\}
   \Biggr]\ ,
\nonumber\\
\label{Bt A on-shell}
\\
 &&\overline{B}^s
  (p_0 = \bar{p}+i\epsilon,\bar{p};M_\rho,0;T) \nonumber\\
 &=& \int\,\frac{d^3 k}{(2\pi)^3}
   \Biggl[ \frac{1}{2\omega_\rho}
           \frac{1}{e^{\omega_\rho/T}-1} \nonumber\\ 
 &&\times
    \Biggl\{ \frac{(2\vec{k}\cdot\vec{p}-\bar{p}^2)^2}{\bar{p}^2}
     \Biggl( \frac{1}{(\omega_\rho - \bar{p})^2 - (\omega_\pi^p)^2 -
                       2i\epsilon (\omega_\rho - \bar{p})} +
              \frac{1}{(\omega_\rho + \bar{p})^2 - (\omega_\pi^p)^2 +
                       2i\epsilon (\omega_\rho + \bar{p})}
     \Biggr) \nonumber\\ 
  &&{}-\frac{\vec{p}\cdot (2\vec{k}-\vec{p})}{\bar{p}}
     \Biggl( \frac{2\omega_\rho - \bar{p}}
                   {(\omega_\rho - \bar{p})^2 - (\omega_\pi^p)^2 -
                      2i\epsilon (\omega_\rho - \bar{p})}-
              \frac{2\omega_\rho + \bar{p}}
                   {(\omega_\rho + \bar{p})^2 - (\omega_\pi^p)^2 +
                     2i\epsilon (\omega_\rho + \bar{p})}
     \Biggr)
    \Biggr\} \nonumber\\
  &&\qquad\quad{}+\frac{1}{2\omega_\pi^p}
           \frac{1}{e^{\omega_\pi^p/T}-1} \nonumber\\ 
  &&\times
    \Biggl\{ \frac{(2\vec{k}\cdot\vec{p}-\bar{p}^2)^2}{\bar{p}^2}
     \Biggl( \frac{1}{(\omega_\pi^p - \bar{p})^2 - \omega_\rho^2 -
                        2i\epsilon (\omega_\pi^p - \bar{p})} +
              \frac{1}{(\omega_\pi^p + \bar{p})^2 - \omega_\rho^2 +
                        2i\epsilon (\omega_\pi^p + \bar{p})}
     \Biggr) \nonumber\\ 
  &&{}+\frac{\vec{p}\cdot (2\vec{k}-\vec{p})}{\bar{p}}
     \Biggl( \frac{2\omega_\pi^p - \bar{p}}
                   {(\omega_\pi^p - \bar{p})^2 - \omega_\rho^2 -
                     2i\epsilon (\omega_\pi^p - \bar{p})}-
              \frac{2\omega_\pi^p + \bar{p}}
                   {(\omega_\pi^p + \bar{p})^2 - \omega_\rho^2 +
                     2i\epsilon (\omega_\pi^p + \bar{p})}
     \Biggr)
    \Biggr\}
   \Biggr]\ . 
\nonumber\\
\label{Bs A on-shell}
\end{eqnarray}
We further take the $M_\rho\rightarrow0$ limits of the above
expressions.
The limit of 
$\overline{B}_{0}(p_0 = \bar{p}+i\epsilon, \bar{p};M_\rho,0;T)$
includes the infrared logarithmic divergence $\ln M_\rho^2$.
However, it appears multiplied by $M_\rho^2$ in 
$\overline{\Pi}_\perp^t$ and $\overline{\Pi}_\perp^s$, and
the product
$M_\rho^2
\overline{B}_{0}(p_0 = \bar{p}+i\epsilon, \bar{p};M_\rho,0;T)$
vanishes at $M_\rho\rightarrow0$ limit:
\begin{equation}
\lim_{M_\rho \rightarrow 0}
M_\rho^2
\overline{B}_{0}(p_0 = \bar{p}+i\epsilon, \bar{p};M_\rho,0;T)
= 0
\ .
\label{B0 A on-shell VM}
\end{equation}
As for
$\overline{B}^{t,s}(p_0=\bar{p}+i\epsilon,\bar{p};M_\rho,0;T)$,
we obtain
\begin{eqnarray}
 \lim_{M_\rho \to 0}
 \overline{B}^t(p_0=\bar{p}+i\epsilon,\bar{p};M_\rho,0;T)
 &=& -2 \tilde{I}_{2}(T)\ , \nonumber\\
 \lim_{M_\rho \to 0}
 \overline{B}^s(p_0=\bar{p}+i\epsilon,\bar{p};M_\rho,0;T)
 &=& -2 \tilde{I}_{2}(T)\ .
\label{Bts A on-shell VM}
\end{eqnarray}

In the static limit ($p_0 \to 0$), the functions
$M_\rho^2 \overline{B}_{0}(p_0,\bar{p};M_\rho,0;T)$ and
$\overline{B}^t(p_0,\bar{p};M_\rho,0;T)-
\overline{B}^L(p_0,\bar{p};M_\rho,0;T)$ become
\begin{eqnarray}
 &&\lim_{p_0 \to 0}
 M_\rho^2 \overline{B}_{0}(p_0,\bar{p};M_\rho,0;T) \nonumber\\
 &&\qquad\qquad = M_\rho^2 \int\frac{d^3 k}{(2\pi)^3}
   \frac{-1}{\omega_\rho^2 - (\omega_\pi^p)^2}
   \Biggl[ \frac{1}{\omega_\rho}\frac{1}{e^{\omega_\rho/T}-1} -
           \frac{1}{\omega_\pi^p}\frac{1}{e^{\omega_\pi^p/T}-1}
   \Biggr],
\label{B0B_static} \\
 &&\lim_{p_0 \to 0}
 \Bigl[
  \overline{B}^t(p_0,\bar{p};M_\rho,0;T) -
  \overline{B}^L(p_0,\bar{p};M_\rho,0;T)
 \Bigr] \nonumber\\
 &&\qquad\qquad = \int\frac{d^3 k}{(2\pi)^3}
   \frac{-4}{\omega_\rho^2 - (\omega_\pi^p)^2}
   \Biggl[ \frac{\omega_\rho}{e^{\omega_\rho/T}-1} -
           \frac{\omega_\pi^p}{e^{\omega_\pi^p/T}-1}
   \Biggr].
\label{Bt-BL static}
\end{eqnarray}
Taking the low momentum limits of these expressions, we obtain
\begin{eqnarray}
 &&\lim_{\bar{p} \to 0}
 M_\rho^2 \overline{B}_{0}(p_0=0,\bar{p};M_\rho,0;T)
  = \tilde{I}_{2}(T) - \tilde{J}_{1}^2(M_\rho;T), 
\label{B0B_staticL} \\
 &&\lim_{\bar{p} \to 0}
 \Bigl[
  \overline{B}^t(p_0=0,\bar{p};M_\rho,0;T) -
  \overline{B}^L(p_0=0,\bar{p};M_\rho,0;T)
 \Bigr] \nonumber\\
 &&\qquad\qquad = -\frac{4}{M_\rho^2}\Bigl[
     \tilde{J}_{-1}^2(M_\rho;T) - \tilde{I}_{4}(T) \Bigr].
\label{Bt-BL staticL}
\end{eqnarray}
Moreover, in the VM limit ($M_\rho \to 0$), these are reduced to
\begin{eqnarray}
 &&\lim_{\bar{p} \to 0}
 M_\rho^2 \overline{B}_{0}(p_0=0,\bar{p};M_\rho,0;T)
 \stackrel{M_\rho \to 0}{\to} 0, \nonumber\\
 &&\lim_{\bar{p} \to 0}
 \Bigl[
  \overline{B}^t(p_0=0,\bar{p};M_\rho,0;T) -
  \overline{B}^L(p_0=0,\bar{p};M_\rho,0;T)
 \Bigr]
 \stackrel{M_\rho \to 0}{\to} 2\tilde{I}_{2}(T)\ .
\label{B0B Bt-BL staticLM}
\end{eqnarray}

We consider the functions $\overline{B}_0$ and 
$\overline{B}^{\mu\nu}$ appearing in the two-point functions of
$\overline{\cal V}_\mu$ and $\overline{V}_\mu$.
$\overline{B}^t$ and $\overline{B}^s$ are
calculated as
\begin{eqnarray}
 &\overline{B}^{t}(p_0,\bar{p};0,0;T)
 =\overline{B}^{s}(p_0,\bar{p};0,0;T)
 = - 2 \overline{A}_{0}(0;T)
 = -2 \tilde{I}_{2}(T),& \nonumber\\
 &\overline{B}^{t}(p_0,\bar{p};M_\rho ,M_\rho ;T)
 = \overline{B}^{s}(p_0,\bar{p};M_\rho ,M_\rho ;T)
 = - 2 \overline{A}_{0}(M_\rho;T)
 = -2 \tilde{J}_{1}^2(M_\rho;T).&
\nonumber\\ 
\label{B.8}
\end{eqnarray}

The function $\overline{B}_{0}$ is expressed as
\begin{eqnarray}
 &&\overline{B}_{0}(p_0,\bar{p};M_\rho,M_\rho;T) \nonumber\\
 &=& \int\,\frac{d^3 k}{(2\pi)^3}
   \Biggl[ \frac{-1}{2\omega_\rho}\frac{1}{e^{\omega_\rho/T}-1}
    \Biggl\{ \frac{1}{(\omega_\rho - p_0)^2 - (\omega_\rho^p)^2} +
             \frac{1}{(\omega_\rho + p_0)^2 - (\omega_\rho^p)^2}
    \Biggr\}\nonumber\\ 
 &&\qquad\quad {}+
           \frac{-1}{2\omega_\rho^p}\frac{1}{e^{\omega_\rho^p/T}-1}
    \Biggl\{ \frac{1}{(\omega_\rho^p - p_0)^2 - \omega_\rho^2} +
             \frac{1}{(\omega_\rho^p + p_0)^2 - \omega_\rho^2}
    \Biggr\}
   \Biggr]\ , 
\end{eqnarray}
where we define 
\begin{equation}
\omega_\rho^p = \sqrt{ |\vec{k}-\vec{p}|^2 + M_\rho^2 }
\ .
\end{equation}
In the static limit $(p_0 \to 0)$, the functions $\overline{B}_0$ and
$\overline{B}^L$ are expressed as
\begin{eqnarray}
 &&\lim_{p_0 \to 0}
 M_\rho^2 \overline{B}_{0}(p_0,\bar{p};M_\rho ,M_\rho ;T)
\nonumber\\
 &&\qquad\qquad = M_\rho^2 \int \frac{d^3 k}{(2\pi)^3}
   \frac{-1}{\omega_\rho^2 - (\omega_\rho^p)^2}
   \Biggl[ \frac{1}{\omega_\rho}\frac{1}{e^{\omega_\rho/T}-1} -
           \frac{1}{\omega_\rho^p}\frac{1}{e^{\omega_\rho^p/T}-1}
   \Biggr], \\
 &&\lim_{p_0 \to 0}
 \overline{B}^L(p_0,\bar{p};M_\rho,M_\rho;T) \nonumber\\
 &&\qquad\qquad= \int \frac{d^3 k}{(2\pi)^3}
   \frac{1}{\vec{p}\cdot(2\vec{k}-\vec{p})}
   \Biggl[ \frac{1}{\omega_\rho}\frac{1}{e^{\omega_\rho/T}-1}
    \Bigl\{ 4\omega_\rho^2 - \vec{p}\cdot(2\vec{k}-\vec{p})
    \Bigr\} \nonumber\\
 &&\qquad\qquad\qquad\qquad\qquad\qquad\quad
      {}-  \frac{1}{\omega_\rho^p}\frac{1}{e^{\omega_\rho^p/T}-1}
    \Bigl\{ 4(\omega_\rho^p)^2 + \vec{p}\cdot(2\vec{k}-\vec{p})
    \Bigr\}
   \Biggr].
\end{eqnarray}
Taking the low momentum limit $(\bar{p} \to 0)$, 
these expressions become
\begin{eqnarray}
 &&\lim_{\bar{p} \to 0}
  M_\rho^2 \overline{B}_{0}(p_0=0,\bar{p};M_\rho ,M_\rho ;T)
  = \frac{1}{2}\Bigl[ \tilde{J}_{-1}^0(M_\rho;T) -
    \tilde{J}_{1}^2(M_\rho;T) \Bigr], \\
 &&\lim_{\bar{p} \to 0}
  \overline{B}^L(p_0=0,\bar{p};M_\rho,M_\rho;T)
  = {}-2\Bigl[ M_\rho^2 \tilde{J}_{1}^0(M_\rho;T) +
     2\tilde{J}_{1}^2(M_\rho;T) \Bigr].
\end{eqnarray}
In the VM limit $(M_\rho \to 0)$, these are reduced to
\begin{eqnarray}
 &&\lim_{\bar{p} \to 0}
  M_\rho^2 \overline{B}_{0}(p_0=0,\bar{p};M_\rho ,M_\rho ;T)
  \stackrel{M_\rho \to 0}{\to} 0, \\
 &&\lim_{\bar{p} \to 0}
  \overline{B}^L(p_0=0,\bar{p};M_\rho,M_\rho;T)
  \stackrel{M_\rho \to 0}{\to} {}-4\tilde{I}_2(T).
\end{eqnarray}

On the other hand, the functions 
$\overline{B}_0, \overline{B}^L-\overline{B}^s$
and $\overline{B}^T-\overline{B}^s$ in the low momentum limit $(\bar{p} \to 0)$
are given by
\begin{eqnarray}
 &&\lim_{\bar{p} \to 0}
 \overline{B}_{0}(p_0,\bar{p};M_\rho ,M_\rho ;T) 
 = \frac{1}{2}\tilde{F}^2_{3}(p_0;M_\rho ;T) \ ,
\label{B0B rest} \\
 &&\lim_{\bar{p} \to 0}\Bigl[
 \overline{B}^L(p_0,\bar{p};M_\rho,M_\rho;T)-
 \overline{B}^s(p_0,\bar{p};M_\rho,M_\rho;T)
 \Bigr]
  = \frac{2}{3}\tilde{F}_{3}^4(p_0;M_\rho;T), 
\label{BL rest}\\
 &&\lim_{\bar{p} \to 0}\Bigl[
 \overline{B}^T(p_0,\bar{p};M_\rho,M_\rho;T)-
 \overline{B}^s(p_0,\bar{p};M_\rho,M_\rho;T)
 \Bigr]
 = \frac{2}{3}\tilde{F}_{3}^4(p_0;M_\rho;T), 
  \label{BT rest}
\end{eqnarray}
where the function $\tilde{F}_{3}^n$ is
defined in Appendix~\ref{app:Functions}.


\setcounter{equation}{0}
\chapter{Quantum Corrections}
\label{app:QC}

In this appendix we briefly summarize the quantum corrections to 
the two-point functions.

Let us define
the functions $B_0^{\rm(vac)}$, $B^{{\rm(vac)}\mu\nu}$ and
$A_0^{\rm(vac)}$
by the following integrals~\cite{HY:PRep}:
\begin{eqnarray}
  A_0^{\rm(vac)}(M)
   &=& \int \frac{d^n k}{i(2\pi)^4}
      \frac{1}{M^2 - k^2}\ , 
\label{A0vac def app}
\\
  B_0^{\rm(vac)}(p;M_1,M_2)
   &=& \int \frac{d^n k}{i(2\pi)^4}
      \frac{1}{[M_1^2-k^2][M_2^2-(k-p)^2]}\ , 
\label{B0vac def app}
\\
  B^{{\rm(vac)}\mu\nu}(p;M_1,M_2)
   &=& \int \frac{d^n k}{i(2\pi)^4}
      \frac{(2k-p)^\mu (2k-p)^\nu }{[M_1^2-k^2][M_2^2-(k-p)^2]}\ .
\label{Bmnvac def app}
\end{eqnarray}
In the present analysis it is important to include the quadratic
divergences to obtain the RGEs in the Wilsonian sense.
Here, following 
Refs.~\cite{HY:conformal,HY:WM,HY:PRep}, we adopt the
dimensional regularization and identify the quadratic divergences with
the presence of poles of ultraviolet origin at $n=2$~\cite{Veltman}.
This can be done by the following replacement in the Feynman
integrals:
\begin{equation}
\int \frac{d^n k}{i (2\pi)^n} \frac{1}{-k^2} \rightarrow 
\frac{\Lambda^2} {(4\pi)^2} \ ,
\qquad
\int \frac{d^n k}{i (2\pi)^n} 
\frac{k_\mu k_\nu}{\left[-k^2\right]^2} \rightarrow 
- \frac{\Lambda^2} {2(4\pi)^2} g_{\mu\nu} \ .
\label{quad repl app}
\end{equation}
On the other hand, 
the logarithmic divergence is identified with the pole at 
$n=4$:
\begin{equation}
\frac{1}{\bar{\epsilon}} + 1 \rightarrow
\ln \Lambda^2
\ ,
\label{logrepl:2 app}
\end{equation}
where
\begin{equation}
\frac{1}{\bar{\epsilon}} \equiv
\frac{2}{4 - n } - \gamma_E + \ln (4\pi)
\ ,
\end{equation}
with $\gamma_E$ being the Euler constant.

By using the replacements in Eqs.~(\ref{quad repl app}) and
(\ref{logrepl:2 app}), the integrals in
Eqs.~(\ref{A0vac def app}), (\ref{B0vac def app}) and (\ref{Bmnvac def app})
are evaluated as
\begin{eqnarray}
&&
A_0^{\rm(vac)}(M) 
=
\frac{\Lambda^2} {(4\pi)^2}
- \frac{M^2}{(4\pi)^2} \ln \frac{\Lambda^2}{M^2}
\ ,
\label{A0vac}
\\
&&
  B_0^{\rm(vac)}(p^2;M_1,M_2) 
=
\frac{1}{(4\pi)^2}
\left[
  \ln \Lambda^2 - 1 - F_0(p^2;M_1,M_2)
\right]
\ ,
\label{B0vac}
\\
&&
  B^{{\rm(vac)}\mu\nu}(p;M_1,M_2)
\nonumber\\
&& \quad
=
  - g^{\mu\nu} \frac{1}{(4\pi)^2}
  \left[ 
    2 \Lambda^2 - M_1^2 \ln \frac{\Lambda^2}{M_1^2}
   - M_2^2 \ln \frac{\Lambda^2}{M_2^2}
   - (M_1^2-M_2^2) F_A(p^2;M_1,M_2)
  \right]
\nonumber\\
&& \qquad
  {}- \left( g^{\mu\nu}p^2 - p^\mu p^\nu \right) 
    \frac{1}{(4\pi)^2} 
  \left[ 
    \frac{1}{3} \ln \Lambda^2 - F_0(p^2;M_1,M_2)
    + 4 F_3(p^2;M_1,M_2)
  \right]
\ ,
\label{div:Bmunu 2}
\end{eqnarray}
where $F_0$, $F_A$ and $F_3$ are defined by
\begin{eqnarray}
F_0 (s;M_1,M_2) &=& \int^1_0 dx \ln 
\left[ (1-x) M_1^2 + x M_2^2 - x (1-x) s \right] \ , \nonumber\\
F_A (s;M_1,M_2) &=& \int^1_0 dx\,(1-2x)\, \ln 
\left[ (1-x) M_1^2 + x M_2^2 - x (1-x) s \right] \ , \nonumber\\
F_3 (s;M_1,M_2) &=& \int^1_0 dx \, x (1-x) 
\ln \left[ (1-x) M_1^2 + x M_2^2 - x (1-x) s \right] 
\ .
\end{eqnarray}

We consider the quantum corrections denoted by $\Pi^{{\rm (vac)}\mu\nu}$
[see Eq.~(\ref{eq:TPart})].
At tree level
the two-point functions of $\overline{\cal A}_\mu$,
$\overline{\cal V}_\mu$ and $\overline{V}_\mu$ are given by
\begin{eqnarray}
 \Pi_\perp^{\rm{(tree)}\mu\nu}(p)
  &=& F_{\pi, \rm{bare}}^2 g^{\mu\nu} +
     2z_{2, \rm{bare}}(p^2 g^{\mu\nu} - p^\mu p^\nu)\ , \nonumber\\
 \Pi_\parallel^{\rm{(tree)}\mu\nu}(p)
  &=& F_{\sigma, \rm{bare}}^2 g^{\mu\nu} +
     2z_{1, \rm{bare}}(p^2 g^{\mu\nu} - p^\mu p^\nu)\ , \nonumber\\
 \Pi_V^{\rm{(tree)}\mu\nu}(p)
  &=& F_{\sigma, \rm{bare}}^2 g^{\mu\nu} -
     \frac{1}{g_{\rm{bare}}^2}
     (p^2 g^{\mu\nu} - p^\mu p^\nu)
\ ,\nonumber\\
 \Pi_{V\parallel}^{\rm{(tree)}\mu\nu}(p)
  &=& {}- F_{\sigma, \rm{bare}}^2 g^{\mu\nu} +
     z_{3, \rm{bare}}(p^2 g^{\mu\nu} - p^\mu p^\nu)
\ .
\label{TP tree app}
\end{eqnarray}
Thus the one-loop contributions to
$\Pi_\perp^{{\rm (vac)}\mu\nu}$
give the quantum corrections to $F_\pi ^2$ and $z_2$.
Similarly,
each of the one-loop contributions to
$\Pi_\parallel^{{\rm (vac)}\mu\nu}$, $\Pi_V^{{\rm (vac)}\mu\nu}$ and
$\Pi_{V\parallel}^{{\rm (vac)}\mu\nu}$ includes the quantum corrections to 
two parameters shown above.
For distinguishing the quantum corrections to two parameters included
in the two-point function,
it is convenient to decompose each two-point function as
\begin{equation}
 \Pi^{{\rm (vac)}\mu\nu}(p)=\Pi^{{\rm (vac)}S}(p)g^{\mu\nu} +
                 \Pi^{{\rm (vac)}LT}(p)(g^{\mu\nu}p^2 - p^\mu p^\nu).
\label{decomp T0 app}
\end{equation}
It should be noticed that, since we use the Lagrangian with Lorentz
invariance, the form of the quantum corrections is Lorentz invariant.
Then, the relation between four components given in 
Eq.~(\ref{Pi perp decomp}) and two components
shown above are given by
\begin{eqnarray}
 &&\Pi^{{\rm (vac)}t} = \Pi^{{\rm (vac)}s} = \Pi^{{\rm (vac)}S},
\nonumber\\
 &&\Pi^{{\rm (vac)}L} = \Pi^{{\rm (vac)}T} = \Pi^{{\rm (vac)}LT}.
\end{eqnarray}
Using the decomposition in Eq.~(\ref{decomp T0 app}), we identify 
$\Pi^{{\rm (vac)}S}_{\perp\mbox{\scriptsize(1-loop)}}(p^2)$ with the quantum
correction to $F_\pi^2$, 
$\Pi^{{\rm (vac)}LT}_{\perp\mbox{\scriptsize(1-loop)}}(p^2)$ 
with that to $z_2$,
and so on.
It should be noticed that
the following relation is satisfied~\cite{HY:WM,HY:PRep}:
\begin{equation}
\Pi_V^{{\rm (vac)}S}(p^2) = \Pi_\parallel^{{\rm (vac)}S}(p^2) =
- \Pi_{V\parallel}^{{\rm (vac)}S}(p^2) \ .
\label{Pi V S equal app}
\end{equation}
Then the quantum correction to $F_\sigma^2$ can be extracted from
any of $\Pi_\parallel^{\mu\nu}$,
$\Pi_V^{\mu\nu}$ and $\Pi_{V\parallel}^{\mu\nu}$.

We note that in Refs.~\cite{HY:WM,HY:PRep} the finite corrections
of $\Pi^{{\rm (vac)}\mu\nu}_{\mbox{\scriptsize(1-loop)}}$ are neglected
by assuming that they are small.
In this paper
we include these finite contributions
in addition to the divergent corrections.
As in Refs.~\cite{HShibata,HS:VM},
we adopt the on-shell renormalization condition.
They are expressed as
\footnote{
  In the framework of the ChPT with HLS, 
  the renormalization point $\mu$ should be taken 
  as $\mu \geq M_\rho$
  since the vector meson decouples at $\mu = M_\rho$.
  Below the scale $M_\rho$ the parameter $F_\pi$, which is expressed 
  as $F_\pi^{(\pi)}(\mu)$ in Refs.~\cite{HY:WM,HY:PRep}, runs by the 
  quantum correction from the pion loop 
  alone.
  Then
  $F_\pi^2(\mu=0)$ 
  in the right-hand-side of the renormalization condition 
  Eq.~(\ref{Fpi renorm cond}) is defined by
  $[F_\pi^{(\pi)}(\mu=0)]^2$.
  Due to the presence of the quadratic divergence $F_\pi(M_\rho)$
  is not connected smoothly to $F_\pi^{(\pi)}(M_\rho)$.
  The relation between them is expressed as~\cite{HY:WM,HY:PRep}
  $ [F_\pi^{(\pi)}(M_\rho)]^2 = F_\pi^2(M_\rho) +
    [ N_f/(4\pi)^2] [ a(M_\rho)/2 ] M_\rho^2$,
  where $a(\mu)$ for $\mu>M_\rho$ is defined as
  $ a(\mu) \equiv F_\sigma^2(\mu)/ F_\pi^2(\mu)$.
  By using this relation, the renormalization condition
  (\ref{Fpi renorm cond}) determines the condition
  for $F_\pi^2(M_\rho)$.
  Strictly speaking,
  we should use $F_\pi(M_\rho)$ in the calculations in the present analysis.
  However, the difference between $F_\pi(0)$ and $F_\pi(M_\rho)$ 
  inside the loop correction coming from the finite renormalization
  effect is of higher order, and we use $F_\pi(0)$ 
  inside one-loop corrections in this paper.
}:
\begin{eqnarray}
 \mbox{Re}\Bigl[\Pi_\perp^{{\rm (vac)}S} (p^2=0)\Bigr]
  &=& F_\pi^2 (\mu = 0)\ , 
\label{Fpi renorm cond}
\\
 \mbox{Re}\Bigl[\Pi_V^{{\rm (vac)}S} (p^2=M_\rho^2)\Bigr]
  &=& F_\sigma^2 (\mu = M_\rho), 
\label{Fs ren cond}
\\
 \mbox{Re}\Bigl[\Pi_V^{{\rm (vac)}LT} (p^2=M_\rho^2)\Bigr]
  &=& - \frac{1}{g^2(\mu = M_\rho)},
\label{g ren cond}
\end{eqnarray}
where $\mu$ denotes the renormalization point.
Using the above renormalization conditions,
we can write the two-point functions as
\begin{eqnarray}
\Pi_\perp^{{\rm (vac)}S}(p^2) &=&
  F_\pi^2(0) + \widetilde{\Pi}_\perp^S(p^2) \ ,
\nonumber\\
\Pi_V^{{\rm (vac)}S}(p^2) &=& 
  F_\sigma^2(M_\rho) + \widetilde{\Pi}_V^S(p^2) \ ,
\nonumber\\
\Pi_V^{{\rm (vac)}LT}(p^2) &=& 
  - \frac{1}{g^2(M_\rho)} + \widetilde{\Pi}_V^{LT}(p^2) \ ,
\label{finite renormalization effects}
\end{eqnarray}
where $\widetilde{\Pi}_\perp^S(p^2)$,
$\widetilde{\Pi}_V^S(p^2)$ and $\widetilde{\Pi}_V^{LT}(p^2)$
denote the finite renormalization effects.
Their explicit forms are listed below.
{}From the above renormalization conditions
they satisfy
\begin{equation}
\widetilde{\Pi}_\perp^S(p^2=0)
=
\mbox{Re}\,\widetilde{\Pi}_V^S(p^2=M_\rho^2)
=
\mbox{Re}\,\widetilde{\Pi}_V^{LT}(p^2=M_\rho^2)
= 0
\ .
\label{zero FRE}
\end{equation}
Thus in the present renormalization scheme,
all the quantum effects for the on-shell parameters at leading order
are included through the renormalization group equations.

In the following, using the above functions, we summarize
the quantum corrections to two components
$\Pi^{{\rm (vac)}S}$ and $\Pi^{{\rm (vac)}LT}$ of the two-point functions
$\Pi_\perp^{{\rm (vac)}\mu\nu},\,\Pi_\parallel^{{\rm (vac)}\mu\nu},\,
\Pi_V^{{\rm (vac)}\mu\nu}$
and $\Pi_{V\parallel}^{{\rm (vac)}\mu\nu}$ which are defined in 
Eq.~(\ref{decomp T0 app}).
For $\overline{\cal A}_\mu$-$\overline{\cal A}_\nu$ two-point function,
we obtain
\begin{eqnarray}
 \Pi_{\perp\mbox{\scriptsize(1-loop)}}^{{\rm (vac)}S}(p^2)
  &=& - \frac{N_f}{(4\pi)^2}
    \Bigl[ \frac{2-a}{2}\Lambda^2 + \frac{3}{4}a\,M_\rho^2 \ln\Lambda^2 
     {}+ \frac{1}{4}a\,M_\rho^2 \ln M_\rho^2 \nonumber\\
  &&\qquad\qquad{}- a\,M_\rho^2 \Bigl\{
    1 + F_0(p^2;M_\rho,0) + \frac{1}{4}F_A(p^2;M_\rho,0)
    \Bigr\} \Bigr], \nonumber\\
 \Pi_{\perp\mbox{\scriptsize(1-loop)}}^{{\rm (vac)}LT}(p^2)
  &=& - \frac{N_f}{(4\pi)^2}\frac{a}{4}
    \Bigl[ \frac{1}{3}\ln\Lambda^2 - F_0(p^2;M_\rho,0) + 
      4F_3(p^2;M_\rho,0) \Bigr]. 
\end{eqnarray}
Corrections to
$\overline{\cal V}_\mu$-$\overline{\cal V}_\nu$ two-point function
are given by
\begin{eqnarray}
 \Pi_{\parallel\mbox{\scriptsize(1-loop)}}^{{\rm (vac)}S}(p^2)
  &=& -\frac{N_f}{(4\pi)^2}
   \Bigl[ \frac{a^2 + 1}{4}\Lambda^2 + \frac{3}{4}M_\rho^2\ln M_\rho^2
    {}+ M_\rho^2 \Bigl\{ \frac{1}{4}\ln M_\rho^2 - 1 -
    F_0(p^2;M_\rho,M_\rho) \Bigr\}\Bigr], \nonumber\\
 \Pi_{\parallel\mbox{\scriptsize(1-loop)}}^{{\rm (vac)}LT}(p^2)
  &=& -\frac{N_f}{(4\pi)^2}\frac{1}{8}
   \Bigl[ \frac{a^2 - 4a + 5}{3}\ln\Lambda^2 - F_0(p^2;M_\rho,M_\rho)
\nonumber\\ 
  &&\qquad\qquad\qquad\qquad{}+ 4F_3(p^2;M_\rho,M_\rho) - 
    4(2-a)^2 \ln M_\rho^2 \Bigr]. 
\end{eqnarray}
As for $\overline{V}_\mu$-$\overline{V}_\nu$ we have
\begin{eqnarray}
 \Pi_{V\mbox{\scriptsize(1-loop)}}^{{\rm (vac)}S}(p^2) 
  &=& \Pi_{\parallel\mbox{\scriptsize(1-loop)}}^{{\rm (vac)}S}(p^2), 
\nonumber\\
 \Pi_{V\mbox{\scriptsize(1-loop)}}^{{\rm (vac)}LT}(p^2)
  &=& -\frac{N_f}{(4\pi)^2}
   \Bigl[ \frac{a^2 - 87}{24}\ln\Lambda^2 + 4 + 
     \frac{23}{8}F_0(p^2;M_\rho,M_\rho) \nonumber\\ 
  &&\qquad\qquad{}+ 
     \frac{9}{2}F_3(p^2;M_\rho,M_\rho) -
     \frac{a^2}{8} \Bigl\{ F_0(p^2;0,0) - 4F_3(p^2;0,0) \Bigr\}\Bigr].
\end{eqnarray}
Finally, corrections to $\overline{V}_\mu$-$\overline{\cal V}_\nu$
two-point function are expressed as
\begin{eqnarray}
 \Pi_{V\parallel\mbox{\scriptsize(1-loop)}}^{{\rm (vac)}S}(p^2)
  &=& - \Pi_{\parallel\mbox{\scriptsize(1-loop)}}^{{\rm (vac)}S}(p^2), 
\nonumber\\
 \Pi_{V\parallel\mbox{\scriptsize(1-loop)}}^{{\rm (vac)}LT}(p^2)
  &=& -\frac{N_f}{(4\pi)^2}\frac{1}{8}
   \Bigl[ \frac{1 + 2a - a^2}{3} \ln\Lambda^2 - a(2-a)\ln M_\rho^2
\nonumber\\
  &&\qquad\qquad\qquad\qquad {}- F_0(p^2;M_\rho,M_\rho) +
   4F_3(p^2;M_\rho,M_\rho) \Bigr]. 
\end{eqnarray}

The renormalization conditions in Eqs.~(\ref{Fpi renorm cond})-
(\ref{g ren cond}) lead to the following relations among the bare and
renormalized parameters:
\begin{eqnarray}
 &&F_{\pi,\rm{bare}}^2-\frac{N_f}{4(4\pi)^2}
 \bigl[ 2(2-a)\Lambda^2 + 3aM_\rho^2 \ln\Lambda^2
 \bigr] \nonumber\\
 &&\qquad = F_\pi^2(0) - \frac{N_f}{4(4\pi)^2}aM_\rho^2
    \bigl[ 3\ln M_\rho^2 + \frac{1}{2}
    \bigr], \\
 &&F_{\sigma,\rm{bare}}^2 - \frac{N_f}{4(4\pi)^2}
  \bigl[ (a^2 + 1)\Lambda^2 + 3M_\rho^2 \ln\Lambda^2
  \bigr] \nonumber\\
 &&\qquad = F_\sigma^2(M_\rho) - \frac{N_f}{4(4\pi)^2}M_\rho^2
    \bigl[ 3\ln M_\rho^2 - 4(1-\sqrt{3}\tan^{-1}\sqrt{3})
    \bigr], \\
 &&\frac{1}{g_{\rm{bare}}^2} - \frac{N_f}{(4\pi)^2}
  \frac{87-a^2}{24}\ln\Lambda^2 \nonumber\\
 &&\qquad = \frac{1}{g^2(M_\rho)} - \frac{N_f}{8(4\pi)^2}
    \bigl[ \frac{87-a^2}{3}\ln M_\rho^2 - \frac{147-5a^2}{9}+
           41\sqrt{3}\tan^{-1}\sqrt{3}
    \bigr]. 
\end{eqnarray}

{}From these relations,
we obtain the RGEs for the parameters $F_\pi,\, g$ and $a$ 
as~\cite{HY:WM}
\begin{eqnarray}
 \mu \frac{d{F_\pi}^2}{d\mu} 
 &=& \frac{N_f}{2(4\pi)^2}
     \Bigl[ 3a^2 g^2 {F_\pi}^2 + 2(2-a){\mu}^2\Bigr],
 \label{eq:RGEF}\\
 \mu \frac{da}{d\mu} 
 &=& -\frac{N_f}{2(4\pi)^2}(a-1)
     \Bigl[ 3a(a+1)g^2 - (3a-1)\frac{{\mu}^2}{{F_\pi}^2} \Bigr],
 \label{eq:RGEa}\\
 \mu \frac{dg^2}{d\mu} 
 &=& -\frac{N_f}{2(4\pi)^2}\frac{87-a^2}{6}g^4.
 \label{eq:RGEg}
\end{eqnarray}

The finite renormalization effects of the two-point functions
are expressed as
\begin{eqnarray}
 \tilde{\Pi}_\perp^S(p^2)
 &=& \frac{N_f}{(4\pi)^2}a M_\rho^2 
    \Bigl[ -\Bigl( 1-\frac{M_\rho^2}{4p^2} \Bigr)
     \Bigl\{ 1-\Bigl( 1-\frac{M_\rho^2}{p^2} \Bigr)
      \ln \Bigl( 1-\frac{p^2}{M_\rho^2} \Bigr)\Bigr\} -
     \frac{1}{8} \Bigr] ,\\
 \tilde{\Pi}_V^S(p^2)
 &=& \frac{N_f}{(4\pi)^2}M_\rho^2
    \Bigl[ -\sqrt{3}\tan^{-1}\sqrt{3} +
     2\sqrt{\frac{4M_\rho^2 - p^2}{p^2}}
      \tan^{-1}\sqrt{\frac{p^2}{4M_\rho^2 - p^2}} \Bigr]\ ,
\label{C.2}
\\
 \tilde{\Pi}_V^{LT}(p^2)
 &=& \frac{N_f}{8(4\pi)^2}
    \Bigl[ \frac{a^2}{3}\ln \Bigl( \frac{-p^2}{M_\rho^2} \Bigr)
      {}- 24\Bigl( 1-\frac{M_\rho^2}{p^2} \Bigr) +
     41\sqrt{3}\tan^{-1}\sqrt{3} 
\nonumber \\
 &&\qquad\qquad 
    {}- \frac{2(12M_\rho^2 + 29p^2)}{p^2}
        \sqrt{\frac{4M_\rho^2 - p^2}{p^2}}
        \tan^{-1}\sqrt{\frac{p^2}{4M_\rho^2 - p^2}} \Bigr] .
\label{C.3}
\end{eqnarray}


\setcounter{equation}{0}
\chapter{Functions}
\label{app:Functions}

In this appendix, we list the integral forms of the
functions which appear in 
the expressions of the physical quantities and the
several limits of the loop integrals shown in
Appendix~\ref{app:LIFT}.
The functions $\tilde{I}_{n}(T)$ and
$\tilde{J}^n_{m}(M ;T)$ ($n$, $m$: integers)
are given by
 \begin{eqnarray}
  &&\qquad\quad \tilde{I}_{n}(T) 
   = \int \frac{d^3 k}{(2\pi)^3}\frac{|\vec{k}|^{n-3}}{e^{k/T}-1}
   = \frac{1}{2\pi^2}\hat{I}_{n} T^n \ , \nonumber\\
  &&\qquad\qquad\qquad \hat{I}_{2} = \frac{{\pi}^2}{6},\quad
  \hat{I}_{4} = \frac{{\pi}^4}{15}\ , 
\label{I fun}
\\
  &&\quad \tilde{J}^n_{m}(M ;T) 
   = \int \frac{d^3 k}{(2\pi)^3} \frac{1}{e^{\omega /T}-1}
      \frac{|\vec{k}|^{n-2}}{{\omega}^m}\ , 
\label{J fun}
\end{eqnarray}
where
\begin{equation}
  \omega = \sqrt{k^2 + {M}^2}\ .
\end{equation}
The functions $\tilde{F}^n_{3}(p_0;M;T)$ and
$\tilde{G}_{n}(p_0;T)$, which appear in the vector meson pole
mass in section~\ref{sec:VMM}, are defined as
\begin{eqnarray}
  &&\tilde{F}^n_{3}(p_0;M;T) 
   = \int \frac{d^3 k}{(2\pi)^3}\frac{1}{e^{\omega /T}-1}
      \frac{4|\vec{k}|^{n-2}}{\omega (4{\omega}^2 - {p_0}^2)}\ , 
\nonumber\\
  &&\quad \tilde{G}_{n}(p_0;T) 
   = \int \frac{d^3 k}{(2\pi)^3}\frac{|\vec{k}|^{n-3}}{e^{k/T}-1}
       \frac{4|\vec{k}|^2}{4|\vec{k}|^2 - {p_0}^2}\ .
\end{eqnarray}


\setcounter{equation}{0}
\chapter{Temporal and Spatial Parts of Two-point Function}
\label{app:TPF}

In this appendix, we show that the temporal and spatial components
of the two-point function $\Pi_\perp^{{\rm (vac)}\mu\nu}$ are
independent of the external momentum in the VM limit.

{}From Eq.~(\ref{Pi-AA-vac-b}),
$\Pi_\perp^{{\rm (vac)(b)}\mu\nu}$ is rewritten as
\begin{eqnarray}
 \Pi_\perp^{\rm{(vac)(b)}\mu\nu}(p_0,\bar{p})
&=&  N_f\,\frac{a^t}{4} X_{\bar{\mu}}^\mu X_{\bar{\nu}}^\nu
     B^{\rm{(vac)}\bar{\mu}\bar{\nu}}(p_0,\bar{p};M_\rho,0) \nonumber\\
&\equiv&
     N_f\,\frac{a^t}{4} 
     \tilde{B}^{\rm{(vac)}\mu\nu}(p_0,\bar{p};M_\rho,0),
\end{eqnarray}
where we define
\begin{equation}
\qquad
 X_{\bar{\mu}}^\mu
 = u^\mu u_{\bar{\mu}} + V_\sigma^2 (g_{\bar{\mu}}^\mu - 
   u^\mu u_{\bar{\mu}}).
\end{equation}
In the above expression,
we define the function $B^{(\rm vac)\mu\nu}$ contributed to 
the diagram (b) in Fig.~\ref{fig:AA} as
\begin{eqnarray}
 &&B^{\rm{(vac)}\mu\nu}(p_0,\bar{p};M_\rho,0)
\nonumber\\
&&\qquad
 = \int\frac{dk_0}{i(2\pi)}\int\frac{d^3\bar{k}}{(2\pi)^3}
   \frac{(2k-p)^\mu (2k-p)^\nu}{[k_0^2 - \omega_\pi^2]
    [(k_0 - p_0)^2 - (\omega_\rho^p)^2]},
\end{eqnarray}
with
\begin{eqnarray}
\qquad\quad
 \omega_\pi^2 &=& V_\pi^2 \bar{k}^2, \nonumber\\
 (\omega_\rho^p)^2 &=& V_\sigma^2 |\vec{k}-\vec{p}|^2 + M_\rho^2.
\end{eqnarray}
In terms of each components of $B^{(\rm vac)\mu\nu}$,
the temporal and spatial parts of $\tilde{B}^{\mu\nu}$ are given by
\begin{eqnarray}
 &&\tilde{B}^{\rm{(vac)}t}(p_0,\bar{p};M_\rho,0)
\nonumber\\
&&\qquad
   = \Bigl[ B^{\rm{(vac)}t}(p_0,\bar{p};M_\rho,0) 
   {}+
     (1-V_\sigma^2)\frac{\bar{p}^2}{p^2}
      B^{\rm{(vac)}L}(p_0,\bar{p};M_\rho,0)\Bigr], \nonumber\\
 &&\tilde{B}^{\rm{(vac)}s}(p_0,\bar{p};M_\rho,0)
\nonumber\\
&&\qquad
   = V_\sigma^4 \Bigl[ B^{\rm{(vac)}s}(p_0,\bar{p};M_\rho,0) 
   {}+
     \frac{1-V_\sigma^2}{V_\sigma^2}\,\frac{p_0^2}{p^2}
      B^{\rm(vac)L}(p_0,\bar{p};M_\rho,0)\Bigr].
\label{ef}
\end{eqnarray}
By using the expressions in Appendix~\ref{app:LIFT},
the componets $B^{\rm (vac)t}, B^{\rm (vac)s}$ and $B^{\rm (vac)L}$ 
take the following forms:
\begin{eqnarray}
 &&
 B^{\rm (vac)t}(p_0,\bar{p};M_\rho,0)
\nonumber\\
&&\quad
  = \int\,\frac{d^3 k}{(2\pi)^3}
   \Biggl[ \frac{-1}{2\omega_\pi}
    \frac{(2\omega_\pi - p_0)^2}
         {(\omega_\pi - p_0)^2 - (\omega_\rho^p)^2}
  {}+
    \frac{-1}{2\omega_\rho^p}
    \frac{(2\omega_\rho^p + p_0)^2}
         {(\omega_\rho^p + p_0)^2 - \omega_\pi^2} \nonumber\\
 &&\quad\qquad
  {}-\frac{\vec{p}\cdot(2\vec{k}-\vec{p})}{p_0}
   \Biggl\{ \frac{1}{2\omega_\pi}
     \frac{2\omega_\pi - p_0}
          {(\omega_\pi - p_0)^2 - (\omega_\rho^p)^2} 
  {}+
     \frac{1}{2\omega_\rho^p}
     \frac{2\omega_\rho^p + p_0}
          {(\omega_\rho^p + p_0)^2 - \omega_\pi^2}
   \Biggr\}
  \Biggr],
\nonumber\\
 &&
 B^{\rm (vac)s}(p_0,\bar{p};M_\rho,0) \nonumber\\
 &&\quad
  = \int\,\frac{d^3 k}{(2\pi)^3}
   \Biggl[ \frac{(2\vec{k}\cdot\vec{p}-\bar{p}^2)^2}{\bar{p}^2}
    \Biggl\{ 
     \frac{1}{2\omega_\pi}
     \frac{1}{(\omega_\pi - p_0)^2 - (\omega_\rho^p)^2} 
  {}+
     \frac{1}{2\omega_\rho^p}
     \frac{1}{(\omega_\rho^p + p_0)^2 - \omega_\pi^2} 
    \Biggr\} \nonumber\\
 &&\quad\qquad
  {}+\frac{p_0\vec{p}\cdot (2\vec{k}-\vec{p})}{\bar{p}^2}
   \Biggl\{ \frac{1}{2\omega_\pi}
     \frac{2\omega_\pi - p_0}
          {(\omega_\pi - p_0)^2 - (\omega_\rho^p)^2} 
  {}+
     \frac{1}{2\omega_\rho^p}
     \frac{2\omega_\rho^p + p_0}
          {(\omega_\rho^p + p_0)^2 - \omega_\pi^2}
   \Biggr\}
  \Biggr],
\nonumber\\
 &&
 B^{\rm (vac)L}(p_0,\bar{p};M_\rho,0) \nonumber\\
 &&\quad
  = \int\,\frac{d^3 k}{(2\pi)^3}
  \frac{p^2 \vec{p}\cdot (2\vec{k}-\vec{p})}{p_0 \bar{p}^2}
  \Biggl[ 
     \frac{1}{2\omega_\pi}
     \frac{2\omega_\pi - p_0}
          {(\omega_\pi - p_0)^2 - (\omega_\rho^p)^2} 
  {}+
     \frac{1}{2\omega_\rho^p}
     \frac{2\omega_\rho^p + p_0}
          {(\omega_\rho^p + p_0)^2 - \omega_\pi^2}
  \Biggr].
\end{eqnarray}

Now we take the VM limit.
Note that the VM condition for $a^t$ and $a^s$ implies 
that $\sigma$ velocity becomes equal to the pion velocity, 
$V_\sigma \to V_\pi$ for $T \to T_c$ as shown in Eq.~(\ref{vp=vs}).
The components $B^{\rm{(vac)}t}, B^{\rm{(vac)}s}$ and $B^{\rm{(vac)}L}$
are calculated as follows:
\begin{eqnarray}
 &&\lim_{\rm VM}B^{\rm{(vac)}t}(p_0,\bar{p};M_\rho,0)
\nonumber\\
&&\quad\quad
   = \int\frac{d^3 k}{(2\pi)^3} \frac{-1}{\omega_\pi}
     \frac{{\cal I}(\bar{k};p_0,\bar{p}) + \vec{p}\cdot(2\vec{k}-\vec{p})
      {\cal J}(\bar{k};p_0,\bar{p})}
      {[(\omega_\pi - p_0)^2 - (\omega_\pi^p)^2]
       [(\omega_\pi + p_0)^2 - (\omega_\pi^p)^2]}, \label{t}
\\
 &&\lim_{\rm VM}B^{\rm{(vac)}s}(p_0,\bar{p};M_\rho,0)
\nonumber\\
&&\quad\quad
   = \int\frac{d^3 k}{(2\pi)^3} \frac{1}{\omega_\pi}
     \frac{1}{\bar{p}^2}
     \frac{1}
      {[(\omega_\pi - p_0)^2 - (\omega_\pi^p)^2]
       [(\omega_\pi + p_0)^2 - (\omega_\pi^p)^2]}
\nonumber\\
&&\quad\qquad\times
 \Bigl[
 \Bigl( \vec{p}\cdot(2\vec{k}-\vec{p})\bigr)^2 
      {\cal K}(\bar{k};p_0,\bar{p}) 
  {}+ 
      p_0^2 \vec{p}\cdot(2\vec{k}-\vec{p})
      {\cal J}(\bar{k};p_0,\bar{p})
 \Bigr], \label{s}
\\
 &&\lim_{\rm VM}B^{\rm{(vac)}L}(p_0,\bar{p};M_\rho,0)
\nonumber\\
&&\quad\quad
   = \int\frac{d^3 k}{(2\pi)^3} 
     \frac{1}{\omega_\pi}
     \frac{p^2}{\bar{p}^2}
     \frac{\vec{p}\cdot(2\vec{k}-\vec{p}) {\cal J}(\bar{k};p_0,\bar{p})}
      {[(\omega_\pi - p_0)^2 - (\omega_\pi^p)^2]
       [(\omega_\pi + p_0)^2 - (\omega_\pi^p)^2]}, \label{L}
\end{eqnarray}
where we define the functions as
\begin{eqnarray}
&&
 {\cal I}(\bar{k};p_0,\bar{p})
 = \omega_\pi^2 [ 4\omega_\pi^2 - 4(\omega_\pi^p)^2 - 3p_0^2 ] +
     p_0^2[ p_0^2 - (\omega_\pi^p)^2 ], \nonumber\\
&&
 {\cal J}(\bar{k};p_0,\bar{p})
 = -p_0[ -3\omega_\pi^2 + p_0^2 - (\omega_\pi^p)^2 ], \nonumber\\
&&
 {\cal K}(\bar{k};p_0,\bar{p})
 = \omega_\pi^2 + p_0^2 - (\omega_\pi^p)^2.
\end{eqnarray}
Using Eqs.~(\ref{t})-(\ref{L}) with these functions,
we obtain $\tilde{B}^{\rm{(vac)}t,s}$ as
\begin{eqnarray}
 &&\lim_{\rm VM}\tilde{B}^{\rm{(vac)}t}(p_0,\bar{p};M_\rho,0)
\nonumber\\
&&\qquad
   = \int\frac{d^3 k}{(2\pi)^3} \frac{-1}{\omega_\pi}
     \frac{{\cal I}(\bar{k};p_0,\bar{p}) + 
       V_\pi^2\vec{p}\cdot(2\vec{k}-\vec{p})
      {\cal J}(\bar{k};p_0,\bar{p})}
      {[(\omega_\pi - p_0)^2 - (\omega_\pi^p)^2]
       [(\omega_\pi + p_0)^2 - (\omega_\pi^p)^2]}, \label{tilde-t}
\\
 &&\lim_{\rm VM}\tilde{B}^{\rm{(vac)}s}(p_0,\bar{p};M_\rho,0)
\nonumber\\
&&\qquad
   = \int\frac{d^3 k}{(2\pi)^3} \frac{1}{\omega_\pi}
     \frac{1}{V_\pi^2 \bar{p}^2}
     \frac{1}
      {[(\omega_\pi - p_0)^2 - (\omega_\pi^p)^2]
       [(\omega_\pi + p_0)^2 - (\omega_\pi^p)^2]}
\nonumber\\
&&\quad\qquad\times
  \Bigl[
  \Bigl( V_\pi^2 \vec{p}\cdot(2\vec{k}-\vec{p}) \Bigr)^2 
      {\cal K}(\bar{k};p_0,\bar{p}) 
  {}+ 
      p_0^2 V_\pi^2 \vec{p}\cdot(2\vec{k}-\vec{p})
      {\cal J}(\bar{k};p_0,\bar{p})
  \Bigr]. \label{tilde-s}
\end{eqnarray}
The integrand of the functions $\tilde{B}^{\rm (vac)t}$ and 
$\tilde{B}^{\rm (vac)s}$ in the VM limit are same as
those of $B^{\rm (vac)t}$ and $B^{\rm (vac)s}$ with $V_\pi = 1$
when we make the following replacement in $\tilde{B}^{\rm (vac)t,s}$:
\begin{equation}
\qquad
 V_\pi \bar{k} \to \bar{k}, \qquad
 V_\pi |\vec{k} - \vec{p}| \to |\vec{k} - \vec{p}|.
\label{replacement}
\end{equation}
The functions $B^{\rm (vac)t,s}$ with $M_\rho \to 0$ are
independent of the external momentum $p_0$ and $\bar{p}$
[see Appendix~\ref{app:LIFT}]
~\footnote{
 In Ref.~\cite{HKRS:SUS} where $V_\pi = 1$ was taken,
 it was shown that
 the hadronic corrections $\bar{\Pi}_\perp^{t,s}(p_0,\bar{p};T)$
 at the VM limit are independent of the external momentum $p_0$
 and $\bar{p}$.
 The structure of the integrand in the vacuum part is the same as
 that in the hadronic part except for the absence
 of the Bose-Einstein distribution function.
 Thus the vacuum part is also independent of $p_0$
 and $\bar{p}$.
}.
Thus
we find that the functions $\tilde{B}^{(\rm vac)t}$ and 
$\tilde{B}^{(\rm vac)s}$ in the VM limit are 
obtained independently of the external momentum $p_0$ and $\bar{p}$:
\begin{eqnarray}
&&
 \lim_{\rm VM}\tilde{B}^{\rm{(vac)}t}(p_0,\bar{p};M_\rho,0)
 = -\frac{1}{V_\pi}\frac{\Lambda^2}{8\pi^2},
\nonumber\\
&&
 \lim_{\rm VM}\tilde{B}^{\rm{(vac)}s}(p_0,\bar{p};M_\rho,0)
 = -{V_\pi}\frac{\Lambda^2}{8\pi^2}.
\end{eqnarray}


\setcounter{equation}{0}
\chapter{Current Correlator in Operator Product Expansion}
\label{app:CCOPE}

In this appendix, 
following Ref.~\cite{FLK} where the current correlator
is discussed using the OPE in dense matter,
we determine the Lorentz non-invariant effect in bare pion velocity
$V_{\pi,{\rm bare}}$ through the Wilsonian matching.
The tensor structure of the current correlator with Lorentz
non-scalar operators in dense matter is same as that in hot matter.
Thus in order to obtain the current correlator in terms of OPE variables,
we simply make a replacement of the matrix element in dense matter
with the one in hot matter.

The axial-vector current correlator $G_A^{\mu\nu}$ obtained from 
the OPE in dense matter 
is given by~\cite{FLK}
\begin{eqnarray}
 &&
 G_A^{\mu\nu}(q_0,\bar{q})
 = (q^\mu q^\nu - g^{\mu\nu}q^2)\frac{-1}{4}
   \Biggl[ \frac{1}{2\pi^2}
   \Biggl( 1+\frac{\alpha_s}{\pi} \Biggr) \ln \Biggl( \frac{Q^2}{\mu^2}
   \Biggr) + \frac{1}{6Q^4} \Big\langle \frac{\alpha_s}{\pi} G^2 
   \Big\rangle_\rho \nonumber\\ 
 &&\qquad\qquad{}- \frac{2\pi\alpha_s}{Q^6} \Big\langle 
   \Bigl( \bar{u}\gamma_\mu \gamma_5 \lambda^a u - 
   \bar{d}\gamma_\mu \gamma_5 \lambda^a d \Bigr)^2 \Big\rangle_\rho 
\nonumber\\
 &&\qquad\qquad{}- \frac{4\pi\alpha_s}{9Q^6}
   \Big\langle 
   \Bigl( \bar{u}\gamma_\mu\lambda^a u + \bar{d}\gamma_\mu\lambda^a d 
   \Bigr)\sum_q^{u,d,s}\bar{q}\gamma_\mu\lambda^a q \Big\rangle_\rho
   \Biggr] \nonumber\\
 &&\qquad{}+
   \sum_{\tau = 2}\sum_{k = 1}
   \bigl[ -g^{\mu\nu}q^{\mu_1}q^{\mu_2} + g^{\mu\mu_1}q^\nu q^{\mu_2}
    + q^\mu q^{\mu_1}g^{\nu\mu_2} + g^{\mu\mu_1}g^{\nu\mu_2}Q^2
   \bigr] \nonumber\\
 &&\qquad\qquad \times
   q^{\mu_3}\cdots q^{\mu_{2k}}
   \frac{2^{2k}}{Q^{4k + \tau - 2}} 
   A_{\mu_1 \cdots \mu_{2k}}^{2k + \tau, \tau} \nonumber\\
 &&\qquad{}+
  \sum_{\tau = 2}\sum_{k=1}\Bigl[ g^{\mu\nu} - \frac{q^\mu q^\nu}{q^2}
   \Bigr] q^{\mu_1}\cdots q^{\mu_{2k}}
   \frac{2^{2k}}{Q^{4k + \tau -2}}
   C_{\mu_1 \cdots \mu_{2k}}^{2k + \tau, \tau}.
\label{app:acc-ope}
\end{eqnarray}
with $Q^2 = -q^2$.
$\tau = d - s$ denotes the twist, where $s = 2k$ is the number of spin
indices of the operator of dimension $d$.
When we consider the operators of $(\tau,s)=(2,2)$ and $(2,4)$,
$G_A^{\mu\nu}$ has $A_{\alpha\beta}^{4,2}$ and 
$A_{\alpha\beta\lambda\sigma}^{6,2}$ defined by~\footnote{
 Eq.~(8) in Ref.~\cite{FLK} does not have the factor
 $\frac{\rho}{2m}$.
 In the following, we explain the reason why this factor 
 $\frac{\rho}{2m}$ appears in Eq.~(\ref{app:tensor}):
 In the linear density approximation, the matrix element of an
 operator $\hat{A}$ is expressed as $\langle G | \hat{A} | G \rangle
 = \langle 0 | \hat{A} | 0 \rangle + \frac{\rho}{2m}
 \langle p | \hat{A} | p \rangle$ where $| G \rangle$ denotes the ground
 state of nuclear matter~\cite{FLK}.
 The first term on the right-hand-side is the vacuum expectation value,
 which vanishes for operators with spin.
 In the second term $| p \rangle$ denotes a nucleon state with momentum
 $p$.
 In the present analysis, 
 we consider the operators with spin 2 and spin 4 which
 are dominant to the Lorentz non-invariant effect.
 The matrix elements of these operators with spin $s$ are expressed as
 $\tilde{A}_s \equiv \langle G | \hat{A}_s | G \rangle 
 = \frac{\rho}{2m} \langle p | \hat{A}_s | p \rangle 
 \equiv \frac{\rho}{2m}A_s$.
 $A_{2,4}$ in Eq.~(8) in Ref.~\cite{FLK} correspond to
 $\tilde{A}_{2,4}$.
 In terms of $A_s$, $A_{\alpha\beta}$ is expressed as
 $A_{\alpha\beta}^{4,2} = (p_\alpha p_\beta - \frac{1}{4}
 g_{\alpha\beta}p^2)\tilde{A}_2^{4,2} = (p_\alpha p_\beta - 
 \frac{1}{4}g_{\alpha\beta}p^2)\frac{\rho}{2m}A_2^{4,2}$
 as showen in Eq.~(\ref{app:tensor}).
}
\begin{eqnarray}
 &&
 A_{\alpha\beta}^{4,2}
 = \Bigl( p_\alpha p_\beta - \frac{1}{4}g_{\alpha\beta}p^2 \Bigr)
    \frac{\rho}{2m} A_2^{4,2},
\nonumber\\
 &&
 A_{\alpha\beta\lambda\sigma}^{6,2}
 = \Bigl[ p_\alpha p_\beta p_\lambda p_\sigma
    {}- \frac{p^2}{8}\bigl( p_\alpha p_\beta g_{\lambda\sigma} 
     {}+ p_\alpha p_\lambda g_{\beta\sigma}
     {}+ p_\alpha p_\sigma g_{\lambda\beta}
     {}+ p_\beta p_\lambda g_{\alpha\sigma}
\nonumber\\
 &&\qquad\qquad
     {}+ p_\beta p_\sigma g_{\alpha\lambda}
     {}+ p_\lambda p_\sigma g_{\alpha\beta} \bigr)
    {}+ \frac{p^4}{48}\bigl( g_{\alpha\beta}g_{\lambda\sigma}
     {}+ g_{\alpha\lambda}g_{\beta\sigma}
     {}+ g_{\alpha\sigma}g_{\beta\lambda} \bigr) \Bigr]
    \frac{\rho}{2m} A_4^{6,2},
\label{app:tensor}
\end{eqnarray}
where $\rho$ denotes the nuclear density and $m$ the nucleon mass, and
$p^\mu$ is the nucleon momentum and we take 
$p^\mu = m u^\mu = m (1,\vec{0})$.
{}From Eq.~(\ref{app:tensor}), we have the following expressions of
$G_A^{L(0)}$ and $G_A^{L(1)}$~\cite{FLK}~\footnote{
  The expression of Eq.~(\ref{app:OPE-L(1)}) is obtained from Eqs.~(11)
  and (14) in Ref.~\cite{FLK}, where we neglect the ${\cal O}(\alpha_s)$
  terms in $A_{\alpha\beta}^{4,2}$ and 
  $A_{\alpha\beta\lambda\sigma}^{6,2}$.
}:
\begin{eqnarray}
 G_A^{{\rm (OPE)}L(0)}(-q^2)
  &=& \frac{-1}{3}g^{\mu\nu}G_{A,\mu\nu}^{{\rm (OPE)}(0)}
\nonumber\\
  &=& \frac{-1}{4}
   \Biggl[ \frac{1}{2\pi^2}
   \Biggl( 1+\frac{\alpha_s}{\pi} \Biggr) \ln \Biggl( \frac{Q^2}{\mu^2}
   \Biggr) + \frac{1}{6Q^4} \Big\langle \frac{\alpha_s}{\pi} G^2 
   \Big\rangle_\rho \nonumber\\ 
 &&{}- \frac{2\pi\alpha_s}{Q^6} \Big\langle 
   \Bigl( \bar{u}\gamma_\mu \gamma_5 \lambda^a u - 
   \bar{d}\gamma_\mu \gamma_5 \lambda^a d \Bigr)^2 \Big\rangle_\rho 
\nonumber\\
 &&{}- \frac{4\pi\alpha_s}{9Q^6}
   \Big\langle 
   \Bigl( \bar{u}\gamma_\mu\lambda^a u + \bar{d}\gamma_\mu\lambda^a d 
   \Bigr)\sum_q^{u,d,s}\bar{q}\gamma_\mu\lambda^a q \Big\rangle_\rho
   \Biggr]
\nonumber\\
 &&{}+ \frac{\rho m}{(-q^2)^2}A_2^{\rm u+d}
   {}- \frac{5}{3}\frac{\rho m^3}{(-q^2)^3}A_4^{\rm u+d},
\label{app:OPE-L(0)}
\\
 G_A^{{\rm (OPE)}L(1)}(-p^2)
  &=& 2 \rho m^3 \frac{A_4^{u+d}}{q^8}.
\label{app:OPE-L(1)}
\end{eqnarray}

Following the same procedure done in section~\ref{ssec:WMCTf},
we obtain the matching condition on the bare pion velocity:
\begin{equation}
 1 - V_{\pi,{\rm bare}}^2
 = \frac{1}{G_0}
   \frac{2 \rho m^3}{\Lambda^6}A_4^{\rm u+d},
\label{app:deviation-rho}
\end{equation}
where we define $G_0$ as
\begin{eqnarray}
 &&
 G_0 \equiv \frac{1}{8\pi^2}
   \Biggl[ \Bigl( 1 + \frac{\alpha_s}{\pi} \Bigr) +
   \frac{2\pi^2}{3}\frac{\big\langle \frac{\alpha_s}{\pi}G^2
   \big\rangle_\rho }{\Lambda^4} + \pi^3 \frac{1408}{27}\frac{\alpha_s
   \langle \bar{q}q \rangle^2_\rho }{\Lambda^6} \Biggr]
\nonumber\\
 &&\qquad\qquad{}+
   \frac{2 \rho m}{\Lambda^4}A_2^{\rm u+d} -
   \frac{5 \rho m^3}{\Lambda^6}A_4^{\rm u+d}.
\end{eqnarray}

In the above analysis, we consider the OPE in dense matter.
The quantity in hot matter corresponding to $\rho m^3$ in 
Eq.~(\ref{app:deviation-rho}) is determined as follows:
{}From Eq.~(\ref{app:acc-ope}), the term of $(\tau,s)=(2,4)$ is 
expressed as~\cite{FLK}
\begin{eqnarray}
 \frac{1}{q^2}{{G_A^{(1)}}_\mu}^\mu |_{\tau = 2, s = 4}
 &=& - G_A^{L(1)}|_{\tau = 2, s = 4} - 2G_A^{T(1)}|_{\tau = 2, s = 4}
\nonumber\\
 &\simeq& - \frac{20 \rho m^3}{(-q^2)^4} A_4^{\rm u+d}.
\label{app:A_4-rho}
\end{eqnarray}
On the other hand, the same matrix element in hot matter is
expressed as~\cite{HKL}
\begin{eqnarray}
 \frac{1}{q^2}{{G_A}_\mu}^\mu |_{\tau = 2, s = 4}
 &=& -8i \frac{q^\mu q^\nu q^\lambda q^\sigma}{Q^{10}}
   \Big\langle \Bigl( \bar{u}\gamma_\mu D_\nu D_\lambda D_\sigma u +
   \bar{d}\gamma_\mu D_\nu D_\lambda D_\sigma d \Bigr) \Big\rangle_T
\nonumber\\
 &=& \frac{16 \cdot 3}{5}\Biggl[ \frac{16\pi^4 T^6}{63}
  B_3 \Bigl( \frac{m_\pi}{T} \Bigr) - \frac{\pi^2 m_\pi^2 T^4}{10}
  B_2 \Bigl( \frac{m_\pi}{T} \Bigr) + \frac{m_\pi^4 T^2}{48} 
  B_1 \Bigl( \frac{m_\pi}{T} \Bigr) \Biggr]
\nonumber\\
 &&\quad\times
  \frac{q^\mu q^\nu q^\lambda q^\sigma}{Q^{10}}
   \Bigl( u_\mu u_\nu u_\lambda u_\sigma  - \mbox{traces} \Bigr)
   A_4^{\rm u+d},
\label{app:A_4-T}
\end{eqnarray}
where the functions $B_1$, $B_2$ and $B_3$ are defined by
\begin{eqnarray}
 &&B_1(z)=\frac{6}{\pi^2}\int_z^\infty dy\, \sqrt{y^2 - z^2}
   \frac{1}{e^y - 1}, \nonumber\\
 &&B_2(z)=\frac{15}{\pi^4}\int_z^\infty dy\, y^2 \,\sqrt{y^2 - z^2}
   \frac{1}{e^y - 1}, \nonumber\\
 &&B_3(z)=\frac{63}{8\pi^6}\int_z^\infty dy\, y^4 \,\sqrt{y^2 - z^2}
   \frac{1}{e^y - 1}, \nonumber\\
 &&A_4^{\rm u+d}(\mu^2 = 1 \,\mbox{GeV}^2)
    = 0.255\,.
\end{eqnarray}
The momentum dependent part of Eq.~(\ref{app:A_4-T}) is 
expanded around $\bar{q}=0$ as
\begin{equation}
  \frac{q^\mu q^\nu q^\lambda q^\sigma}{Q^{10}}
  \Bigl( u_\mu u_\nu u_\lambda u_\sigma  - \mbox{traces} \Bigr)
 = \frac{5}{16}\frac{1}{(-q^2)^3} - \frac{5}{4}\frac{1}{(-q^2)^4}
   \bar{q}^2 + \cdots.
\label{app:p-dep}
\end{equation}
{}From Eqs.~(\ref{app:A_4-T}) and (\ref{app:p-dep}), 
${{G_A^{(1)}}_\mu}^\mu$ with $(\tau,s)=(2,4)$ takes the form
\begin{eqnarray}
 &&
 \frac{1}{q^2}{{G_A^{(1)}}_\mu}^\mu |_{\tau = 2, s = 4} \nonumber\\
 &&
  = - \frac{12}{(-q^2)^4}
   \Biggl[ \frac{16\pi^4 T^6}{63}
  B_3 \Bigl( \frac{m_\pi}{T} \Bigr) - \frac{\pi^2 m_\pi^2 T^4}{10}
  B_2 \Bigl( \frac{m_\pi}{T} \Bigr) + \frac{m_\pi^4 T^2}{48} 
  B_1 \Bigl( \frac{m_\pi}{T} \Bigr) \Biggr]
  A_4^{\rm u+d}.
\end{eqnarray}
Comparing this expression with Eq.~(\ref{app:A_4-rho}),
we find the following correspondence:
\begin{equation}
 \rho m^3 \quad \Leftrightarrow \quad
 \frac{3}{5}\Biggl[ \frac{16\pi^4 T^6}{63}
  B_3 \Bigl( \frac{m_\pi}{T} \Bigr) - \frac{\pi^2 m_\pi^2 T^4}{10}
  B_2 \Bigl( \frac{m_\pi}{T} \Bigr) + \frac{m_\pi^4 T^2}{48} 
  B_1 \Bigl( \frac{m_\pi}{T} \Bigr) \Biggr].
\end{equation}
Thus the expression in hot matter corresponding to
Eq.~(\ref{app:deviation-rho}) is obtained as
\begin{eqnarray}
 &&
 1 - V_{\pi,{\rm bare}}^2 \nonumber\\
 &&\quad=
 \frac{1}{G_0}\frac{6}{5}\Biggl[ \frac{16\pi^4 T^6}{63}
  B_3 \Bigl( \frac{m_\pi}{T} \Bigr) - \frac{\pi^2 m_\pi^2 T^4}{10}
  B_2 \Bigl( \frac{m_\pi}{T} \Bigr) + \frac{m_\pi^4 T^2}{48} 
  B_1 \Bigl( \frac{m_\pi}{T} \Bigr) \Biggr]
 \frac{1}{\Lambda^6}A_4^{\rm u+d}.
\label{app:deviation-T}
\end{eqnarray}
Taking $m_\pi = 0$, Eq.~(\ref{app:deviation-T}) is rewritten into
the same form as Eq.~(\ref{deviation-rho}).


\setcounter{equation}{0}
\chapter{Hadronic Thermal Corrections to Susceptibilities}
\label{app:HTC}

In this appendix we summarize the
hadronic thermal corrections to the two-point functions of
$\overline{\cal A}_\mu$-$\overline{\cal A}_\nu$,
$\overline{V}_\mu$-$\overline{V}_\nu$, $\overline{\cal
V}_\mu$-$\overline{\cal V}_\nu$ and
$\overline{V}_\mu$-$\overline{\cal V}_\nu$.

The four components of the hadronic thermal corrections to the two
point function of $\overline{\cal A}_\mu$-$\overline{\cal A}_\nu$,
$\Pi_\perp$, are expressed as
\begin{eqnarray}
\bar{\Pi}_{\perp}^t(p_0,\bar{p};T)
&=&
  N_f (a-1) \bar{A}_{0}(0,T)
  - N_f a M_\rho^2 \bar{B}_{0}(p_0,\bar{p};M_\rho,0;T)
\nonumber\\
&&
  {}+ N_f \frac{a}{4} \bar{B}^t(p_0,\bar{p};M_\rho,0;T)
\ ,
\label{AA t}
\\
\bar{\Pi}_{\perp}^s(p_0,\bar{p};T)
&=&
  N_f (a-1) \bar{A}_{0}(0,T)
  - N_f a M_\rho^2 \bar{B}_{0}(p_0,\bar{p};M_\rho,0;T)
\nonumber\\
&&
  {}+ N_f \frac{a}{4} \bar{B}^s(p_0,\bar{p};M_\rho,0;T)
\ ,
\label{AA s}
\\
  \bar{\Pi}_{\perp}^L(p_0,\bar{p};T)
&=&
  N_f \frac{a}{4} \bar{B}^L(p_0,\bar{p};M_\rho,0;T)
\ ,
\label{AA L}
\\
\bar{\Pi}_{\perp}^T(p_0,\bar{p};T)
&=&
  N_f \frac{a}{4} \bar{B}^T(p_0,\bar{p};M_\rho,0;T)
\ .
\label{AA T}
\end{eqnarray}

The two components $\bar{\Pi}^t$ and $\bar{\Pi}^s$ of hadronic
thermal corrections to the two-point functions of
$\overline{V}_\mu$-$\overline{V}_\nu$, $\overline{\cal
V}_\mu$-$\overline{\cal V}_\nu$ and
$\overline{V}_\mu$-$\overline{\cal V}_\nu$ are written as
\begin{eqnarray}
&&
\bar{\Pi}_{V}^t(p_0,\bar{p};T)
=
\bar{\Pi}_{V}^s(p_0,\bar{p};T)
\nonumber\\
&& \
=
\bar{\Pi}_{\parallel}^t(p_0,\bar{p};T)
=
\bar{\Pi}_{\parallel}^s(p_0,\bar{p};T)
\nonumber\\
&& \
=
- \bar{\Pi}_{V\parallel}^t(p_0,\bar{p};T)
=
- \bar{\Pi}_{V\parallel}^s(p_0,\bar{p};T)
\nonumber\\
&& \quad
=
- N_f \frac{1}{4}
  \left[ \bar{A}_{0}(M_\rho;T) + a^2 \bar{A}_{0}(0;T) \right]
- N_f M_\rho^2 \bar{B}_{0}(p_0,\bar{p};M_\rho,M_\rho;T)
\ .
\label{rr vv rv ts}
\end{eqnarray}
Among the remaining components only $\bar{\Pi}_\parallel^L$
is relevant to the present analysis.
This is given by
\begin{eqnarray}
\bar{\Pi}_{\parallel}^L(p_0,\bar{p};T)
&=&
N_f \frac{1}{8} \bar{B}^L(p_0,\bar{p};M_\rho,M_\rho;T)
+ N_f \frac{(2-a)^2}{8} \bar{B}^L(p_0,\bar{p};0,0;T)
\ .
\label{vv L}
\end{eqnarray}

For obtaining the pion decay constants and velocity in
subection~\ref{ssec:PP-lowT} we need the limit of $p_0 = \bar{p}$ of
$\bar{\Pi}_\perp^t$ and $\bar{\Pi}_\perp^s$ in Eqs.~(\ref{AA t})
and (\ref{AA s}).
With Eq.~(\ref{B0 Bts VM limits}),
$\bar{\Pi}_\perp^t$ and $\bar{\Pi}_\perp^s$ reduce to the
following forms in the limit $M_\rho \rightarrow0$ and
$a\rightarrow1$:
\begin{eqnarray}
&&
\bar{\Pi}_\perp^t(p_0=\bar{p}+i\epsilon,\bar{p};T)
\ \mathop{\longrightarrow}_{M_\rho \rightarrow 0,\, a\rightarrow1} \
- \frac{N_f}{2} \widetilde{J}_{1}^2(0;T)
= - \frac{N_f}{24} T^2
\ ,
\nonumber\\
&&
\bar{\Pi}_\perp^s(p_0=\bar{p}+i\epsilon,\bar{p};T)
\ \mathop{\longrightarrow}_{M_\rho \rightarrow 0,\, a\rightarrow1} \
- \frac{N_f}{2} \widetilde{J}_{1}^2(0;T)
= - \frac{N_f}{24} T^2
\ .
\label{PiA ts Tc app}
\end{eqnarray}

{}In the static--low-momentum limits of the functions listed in
Eq.~(\ref{JB SL}), the $(\bar{\Pi}_\perp^t - \bar{\Pi}_\perp^L)$
appearing in the axial-vector susceptibility
becomes
\begin{eqnarray}
&&
\lim_{\bar{p}\rightarrow0}
\lim_{p_0\rightarrow0}
\left[
  \bar{\Pi}_\perp^t(p_0,\bar{p};T) - \bar{\Pi}_\perp^L(p_0,\bar{p};T)
\right]
\nonumber\\
&& \quad
=
- N_f \widetilde{J}_{1}^2(0;T)
+ N_f a\, \widetilde{J}_{1}^2(M_\rho;T)
- N_f \frac{a}{M_\rho^2}
\left[
  \widetilde{J}_{-1}^2(M_\rho;T)
  - \widetilde{J}_{-1}^2(0;T)
\right]
\ .
\label{PiA tmL SL}
\end{eqnarray}
For the functions appearing in the vector susceptibility
relevant to the present analysis
we have
\begin{eqnarray}
\lim_{\bar{p}\rightarrow0}
\lim_{p_0\rightarrow0}
\left[ \bar{\Pi}_V^t(p_0,\bar{p};T) \right]
&=&
- \frac{N_f}{4} \left[
  2 \widetilde{J}_{-1}^0(M_\rho;T)
  - \widetilde{J}_{1}^2(M_\rho;T)
  + a^2 \, \widetilde{J}_{1}^2(0;T)
\right]
\ ,
\label{Pir t SL}
\\
\lim_{\bar{p}\rightarrow0}
\lim_{p_0\rightarrow0}
\left[ \bar{\Pi}_{\parallel}^L(p_0,\bar{p};T) \right]
&=&
- N_f \frac{1}{4} \left[
  M_\rho^2 \widetilde{J}_{1}^0(M_\rho;T)
  + 2 \widetilde{J}_{1}^2(M_\rho;T)
\right]
\nonumber\\
&&
{}- N_f \frac{(2-a)^2}{2} \widetilde{J}_{1}^2(0;T)
\ .
\label{Piv L SL}
\end{eqnarray}


In the following, we list the explicit forms of the
functions that figure in the hadronic thermal corrections,
$\bar{A}_{0}$, $\bar{B}_{0}$ and
$\bar{B}^{\mu\nu}$ in various limits relevant to the
present analysis.

The function
$\bar{A}_{0}(M;T)$ is
expressed as
\begin{eqnarray}
&&
\bar{A}_{0}(M;T) = \widetilde{J}_{1}^2(M;T)
\ ,
\label{A0 J}
\end{eqnarray}
where $\tilde{J}_{1}^2(M;T)$ is defined by
\begin{eqnarray}
&&
\widetilde{J}_{l}^n(M;T)
= \int \frac{d^3\vec{k}}{(2\pi)^3}
\frac{1}{ e^{\omega(\vec{k};M)/T} - 1 }
\frac{ \vert \vec{k} \vert^{n-2} }{[ \omega(\vec{k};M) ]^l }
\ ,
\end{eqnarray}
with $l$ and $n$ being integers and $\omega(\vec{k};M) \equiv
\sqrt{ M^2 + \vert\vec{k}\vert^2 }$. In the massless limit $M=0$,
the above integration can be performed analytically.
Here we list the result relevant to the present analysis:
\begin{eqnarray}
&&
\widetilde{J}_{1}^2(0;T) =
\widetilde{J}_{-1}^0(0;T) = \frac{1}{12} T^2 \ .
\end{eqnarray}

It is convenient to decompose $\bar{B}^{\mu\nu}$
into four components as done for $\Pi_\perp^{\mu\nu}$
in Eq.~(\ref{Pi perp decomp}):
\begin{equation}
\bar{B}^{\mu\nu}
 =u^\mu u^\nu \bar{B}^t +
   (g^{\mu\nu}-u^\mu u^\nu) \bar{B}^s +
   P_L^{\mu\nu} \bar{B}^L + P_T^{\mu\nu}\bar{B}^T
\ .
\label{Bmn decomp app}
\end{equation}
We note here that, by explicit computations, the following
relations are satisfied:
\begin{eqnarray}
&&
\bar{B}^t(p_0,\bar{p};M,M;T) =
\bar{B}^s(p_0,\bar{p};M,M;T) =
- 2 \bar{A}_{0}(M;T) = -2 \widetilde{J}_{1}^2(M;T)
\ .
\label{Bts rel}
\end{eqnarray}

To obtain the pion decay constants and velocity in
Section~\ref{ssec:PP-lowT} we need the limit of $p_0 = \bar{p}$ of the
functions in Eqs.~(\ref{AA t}) and (\ref{AA s}).
As for the functions $M_\rho^2\bar{B}_{0}$,
$\bar{B}^t$ and $\bar{B}^s$ appearing
in Eqs.~(\ref{AA t}) and (\ref{AA s}),
we find
that, in the limit of $M_\rho$ going to zero,
they reduce to
\begin{eqnarray}
&&
M_\rho^2 \bar{B}_0(p_0=\bar{p} + i \epsilon,\bar{p};M_\rho,0;T)
\ \mathop{\longrightarrow}_{M_\rho \rightarrow 0} \
0 \ ,
\nonumber\\
&&
\bar{B}^t(p_0=\bar{p}+ i\epsilon,\bar{p};M_\rho,0;T)
\ \mathop{\longrightarrow}_{M_\rho \rightarrow 0} \
- 2 \widetilde{J}_{1}^2(0;T) = - \frac{1}{6} T^2 \ ,
\nonumber\\
&&
\bar{B}^s(p_0=\bar{p}+ i\epsilon,\bar{p};M_\rho,0;T)
\ \mathop{\longrightarrow}_{M_\rho \rightarrow 0} \
- 2 \widetilde{J}_{1}^2(0;T) = - \frac{1}{6} T^2 \ .
\label{B0 Bts VM limits}
\end{eqnarray}

The static--low-momentum limits of the functions appearing in the
corrections to the axial-vector and vector susceptibility
are summarized as
\begin{eqnarray}
&&
\lim_{\bar{p}\rightarrow0}
\lim_{p_0\rightarrow0}
\left[ M_\rho^2 \bar{B}_{0}(p_0,\bar{p};M_\rho,0;T) \right]
=
- \widetilde{J}_{1}^2(M_\rho;T) + \widetilde{J}_{1}^2(0;T)
\ ,
\nonumber\\
&&
\lim_{\bar{p}\rightarrow0}
\lim_{p_0\rightarrow0}
\left[
  \bar{B}^t(p_0,\bar{p};M_\rho,0;T)
  - \bar{B}^L(p_0,\bar{p};M_\rho,0;T)
\right]
\nonumber\\
&& \qquad
=
\frac{-4}{M_\rho^2} \left[
  - \widetilde{J}_{-1}^2(M_\rho;T)
  + \widetilde{J}_{-1}^2(0;T)
\right]
\ ,
\nonumber\\
&&
\lim_{\bar{p}\rightarrow0}
\lim_{p_0\rightarrow0}
\left[ M_\rho^2 \bar{B}_{0}(p_0,\bar{p};M_\rho,M_\rho;T) \right]
=
\frac{1}{2}
\left[
  \widetilde{J}_{-1}^0(M_\rho;T)
  - \widetilde{J}_{1}^2(M_\rho;T)
\right]
\ ,
\nonumber\\
&&
\lim_{\bar{p}\rightarrow0}
\lim_{p_0\rightarrow0}
\left[ \bar{B}^L(p_0,\bar{p};M_\rho,M_\rho;T) \right]
=
- 2 M_\rho^2 \widetilde{J}_{1}^0(M_\rho;T)
- 4 \widetilde{J}_{1}^2(M_\rho;T)
\ ,
\nonumber\\
&&
\lim_{\bar{p}\rightarrow0}
\lim_{p_0\rightarrow0}
\left[ \bar{B}^L(p_0,\bar{p};0,0;T) \right]
=
- 4 \widetilde{J}_{1}^2(0;T)
\ .
\label{JB SL}
\end{eqnarray}


\setcounter{equation}{0}
\chapter{Explicit Calculation of Quantum Correction to $\Delta M$}
\label{app:ECQC}

In this appendix, we compute the quantum effects on the masses of
$0^-$ $(P)$ and $0^+$ $(Q^\ast)$ heavy-light ${\cal M}$-mesons by
calculating the one-loop corrections to the two-point functions of
$P$ and $Q^\ast$ denoted by $\Pi_{PP}$ and $\Pi_{Q^\ast Q^\ast}$.
Here we adopt the following regularization method to identify the
power divergences: We first perform the integration over the
temporal component of the integration momentum, and then in the
remaining integration over three-momentum we make the replacements
given by
\begin{equation}
\int^\Lambda \frac{d^3\vec{k}}{(2\pi)^3} \frac{1}{\bar{k}^2}
\ \rightarrow \ \frac{\Lambda}{2\sqrt{2} \pi^2}
\ ,
\quad
\int^\Lambda \frac{d^3\vec{k}}{(2\pi)^3} \frac{1}{\bar{k}}
\ \rightarrow \ \frac{\Lambda^2}{8 \pi^2}
\ ,
\quad
\int^\Lambda \frac{d^3\vec{k}}{(2\pi)^3}
\ \rightarrow\  \frac{\Lambda^3}{12\sqrt{2}\pi^2}
\ .
\label{p divs}
\end{equation}
Here we use the t'Hooft-Feynman gauge for fixing the gauge of the
HLS.

\begin{figure}
\begin{center}
 \includegraphics
  [width = 13cm]
  {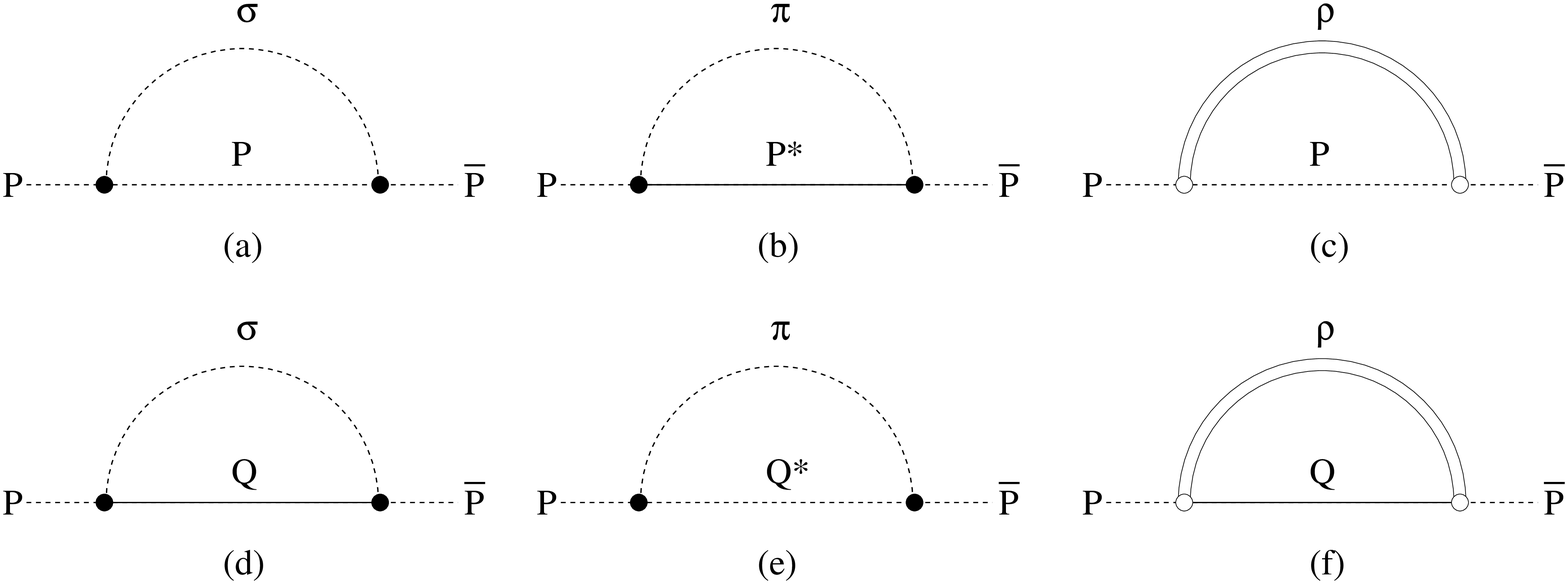}
\end{center}
\caption{Diagrams contributing to $P$-$P$ two point function.}
\label{fig:PP}
\end{figure}

The diagrams contributing to $\Pi_{PP}$ are shown in
Fig.~\ref{fig:PP}.
In the limit of zero external momentum, the divergent parts of
these contributions are given by
\begin{eqnarray}
&& \biggl. \Pi_{PP}^{(a)[\sigma P]} \biggr\vert_{\rm div}
  = \frac{2k^2}{F_\sigma^2}
  \left[
    - \frac{M_H}{(4\pi)^2}
      \left( \Lambda^2 - 2 M_\rho \ln \Lambda \right)
    + \frac{M_H^2}{4\pi^2}
      \left( \frac{\Lambda}{\sqrt{2}} - M_H \ln \Lambda \right)
  \right]
\ ,
\nonumber\\
&& \biggl. \Pi_{PP}^{(b)[\pi P^\ast]} \biggr\vert_{\rm div}
  = \frac{2k^2}{F_\pi^2}
  \left[
    \frac{\Lambda^3}{24\sqrt{2}\pi^2}
    - \frac{M_H}{(4\pi)^2} \Lambda^2
    + \frac{M_H^2}{4\pi^2}
      \left( \frac{\Lambda}{\sqrt{2}} - M_H \ln \Lambda \right)
  \right]
\ ,
\nonumber\\
&& \biggl. \Pi_{PP}^{(c)[\rho P]} \biggr\vert_{\rm div}
  = \frac{g^2}{2\pi^2}  \left( 1 - k \right)^2
  \left( \frac{\Lambda}{\sqrt{2}} - M_H \ln \Lambda \right)
\ ,
\nonumber\\
&& \biggl. \Pi_{PP}^{(d)[\sigma Q]} \biggr\vert_{\rm div}
  = \frac{2k^2}{F_\sigma^2}
  \left[
    \frac{\Lambda^3}{24\sqrt{2}\pi^2}
    - \frac{M_G}{(4\pi)^2}
      \left( \Lambda^2 - 2 M_\rho^2 \ln \Lambda \right)
    + \frac{M_G^2-M_\rho^2}{4\pi^2}
      \left( \frac{\Lambda}{\sqrt{2}} - M_G \ln \Lambda \right)
  \right]
\ ,
\nonumber\\
&& \biggl. \Pi_{PP}^{(e)[\pi Q^\ast]} \biggr\vert_{\rm div}
  = \frac{2k^2}{F_\pi^2}
  \left[
    - \frac{M_G}{(4\pi)^2} \Lambda^2
    + \frac{M_G^2}{4\pi^2}
      \left( \frac{\Lambda}{\sqrt{2}} - M_G \ln \Lambda \right)
  \right]
\ ,
\nonumber\\
&& \biggl. \Pi_{PP}^{(f)[\rho Q]} \biggr\vert_{\rm div}
  = \frac{3 g^2}{2\pi^2} k^2
  \left( \frac{\Lambda}{\sqrt{2}} - M_G \ln \Lambda \right)
\ . \label{Pi PP}
\end{eqnarray}
The particles that figure in the loop are indicated by the suffix
in square bracket; e.g., $[\pi P^\ast]$ indicates that $\pi$ and
$P^\ast$ enter in the internal lines. Here and henceforth, we
suppress, for notational simplification, the group factor
${\mathcal C}_2(N_f)$ defined as $(T_a)_{ij}(T_a)_{jl} = {\mathcal
C}_2(N_f)\delta_{il}$.

\begin{figure}
\begin{center}
 \includegraphics
  [width = 13cm]
  {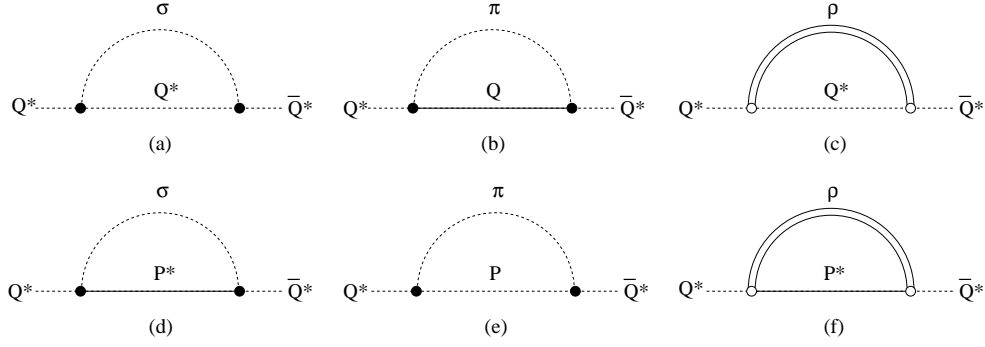}
\end{center}
\caption{Diagrams contributing to $Q^\ast$-$Q^\ast$ two point
function.}
\label{fig:QQ}
\end{figure}

The relevant diagrams contributing to
$\Pi_{Q^\ast Q^\ast}$ are
shown in Fig.~\ref{fig:QQ}.
The divergent parts of
these contributions in the low-energy limit are expressed as
\begin{eqnarray}
&& \biggl. \Pi_{Q^\ast Q^\ast}^{(a)[\sigma Q^\ast]}
  \biggr\vert_{\rm div}
  = \frac{2k^2}{F_\sigma^2}
  \left[
    - \frac{M_G}{(4\pi)^2}
      \left( \Lambda^2 - 2 M_\rho^2 \ln \Lambda \right)
    + \frac{M_G^2}{4\pi^2}
      \left( \frac{\Lambda}{\sqrt{2}} - M_G \ln \Lambda \right)
  \right]
\ ,
\nonumber\\
&& \biggl. \Pi_{Q^\ast Q^\ast}^{(b)[\pi Q]} \biggr\vert_{\rm div}
  = \frac{2k^2}{F_\pi^2}
  \left[
    \frac{\Lambda^3}{24\sqrt{2}\pi^2}
    - \frac{M_G}{(4\pi)^2} \Lambda^2
    + \frac{M_G^2}{4\pi^2}
      \left( \frac{\Lambda}{\sqrt{2}} - M_G \ln \Lambda \right)
  \right]
\ ,
\nonumber\\
&& \biggl. \Pi_{Q^\ast Q^\ast}^{(c)[\rho Q^\ast]}
  \biggr\vert_{\rm div}
  = \frac{g^2}{2\pi^2}  \left( 1 - k \right)^2
  \left( \frac{\Lambda}{\sqrt{2}} - M_G \ln \Lambda \right)
\ ,
\nonumber\\
&& \biggl. \Pi_{Q^\ast Q^\ast}^{(d)[\sigma P^\ast]}
  \biggr\vert_{\rm div}
  = \frac{2k^2}{F_\sigma^2}
  \left[
    \frac{\Lambda^3}{24\sqrt{2}\pi^2}
    - \frac{M_H}{(4\pi)^2}
      \left( \Lambda^2 - 2 M_\rho^2 \ln \Lambda \right)
    + \frac{M_H^2-M_\rho^2}{4\pi^2}
      \left( \frac{\Lambda}{\sqrt{2}} - M_H \ln \Lambda \right)
  \right]
\ ,
\nonumber\\
&& \biggl. \Pi_{Q^\ast Q^\ast}^{(e)[\pi P]} \biggr\vert_{\rm div}
  = \frac{2k^2}{F_\pi^2}
  \left[
    - \frac{M_H}{(4\pi)^2} \Lambda^2
    + \frac{M_H^2}{4\pi^2}
      \left( \frac{\Lambda}{\sqrt{2}} - M_H \ln \Lambda \right)
  \right]
\ ,
\nonumber\\
&& \biggl. \Pi_{Q^\ast Q^\ast}^{(f)[\rho P^\ast]} \biggr\vert_{\rm div}
  = \frac{3 g^2}{2\pi^2} k^2
  \left( \frac{\Lambda}{\sqrt{2}} - M_H \ln \Lambda \right)
\ .
\label{Pi QQ}
\end{eqnarray}

Now, let us compute the difference of $\Pi_{Q^\ast Q^\ast} - \Pi_{PP}$.

It is easy to show that $\left. \Pi_{PP}^{(b+e)}\right\vert_{\rm
div}$ exactly cancels with $\left. \Pi_{Q^\ast Q^\ast}^{(b+e)}
\right\vert_{\rm div}$. {}From the explicit forms given in
Eqs.~(\ref{Pi PP}) and (\ref{Pi QQ}), we have
\begin{eqnarray}
&& \biggl. \Pi_{Q^\ast Q^\ast}^{(b)[\pi Q]}
   - \Pi_{PP}^{(e)[\pi Q^\ast]} \biggr\vert_{\rm div}
  = \frac{2k^2}{F_\pi^2} \frac{\Lambda^3}{24\sqrt{2}\pi^2}
\ ,
\nonumber\\
&& \biggl. \Pi_{Q^\ast Q^\ast}^{(e)[\pi P]}
   - \Pi_{PP}^{(b)[\pi P^\ast]} \biggr\vert_{\rm div}
  = - \frac{2k^2}{F_\pi^2} \frac{\Lambda^3}{24\sqrt{2}\pi^2}
\ .
\end{eqnarray}
Note that the logarithmic, linear and quadratic divergences in
$\Pi_{Q^\ast Q^\ast}$ are exactly canceled by those in $\Pi_{PP}$.
This cancellation simply reflects that the external particles are
chiral partners. This immediately leads to
\begin{equation}
\biggl. \Pi_{Q^\ast Q^\ast}^{(b+e)} - \Pi_{PP}^{(b+e)}\biggr\vert_{\rm div} = 0
\ .
\label{QQPP be}
\end{equation}
The cubic divergence in $\Pi_{Q^\ast Q^\ast}$ is exactly canceled
by that in $\Pi_{PP}$, reflecting that the internal particles are
chiral partners to each other.

In a similar way, a partial cancellation takes place between
$\Pi_{Q^\ast Q^\ast}^{(a)}$ and $\Pi_{PP}^{(d)}$ as well as
between $\Pi_{Q^\ast Q^\ast}^{(d)}$ and $\Pi_{PP}^{(a)}$:
\begin{eqnarray}
&& \biggl.
     \Pi_{Q^\ast Q^\ast}^{(a)[\sigma Q^\ast]}
     - \Pi_{PP}^{(d)[\sigma Q]}
   \biggr\vert_{\rm div}
  = \frac{2k^2}{F_\sigma^2}
    \left[
      - \frac{\Lambda^3}{24\sqrt{2}\pi^2}
      + \frac{M_\rho^2}{4\pi^2}
        \left( \frac{\Lambda}{\sqrt{2}} - M_G \ln \Lambda \right)
    \right]
\ ,
\nonumber\\
&& \biggl.
     \Pi_{Q^\ast Q^\ast}^{(d)[\sigma P^\ast]}
     - \Pi_{PP}^{(a)[\sigma P]}
   \biggr\vert_{\rm div}
  = \frac{2k^2}{F_\sigma^2}
    \left[
      \frac{\Lambda^3}{24\sqrt{2}\pi^2}
      - \frac{M_\rho^2}{4\pi^2}
        \left( \frac{\Lambda}{\sqrt{2}} - M_H \ln \Lambda \right)
    \right]
\ .
\end{eqnarray}
These lead to
\begin{equation}
\biggl. \Pi_{Q^\ast Q^\ast}^{(a+d)}
  - \Pi_{PP}^{(a+d)} \biggr\vert_{\rm div}
  = - g^2 \frac{k^2}{2\pi^2} ( M_G - M_H ) \ln \Lambda
\ ,
\label{QQPP ad}
\end{equation}
where we used $M_\rho^2 = g^2 F_\sigma^2$. The remaining
contributions sum to
\begin{equation}
\biggl. \Pi_{Q^\ast Q^\ast}^{(c+f)}
  - \Pi_{PP}^{(c+f)} \biggr\vert_{\rm div}
  = - g^2 \frac{1 - 2k - 2k^2 }{2\pi^2} ( M_G - M_H ) \ln \Lambda
\ .
\label{QQPP cf}
\end{equation}

By summing up the contributions in Eqs.~(\ref{QQPP be}),
(\ref{QQPP ad}) and (\ref{QQPP cf}),  we obtain the divergent part
of the correction to the mass difference:
\begin{equation}
\biggl. \Pi_{Q^\ast Q^\ast} - \Pi_{PP} \biggr\vert_{\rm div}
 = - {\mathcal C}_2(N_f)\,
   \frac{g^2}{2\pi^2}\Bigl( 1 - 2k - k^2 \Bigr)
   (M_G - M_H)\ln\Lambda
\ ,
\label{quantum correction app}
\end{equation}
where we reinstated the group factor ${\mathcal C}_2(N_f)$. The
logarithmic divergence in the above expression is renormalized by
the bare contribution given by
\begin{equation}
\biggl. \Pi_{Q^\ast Q^\ast} - \Pi_{PP} \biggr\vert_{\rm bare}
  = \Delta M_{\rm bare} \ .
\end{equation}
Thus the RGE for the mass difference
$\Delta M = M_G - M_H$ has the following form:
\begin{equation}
 \mu\frac{d\,\Delta M}{d\mu}
 =  {\mathcal C}_2(N_f)\,
   \frac{g^2}{2\pi^2}\Bigl( 1 - 2k - k^2 \Bigr)\Delta M.
\label{rge}
\end{equation}
We should stress that this RGE is exactly the same as the one
in Eq.~(\ref{rge a=1}) obtained by setting $a=1$, i.e., $F_\sigma = F_\pi$.

Note that the complete cancellation of $\pi$- and $\sigma$-loop
contributions occurs even when we take into account finite 
renormalization effects.

\newpage

\renewcommand{\bibname}{References}

\end{document}